\documentclass[11pt,twoside,final,reqno]{atmp}

\usepackage{amsmath,amssymb,amsxtra}

\usepackage[all]{xy}
\usepackage{color}

\usepackage{epsfig}
\usepackage{calc}
\usepackage[ansinew]{inputenc}
\usepackage{graphicx}

\numberwithin{equation}{section}
\numberwithin{figure}{section}
\numberwithin{table}{section}

\def\IC{\mathbb{C}}
\def\IF{\mathbb{F}}
\def\IP{\mathbb{P}}
\def\IR{\mathbb{R}}
\def\IZ{\mathbb{Z}}
\def\Oc{\mathcal{O}}
\def\Ic{\mathcal{I}}
\def\Uc{\mathcal{U}}
\def\tilde{\widetilde}

\def\eg{{\it e.g.\ }}
\def\half{\frac{1}{2}}

\def\Bl#1{{\rm Bl}_{#1}}  
\def\zf#1#2{z^{#1}_{{\rm fixed},#2}}
\newcommand{\shortexactseq}[5]{0 \longrightarrow #1 \stackrel{#2}{\longrightarrow} #3 \stackrel{#4}{\longrightarrow} #5 \longrightarrow 0}
\DeclareMathOperator{\ch}{c}
\def\Dt{\widetilde{D}}
\def\Et{\widetilde{E}}
\newcommand{\ibar}{\bar{\imath}}
\newcommand{\jbar}{\bar{\jmath}}
\newcommand{\kbar}{\bar{k}}

\begin{document}

\title[Toroidal Orbifolds and Orientifolds]
{Resolved Toroidal Orbifolds and their Orientifolds}

\arxurl{hep-th/0609014}

\author[L\"ust, Reffert, Scheidegger, Stieberger]{D.~L\"ust$^{1,2}$, S.~Reffert$^2$, E.~Scheidegger$^3$, and\ 
S.~Stieberger$^1$}

\address{
$^1$ Arnold--Sommerfeld--Center for Theoretical Physics,\\
Department f\"ur Physik, Ludwig--Maximilians--Universit\"at M\"unchen,\\
Theresienstra\ss e 37, 80333 M\"unchen, Germany\\
\vskip7pt
$^2$ Max--Planck--Institut f\"ur Physik,\\
F\"ohringer Ring 6, 80805 M\"unchen, Germany\\
\vskip7pt
$^3$Universita del Piemonte Orientale "A. Avogadro",\\
Dip. Scienze e Tecnologie Avanzate,\\
Via Bellini 25/g, I-15100 Alessandria, Italy\\
INFN -- Sezione di Torino, Italy\\
\vskip7pt
}  
\addressemail{{\it Email:} luest, sreffert, stieberg @theorie.physik.uni-muenchen.de,\\
esche @mfn.unipmn.it}

\begin{abstract}
We discuss the resolution of toroidal orbifolds. For the resulting smooth Calabi--Yau manifolds, we calculate the intersection ring and determine the divisor topologies. In a next step, the orientifold quotients are constructed.
\end{abstract}

\maketitle

\break
{\bf \tableofcontents}

\break
\section{Introduction}
\label{sec:Intro}

Orbifold compactifications have played a prominent role in string theory ever since their advent in~\cite{Dixon:1985jw},~\cite{Dixon:1986jc}. Over the years, toroidal orbifolds have served as simple models to test new structures and phenomena in string theory such as dualities. Moreover, they provide a rich playground for phenomenologial studies of semi--realistic string compactifications. It is therefore natural to investigate more recent issues such as properties of orientifold compactifications and the ensuing $N=1$ low energy effective field theories as well as the stabilization of the moduli in these theories.

The initial impulse for the present paper arose from the search for orientifold compactifications in type IIB string theory which allow the complete stabilization of all K\"ahler moduli. The occurrence of a large number of moduli, i.e. free parameters in the low energy effective field theory, is a malady that most commonly used string compactifications suffer from. In~\cite{Kachru:2003aw}, a mechanism (KKLT) was proposed that stabilizes the dilaton and all geometrical moduli and moreover leads to a meta--stable de Sitter vacuum. The complex structure moduli and the dilaton are stabilized via background 3--form fluxes, whereas the K\"ahler moduli are fixed via non--perturbative effects: a superpotential is generated either via Euclidean D3--brane instantons wrapping a divisor in the compactification manifold~\cite{Witten:1996bn},~\cite{Kallosh:2005gs},~\cite{Bergshoeff:2005yp}, or by a gaugino condensate in the world--volume of D7--branes wrapping such a divisor~\cite{Katz:1996th},~\cite{Gorlich:2004qm}. For both mechanisms, the divisors in question must fulfill certain topological properties: To be able to decide whether the K\"ahler moduli can be stabilized for a given compactification manifold, the knowledge of the topology of a full set of divisors, as well as the intersection ring, which enters the tree--level K\"ahler potential are needed. Moreover, we need to know which K\"ahler moduli become non--geometric and which complex structure moduli disappear after the orientifold quotient, respectively.
 
To date, only few explicit examples which realize the proposed mechanism of moduli stabilization exist~\cite{Denef:2004dm},~\cite{Denef:2005mm},~\cite{Aspinwall:2005ad}. In~\cite{Denef:2005mm}, the $T^6/\IZ_2\times\IZ_2$ orbifold was resolved and the F--theory lift of the correspondinging orientifold quotient provided indeed a working example for moduli stabilization \`a la KKLT. A natural question then is whether the results of~\cite{Denef:2005mm} extend to more general toroidal orbifolds, which are generically less symmetric and carry more non--trivial features than $T^6/\IZ_2\times\IZ_2$. To answer this question we have developed geometric methods for analyzing the topology of resolved orbifolds. Since these methods are of general interest and lend themselves to many more applications than only those given by the initial motivation discussed above, a study dedicated to this subject exclusively is justified. An outline of this program was given in~\cite{Reffert:2005mn}. Its application to the issue of moduli stabilization and in particular to the answer to the above question is given in our companion paper~\cite{pheno}. Our paper builds the bridge between the moduli stabilization at the orbifold point in~\cite{Lust:2005dy} and at the large volume limit in~\cite{pheno}. In the latter, it is in particular shown how to stabilize the non--geometric K\"ahler moduli.

With the methods discussed in the present paper, a given toroidal orbifold can be resolved to yield a smooth Calabi--Yau manifold. While the methods of toric geometry allow us to resolve the singularities locally, the knowledge of the structure of the covering space, i.e. the $T^6$, allows us to glue the resolved patches together in the correct fashion. We are able to determine the topology of all divisors that arise naturally in our construction. Moreover we explain how to determine the full intersection ring of the resulting Calabi--Yau manifold. In a second step, we provide the transition to the orientifold quotient. This quotient corresponds from a geometric point of view to a $\IZ_2$ quotient of the smooth Calabi--Yau manifold.

As mentioned in the beginning, toroidal orbifolds have already lent themselves to numerous applications in string theory in the past. One one hand, this is due to their simple structure which they inherit from the covering torus, while on the other hand they exhibit the non--trivial nature of a generic Calabi--Yau manifold. The lesson is that the singular orbifold quotient is merely a special, degenerate point in the moduli space of a smooth Calabi--Yau manifold and that this smooth manifold is actually the object of interest. It is the construction of this object that we focus on, as well as the transition to the orientifold quotient. The orientifold quotient is the relevant object for all models in type II string theories featuring D--branes. We provide explicit examples of models where the number of geometric moduli is reduced after taking the quotient, i.e. models with $h^{1,1}_{-}\neq0$ and $h^{2,1}_{+} \neq 0$. 

The plan of the paper is as follows. In Section~\ref{sec:Orbifolds}, we briefly discuss the different possible orbifold constructions. In Section~\ref{sec:local}, the resolution of the orbifold singularities in a local set-up using the methods of toric geometry is treated. In Section~\ref{sec:toric}, the necessary background in toric geometry is given and the resolution process explained. In Section~\ref{sec:ExcepTop}, we state how to determine the topology of the divisors in the local resolved geometries. In Section~\ref{sec:LocalExamples}, we apply these techniques to explicitly perform the resolution of the singularity of $\IC^3/\IZ_{6-I}$. 

While the material in the preceding sections is mostly standard, Sections~\ref{sec:Global} and~\ref{sec:orientifolds} form the core part of the new material presented in this paper. In Section~\ref{sec:Global}, we give the general method of constructing a smooth Calabi--Yau manifold from the resolved local patches discussed in the previous section. In Section~\ref{sec:FixedSets}, the fixed point configurations of the singular orbifold and the counting of the exceptional divisors arising from the resolution are reviewed. In Section~\ref{sec:IntersectionRing}, we show how to obtain the linear relations among the divisors in the global model and present the method to calculate the full intersection ring of the smooth Calabi--Yau manifold, albeit not necessarily in an integral basis. With the knowledge of the preceding section we can demonstrate in Section~\ref{sec:Topology} how to determine the topology of these divisors. In Section~\ref{sec:TwistedCplx}, we shed light on the origin of the twisted complex structure moduli, a topic on which so far only very little is known. As in Section~\ref{sec:local} we apply these general methods in Section \ref{sec:GlobalExamples} to the examples of $T^6/\IZ_{6-I}$ on the root lattices of $G_2\times SU(3)^2$ and $G_2^2\times SU(3)$ and two $T^6/\IZ_2\times\IZ_2$ shift orbifolds. 

The transition from the smooth Calabi--Yau manifold to its orientifold quotient is made in Section~\ref{sec:orientifolds}. In Section \ref{sec:Actions}, we consider the global orientifold involution and its relation to the involutions on the local patches. Furthermore, we discuss the occurrence of divisors which are not invariant under the global involution, leading to $h^{1,1}_{-}\neq 0$, and of complex structure moduli which are invariant, leading to $h^{2,1}_{+} \neq 0$. Both of them are non--geometric moduli. In Section~\ref{sec:Oring} we explain how the intersection ring of the orientifold quotient is obtained from the one of the Calabi--Yau manifold. The global configuration of the orientifold--plane and the resulting charges, which have to be canceled are discussed in Section~\ref{sec:Tadpoles}. The orientifold quotients of some of the examples of Section~\ref{sec:GlobalExamples} conclude the topic. We end with concluding remarks and a brief outlook in Section~\ref{sec:Conclusions}.

In Appendix~\ref{sec:toricpatches}, the resolutions of all the local orbifold singularities which we encounter in the examples are collected. The remaining appendices give the details of a few more examples, namely $T^6/\IZ_3$, $T^6/\IZ_4$ and $T^6/\IZ_{6-II}$ on several lattices, $T^6/\IZ_2\times\IZ_4$, and $T^6/\IZ_3\times\IZ_3$.


\section{Type IIB orientifolds of toroidal orbifolds}
\label{sec:Orbifolds}

A fundamental problem in string theory is the classification of string compactifications. $N=2$ theories in four dimensions can be obtained as compactifications of type II string theories on Calabi--Yau manifolds. Therefore we need a list of Calabi--Yau manifolds. We will focus on a particular class of such manifolds, namely resolutions of torus orbifolds. In this section, we will review the present content of this list and indicate to which part we will restrict ourselves in the remainder of the article. For compiling this list, we need topological criteria to decide which Calabi--Yau manifolds are equivalent. 


An important but expensive calculational tool to settle the question of possible equivalences is the classification of real six manifolds by C.T.C.~Wall \cite{Wall:1966ab}. Applied to Calabi--Yau manifolds it states that two simply connected manifolds $X$ and $X'$ are of the same topological type, i.e. are diffeomorphic, if besides the Hodge numbers, the triple intersection numbers $S_{abc}=\int_X \omega_a\wedge \omega_b\wedge \omega_c$ and the linear forms $\int_X \ch_2\wedge \omega_a$ are the same, possibly up to an integer linear basis transformation $\omega=M \omega'$ in the K\"ahler cone of $X$ and $X'$. If the bases can only be related by a rational linear transformation, the spaces are rational homotopy equivalent. Only finitely many diffeomorphic types can exist in a rational homotopy equivalence class.

To date there is no complete classification of torus orbifolds and their resolutions in three dimensions. For several simple classes there are partial classifications. Coxeter orbifolds with $G$ abelian and not containing any shifts were classified in~\cite{Erler:1992ki} and~\cite{Font:1988mk}. The latter also take into account discrete torsion. However, there are also examples of non--Coxeter orbifolds, e.g. in~\cite{Casas:1991ac}. For abelian orbifolds including shifts there is a result in~\cite{Oguiso:1999ab}. The classification by Borcea and Voisin in~\cite{Voisin:1993ab} and~\cite{Borcea:1997tq} of Calabi--Yau threefolds with involutions coming from automorphisms of K3s has a partial overlap with the previous two classifications. To our knowledge, there is no classification for $G$ non--abelian with or without shifts. In two dimensions, there is a classification of torus orbifolds without shifts by Fujiki~\cite{Fujiki:1988ab}, see also \cite{Wendland:2000ry}.

The statements of these partial classifications are of differing richness. While all of them state the allowed groups and their actions, some of them also determine topological quantities like the Euler number, the Hodge numbers, or even the intersection ring. \cite{Erler:1992ki}, \cite{Font:1988mk} and~\cite{Borcea:1997tq} provide the Hodge numbers, while~\cite{Oguiso:1999ab} does not. Voisin~\cite{Voisin:1993ab} even explicitly calculates the intersection ring of the manifolds in her list. In the following two sections we present a general method to determine the topology of resolutions of abelian torus orbifolds, including the intersection ring, the linear forms $\ch_2$ and the topology of the divisors. For a discussion of the relation between the vanishing of $\ch_2$ and torus orbifolds see~\cite{Wilson:1997ab}.

The orbifolds which will be considered here are listed in the following table which is taken from~~\cite{Erler:1992ki}.
\begin{table}[h!]
\begin{center}
\begin{center}
{\small
\begin{tabular}{|l|c|c|c|c|c|c|}
\hline
$\ G$&Lattice &$h^{1,1}_{\rm untw.}$&$h^{2,1}_{\rm untw.}$&
$h^{1,1}_{\rm twist.}$&$h^{2,1}_{\rm twist.}$ \\ [2pt]
\hline
$\ \IZ_3 $    &\  $SU(3)^3 $          &9 & 0 & 27 & 0\cr
$\  \IZ_4  $   &\ $SU(4)^2  $      &5 & 1 & 20 & 0\cr
$\  \IZ_4 $    &\  $SU(2)\times SU(4)\times SO(5)$ &5 & 1 & 22 & 2\cr
$\  \IZ_4$      &\  $SU(2)^2\times SO(5)^2 $ &5 & 1 & 26 & 6\cr
$\  \IZ_{6-I} $   & $(G_2\times SU(3)^{2})^{\flat}$  &5 & 0 & 20 & 1\cr
$\  \IZ_{6-I}  $  &$ SU(3)\times G_2^2$ &5 & 0 & 24 & 5\cr
$\  \IZ_{6-II} $  &$ SU(2)\times SU(6)$    &3 & 1 & 22 & 0\cr
$\  \IZ_{6-II} $  &$SU(3)\times SO(8) $&3 & 1 & 26 & 4\cr
$\  \IZ_{6-II} $  &$(SU(2)^2\times SU(3)\times SU(3))^{\sharp} $ &3 & 1 & 28 & 6\cr
$\  \IZ_{6-II}  $ &$ SU(2)^2\times SU(3)\times G_2$  &3 & 1 & 32 & 10\cr
$\  \IZ_7  $      &$ SU(7) $                             &3 & 0 & 21 & 0\cr
$\  \IZ_{8-I}  $  &$ (SU(4)\times SU(4))^*$                              &3 & 0 & 21 & 0\cr
$\  \IZ_{8-I}  $  &$SO(5)\times SO(9)    $     &3 & 0 & 24 & 3\cr
$\  \IZ_{8-II} $  &$ SU(2)\times SO(10)  $     &3 & 1 & 24 & 2\cr
$\  \IZ_{8-II}  $   &$ SO(4)\times SO(9)$   &3 & 1 & 28 & 6\cr
$\  \IZ_{12-I}$   &$ E_6 $  &3 & 0 & 22 & 1\cr
$\  \IZ_{12-I} $  &$ SU(3)\times F_4$  &3 & 0 & 26 & 5\cr
$\  \IZ_{12-II} $  &$SO(4)\times F_4$  &3 & 1 & 28 & 6\cr
$\  \IZ_2 \times\IZ_2$     &$SU(2)^6$  &3 & 3 & 48 & 0\cr
$\  \IZ_2 \times\IZ_4$    &$SU(2)^2\times SO(5)^2$ &3 & 1& 58 & 0\cr
$\  \IZ_2 \times\IZ_6 $    &$ SU(2)^2\times SU(3)\times G_2$&3 & 1 & 48 & 2\cr
$\  \IZ_2 \times\IZ_{6'}$&$ SU(3)\times G_2^2$&3 & 0 & 33 & 0\cr
$\  \IZ_3 \times\IZ_3  $   &$SU(3)^3$   &3 & 0 & 81 & 0\cr
$\  \IZ_3 \times\IZ_6 $    &$SU(3)\times G_2^2 $&3 & 0 & 70 & 1\cr
$\  \IZ_4 \times\IZ_4 $    &$SO(5)^3 $ &3 & 0 & 87 & 0\cr
$\  \IZ_6 \times\IZ_6  $   &$G_2^3$ &3 & 0 & 81 & 0\cr
\hline 
\end{tabular}}
\end{center}
\caption{Twists, lattices and Hodge numbers.}
\label{tab:ZN}
\end{center}
\end{table}
The table give the torus lattices and the twisted and untwisted Hodge numbers.
The lattices marked with $\flat$, $\sharp$, and $*$ are realized as so--called generalized Coxeter twists, the automorphism being in the first and second case $S_1S_2S_3S_4P_{36}P_{45}$ and in the third  $S_1S_2S_3P_{16}P_{25}P_{34}$, where the $S_i$ are Weyl reflections and the $P_{ij}$ transpositions of roots.

In addition, we will consider two orbifolds with $G=\IZ_2\times\IZ_2$ with one, respectively two shifts included which were introduced in~\cite{Morrison:1996pp}. They are related to the Schoen Calabi--Yau manifold~\cite{Schoen:1988ab} (see also~\cite{Braun:2004xv}) with $h^{1,1}=h^{2,1}=19$, and to the Enriques Calabi--Yau manifold~\cite{Ferrara:1995yx},~\cite{Klemm:2005pd} with $h^{1,1}=h^{2,1}=11$.


\section{The local models}
\label{sec:local}

In this section, we examine the resolution of the singularities of toroidal orbifolds. Close to an orbifold fixed point, which is a quotient singularity, the geometry looks like $\IC^3/G$. This non-compact local geometry is conveniently described as a toric variety. We will not give the full definitions here but refer the reader to the literature, Chapter 7 of~\cite{Hori:2003ab} for example gives an easily accessible introduction, and~\cite{Fulton:1993ab} for the necessary details. 

We will give the description of these local orbifolds in terms of their fans, resolve the singularities by blowing up the orbifold fixed points and give the intersection properties along the lines of~\cite{DelaOssa:2001xk}.


\subsection{Toric geometry and resolution of singularities}
\label{sec:toric}

A systematic way to resolve abelian orbifold singularities is toric geometry~\cite{Fulton:1993ab}. In the following paragraphs, we summarize some of the basic facts about toric geometry. An $n$--dimensional toric variety takes the form
\begin{equation}
  \label{eq:toricvariety}
  X_\Sigma=(\IC^d\setminus F_\Sigma)/(\IC^*)^r,
\end{equation}
where $n=d-r$, and the algebraic torus $(\IC^*)^r$ acts by coordinatewise multiplication. The set $F_\Sigma$ is a subset that remains fixed under a continuous subgroup of $(\IC^*)^r$ and must be subtracted for the variety to be well defined. The action of $(\IC^*)^r$ is encoded in a lattice $N$ which is isomorphic to $\IZ^d$ and by its fan $\Sigma$. A fan is a collection of strongly convex rational cones in $N\otimes_\IZ \IR$ with the property that each face of a cone in $\Sigma$ is also a cone in $\Sigma$ and the intersection of two cones in $\Sigma$ is a face of each. The $k$--dimensional cones in $\Sigma$ are in one--to--one correspondence with the codimension $k$--submanifolds of $X_\Sigma$. In particular, the one--dimensional cones correspond to the divisors in $X_\Sigma$. The fan $\Sigma$ can be encoded by the generators of its edges or one--dimensional cones, i.e. by vectors $v_i\in N$. To each $v_i$ we associate a homogeneous coordinate $z_i$ of $X_\Sigma$. The $(\IC^*)^r$ action is encoded on the $v_i$ in $r$ linear relations
\begin{equation}
  \label{eq:linrels}
  \sum_{i=1}^d l^{(a)}_i v_i = 0, \qquad a=1,\dots,r, \quad l^{(a)}_i \in \IZ.
\end{equation}
To each $v_i$ we assign an invariant monomial $U^i = \prod_{i=1}^d z_i^{\langle v_i,m \rangle}$, where $m\in M$ is an element of the lattice dual to $N$. These monomials are the local coordinates of $X_\Sigma$. 

We are only interested in Calabi--Yau orbifolds $\IC^m/G$ of dimensions $m=2,\,3$, so we require $X_\Sigma$ to have trivial canonical class. This translates to demanding that all but one of the $v_i$ lie in the same affine hyperplane one unit away from the origin $v_0$. This means that the last component of all the $v_i$ (except $v_0$) equals one. Thus, the $v_i$ form a cone $C_{\Delta^{(m)}}$ over the polyhedron $\Delta^{(m)}$ spanned by the $v_i$, $i\not=0$ with apex $v_0$. This allows us to draw toric diagrams in two (one) dimensions instead of $m=3$ ($m=2$). 

The fan $\Sigma$ associated to the singularity $\IC^3/G$ is obtained as follows: We have a single three--dimensional cone in $\Sigma$, generated by $v_1,\,v_2,\,v_3$. A generator $\theta$ of $G$ of order $n$ acts on the coordinates of $\IC^3$ by
\begin{equation}
  \label{eq:twistc}
  \theta:\ (z^1,\, z^2,\, z^3) \to (\varepsilon\, z^1, \varepsilon^{n_1}\, z^2, \varepsilon^{n_2}\, z^3),\quad \varepsilon=e^{2 \pi i/n},
\end{equation} 
For such an action we will adopt the shorthand notation $\frac{1}{n}(1,n_1,n_2)$. The Calabi-Yau condition is trivially fulfilled as the orbifold actions are chosen such that $1+n_1+n_2=n$ and $\varepsilon^n=1$. Then the local coordinates of $X_\Sigma$ are $U^a=(z^1)^{(v_1)_a}(z^2)^{(v_2)_a}(z^3)^{(v_3)_a}$. To find the coordinates in $N$ of the generators $v_i$ of the fan, we require the $U^k$ to be invariant under the action of $\theta$. This results in finding two linearly independent solutions to the equation
\begin{equation}
  \label{eq:vi}
  (v_1)_a+n_1\,(v_2)_a+n_2\,(v_3)_a= 0\ \mod\, n.
\end{equation}
The divisors corresponding to the 1--dimensional cones $v_i$ will be denoted by $D_i$.

$X_\Sigma$ is smooth if all the top--dimensional cones in $\Sigma$ have volume one. Here, there is only one such cone whose volume is $|G|$, hence $X_\Sigma$ is singular. There is a standard procedure for resolving singularities of toric varieties. It consists of adding all lattice points in $N$ which lie in the polyhedron $\Delta^{(m)}$ in the affine hyperplane at distance one which is spanned by the generators $v_i$. Adding points only in this hyperplane ensures that the the canonical class of the variety is not affected, i.e. the resulting manifold is still Calabi--Yau.

For the fan $\Sigma$ this means that the corresponding one--dimensional generators $w_i$ are added to it and that it has to be subdivided accordingly. We denote the refined fan by $\widetilde\Sigma$. In~\cite{Aspinwall:1994ev} it is shown that these new generators can be related (in the case $m=3$) to the group elements of $G$ as follows:   
\begin{equation}
  \label{eq:criterion}
  w_i=g^{(i)}_1\,v_1+g^{(i)}_2\,v_2+g^{(i)}_3\,v_3, \qquad \qquad\sum_{k=1}^3 g^{(i)}_k=1,\quad  0\leq g^{(i)}_k<1.
\end{equation}
where $g^{(i)}=(g^{(i)}_1,\,g^{(i)}_2,\,g^{(i)}_3)\in (\IZ_n)^3$ represents the corresponding group element $\theta_i$. The corresponding exceptional divisors are denoted $E_i$. To each of the new generators we associate a new homogeneous coordinate which we denote by $y^i$, as opposed to the $z^i$ we associate to the original $v_i$. Let us pause for a moment to think about what this method of resolution means. The obvious reason for enforcing the criterion~(\ref{eq:criterion}) is that group elements which do not respect it fail to fulfill the Calabi--Yau condition: Their third component is no longer equal to one. But what is the interpretation of these group elements that do not contribute ? Another way to phrase the question is: Why do not all twisted sectors contribute exceptional divisors ? A closer look at the group elements shows that all those elements of the form $ \frac{1}{n}\,(1,a,b)$ which fulfill~(\ref{eq:criterion}) give rise to inner points of the toric diagram. Those of the form $\frac{1}{n}\, (1,0,b)$ lead to points on the edge of the diagram. They always  fulfill~(\ref{eq:criterion}) and each element which belongs to such a subgroup contributes a divisor to the respective edge, therefore  there will be $n-1$ points on it. The elements which do not fulfill~(\ref{eq:criterion}) are in fact anti--twists, i.e. they have the form $ \frac{1}{n}\,(n-1, n-a, n-b)$. Since the anti--twist does not carry any information which was not contained already in the twist, there is no need to take it into account separately, hence also from  this point of view it makes sense that it does not contribute an exceptional divisor to the resolution. 

The case $m=2$ is even simpler. The singularity $\IC^2/\IZ_n$ is called a rational double point of type $A_{n-1}$ and its resolution is called a Hirzebruch--Jung sphere tree consisting of $n-1$ exceptional divisors intersecting themselves according to the Dynkin diagram of $A_{n-1}$. The corresponding polyhedron $\Delta^{(1)}$ consists of a single edge joining two vertices $v_1$ and $v_2$ with $n-1$ equally spaced lattice points $w_1,\dots,w_{n-1}$ in the interior of the edge. 

After having added the generators, we have to subdivide the fan. The subdivision of the fan $\Sigma$ into $\widetilde\Sigma$ corresponds to a triangulation of the toric diagram. In general, there are several triangulations, and therefore several possible resolutions. They are all related via birational transformations. The diagram of the resolution $X_{\widetilde\Sigma}$ of $X_{\Sigma}=\IC^m/G$ contains $|G|$ simplices, yielding $|G|$ three-dimensional cones of volume one. Hence $X_{\widetilde\Sigma}$ is smooth. This variety can, of course, also be written in the form~(\ref{eq:toricvariety}) where $r$ is now the number of additional generators $w_i$. Their defining relations~(\ref{eq:criterion}) are, after clearing the denominators, precisely the $r$ linear relations~(\ref{eq:linrels}) which encode the $\left(\IC^*\right)$ action. The divisors corresponding to such a linear combination are ``sliding divisors'' in the compact geometry, as we will discuss in great detail in Section~\ref{sec:IntersectionRing}. Finally, the excluded set $F_\Sigma$ is obtained as follows: Take the set of all combinations of generators $v_i$ and $w_i$ of one--dimensional cones in $\Sigma$ that {\it do not span a cone} in $\Sigma$ and define for each such combination a linear space by setting the coordinates associated to the $v_i$ to zero. $F_\Sigma$ is the union of these linear spaces, i.e. the set of simultaneous zeros of coordinates not belonging to the same cone. In the case of several possible triangulations, it is the excluded set $F_{\tilde\Sigma}$ that distinguishes the different resulting geometries.

In the dual diagram, the geometry and intersection properties of a toric manifold are often easier to grasp than in the original toric diagram. The divisors, which are represented by vertices in the original toric diagram become faces in the dual diagram, the curves marking the intersections of two divisors remain curves and the intersections of three divisors which are represented by the faces of the original diagram become vertices. In the dual graph, it is immediately clear which of the curves are compact. The curves at the intersection of two exceptional divisors are the exceptional curves.

It is convenient to introduce a matrix $(\,P\, |\,Q\,)$. $P$ contains as its rows the vectors $v_i$ and $w_i$; the columns of $Q$ contain the linear relations~(\ref{eq:linrels}) between the divisors, i.e. $Q = \left(l_i^{(a)}\right)$. From the rows of $Q$, which we denote by $C_i,\ i=1,\dots,n$, we can read off the linear equivalences between the divisors. (Note that two divisors $\sum b_i D_i$ and $\sum b_i' D_i$ are linearly equivalent if and only if they are homologically equivalent.) For most applications, it is most convenient to choose the $C_i$ to be the generators of the Mori cone. The Mori cone is the space of effective curves, i.e. the space of all curves $C\in X_\Sigma$ with $C\cdot D\geq 0$ for all divisors $D\in X_\Sigma$. It is dual to the K\"ahler cone. In our cases, the Mori cone is spanned by curves corresponding to two--dimensional cones. The generators for the Mori cone correspond to those linear relations with which all others can be expressed as positive, integer linear combinations. We will briefly review the method of finding the generators of the Mori cone. It can be found e.g. in~\cite{Berglund:1995gd}. We present it here adapted to our context.
\begin{enumerate}
  \item In a given triangulation, take the three-dimensional simplices $S_k$ (corresponding to the three-dimensional cones). Take those pairs of simplices $(S_l,S_k)$ that share a two--dimensional simplex $S_k\cap S_l$.
  \item For each such pair find the unique linear relation among the vertices in $S_k\cup S_l$ such that 
  \begin{enumerate}
    \item the coefficients are minimal integers and 
    \item the coefficients for the points in $(S_k \cup S_l) \setminus (S_k \cap S_l)$ are positive.
  \end{enumerate}
  \item Find the minimal integer relations among those obtained in step 2 such that each of them can be expressed as a positive integer linear combination of them. 
\end{enumerate}

Next, we want to investigate the intersection properties of the divisors of the resolved variety $X_{\widetilde\Sigma}$. 
The general rule for triple intersections is that the intersection number of three distinct divisors is 1 if they belong to the same cone and 0 otherwise. The set of collections of divisors which do not intersect because they do not lie in the same come forms a further characteristic quantity of a toric variety, the Stanley--Reisner ideal. It contains the same information as the exceptional set $F_{\Sigma}$. Intersection numbers for triple intersections of the form $D_i^2 D_j$ or $E_k^3$ can be obtained by making use of the linear equivalences between the divisors. Since we are working here with non--compact varieties at least one compact divisor has to be involved. For intersections in compact varieties there is no such condition. The intersection ring of a toric variety is -- up to a global normalization -- completely determined by the linear relations and the Stanley--Reisner ideal. The normalization is fixed by one intersection number of three distinct divisors.

The matrix elements of $Q$ are the intersection numbers between the curves $C_i$ and the divisors $D_i,\,E_i$. We can use this to determine how the compact curves of our blown--up geometry are related to the $C_i$.


\subsection{Topology of the exceptional divisors}
\label{sec:ExcepTop}

There are two types of exceptional divisors depending on whether the the corresponding point lies in the interior or sits on the boundary of the toric diagram. The latter case can easily be discussed by looking at the dual toric diagram.We see that it corresponds to the two--dimensional situation with an extra non--compact direction, hence it has the topology of $\IC \times \IP^1$.

We therefore turn to the divisors corresponding to points in the interior of the toric diagram. For that purpose we recall the notion of the star of a cone $\sigma$, denoted ${\rm Star}(\sigma)$, which is the set of all cones $\tau$ in the fan $\Sigma$ containing $\sigma$. In our situation, the topology of an exceptional divisor $E_i$ is then determined in terms of ${\rm Star}(\sigma_{w_i})$. This means that we simply remove from the fan $\Sigma$ all cones, i.e. points and lines in the toric diagram, which do not contain $w_i$. The diagram of the star does not need to be convex anymore. Then we compute the linear relations and the Mori cone for the star. This means in particular that we drop all the simplices $S_k$ in the induced triangulation of the star which do not lie in its toric diagram. As a consequence, certain linear relations of the full diagram will be removed in the process of determining the Mori cone.  The generators of the Mori cone of the star will in general be different from those of $\Sigma$. 

Once we have obtained the Mori cone of the star we can rely on the classification of compact toric surfaces. This tells us that any toric surface is either a $\IP^2$, a Hirzebruch surface $\IF_n$, or a toric blow-up thereof. The generator of the Mori cone of $\IP^2$ is of the form $$Q^T = \left(\begin{array}{cccc}-3& 1& 1& 1\end{array}\right).$$ For $\IF_n$ the generators take the form $$Q^T = \left(\begin{array}{ccccc}-2 & 1 & 1 & 0 & 0\cr-n-2 & 0 & n & 1 & 1\end{array}\right) \qquad {\rm or} \qquad  Q^T = \left(\begin{array}{ccccc} -2 & 1 & 1 & 0 & 0\cr n-2 & 0 & -n & 1 & 1\end{array}\right)$$ since $\IF_{-n}$ is isomorphic to $\IF_n$. Finally, every toric blow-up of a point adds an additional independent relation whose form is $$Q^T = \left(\begin{array}{cccccc}0 & ... & 0 &1 &1 &-2\cr\end{array}\right).$$ We will denote the blow-up of a surface $S$ in $n$ points by $\Bl{n}S$. 

Also the exceptional divisors corresponding to points on the boundary of the toric diagram can be dealt with using the star. Since the geometry is effectively reduced by one dimension, the only compact toric manifold in one dimension is $\IP^1$ and the corresponding generator is $$Q^T = \left(\begin{array}{cccc}-2& 1& 1& 0\cr\end{array}\right),$$ where the 0 corresponds to the non-compact factor $\IC$.  

However, this is not yet the full story, since our toric variety $X_{\widetilde{\Sigma}}$ is actually three-dimensional. In particular, the stars are in fact cones over a polygon. Therefore, we have an additional possibility for a toric blow-up. We can add a point to the polygon such that the corresponding relation is of the form $$Q^T = \left(\begin{array}{ccccccc}0 & ... & 0 &1 &1 &-1 &-1\end{array}\right).$$ This corresponds to adding a cone over a lozenge and is well-known from the resolution of the conifold singularity. The lozenge has to be subdivided into two simplices, and there are two ways of doing this. The process of going from one way to the other is known as a flop and reverses the signs of the corresponding relation. It also affects some of the other relations. The curve $C_-$ that is flopped is the intersection of two divisors, say $E_1$ and $E_2$. If any other curve $C$ intersects one of these two divisors, i.e. $C\cdot E_i \not=0$, the new relation corresponding to $C$ is the sum of the relation of $C_-$ and $C$. Topologically, this means that an exceptional curve $C_-$ is blown down and another one, $C_+$, is blown up. As a consequence, we have to include these blow-ups in the list of surfaces given above. In addition, we have to include topologies that can be obtained by flopping a curve in a surface of this enlarged list of surfaces.


\subsection{Examples}
\label{sec:LocalExamples}

We shall now put the methods discussed above to use by studying some examples. The resolutions of $\IC^3/\IZ_{6-I}$ and $\IC^3/\IZ_{2}\times\IZ_6$ are discussed in the main text, whereas the remaining ones can be found in the appendix.


\subsubsection{Resolution of $\IC^3/\IZ_{6-I}$}
\label{sec:Z6-I}

\begin{figure}[h!]
  \begin{center}
    \includegraphics[width=100mm]{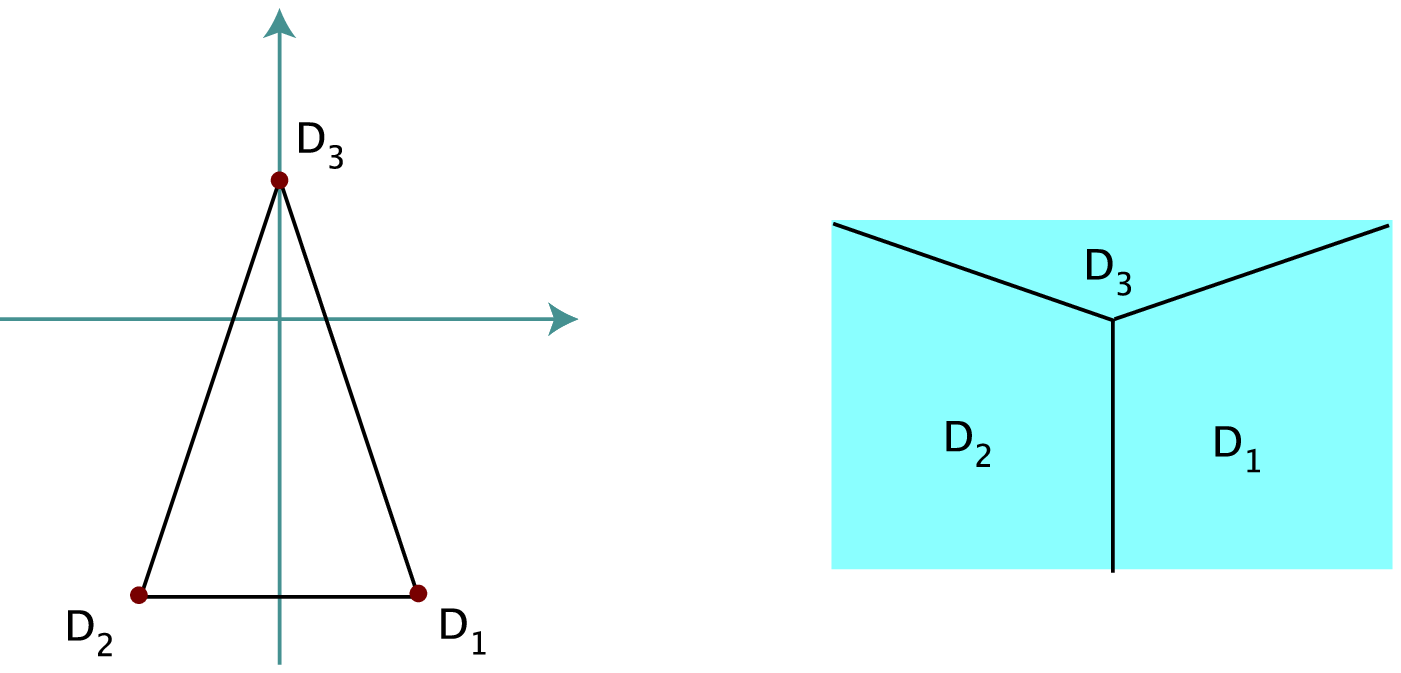}
    \caption{Toric diagram of $\IC^3/\IZ_{6-I}$ and dual graph}
    \label{fig:sixib}
  \end{center}
\end{figure}

The group $\IZ_{6-I}$ acts as follows on $\IC^3$:
\begin{equation}
  \label{eq:twistsixi}
  \theta:\ (z^1,\, z^2,\, z^3) \to (\varepsilon\, z^1, \varepsilon\, z^2, \varepsilon^4\, z^3),\quad \varepsilon=e^{2 \pi i/6}.
\end{equation}
To find the components of the $v_i$, we have to solve $(v_1)_i+(v_2)_i+4\,(v_3)_i=0\ \mod\,6$. This leads to the following three generators of the fan (or some other linear combination thereof):
\begin{equation}
  v_1=(1,-2,1),\ v_2=(-1,-2,1),\ v_3=(0,1,1).
\end{equation}
The toric diagram and its dual of ${\IC}^3/\IZ_{6-I}$ are depicted in Figure \ref{fig:sixib}.

\begin{figure}[h!]
  \begin{center}
    \includegraphics[width=100mm]{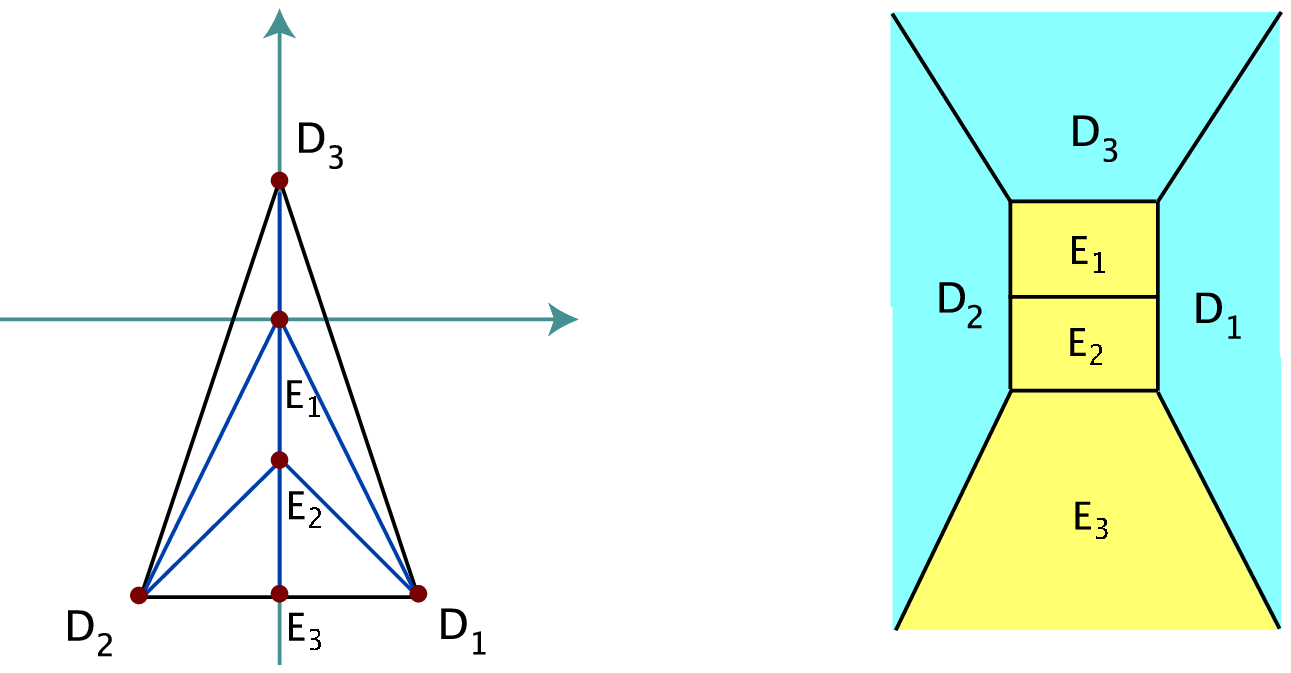}
    \caption{Toric diagram of the resolution of $\IC^3/\IZ_{6-I}$ and dual graph}
    \label{fig:fsixi}
  \end{center}
\end{figure}

To resolve the singularity, we find that $\theta,\,\theta^2$ and $\theta^3$ fulfill~(\ref{eq:criterion}). This leads to the following new generators:
\begin{equation*}
  \begin{array}{rclrl}
    \medskip
    w_1&=& \frac{1}{6}\,v_1+\frac{1}{6}\,v_2+\frac{4}{6}\,v_3 &=&(0,0,1), \\
    \medskip
    w_2&=& \frac{3}{6}\,v_1+\frac{2}{6}\,v_2+\frac{2}{6}\,v_3 &=&(0,-1,1), \\
    w_3&=& \frac{3}{6}\,v_1+\frac{3}{6}\,v_2                  &=&(0,-2,1).
  \end{array}
\end{equation*}
In this case, the triangulation is unique. Figure \ref{fig:fsixi} shows the corresponding toric diagram and its dual graph.
The $\tilde U_i$ are
\begin{eqnarray}
  \label{eq:tildeUsixi}
  \tilde U_1&=&{z^1\over z^2},\quad \tilde U_2={z^3\over(z^1)^2(z^2)^2y^2(y^3)^2},\quad \tilde U_3=z^1z^2z^2y^1y^2y^3.
\end{eqnarray}
According to (\ref{eq:toricvariety}), the new blown--up geometry is
\begin{equation}
  \label{eq:blowupsixi}
  X_{\tilde\Sigma}=\,({\IC}^{6}\setminus F_{\tilde\Sigma})/({\IC}^*)^3,
\end{equation}
where the action of $(\IC^*)^3$ will be given below~(\ref{eq:lineqsixI}). The excluded set is
\begin{equation*}
  \label{eq:excludesixi}
  F_{\tilde\Sigma}=\{(z_3,y_2)=0,\,(z_3,y_3)=0,\,(y_1,y_3)=0,\,(z_1,z_2)=0\,\}.
\end{equation*}
As can readily be seen in the dual graph, we have 7 compact curves in $X_{\tilde\Sigma}$. Two of them, $\{y^1=y^2=0\}$ and $\{y^2=y^3=0\}$ are exceptional. They both have the topology of ${\IP}^1$. Take for example $C_1$:  To avoid being on the excluded set, we must have $y^3\neq0,\ z^3\neq0$ and $(z^1,z^2)\neq0$. Therefore $C_1=\{(z^1,z^2, 1,0,0,1), (z^1,z^1)\neq0\}/\langle (z^1,z^2)\sim (\lambda z^1,\lambda z^2)\rangle$, which corresponds to a ${\IP}^1$.

The fan $\widetilde{\Sigma}$ has six three--dimensional cones: $S_1=\langle D_1,\,E_2,\,E_3\rangle,\ S_2=\langle D_1,\,E_2,\,E_1\rangle,\\ S_3=\langle D_1,\,E_1,\,D_3\rangle,$  $S_4=\langle D_2,\,E_2,\,E_3\rangle,\ S_5=\langle D_2,\,E_2,\,E_1\rangle,$ and $S_6=\langle D_2,\,E_1,\,D_3\rangle$. For this example, we will show the method of working out the Mori generators step by step.
We give the pairs, the sets $S_l\cup S_k$ (the points underlined are those who have to have positive coefficients) and the linear relations:
\begin{eqnarray}
  \label{eq:Moripairs}
  S_6\cup S_3&=&\{\underline{D_1},\,\underline{D_2},\,D_3,\,E_1\},\quad D_1+D_2+4\, D_3-6\,E_1=0,\nonumber\\
  S_5\cup S_2&=&\{\underline{D_1},\,\underline{D_2},\,E_1,\,E_2\},\quad D_1+D_2+2\, E_1-4\,E_2=0,\nonumber\\
  S_4\cup S_1&=&\{\underline{D_1},\,\underline{D_2},\,E_2,\,E_3\},\quad D_1+D_2-2\,E_3=0,\nonumber\\
  S_3\cup S_2&=&\{D_1,\,\underline{D_3},\,E_1,\,\underline{E_2}\},\quad D_3-2\,E_1+E_2=0,\nonumber\\
  S_2\cup S_1&=&\{D_1,\,\underline{E_1},\,E_2,\,\underline{E_3}\},\quad E_1-2\,E_2+E_3=0,\nonumber\\
  S_6\cup S_5&=&\{D_2,\,\underline{D_3},\,E_1,\,\underline{E_2}\},\quad D_3-2\,E_1+E_2=0,\nonumber\\
  S_5\cup S_4&=&\{D_2,\,\underline{E_1},\,E_2,\,\underline{E_3}\},\quad E_1-2\,E_2+E_3=0.
\end{eqnarray}
We find the following three Mori generators:
\begin{equation}
  \label{eq:Morigensixi}
  C_1=\{1,1,0,0,0,-2\},\quad C_2=\{0,0,1,-2,1,0\},\quad C_3=\{0,0,0,1,-2,1\}.
\end{equation}
With this, we are ready to write down $(P\,|\,Q)$:
\begin{equation}
  \label{eq:PQ}
  (P\,|\,Q)=\left(
  \begin{array}{cccccccc}
  D_1&1&-2&1&|&1&0&0\nonumber\\
  D_2&-1&-2&1&|&1&0&0\nonumber\\
  D_3&0&1&1&|&0&1&0\nonumber\\
  E_1&0&0&1&|&0&\!\!-2&1\nonumber\\
  E_2&0&-1&1&|&0&1&\!\!\!-2\nonumber\\
  E_3&0&-2&1&|&\!\!\!-2&0&1.
  \end{array}\right)
\end{equation}
From the rows of $Q$, we can read off the linear equivalences which we bring into a form which will be relevant in Section~\ref{sec:Z6I_G2xSU3xSU3}.
\begin{eqnarray}
  \label{eq:lineqsixI}
  0 &\sim& 6\,D_{{1}}+E_{{1}}+2\,E_{{2}}+3\,E_{{3}},\nonumber\\
  0 &\sim& 6\,D_{{2}}+E_{{1}}+2\,E_{{2}}+3\,E_{{3}},\nonumber\\
  0 &\sim& 3\,D_{{3}}+2\,E_{{1}}+E_{{2}}.
\end{eqnarray}
From the columns of $Q$ we can read off the action of $\left(\IC^*\right)^3$ in~(\ref{eq:blowupsixi}):
\begin{equation}
 \label{eq:rescalessixi}
 (z^1,\,z^2,\,z^3,\,y^1,\,y^2,\,y^3) \to (\lambda_1\,z^1,\,\lambda_1\,z^2,\,\lambda_2\,z^3,\, \frac{\lambda_3}{\lambda_2^2}\,y^1,\,\frac{\lambda_2}{\lambda_3^2}\,y^2,\,\frac{\lambda_3}{\lambda_1^2}\,y^3).
\end{equation}
which leaves the $\tilde{U_i}$ invariant.
The matrix elements of $Q$ contain the intersection numbers of the $C_i$ with the $D_1,\,E_1$, \eg $E_1\cdot C_2=-2,\ D_3\cdot C_3=0$, etc.
We know that $E_1\cdot E_3=0$. From the linear equivalences between the divisors, we find the following relations between the curves $C_i$ and the seven compact curves of our geometry: 
$C_1=E_2\cdot E_3,\ C_2=D_1\cdot E_1=D_2\cdot E_1,\ C_3=D_1\cdot E_2=D_2\cdot E_2,\ E_1\cdot E_2=C_1+2\,C_3,\ D_3\cdot E_1=C_1+4\,C_2+2\,C_3$. From these relations and $(P\,|\,Q)$, we can get all triple intersection numbers, \eg $D_3 E_1^2=C_1\cdot E_1+4\,C_2\cdot E_1+2\,C_3\cdot E_1=-6$.
To be very explicit, we will give the full table of triple intersections for this example:
\begin{table}
  \begin{center} 
    \begin{tabular}{|c|cccccc|}\hline
      {\rm Curve}&$D_1$&$D_2$&$D_3$&$E_1$&$E_2$&$E_3$\cr
      \noalign{\hrule}\noalign{\hrule}
      $E_1\cdot E_2$&1&1&0 &2&\!\!\!-4&0\cr
      $E_2\cdot E_3$&1&1&0&0&0&\!\!\!-2\cr
      $D_1\cdot E_1$&0&0&1&\!\!\!-2&1&0\cr
      $D_1\cdot E_2$&0&0&0&1&\!\!\!-2&1\cr
      $D_2\cdot E_1$&0&0&1&\!\!\!-2&1&0\cr
      $D_2\cdot E_2$&0&0&0&1&\!\!\!-2&1\cr
      $D_3\cdot E_1$&1&1&4&\!\!\!-6&0&0\cr
      \noalign{\hrule}
    \end{tabular}
  \end{center}
  \caption{Triple intersection numbers of the blow-up of $\IZ_{6-II}$}
  \label{tab:inter}
\end{table}
Using the linear equivalences, we can also find the triple self--intersection, \eg $E_1^3=8$.

From the intersection numbers in $Q$, we find that $\{D_2, D_3, E_1+2\,D_3\}$ form a basis of the K\"ahler cone which is dual to the basis $\{C_1,C_2,C_3\}$ of the Mori cone.

Finally, we determine the topology of the exceptional divisors $E_1$, $E_2$, and $E_3$.
\begin{figure}[h!]
\begin{center}
\includegraphics[width=140mm]{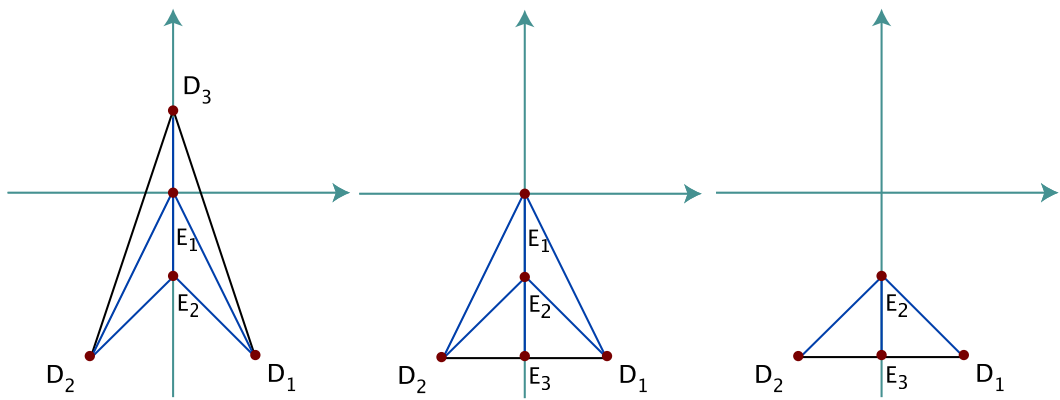}
\caption{The stars of the exceptional divisors $E_1$, $E_2$, and $E_3$, respectively.}
\label{fig:fstarsixi}
\end{center}
\end{figure}
As explained above, we need to look at the respective stars which are displayed in Figure \ref{fig:fstarsixi}.
In order to determine the Mori generators for the star of $E_1$, we have to drop the cones involving $E_3$ which are $S_1$ and $S_4$. From the seven relations in (\ref{eq:Moripairs}) only four of them remain, corresponding to $C_2$, $C_1+2\,C_3$ and $C_1+4\,C_2+2\,C_3$. These are generated by $C_1+2\,C_3=(1,1,0,2,-4,0)$ and $C_2=(0,0,1,-2,1,0)$ which are the Mori generators of $\IF_4$. Similarly, for the star of $E_2$ only the relations not involving $S_3$ and $S_6$ remain. These are generated by $C_1$ and $C_3$, and from (\ref{eq:Morigensixi}) we recognize them to be the Mori generators of $\IF_2$. Finally, the star of $E_3$ has only the relation corresponding to $C_2$. Hence, the topology of $E_3$ is $\IP^1\times \IC$, as it should be, since the point sits on the boundary of the toric diagram of $X_\Sigma$.


\section{The global models}
\label{sec:Global}


\subsection{The fixed point sets and the divisors}
\label{sec:FixedSets}

To be able to glue the blown--up local models together in the correct fashion to obtain the compact smooth Calabi--Yau manifold, we have to study the fixed sets of $T^6$ under the orbifold group, since these are the singular loci of the orbifold.

A point $z_{\rm fixed}$ is fixed under $\theta\in G$ if it fulfills 
\begin{equation}
  \label{eq:fix}
  \theta\,z_{\rm fixed}= z_{\rm fixed}+a,\quad a\in \Lambda,
\end{equation}
where $a$ is a vector of the torus lattice $\Lambda$. In the real lattice basis, $z^i = x^{2i-1} + ix^{2i}$, we have the identification $x^i \sim x^i + 1$. The fixed sets depend on the particulars of the torus lattice, so the fixed sets of orbifold groups that can live on different lattices  will be different for each lattice. We will denote a fixed point by $z_{\rm fixed}=(\zf{1}{\alpha},\zf{2}{\beta},\zf{3}{\gamma})$ where $\alpha,\,\beta,\,\gamma$ label the point in the $1,\,2,\,3$--direction, respectively.

In this way we obtain the sets that are fixed under the respective element of the orbifold group.
We do not need to look at all elements, since some of them are merely the anti-twists of others and therefore give rise to the same fixed sets. Since the fixed points were determined on the covering space, we have to check if they all give rise to distinct equivalence classes. Except for the prime orbifolds, i.e. $T^6/\IZ_3$ and $T^6/\IZ_7$, this is in general not the case. Some of the fixed points are mapped to other fixed points by some group element and are therefore identified in the quotient. When we will count the number of exceptional divisors, i.e. $h^{1,1}_{\rm twist.}$, below, it is crucial to only count one divisor per equivalence class. In the following, one therefore has to distinguish carefully whether one works in the actual orbifold $T^6/G$ or in the covering space. Most calculations are done in the covering space, so it must be kept in mind that in the quotient, points are identified under the action of the orbifold group.

The prime orbifolds are special for having a very simple fixed point configuration. All twisted sectors give rise to the same set of fixed points. In the non-prime cases, also fixed lines (i.e. fixed tori) appear, since these orbifold groups contain $\IZ_2$ subgroups.

To get an idea of how exactly the local patches will be glued, it is useful to have a schematic picture of the configuration, i.e. the intersection pattern of the singularities. In this picture, each complex coordinate is shown as a real coordinate axis and the opposite faces of the resulting cube of length 1 are identified. The fixed point loci are represented by their projections to the real coordinates.

In some cases, not all information that will be necessary for us later will be captured by looking at the fixed sets under a single group element. The points at the intersections of three $\IZ_2$ fixed lines will be relevant as well, which can be easily identified from the schematic figures we provide. This case is the only instance of intersecting fixed lines where the intersection point itself is not fixed under a single group element, and arises only for $\IZ_n\times\IZ_m$ orbifolds with both $n$ and $m$ even.

We now perform the blowups of the local patches as described in section~\ref{sec:toric} and appendix~\ref{sec:toricpatches}, glue the patches together and figure out the number of divisors in the compact model.

The divisors are either inherited from the divisors of the covering torus or originate from the blow-ups of the fixed loci under the group action. We first discuss the inherited divisors. It is easiest to discuss them in terms of the dual 2-forms. A basis for $H^2(T^6)$ is $h_{i\jbar} = dz^{i} \wedge d\bar{z}^{\jbar},\; i,\,\jbar = 1,\ldots,3$, cf. \S 2 in~\cite{Lust:2005dy}. The orbifold group acts on these forms by $\theta(h_{i\jbar}) = e^{2\pi i (g_i-g_{\jbar})} dz^{i} \wedge d\bar{z}^{\jbar}$. Therefore, we immediately see that the three forms $h_{i\ibar}$ are invariant and descend to forms on the orbifold. If, in addition, it happens that for all generators $\theta^a$ $g_i = g_{\jbar}$ simultaneously for some $i\not=\jbar$, then we get an additional pair of 2--forms on the orbifold. It is straightforward to check that the only cases where this happens are the groups $\IZ_3$ for which $g={1\over 3}(1,1,1)$, cf.~(\ref{eq:twistthree}), $\IZ_4$ for which $g=\frac{1}{4}(1,1,2)$, cf.~(\ref{eq:twistfour}) and $\IZ_{6-I}$ for which $g={1 \over 6}(1,1,4)$, cf.~(\ref{eq:twistsixi}). 

For the exceptional divisors we need to consider the fixed loci which, as we have seen in the previous section, are either fixed lines or fixed points.
The corresponding toric patches of the blown--up fixed loci are all determined along the lines of section 2 and can be found in appendix \ref{sec:toricpatches}. 

The fixed lines come from twists of the form ${1\over m}\,(0,1,m-1)$ or cyclic permutation thereof. The corresponding singularities locally have the form of $\IC^2/\IZ_m$. In the global setting, the singular loci are $T^2/\IZ_k$ curves of such $\IC^2/\IZ_m$ singularities where $mk = |G|$, the order of the orbifold group $G$ in $T^6/G$. The possible values for $k$ are $2,3,4$, and $6$. 

The fixed points come from twists of the form ${1\over n}\,(a,b,c),\ a+b+c=n$ which locally look like  $\IC^3/\IZ_n$. Isolated fixed points correspond to toric diagrams with only internal, compact exceptional divisors. When the fixed point sits on a fixed line of order $k$, its toric diagram has $k-1$ exceptional divisors on one of its boundaries, if the fixed point sits at the intersection of two (three) fixed lines, it has exceptional divisors on two (three) of its boundaries. In our examples, the twists occur either in their standard embeddings or permutations thereof. The permutations merely interchange the coordinates, the number and configuration of exceptional divisors remains the same as in the standard form.

We are now presenting the count of the exceptional divisors for all our examples. Most cases are straightforward. There are only two possible complications: One is that the lines fixed under several group elements lie on the same locus. The correct way of counting them is to always count the fixed line of highest order. The other is that several fixed points under different group elements lie in the same loci. There are actually two different ways of counting them, one of which requires no knowledge whatsoever of the form of the local patches. It is enough to know the loci and conjugacy classes of the fixed points. For this first method, we simply count each fixed point as many times as it arises under the different group elements. This automatically gives the right number. The second method involves the toric patches as given in the above tables. To decide which of the patches of the group elements that leave the point invariant to use, we look if the point sits on any (intersection of) fixed line(s). We choose the patch with the right configuration of exceptional divisors on its boundaries to match the order of the fixed lines it sits on. So if the fixed point sits for example on the intersection of two order two and one order six fixed lines, we choose the $\IC^3/\IZ_2\times\IZ_6$--patch with one, one and five exceptional divisors on its boundaries. The patch contributes exactly the number of compact exceptional divisors one would expect from the first counting scheme, so the two methods are consistent.


\subsection{Linear relations and the intersection ring}
\label{sec:IntersectionRing}

There is a purely combinatorial way to determine the intersection ring of the resolved torus orbifold. This method is completely analogous to the one in Section~\ref{sec:toric}. Recall that there we determined the intersection numbers between three distinct divisors, as well as those divisors which never intersect, and then used the linear relations to compute all the remaining intersection numbers. In the global situation we proceed in the same manner. 

The inherited divisors $R_i$ and the exceptional divisors $E_{k,\alpha,\beta,\gamma}$ together form a basis for the divisor classes of the resolved orbifold. This basis is in general not an integral basis. In addition, there are further natural divisors which will be expressed in terms of linear combinations of these basis elements. These additional divisors are simply planes lying at fixed loci of the orbifold action: $D_{i\alpha} = \{ z^i = \zf{i}{\alpha} \}$, where $\alpha$ runs over the fixed loci in the $i$th direction. In terms of the local toric patches they correspond to the non-compact divisors $D_i$, e.g. in (\ref{eq:PQ}). The linear relations encoded in the matrix $Q$ are ``sliding'' divisors in the compact geometry, and are related to the inherited divisors as follows. Consider the divisors $\{ z^i = c \not = \zf{i}{\alpha} \}$. They ``slide'' in the sense that they can move away from the fixed points. The way they can move is constrained by the linear relations in the local geometry. We need, however, to pay attention whether we use the local coordinates $\tilde z^i$ near the fixed point on the orbifold or the local coordinates $z^i$ on the cover. Locally, the map is $\tilde z^i = \left(z^i\right)^{n_i}$, where $n_i$ is the order of the group element that fixes the plane $D_i$. The divisor $R_i=\{ \tilde z^i = c^{n_i} \}$ on the orbifold lifts to a union of $n_i$ divisors $R_i = \bigcup_{k=1}^{n_i} \{ z^i = \varepsilon^k c\}$ on the cover with $\varepsilon^{n_i} =1$. Consider the local toric patch before blowing up. The fixed point lies at $c=\zf{i}{\alpha}$ and in the limit as $c$ approaches this point, we find the relation between $R_i$ and $D_i$ to be $R_i \sim n_iD_i$. This expresses the fact that at the fixed point the polynomial defining $R_i$ on the cover has a zero of order $n_i$ on $D_i$. In the local toric patch, $R_i \sim 0$, hence $n_iD_i \sim 0$. After blowing up, the $R_i$ and $n_iD_i$ differ by the exceptional divisors $E_k$ introduced in the process of resolution. The difference is expressed precisely by the linear relation in the $i$th direction~(\ref{eq:linrels}) of the resolved toric variety $X_{\widetilde\Sigma}$ and takes the form 
\begin{equation}\label{eq:Reqlocal}
  R_i \sim n_iD_i + \sum_k E_k.
\end{equation}
From this we see that this relation is independent from the chosen resolution. Such a relation holds for every fixed point $\zf{i}{\alpha}$ which adds the labels $\alpha$ to the relation:
\begin{equation}
  \label{eq:Reqglobal}
  R_i \sim n_iD_{i\alpha} + \sum_{k, \beta, \gamma} E_{k\alpha\beta\gamma} \qquad {\rm \;for\; all\; } \alpha {\rm \; and \; all\;} i
\end{equation}
where $n_i$ is the order of the group element that fixes the plane $D_{i\alpha}$. The precise form of the sum over the exceptional divisors depends on the singularities involved. 

In general, an orbifold of the form $T^6/G$ has local singularities of the form $\IC^m/H$, where $H$ is some subgroup of index $p = [G:H]$ in $G$. If $H$ is a strict subgroup of $G$, the above discussion applies in exactly the same way and yields the relations (\ref{eq:Reqlocal}) for divisors $R'_i$ with vanishing orders $n_i'$. In the end, however, we have to take into account that $H$ is a subgroup and have to embed the corresponding relations involving $R_i'$ for the action of $H$ into those involving $R_i$ for the action of $G$. The $R_i'$ are related to the $R_i$ simply by
\begin{equation}
  \label{eq:ReqGH}
  R_i = \frac{|G|}{|H|} R_i' = p\, R_i'.
\end{equation}

There is another effect for sets which are fixed only by a strict subgroup $H$. The fixed point sets are mapped into each other by the generator of the normal subgroup $G/H$. This means that we have to consider equivalence classes of invariant divisors. These are represented by $S = \sum_{\alpha} \widetilde{S}_{\alpha}$, where $\widetilde{S}_{\alpha}$ stands for any divisor $\Dt_{i\alpha}$ or $\Et_{k\alpha\beta}$ on the cover and the sum runs over the $p$ elements of the stabilizer $G/H$. In this case, we can add up the corresponding relations:
$$
  \sum_\alpha R_i' \sim n_i'\sum_{\alpha} \Dt_{i\alpha} + \sum_{k,\beta} \sum_{\alpha} \Et_{k\alpha\beta}.
$$
The left hand side is equal to $p\, R_i' = R_i$, therefore
\begin{equation}
  \label{eq:ReqH}
  R_i \sim n_i' D_i + \sum_{k,\beta} E_{k\beta},
\end{equation}
which is the same as the relation for $R_i'$. We will work with the invariant divisors in order to avoid fractional intersection numbers.

Something special happens if $n_i = n_j = n$ for $i \not= j$. In this situation, there are additional divisors on the cover, $R_{ij} = \bigcup_{k=1}^n \{z^i + \varepsilon^k z^j = \varepsilon^{k+k_0} c^{ij} \}$ for some integer $k_0$ and some constant $c_{ij}$, which descend to divisors on the orbifold. We have $\varepsilon^n = 1$ for even $n$, and $\varepsilon^{2n} = 1$ for odd $n$. Since the natural basis for $H^2(T^6)$ are the forms $h_{i\jbar}$ (see the previous subsection), we have to combine the various components of the $R_{ij}$ in a particular way in order to obtain divisors $R_{i\jbar}$ which are Poincar\'e dual to these forms. If we define the variables
\begin{align}
  \label{eq:zij}
  z^{ij}_{k} &= z^i + \varepsilon^k z^j, 
\end{align}
then 
\begin{align}
  \label{eq:Rij}
  R_{i\jbar} &= \bigcup_{k=1}^n \{z^{ij}_{k} + \bar{z}^{ij}_{k} = c^{ij} \} \cup \{z^{ij}_{k} - \bar{z}^{ij}_{k} = c^{ij} \}.
\end{align}
These divisors again satisfy linear relations of the form~\eqref{eq:Reqglobal}\footnote{We do not know how to determine them explicitly.}
:
\begin{align}
  \label{eq:Rijeqglobal}
  R_{i\jbar} \sim n D_{i\jbar\alpha} + \sum_{k, \beta, \gamma} E_{k\alpha\beta\gamma}.
\end{align}

With the local and global linear relations at our disposal we can proceed to determine the intersection ring: First we compute the intersection numbers including the $R_i$ between distinct divisors as well as the Stanley--Reisner ideal from a local compactification of the blown--up singularity. Then we determine all the other intersection numbers by using the linear relations (\ref{eq:Reqglobal}) and the fact that two divisors at different fixed sets never intersect. 

For the local compactification we fix $\alpha$ and drop it from the notation for the time being. For the compactification of the blow-up of $\IC^3$ we choose $\left( \IP^1 \right)^3$. Then we can again invoke the methods of toric geometry from Section 2.1. We start with a lattice $N \cong \IZ^3$ with basis $f_i = m_i e_i$, where $e_i$ is the standard basis. The $m_i$ are positive integers that have to be chosen such that $m_1m_2m_3 = n_1n_2n_3 /|G|$, the $n_i$ are the same as in~(\ref{eq:Reqlocal}). We construct an auxiliary polyhedron $\Delta^{(3)}$ by taking the cone $C_{\Delta^{(2)}}$ from Section~\ref{sec:toric}, rotating and rescaling it such that the vertices corresponding to the divisors $D_i$ are at $v_{i+3} = n_i f_i$, $i=1,2,3$. Then we add the vertices $v_i = -f_i$ corresponding to the divisors $R_i$, $i=1,2,3$. The points $v_{k+6}$ corresponding to the exceptional divisors $E_k$ then lie on the facet $\langle v_4, v_5, v_6 \rangle$. It is easy to check that the linear relations of the polyhedron $\Delta^{(3)}$ are precisely~(\ref{eq:Reqlocal}). For its triangulation we require that it be a star triangulation, i.e. that all simplices contain the origin, and that the triangulation of the simplex $\langle 0, v_4, v_5, v_6 \rangle$ be induced from the triangulation of the cone $C_{\Delta^{(2)}}$. Computing the intersection numbers for three distinct divisors by determining the volume of the corresponding simplex yields the local intersection numbers of the global orbifold. The local Stanley--Reisner ideal, i.e. the set of those divisors which do not intersect because they belong to different cones, can be immediately read off from the auxiliary polyhedron.

Note that this procedure equally applies to resolutions of fixed points and fixed lines. In the latter case, we start with the two--dimensional cone $C_{\Delta^{(1)}} \subset N'_{\IR} \cong \IR^2$ obtained from the resolution of the fixed line at the intersection of $D_1$ and $D_2$. We extend the underlying lattice to $N = \IZ \oplus N' \cong \IZ^3$. Then we add the generator $v_3 = (1,0,0)$ corresponding to the divisor $D_3$ intersecting the fixed line in a point. (The indices of the $D_i$ have to be permuted according to the global coordinates of the singularity.) In this way, we obtain the cone $C_{\Delta^{(2)}} = \{0\} \times C_{\Delta^{(1)}} \cup v_3$ which is the input for the construction of $\Delta^{(3)}$ above. 

For the local patches corresponding to singularities of the form $\IC^m/H$ with $H$ a strict subgroup of $G$, the auxiliary polyhedron $\Delta^{(3)}_H$ is obtained by modifying the polyhedron $\Delta^{(3)}_G$ for $\IC^3/G$. We observe that the exceptional divisors coming from the resolution of $\IC^m/H$ always form a subset of those coming from the resolution of $\IC^3/G$. Hence, we simply drop those points in $\Delta^{(3)}_G$ which do not correspond to an exceptional divisor coming from the resolution of $\IC^m/H$. 

If the equivalence class corresponding to the divisors $D_{i\alpha}$ or $E_{k\alpha\beta\gamma}$ has $p>1$ elements, we have two possibilities: Either we work with the representatives on the cover and plug in the invariant combination at the end of the calculation or we work with the invariant divisors and modify the polyhedra accordingly. The second possibility reduces the calculations by a large amount, so we concentrate on this one. The modification of the polyhedron is determined by the linear relations~(\ref{eq:Reqlocal}) with $R_i=R_i'$ and $n_i=n_i'$ for the corresponding local singularity $\IC^m/H$. This amounts to dividing the $i$th component of $v_k$, $k \geq 4$, by $p$ such that the modified polyhedron also satisfies~(\ref{eq:Reqlocal}). In addition, if it happens that two or more conjugacy classes of the fixed point set, i.e. two or more exceptional divisor classes $E_{k,\alpha}$ lie at the same locus, we have to take into this into account. For fixed points, we can multiply the corresponding generator $v_{k+3}$ by the number of components. For fixed lines, we have to work with as many copies of the corresponding polyhedron as there are components.

We construct the auxiliary polyhedron $\Delta^{(3)}$ for every equivalence class of the fixed point set, and add the labels $\alpha, \beta, \gamma$ denoting the fixed point set to the divisors $D_i$ and $E_k$. The lattice $N$ is the same for all polyhedra. In this way, we get all the intersection numbers $S_{abc} = S_a\cdot S_b\cdot S_c$ between between distinct divisors in the over-complete set $\{ S_a \} = \{R_i, D_{i,\alpha}, E_{k,\alpha,\beta,\gamma} \}$. The presence of the $R_i$ in all the auxiliary polyhedra ensures the correct relative normalizations of the intersection numbers in the different patches. The choice $m_i$ of the lattice basis fixes the overall normalization. In addition, we have the local Stanley--Reisner ideal. There is global analogue of the Stanley--Reisner ideal. It is the set of all pairs of divisors with indices $i,\alpha$ and $i,\alpha'$ with $\alpha \not= \alpha'$. The divisors in such a pair never intersect since they lie at disjoint fixed point sets $\alpha$ and $\alpha'$, respectively. This is easily read off from the schematic picture of the fixed point loci.

Using the linear relations~(\ref{eq:Reqglobal}) which take the general form $\sum_{a} n_s S_a = 0$, we can build a system of equations for the remaining intersection numbers involving two and three equal divisors, $S_{aab}$ and $S_{aaa}$ respectively, by multiplying the linear relations by all possible products $S_bS_c$. This yields a highly overdetermined system of equations $\sum_a n_a S_{abc} = 0$ whose solution determines all the remaining intersection numbers. The information contained in the local and global Stanley--Reisner ideals simplifies this system greatly, since most of these equations are trivially satisfied after setting the corresponding intersections to zero.

As often the case, there is a more direct but equivalent way to obtain the intersection numbers which does not involve the polyhedra. It goes as follows. The intersections between distinct divisors $D_{i\alpha}$ and $E_{k\alpha\beta\gamma}$ are those computed in the local patch, see Section~\ref{sec:toric}. The intersections between $R_j$ and $D_{i\alpha}$ are easily obtained from their defining polynomials on the cover. The intersection number between $R_1$, $R_2$, and $R_3$ is simply the number of solutions to $\{\left(\widetilde{z}^1\right)^{n_1} = c_1^{n_1}, \left(\widetilde{z}^2\right)^{n_2} = c_2^{n_2}, \left(\widetilde{z}^3\right)^{n_3} = c_3^{n_3}\}$ which is $n_1n_2n_3$. Taking into account that we calculated this on the cover, we need to divide by $|G|$ in order to get the result on the orbifold. Similarly, the divisors $D_{i\alpha}$ are defined by linear equations in the $\widetilde{z}^i$, hence we set the corresponding $n_i$ to 1. Therefore,
\begin{align}
  \label{eq:R1R2R3}
  R_1R_2R_3 &= \frac{1}{|G|} n_1n_2n_3, & R_iR_jD_{k\alpha} &= \frac{1}{|G|} n_in_j, & R_iD_{j\alpha}D_{k\beta} = \frac{n_i}{|G|},
\end{align}
for $i,\, j,\, k$ pairwise distinct, and all $\alpha$ and $\beta$. Furthermore, $R_i$ and $D_{i\alpha}$ never intersect by definition. The only remaining intersection numbers involving both $R_j$ and $D_{i\alpha}$ are then $R_jD_{i\alpha}E_{k\alpha\beta\gamma}$. These vanish if $D_{i\alpha}$ and $E_{k\alpha\beta\gamma}$ do not intersect in the local toric patch, otherwise they are equal to 1. Finally, there are the intersections between the $R_i$ and the exceptional divisors. If the exceptional divisor lies in the interior of the toric diagram or on the boundary adjacent to $D_{i\alpha}$, then it never intersects $R_i$. Also, $R_iR_jE_{k\alpha\beta\gamma} = 0$. 

Using this procedure it is also straightforward to compute the intersection numbers involving the divisors $R_{i\jbar}$ and $D_{i\jbar}$. From the defining polynomials in~\eqref{eq:Rij} we find that the only non--vanishing intersection numbers are
\begin{align}
  \label{eq:R1R12R21}
R_{i\jbar}R_{j\ibar}R_k &= -\frac{1}{|G|} n_i^2n_k, & D_{i\jbar\alpha}R_{j\ibar}R_k &= -\frac{1}{|G|} n_i n_k, & R_{i\jbar}R_{j\ibar}D_{k\alpha} &= -\frac{1}{|G|} n_i^2,\notag\\
  D_{i\jbar\alpha}D_{j\ibar\beta}R_k & = -\frac{1}{|G|} n_k, &  D_{i\jbar\alpha}R_{j\ibar}D_{k\beta} & = -\frac{1}{|G|} n_i, & D_{i\jbar\alpha}D_{j\ibar\beta}D_{k\gamma} &= -\frac{1}{|G|}, \notag\\
  R_{i\jbar}R_{j\kbar}R_{k\ibar} &= \frac{1}{|G|} n_i^3, &  R_{i\jbar}R_{j\kbar}D_{k\ibar\alpha} &= \frac{1}{|G|} n_i^2, &  R_{i\jbar}D_{j\kbar\alpha}D_{k\ibar\beta} &= \frac{1}{|G|} n_i, \notag\\
  D_{i\jbar\alpha}D_{j\kbar\beta}D_{k\ibar\gamma} &= \frac{1}{|G|},
\end{align}
for $i,\, j,\, k$ pairwise distinct, and all $\alpha$, $\beta$, and $\gamma$. The negative signs come from carefully taking into account the orientation reversal due to complex conjugation. The intersection numbers with the exceptional divisors should be determined in a similar way.

\subsection{Divisor topologies}
\label{sec:Topology}

In this section we show how the topology of the divisors is determined. We first discuss the exceptional divisors $E_{k\alpha\beta\gamma}$, then the fixed planes $D_{i\alpha}$ and the generic planes $R_i$ by looking at the structure of the fixed point set. Then we explain how some of this information can be distilled from the intersection ring.

The topology of the exceptional divisors $E$ depends on the structure of the fixed point set they originate from. The following three situations can occur:
\renewcommand{\labelenumi}{E\theenumi)}
\begin{enumerate}
  \item Fixed points
  \label{item:E1}
  \item Fixed lines without fixed points
  \label{item:E2}
  \item Fixed lines with fixed points on top of them
  \label{item:E3}
\end{enumerate}
In addition, we have to take into account whether the equivalence class of the fixed point set consists of a single element or more. We first discuss the case of a single element. The topology of the divisors in case~E\ref{item:E1}) has already been discussed in great detail in Section~\ref{sec:ExcepTop}. The local topology of the divisors in the cases~E\ref{item:E2}) and~E\ref{item:E3}) has also been discussed in that section, and found to be (a blow--up of) $\IC\times\IP^1$. The $\IC$ factor is the local description of the $T^2/\IZ_k$ curve on which there were the $\IC^2/\IZ_m$ singularities whose resolution yielded the $\IP^1$ factor. 

\begin{figure}[h!]
\begin{center}
\includegraphics[width=140mm]{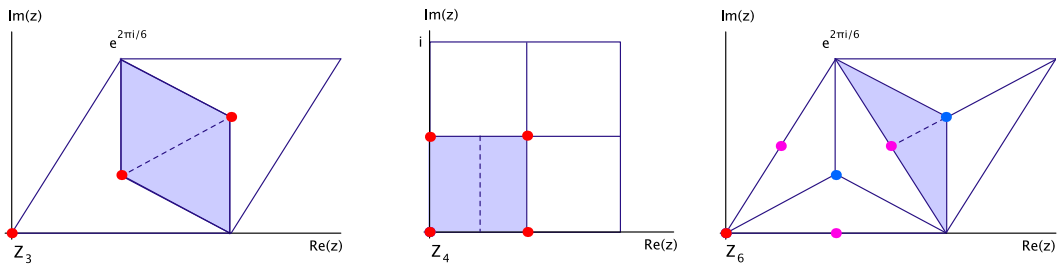}
\caption{The fundamental domains of $T^2/\IZ_k$, $k=3,4,6$. The dashed line indicates the folding.}
\label{fig:fundomains}
\end{center}
\end{figure}

For the determination of the topology of the resolved curves, it is necessary to know the topology of $T^2/\IZ_k$. This can be determined from the action of $\IZ_k$ on the respective fundamental domains. For $k=2$, there are four fixed points at $0, 1/2, \tau/2$, and $(1+\tau)/2$ for arbitrary $\tau$. The fundamental domain for the quotient can be taken to be the rhombus $[0,\tau,\tau+1/2,1/2]$ and the periodicity folds it along the line $[\tau/2,(1+\tau)/2]$. Hence, the topology of $T^2/\IZ_2$ without its singularities is that of a $\IP^1$ minus 4 points. For $k=3,4,6$ the value of $\tau$ is fixed to be $i,\, \exp(\frac{2\pi i}{3}),\, \exp(\frac{2\pi i}{6})$, respectively, and the fundamental domains are shown in Figure~\ref{fig:fundomains}. 
From this figure, we see that the topology of $T^2/\IZ_k$ for $k=3,4,6$ is that of a $\IP^1$ minus 3, 2, 3 points, respectively. In the case~E\ref{item:E2}), there are no further fixed points, so the blow--up procedure glues in points in this $\IP^1$. The topology of such an exceptional divisor is therefore the one of $\IF_0 = \IP^1 \times \IP^1$. In the case~E\ref{item:E3}), the topology further depends on the possible fixed points lying on these fixed lines. This depends on the choice of the root lattice for $T^6/G$, and can therefore only be discussed case by case. This will be done in the examples in Section~\ref{sec:GlobalExamples} and in all the appendices. The general procedure consists of looking at the corresponding toric diagram. There will always be an exceptional curve whose line ends in the point corresponding to the exceptional divisor. (In the example $\IC^3/\IZ_{6-I}$ discussed in Section~\ref{sec:Z6-I}, this is the curve $C_2=E_2\cdot E_3$ in the diagram on the left hand side of Figure~\ref{fig:fstarsixi}.) This exceptional curve meets the $\IP^1$ (minus some points) we have just discussed in the missing points and therefore the blow--up adds in the missing points. Any further lines ending in that point of the toric diagram correspond to additional blow--ups, i.e. additional $\IP^1$s that are glued in at the missing points. Therefore, for each fixed point lying on the fixed line and each additional line in the toric diagram, there will be a blow--up of $\IF_0=\IP^1 \times \IP^1$.

If there are $p$ elements in the equivalence class of the fixed line, the topology in the case~E\ref{item:E2}) is quite different. This is because the $p$ different $T^2/\IZ_k$s are mapped into each other by the corresponding generator in such a way that the different singular points are permuted. When forming the invariant combination by summing over all representatives, the singularities disappear and we are left with a $T^2$. Hence, in the case~E\ref{item:E2}) without fixed points, the topology of $E= \sum_{\alpha=1}^k \Et_\alpha$ is $\IP^1 \times T^2$. 

Similarly, the topology of the divisors $D_{i\alpha}$ depends on the structure of the fixed point sets lying in the divisor. Recall that these divisors are defined by $D_{i\alpha} = \{z^i = \zf{i}{\alpha}\}$. The orbifold group $G$ acts on these divisors by $(z_j,z_k) \to (\varepsilon^{n_j}z_j,\varepsilon^{n_k} z_k)$ for $(z_j,z_k)\in D_{i\alpha}$ and $j\not=i\not=k$. Since $n_j+n_k = n - n_i < n$, the resolved space will not be a Calabi--Yau manifold anymore, but a rational surface or a ruled surface over a $T^2$. This is because for resolutions of this type of actions, the canonical class cannot be preserved. (In more mathematical terms, no crepant resolution exists.) In order to determine the topology, we will use a simplicial cell decomposition, remove the singular sets, glue in the smoothening spaces, i.e. perform the blow--ups, and use the additivity of the Euler number. This has to be done case by case. If, in particular, the fixed point set contains points, there will be a blow--up for each fixed point and for each line in the toric diagram of the fixed point which ends in the point corresponding to $D_i$. Another possibility is to apply the techniques of toric geometry in Section~\ref{sec:toric} to singularities of the form $\IC^2/\IZ_n$ for which $n_1+n_2 \not= n$. As before, we also have to take into account whether the equivalence class of the fixed point set defining $D$ consists of a single element or more.

Note that when embedding the divisor $D$ into a (Calabi--Yau) manifold $X$ in general, not all the divisor classes of $D$ are realized as classes in $X$. In the case of resolved torus orbifolds, this is because the underlying lattice of $D$ is not necessarily a sublattice of the underlying lattice of $X$. This means that the fixed point set of $D$ as a $T^4$ orbifold can be larger than the restriction of the fixed point set of the $T^6$ orbifold to $D$. In order to determine the topology of $D$ we have to work with the larger fixed point set of $D$ as a $T^4$ orbifold. It turns out that there is always a lattice defining a $T^6$ orbifold for which all divisor classes of $D$ are realized in $X$. In fact, we observe that the topology of all those divisors which are present in several different lattices is independent of the lattice. 

The divisors $R_i$ contain by definition no component of the fixed point set. However, if fixed lines are present, they can intersect fixed lines in points. If there are no fixed lines piercing them, the action of the orbifold group is free and their topology is that of a $T^4$. Otherwise, the intersection points have to be resolved in the same way as for the divisors $D_{i\alpha}$. In this case, their topology is always that of a K3 surface.

We can also use the intersection ring to study the topology of these divisors. If we describe the divisor $S$ (which can be of any type, i.e. $R$, $E$ or $D$ above) of the Calabi--Yau manifold $X$ by an embedding $i:S\longrightarrow X$ then we have the associated short exact sequence for the tangent bundles $T_S$ and $T_X$ and the normal bundle $N_{S/X}$ of $S$ in $X$:
\begin{equation}
  \label{eq:embedding}
  \shortexactseq{T_S}{}{T_X|_S}{}{N_{S/X}}.
\end{equation}
By the adjunction formula~\cite{Griffiths:1994ab} $N_{S/X} \cong \Oc(S)|_S$, we can relate the topology of $S$ to that of $X$. Expanding $\ch(T_X) = \ch(T_S)\ch(N_{S/X})$ and using the restriction formula $\int_S \omega = \int_X \omega \wedge S$,
we obtain
\begin{align}
  \label{eq:c1(S)}
  \ch_1(S) &= -S, \\
  \label{eq:c2(S)}
  \ch_1(N_{S/X})^2 &= S^2 \quad = \ch_2(X) - \ch_2(S), \\
  \label{eq:c2.S}
  \ch_2(X)\cdot S + S^3 &= \chi(S),
\end{align}
which gives the relation between the Chern classes of $S$ and the topological numbers of $X$. Furthermore, we have the holomorphic Euler characteristic of $S$
\begin{equation}
  \label{eq:holoEuler}
  \chi(\Oc_S) = 1 - h^{1,0}(S) + h^{2,0}(S).
\end{equation}
Noether's formula~\cite{Griffiths:1994ab} relates $\chi(\Oc_S)$ to the Chern classes of $S$:
\begin{equation}
  \label{eq:Noether}
  \chi(\Oc_S) = \frac{1}{12} \int_S \left(\ch_1(S)^2 + \ch_2(S)\right),
\end{equation}
from which we get
\begin{equation}
  \label{eq:S3}
  12 \chi(\Oc_S) = S^3 + \chi(S).
\end{equation}
This equation can be used in two ways. Since we already determined the topology of the divisors $R_i$, $D_{i\alpha}$ and $E_{k\alpha\beta\gamma}$, i.e. the Euler numbers $\chi(\Oc_S)$ and $\chi(S)$, we can cross--check them with the self--intersection numbers in the intersection ring. We can also explicitly check the number of blow--ups of $S$. On one hand, it is known~\cite{Griffiths:1994ab} that the holomorphic Euler characteristic is a birational invariant, which means that it does not change under blow--ups. On the other hand we know that blowing up a surface adds a 2--cycle to it, hence increases the Euler number $\chi(S)$ by 1. Therefore, the self--intersection number $S^3$ is decreased by 1. Furthermore, $S^3$ restricted to $S$ becomes $S^2 = \ch_1(S)^2 = K_S^2$, where $K_S$ is the canonical divisor of $S$. Like $\chi(\Oc_S)$ and $\chi(S)$, $K_S^2$ is a characteristic quantity of an algebraic surface $S$. We have collected these three numbers for the basic topologies that we found above in the following table:
\begin{equation}
  \label{eq:KS}
  \begin{array}{|c|c|c|c|c|}
    \hline
    S & \chi(S) & \chi(\Oc_S) & K_S^2 & h^{1,0}(S) \\
    \hline
    \IP^2 & 3 & 1 & 9 & 0\\
    \IF_n & 4 & 1 & 8 & 0\\
    \IP^1 \times T^2 & 0 & 0 & 0 & 1\\
    T^4 & 0 & 0 & 0 & 2\\
    \rm{K3} & 24 & 2 & 0 & 0 \\
    \hline
  \end{array}
\end{equation}
The invariants of the blow--ups of these surfaces are then obtained from the above observations. 

The second use of~(\ref{eq:S3}) is to determine $\ch_2\cdot S$ in~(\ref{eq:c2.S}) from the topology of $S$.


\subsection{Twisted complex structure moduli}
\label{sec:TwistedCplx}

There are two types of complex structure deformations. Loosely speaking, we can deform the complex structure of the underlying torus with a given lattice, or we can deform the fixed point set. The former type of deformation corresponds to the untwisted complex structure moduli at the orbifold point while the latter correspond to the twisted complex structure moduli. The deformations of the complex structure of the underlying tori has been studied in great detail in~\cite{Lust:2005dy}. 

To study the twisted complex structure moduli we therefore have to look at the fixed point set. Isolated fixed points do not admit any complex structure deformations. Hence, we need only consider fixed lines. However, if there are fixed points on them, their complex structure again cannot be deformed. So, we are left with fixed lines without fixed points. In the previous section, we have argued that after resolving the singularities they yield exceptional divisors which are ruled surfaces over a $\IP^1$ or a $T^2$. 

These ruled surfaces can also be viewed as an algebraic family of algebraic curves (here rational curves $\IP^1$) parametrized by the base curve $C$. For any smooth complex projective threefold $X$ with such a family of algebraic curves there is a map $\varphi_*: H_1(C) \to H_3(X)$ which sends the 1--cycle $\gamma$ on $C$ to the 3--cycle $\varphi_*(\gamma)$ traced out by the fiber curve $E_t$ as $t$ traces out $\gamma$~\cite{Clemens:1972ab}. Since the fibers $E_t$ are algebraic cycles, the dual map on the cohomology respects the Hodge decomposition and yields a map $\varphi^*: H^{1,0}(C) \to H^{2,1}(X)$. For our varieties $C$ is either a $\IP^1$ or a $T^2$. Since $h^{1,0}(\IP^1) = 0$ and $h^{1,0}(T^2) = 1$, only the ruled surfaces over $T^2$ give such a map. Hence we find that
\begin{equation}
  \label{eq:h21tw}
  h^{2,1}_{\rm twist.}(X) = \sum_{i} (n_i - 1) h^{1,0}(C_i)
\end{equation}
where the sum runs over the curves $C_i$ with topology $T^2$ parametrizing exceptional curves which come from the resolution of $\IC^2/\IZ_{n_i}$ singularities. To reiterate in plain language, each equivalence class of order $n$ fixed lines without fixed points on them contributes as many twisted complex structure moduli as the fixed line has $\IP^1$ components in its resolution, namely $n-1$. This is precisely the situation E\ref{item:E2}) in Section~\ref{sec:Topology}. 

We note that this situation has a well--known analogue for hypersurfaces in toric varieties. In that case the complex structure deformations split into polynomial and nonpolynomial ones. The former correspond to deformation of the hypersurfaces while the latter corresponds to deformations of a certain ambient toric variety. The nonpolynomial deformations also come from curves of $\IC^2/\IZ_n$ singularities. After resolution of the singularities these become families of $\IP^1$s parametrized by the curve, in other words, ruled surfaces and a similar reasoning applies~\cite{Katz:1996ht}.


\subsection{Examples}
\label{sec:GlobalExamples}


\subsubsection{The $\IZ_{6-I}$ orbifold on $G_2\times SU(3)^2$}
\label{sec:Z6I_G2xSU3xSU3}

In order to find the fixed point sets, we need to look at the $\theta$--, $\theta^2$-- and $\theta^3$--twists. $\theta^4$ and $\theta^5$ yield no new information, since they are simply the anti--twists of $\theta^2$ and, $\theta$. The action of the twist $\theta$ on the lattice $G_2\times SU(3)^2$ was given in (A.12) and the resulting complex structure in (A.16) of~\cite{Lust:2005dy}.
The $\IZ_{6-I}$--twist has only one fixed point for each coordinate, namely $\zf{1}{1}=\zf{2}{1}=\zf{3}{1} = 0$. The $\IZ_3$--twist has three fixed points, namely $\zf{1}{\alpha} = \zf{2}{\beta} = 0,1/3, 2/3$ for $\alpha,\beta=1,3,5$ and $\zf{3}{\gamma}=0, 1/\sqrt3 \,e^{\pi i/6}, 1+i/\sqrt3$ for $\gamma=1,2,3$. The $\IZ_2$--twist, which arises in the $\theta^3$-twisted sector, has four fixed points, corresponding to $\zf{1}{\alpha} = 0,\half,\half \tau,\half(1+\tau)$, $\alpha=1,2,4,6$ for the respective modular parameter $\tau$. As a general rule, we shall use red to denote the fixed set under $\theta$, blue to denote the fixed set under $\theta^2$ and pink to denote the fixed set under $\theta^3$. Note that the figure shows the covering space, not the quotient. 

The equivalence classes of the fixed point set are described as follows: We first look at the $z^1$ and $z^2$--directions. The two $\IC^3/\IZ_3$ fixed points at $1/3$ and $2/3$ are mapped to each other by $\theta$ and form orbits of length two. We choose to represent this orbit by $\zf{i}{2}$, $i=1,2$. The three $\IC^3/\IZ_3$ fixed points in the $z^3$--direction each form a separate conjugacy class. Therefore, we obtain the 15 conjugacy classes of $\IC^3/\IZ_3$ fixed points, 5 in each plane $z^3=\zf{3}{\gamma}$, $\gamma=1,2,3$:
\begin{align}
  \label{eq:zthreeconj}
  \mu=1:\; &(0,0, \zf{3}{\gamma}) & & &\notag\\  
  \mu=2:\; &(0,\tfrac{1}{3},\zf{3}{\gamma}),\ (0,\tfrac{2}{3},\zf{3}{\gamma}) & \mu=3:\; & (\tfrac{1}{3},0,\zf{3}{\gamma}),\ (\tfrac{2}{3},0,\zf{3}{\gamma})\notag\\ 
  \mu=4:\; &(\tfrac{1}{3},\tfrac{1}{3},\zf{3}{\gamma}),\ (\tfrac{2}{3},\tfrac{2}{3}, \zf{3}{\gamma}) & \mu=5:\; & (\tfrac{1}{3}, \tfrac{2}{3},\zf{3}{\gamma}),\ (\tfrac{2}{3},\tfrac{1}{3},\zf{3}{\gamma}). 
\end{align}
The fixed points in the $z^1$--direction form the two orbits under $\theta^2$, namely $0$, and $\half \to \half(1+\tau) \to \half \tau$. The corresponding two conjugacy classes will be represented by $\zf{i}{3}$, $i=1,2$.
\begin{table}[h!]
  \begin{center}
  \begin{tabular}{|c|c|c|c|}
    \hline
    Group el.& Order & Fixed Set& Conj. Classes \cr
    \hline 
    \noalign{\hrule}
    $\ \theta  $ & 6 &  3\ {\rm fixed\ points} &\  3\cr
    $\ \theta^2$ & 3 & 27\ {\rm fixed\ points} &\ 15\cr
    $\ \theta^3$ & 2 &  4\ {\rm fixed\  lines} &\  2\cr
    \hline
  \end{tabular}
  \caption{Fixed point sets for $\IZ_{6-I}$ on $G_2\times SU(3)^2$.}
  \label{tab:fssixigsuii}
\end{center} 
\end{table}
Table~\ref{tab:fssixigsuii} summarizes the relevant data of the fixed point set. The invariant subtorus under $\theta^3$ is $(0,0,x^5-x^6,-x^6,x^5,x^6)$, corresponding to the complex $z^3$--coordinate being invariant.
\begin{figure}[h!]
\begin{center}
\includegraphics[width=85mm]{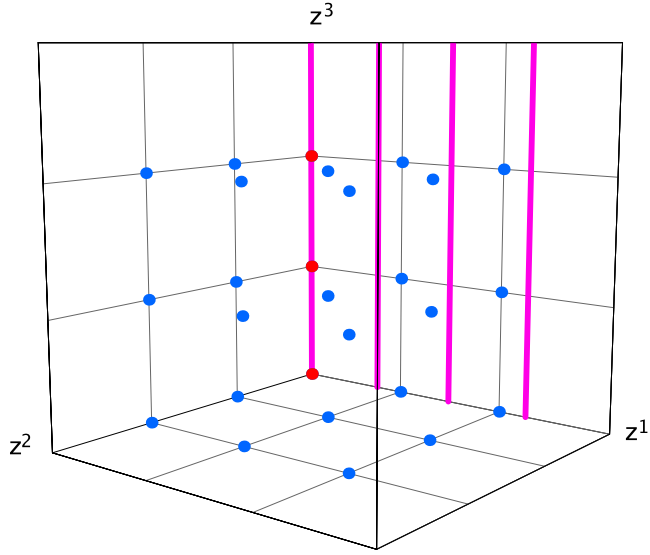}
\caption{Schematic picture of the fixed point set of the $\IZ_{6-I}$ orbifold on $G_2\times SU(3)^2$}
\label{fig:ffixsixi}
\end{center}
\end{figure}
Figure~\ref{fig:ffixsixi} shows the configuration of the fixed point set in a schematic way, where each complex coordinate is shown as a coordinate axis and the opposite faces of the resulting cube of length 1 are identified. 

Next, we resolve all the singularities and determine all the divisors of the blown--up torus orbifold including their topologies. From Table~\ref{tab:fssixigsuii} and Figure~\ref{fig:ffixsixi}, we see that there are three fixed points with a $\IC^3/\IZ_{6-I}$ singularity. Resolving this singularity amounts to replacing the $\IC^3/\IZ_{6-I}$ patch by $X_{\widetilde{\Sigma}}$ in~(\ref{eq:blowupsixi}). By Figure~\ref{fig:fsixi}, each resolution contributes two exceptional divisors $E_{1,\gamma}$, and $E_{2,1,\gamma}$, $\gamma=1,2,3$, from the interior of the diagram and one exceptional divisor from the boundary of the diagram, respectively The latter will be considered in the next paragraph.

Returning to Table~\ref{tab:fssixigsuii} and Figure~\ref{fig:ffixsixi}, we have furthermore 15 conjugacy classes of $\IC^3/\IZ_3$ fixed points. Blowing them up replaces each of them locally by $X_{\widetilde{\Sigma}}$ in~(\ref{eq:blowupthree}) and contributes one exceptional divisor as can be seen from Figure~\ref{fig:frthree}. Since three of these fixed points sit at the location of the $\IC^3/\IZ_{6-I}$ fixed points which we have already taken into account ($E_{2,1,\gamma}$), we only count 12 of them, and denote the resulting divisors by $E_{2,\mu,\gamma},\ \mu = 2,\dots,5, \, \gamma=1,2,3$. The invariant divisors are constructed according to the conjugacy classes in~(\ref{eq:zthreeconj}):
\begin{align}
  \label{eq:E2conj}
  E_{2,2,\gamma} &= \Et_{2,1,2,\gamma} +  \Et_{2,1,3,\gamma}, & E_{2,3,\gamma} &=  \Et_{2,3,1,\gamma} +  \Et_{2,5,1,\gamma},\notag\\
  E_{2,4,\gamma} &= \Et_{2,3,2,\gamma} +  \Et_{2,5,3,\gamma}, & E_{2,5,\gamma} &=  \Et_{2,3,3,\gamma} +  \Et_{2,5,2,\gamma},
\end{align}
where $\Et_{2,\alpha,\beta,\gamma}$ are the representatives on the cover.

Then, we finally have 2 conjugacy classes of fixed lines of the form $\IC^2/\IZ_2$. We see that after the resolution, each class contributes one exceptional divisor $E_{3,\alpha}, \alpha=1,2$. On the fixed line at $\zf{1}{1}=\zf{2}{1}=0$ sit the three $\IC^3/\IZ_{6-I}$ fixed points. The divisor coming from the blow--up of this fixed line, $E_{3,1}$, is identified with the three exceptional divisors corresponding to the points on the boundary of the toric diagram of the resolution of $\IC^3/\IZ_{6-I}$ that we mentioned above. The other exceptional divisor is the invariant combination $E_{3,2} = \sum_{\alpha=2,4,6} \Et_{3,\alpha}$, where $\Et_{3,\alpha}$ are the representatives on the cover. Consequently, $E_{3,1}$ could have a different topology than $E_{3,2}$.

This results in $3\cdot 2+12\cdot 1+2\cdot 1=20$ exceptional divisors.  There is one $\IC^2/\IZ_2$ fixed line without fixed points on it, therefore, by~(\ref{eq:h21tw}), $h^{2,1}_{\rm twist.}=1$.

Furthermore, we have fixed planes $\Dt_{1,\alpha} = \{ z^1=\zf{1}{\alpha}\}$, $\alpha=1,\dots,6$, $\Dt_{2,\beta} = \{z^2=\zf{2}{\beta}\}$, $\beta=1,2,3$, and $\Dt_{3,\gamma} = \{z^3 = \zf{3}{\gamma}\}$, $\gamma=1,2,3$ on the cover. From these we define the invariant combinations
\begin{align*}
  D_{1,1} &= \Dt_{1,1}, & D_{1,2} &= \Dt_{1,2} + \Dt_{1,4} + \Dt_{1,6}, &
  D_{1,3} &= \Dt_{1,3} + \Dt_{1,5}, \\
  D_{2,1} &= \Dt_{2,1}, & D_{2,2} &= \Dt_{2,2} + \Dt_{2,3}, & D_{3,\gamma} &= \Dt_{3,\gamma}.
\end{align*}
Next, we need the global linear relations~(\ref{eq:Reqglobal}) in order to determine the intersection ring. The relation for $D_{1,1}$ is obtained from (\ref{eq:lineqsixI}) :
\begin{equation}
  \label{eq:Z6IrelD11} 
  R_1=6\,D_{{1,1}}+\sum_{\gamma=1}^3E_{{1,\gamma}} +2\, \sum_{\mu=1}^2 \sum_{\gamma=1}^3 E_{{2,\mu,\gamma}}+3\, E_{{3,1}}.
\end{equation}
The divisor $D_{1,2}$ only contains a single equivalence class of $\IC^2/\IZ_2$ fixed lines. 
From the local relations~(\ref{eq:lineqsctwo}), we find the local relation to $R_1$ as in~(\ref{eq:ReqH}) (here, we already changed the labels of the divisors to match  the labels of the $\IC^3/\IZ_{6-I}$ patch):
\begin{equation}
  \label{eq:Z6IrelD12}
  R_1 = 2D_{{1,2}}+\, E_{{3,2}}.
\end{equation}
Next, we look at the divisor $D_{1,3}$, which only contains $\IC^3/\IZ_3$ fixed points. The local linear equivalences~(\ref{eq:lineqthree}) and~(\ref{eq:ReqH}) lead to
\begin{equation}
  \label{eq:Z6Irel13}
  R_1 = 3\,D_{{1,3}}+\sum_{\mu=3}^5 \sum_{\gamma=1}^3 E_{{2,\mu,\gamma}}.
\end{equation}
The linear relations for $D_{2,\beta}$ are the same as those for $D_{1,\alpha}$ except that the one coming from the $\IC^2/\IZ_2$ fixed line is absent: 
\begin{eqnarray}
  \label{eq:Z6Irels2}
  R_2&=&6\,D_{{2,1}}+\sum_{\gamma=1}^3 E_{{1,\gamma}}+2\, \sum_{\mu=1,3} \sum_{\gamma=1}^3 E_{{2,\mu,\gamma}}+3\, \sum_{\alpha=1}^2 E_{{3,\alpha}},\nonumber\\
  R_2&=&3\,D_{{2,2}}+\sum_{\mu=2,4,5} \sum_{\gamma=1}^3 E_{{2,\mu,\gamma}}.
\end{eqnarray}
Finally, the relations for $D_{3,\gamma}$ are again obtained from~(\ref{eq:lineqsixI}):
\begin{equation}
  \label{eq:Z6Irels3}
  R_3=3\,D_{{3,\gamma}}+2\, E_{1,\gamma} + \sum_{\mu=1}^5 E_{{2,\mu,\gamma}} \qquad \gamma=1,\dots,3.
\end{equation}
Now, we are ready to compute the intersection ring. First, we need to determine the basis for the lattice $N$ in which the auxiliary polyhedra will live. From~(\ref{eq:Z6IrelD11}), (\ref{eq:Z6Irels2}), and~(\ref{eq:Z6Irels3}) we see that $n_1=n_2=6$, and $n_3=3$. Hence we can choose $m_1=m_2=3$, and $m_3=2$ and the lattice basis is $f_1=(3,0,0)$, $f_2=(0,3,0)$, $f_3=(0,0,2)$. We start with the polyhedron $\Delta^{(3)}$ for the $\IZ_{6-I}$ fixed points. Its lattice points are
\begin{align}\label{eq:Z6Ipoly}
  v_1 &= (-3,0,0), & v_2 &= (0,-3,0), & v_3 &= (0,0,-2),\notag\\
 v_4 &= (18,0,0), & v_5 &= (0,18,0), & v_6 &= (0,0,6),\\
  v_7 &= (3,3,4), & v_8 &= (6,6,2), & v_9 &= (9,9,0),\notag
\end{align}
corresponding to the divisors $R_1,R_2,R_3,D_1,D_2,D_3,E_1,E_2,E_3$ in that order. The polyhedron is shown in Figure~\ref{fig:Z6I-cpt}. By applying the methods described at the end of Section~\ref{sec:toric} we obtain the following intersection numbers between three distinct divisors:
\begin{figure}[h!]
\begin{center}
\includegraphics[width=140mm]{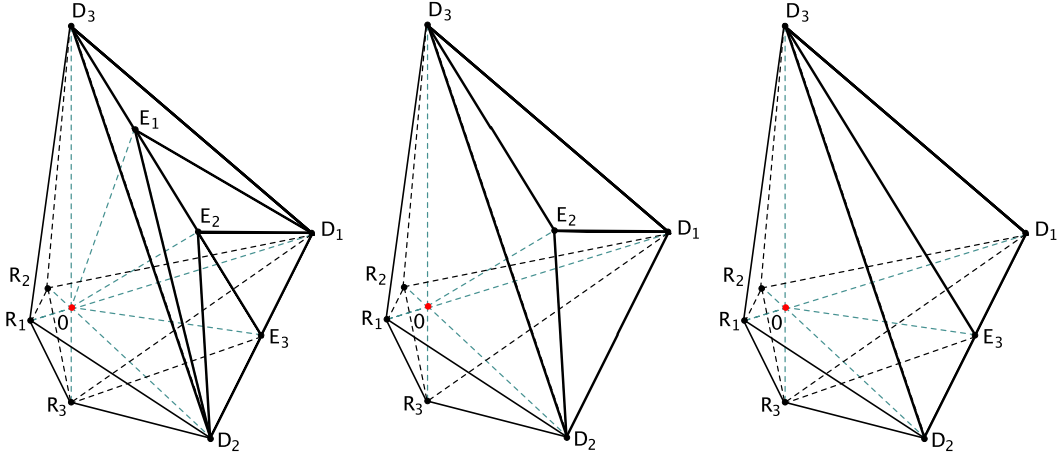}
\caption{The polyhedra $\Delta_H^{(3)}$ describing the local compactification of the resolution of $\IC^3/H$ with $H=G=\IZ_{6-I}$, $H=\IZ_3$, and $H=\IZ_2$.}
\label{fig:Z6I-cpt}
\end{center}
\end{figure}
\begin{align}
  \label{eq:intZ6I}
  R_1R_2R_3 &= 18, & R_1R_2D_3 &= 6, & R_1R_3D_2 &=3, & R_1D_2D_3 &= 1, \notag\\
  R_2R_3D_1 &=  3, & R_2D_1D_3 &= 1, & R_3D_1E_3 &=1, & R_3D_2E_3 &= 1, \notag\\
  D_1E_1D_3 &=  1, & D_1E_1E_2 &= 1, & D_1E_2E_3 &=1, & D_2D_3E_1 &= 1, \notag\\
  D_2E_1E_2 &=  1, & D_2E_2E_3 &= 1,  
\end{align}
and the local Stanley--Reisner ideal
\begin{gather}
  \left\{R_iD_i = 0, R_iE_1 = 0, R_iE_2 = 0, R_1E_3 = 0, R_2E_3 = 0, \right.\notag\\
  \left. D_1D_2 = 0, D_3E_2 = 0, D_3E_3 = 0, E_1E_3 = 0,\; i=1,2,3 \right\}.  
  \label{eq:SRIZ6I}
\end{gather}
Now, we add the labels of the fixed points $\alpha,\beta,\gamma$ to the divisors: $D_i \to D_{i\alpha}$, $E_1 \to E_{1\gamma}$, $E_2 \to E_{2\alpha\beta\gamma}$, $E_3 \to E_{3\alpha}$, and set $\alpha=1, \beta=1, \gamma=1,2,3$. 

As explained in Section~\ref{sec:IntersectionRing}, the polyhedra $\Delta_H$ for the other patches are obtained from $\Delta_G^{(3)}$ by dropping and rescaling some of the points. For the $\IC^3/\IZ_3$ patches we drop $v_7$ and $v_9$, for those at $\mu=2$ we set $v_5 = (0,9,0)$, $v_8 = (6,3,2)$, for those at $\mu=3$ we set $v_4 = (9,0,0)$, $v_8 = (3,6,2)$, and finally for those at $\mu=4,5$ we set $v_4 = (9,0,0), v_5 = (0,9,0), v_8 = (6,6,4)$. For the $\IZ_2$ fixed line at $\alpha=1,\beta=2$ we drop $v_7$ and $v_8$ and set $v_4 = (6,0,0), v_9 = (3,9,0)$. Computing the analogues of~(\ref{eq:intZ6I}) and~(\ref{eq:SRIZ6I}) yields all the local information we need. The global information comes from the linear relations~(\ref{eq:Z6IrelD11}) to~(\ref{eq:Z6Irels3}) and the examination of Figure~\ref{fig:ffixsixi} to determine those pairs of divisors that never intersect. Here, one has to be careful with the divisors $E_{2,\mu,\gamma}$ for $\mu=4,5$. Solving the resulting overdetermined system of linear equations then yields the intersection ring of $X$ in the basis $\{R_i, E_{k\alpha\beta\gamma}\}$ (without the divisors $R_{i\jbar}$):
\begin{align}
  \label{eq:ringZ6I}
  R_1R_2R_3 &= 18, & R_3E_{3,1}^2 &= -2, & R_3E_{3,2}^2 &= -6, & E_{1,\gamma}^3 &= 8, \notag\\
  E_{1,\gamma}^2 E_{2,1,\gamma} &= 2, & E_{1,\gamma}E_{2,1,\gamma}^2 &= -4, & E_{2,1\gamma}^3 &= 8, & E_{2,\mu,\gamma}^3 &= 9, \notag\\
  E_{2,1,\gamma}E_{3,1}^2 &= -2, & E_{3,1}^3 &= 8,
\end{align}
for $\mu=2,\dots,5$, $\gamma=1,2,3$. Here we have given only the nonvanishing intersection numbers. Those involving the $D_{i\alpha}$ can be obtained using the linear relations~(\ref{eq:Z6IrelD11}) to~(\ref{eq:Z6Irels3}).

Next, we discuss the divisor topologies. The topology of the exceptional divisors has been determined in Section~\ref{sec:LocalExamples}: $E_{1,\gamma} = \IF_4$ and $E_{2,1,\gamma} = \IF_2$. By the remark at the end of Appendix~\ref{sec:Z3orb}, the $E_{2,\mu,\gamma}$, $\mu=2,\dots,5$, have the topology of a $\IP^2$. The divisor $E_{3,1}$ is of type~E\ref{item:E3}) and has a single representative, hence the basic topology is that of a $\IF_0$. There are 3 $\IC^3/\IZ_{6-I}$ fixed points on it, but there is only a single line ending in $E_3$ in the toric diagram of Figure~\ref{fig:fsixi}, which corresponds to the exceptional $\IP^1$, therefore there are no further blow--ups. The divisor $E_{3,2}$ is of type~E\ref{item:E2}), but there are 3 representatives, so its topology is that of $\IP^1\times T^2$. 

The topology of $D_{2,1}$ is determined as follows: The fixed point set of the action $\frac{1}{6}(1,4)$ agrees with the restriction of the fixed point set of $T^6/\IZ_{6-I}$ to $D_{2,1}$. The Euler number of $D_{2,1}$ minus the fixed point set is $(0-4\cdot 0-6\cdot 1)/6=-1$. The blow--up procedure glues in 3 $\IP^1\times T^2$s at the $\IZ_2$ fixed lines which does not change the Euler number. The last fixed line is replaced by a $\IP^1 \times T^2$ minus 3 points, upon which there is still a free $\IZ_3$ action. Its Euler number is therefore $(0-3)/3 = -1$. The 6 $\IZ_3$ fixed points fall into 3 equivalence classes, furthermore we see from Figure~\ref{fig:frthree} that there is one line ending in $D_2$. Hence, each of these classes is replaced by a $\IP^1$, and the contribution to the Euler number is $3\cdot 2=6$. Finally, for the 3 $\IC^3/\IZ_{6-I}$ fixed points there are 2 lines ending in $D_2$ in the toric diagram in Figure~\ref{fig:fsixi}. At a single fixed point, the blow--up yields two $\IP^1$s touching in one point whose Euler number is $2\cdot 2 -1=3$. Adding everything up, the Euler number of $D_{2,1}$ is $-1 + 0 -1 + 6 + 3\cdot 3 = 13$ which can be viewed as the result of a blow--up of $\IF_0$ in 9 points. The same is true for $D_{1,1}$, however, there are no $\IC^2/\IZ_2$ fixed lines without fixed points. The topology of each representative of $D_{1,2}$ minus the fixed point set, viewed as a $T^4$ orbifold, is that of a $T^2 \times (T^2/\IZ_2 \setminus \{ 4\ \rm{pts} \})$. (Note that also here the fixed point set is larger than the restriction of the fixed point set to $D_{1,2}$.) They are permuted under the residual $\IZ_3$ action and the 12 points fall into 3 orbits of length 1 and 3 orbits of length 3. Hence, the topology of the class is still that of a $T^2 \times (T^2/\IZ_2 \setminus \{ 4 \ \rm{pts} \})$. After the blow--up it is therefore a $\IP^1 \times T^2$. The topology of the divisors $D_{2,2}$ and $D_{1,3}$ is the same as the topology of $D_{i\alpha}$ in the $\IZ_3$ orbifold which is discussed in detail in Appendix~\ref{sec:Z3orb}. It can be viewed as a blow--up of $\IP^2$ in 12 points. Finally, there are the divisors $D_{3\gamma}$. The action $\frac{1}{6}(1,1)$ on $T^4$ has 24 fixed points, 1 of order 6, 15 of order 2, and 8 of order 3. The $\IZ_2$ fixed points fall into 5 orbits of length 3 under the order three element, and the $\IZ_3$ fixed points fall into 4 orbits of length 2 under the order two element. For each type of fixed point there is a single line ending in $D_3$ in the corresponding toric diagram, therefore the fixed points are all replaced by a $\IP^1$. The Euler number therefore is $(0-24)/6 + (1+5+4)\cdot 2 = 16$. Hence, the $D_{3,\gamma}$ can be viewed as blow--ups of $\IF_0$ in 12 points. Note again, that the $\IZ_2$ fixed points do not belong to the restriction of the fixed point set of $T^6/\IZ_{6-I}$ to $D_{3,\gamma}$. 

The divisors $R_1$ and $R_2$ do not intersect any fixed lines lines, therefore they simply have the topology of $T^4$. The divisor $R_3$ has the topology of a K3. In Table~\ref{tab:TopZ6I} we have summarized the topology of all the divisors. All the Euler numbers and types of surfaces we have determined above together with~(\ref{eq:ringZ6I}) agree with Noethers formula~(\ref{eq:S3}). 
\begin{table}[h!]
  \begin{center}
  $
  \begin{array}{c}
    \begin{array}{|c|c|c|c|c|}
      \hline
      E_{1\gamma} & E_{2,1\gamma} & E_{2\mu\gamma} & E_{3,1} & E_{3,2}         \\  
      \hline
      \IF_4       & \IF_2         & \IP^2          & \IF_0   & \IP^1\times T^2 \\
      \hline
    \end{array}
    \\
    \\
    \begin{array}{|c|c|c|c|c|c|c|c|c|}
      \hline
      D_{1,1}     & D_{1,2}         & D_{1,3}      & D_{2,1}     & D_{2,2}      & D_{3,\gamma} & R_1, R_2 & R_3 \\  
      \hline
      \Bl{9}\IF_n & \IP^1\times T^2 & \Bl{12}\IP^2 & \Bl{9}\IF_n & \Bl{12}\IP^2 & \Bl{12}\IF_n & T^4      & \rm{K3} \\
      \hline
    \end{array}
  \end{array}
  $
  \end{center}
  \caption{Divisor topologies for $\IZ_{6-I}$ on $G_2\times SU(3)^2$}
  \label{tab:TopZ6I}
\end{table}
With the knowledge of the Euler numbers and the intersection ring we can determine the second Chern class $\ch_2$ on the basis $\{R_i,E_{k\alpha\beta\gamma}\}$ by~(\ref{eq:c2.S}):
\begin{align}
  \label{eq:c2Z6I}
  \ch_2\cdot E_{1,\gamma} &= -4, & \ch_2\cdot E_{2,1,\gamma} &= -4, & \ch_2 \cdot E_{2,\mu,\gamma} &= -6, & \ch_2\cdot E_{3,1} & = -4,\notag\\
  \ch_2\cdot E_{3,2} &= 0, & \ch_2 \cdot R_i &= 0, & \ch_2 \cdot R_3 = & 24. 
\end{align}
Since the second Chern class is a linear form on $H^2(X,\IZ)$ we can apply it to each of the linear relations in~(\ref{eq:Z6IrelD11}) to~(\ref{eq:Z6Irels3}) and again find complete agreement.


\subsubsection{The $\IZ_{6-I}$ orbifold on $G_2^2\times SU(3)$}
\label{sec:Z6I_G2xG2xSU3}

\begin{figure}[h!]
  \begin{center}
  \includegraphics[width=85mm]{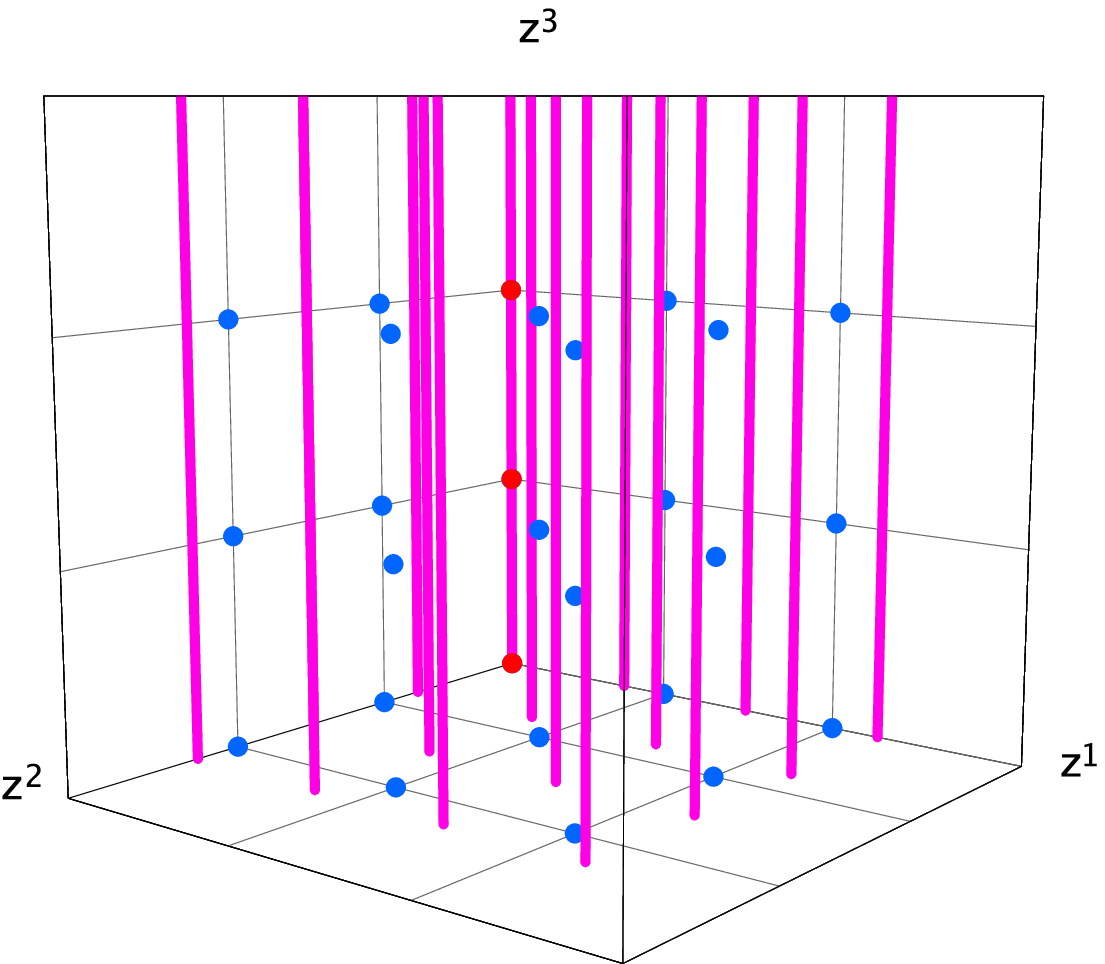}
  \caption{Schematic picture of the fixed set configuration of $\IZ_{6-I}$ on $G_2^2\times SU(3)$}
  \label{fig:ffixedi}
  \end{center}
\end{figure}

Here, the analysis of the fixed point set is very similar to the previous example and we will only point out the differences. The action of the twist $\theta$ on the lattice $G_2^2\times SU(3)$ was given in (A.2) and the resulting complex structure in (A.6) of~\cite{Lust:2005dy}. Unlike the previous example the complex structure obtained from this lattice factorizes into three $T^2$:
\begin{figure}[h!]
  \begin{center}
  \includegraphics[width=140mm]{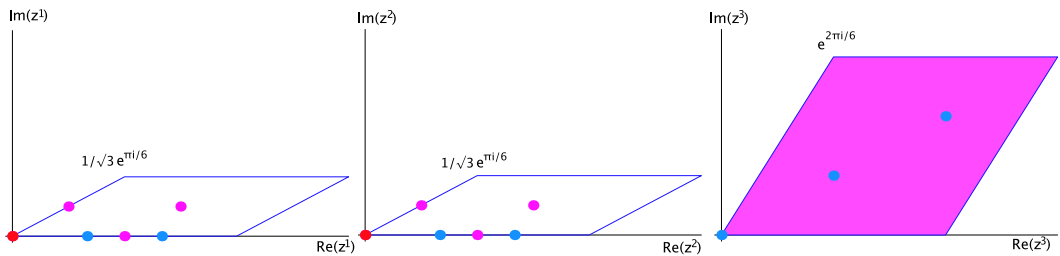}
  \caption{Fundamental regions for the $\IZ_{6-I}$ orbifold}
  \label{fig:fzsixif}
  \end{center}
\end{figure}
Figure~\ref{fig:fzsixif} shows the fundamental regions of the three $T^2$ corresponding to $z^1,\,z^2,\,z^3$ and their fixed points in the different sectors. In addition to the $\IZ_2$ fixed points for the lattice $G_2\times SU(3)^2$ in the $z^1$--direction, there are also fixed points $\zf{2}{\beta} = 0,\half,\half \tau,\half(1+\tau)$, $\beta=1,2,4,6$ in the $z^2$--direction. The corresponding 16 fixed lines $(\zf{1}{\alpha},\zf{2}{\beta},z^3)$ fall into six conjugacy classes under the action of $\theta^2$:
\begin{align}
  \label{eq:ztwoconj}
  \nu=1:\; &(0,0, z^3) \cr
  \nu=2:\; &(\tfrac{1}{2},0,z^3),\ (\tfrac{1}{2}(1+\tau),0,z^3),\ (\tfrac{1}{2}\tau,0,z^3)\cr
  \nu=3:\; &(0,\tfrac{1}{2},z^3),\ (0,\tfrac{1}{2}(1+\tau),z^3),\ (0,\tfrac{1}{2}\tau,z^3) \cr
  \nu=4:\; &(\tfrac{1}{2},\tfrac{1}{2},z^3),(\tfrac{1}{2}(1+\tau),\tfrac{1}{2}(1+\tau),z^3),\ (\tfrac{1}{2}\tau,\tfrac{1}{2}\tau,z^3)\cr
  \nu=5:\; &(\tfrac{1}{2}, \tfrac{1}{2}(1+\tau),z^3),\ (\tfrac{1}{2}(1+\tau),\tfrac{1}{2}\tau,z^3),\ (\tfrac{1}{2}\tau,\tfrac{1}{2},z^3) \cr
  \nu=6:\; &(\tfrac{1}{2}, \tfrac{1}{2}\tau,z^3),\ (\tfrac{1}{2}(1+\tau), \tfrac{1}{2},z^3),\ (\tfrac{1}{2}\tau, \tfrac{1}{2}(1+\tau),z^3).
\end{align}
Table~\ref{tab:fssixi} summarizes the relevant data of the fixed point set. The invariant subtorus under $\theta^3$ is $(0,0,0,0,x^5,x^6)$ which corresponds simply to $z^3$ being invariant.
\begin{table}[h!]
  \begin{center}
    \begin{tabular}{|c|c|c|c|}
    \hline
    Group el.& Order &Fixed Set&Conj. Classes \cr
    \noalign{\hrule}\noalign{\hrule}
    $ \theta$& 6&3 fixed points & 3\cr
    $\theta^2$&3 &27 fixed points &15\cr
    $ \theta^3$&2 &16 fixed  lines &\ 6\cr
    \hline
    \end{tabular}
  \end{center}
  \caption{Fixed point set for $\IZ_{6-I}$ on $G_2^2\times SU(3)$}
  \label{tab:fssixi}
\end{table}
Figure \ref{fig:ffixedi} shows the configuration of the fixed sets in a schematic way.  
Since the fixed points sets of the order 6 and order 3 elements are the same as for the lattice $G_2 \times SU(3)^2$, we have the same exceptional divisors $E_{1,\gamma}$ and $E_{2,\mu,\gamma}$, $\gamma=1,2,3$, $\mu=1,\dots,5$ as in the previous section. In addition, we have 6 conjugacy classes of $\IC^2/\IZ_2$ fixed lines instead of only two. Again, on the fixed line at $\zf{1}{1}=\zf{2}{1}=0$ sit the three $\IC^3/\IZ_{6-I}$ fixed points. The divisor coming from the blow--up of this fixed line, $E_{3,1}$, is identified with the three exceptional divisors corresponding to the points on the boundary of the toric diagram of the resolution of $\IC^3/\IZ_{6-I}$ that we mentioned above. The other exceptional divisors are built as invariant combinations according to the conjugacy classes in~(\ref{eq:ztwoconj}):
\begin{align}
  \label{eq:E3conj}
  E_{3,1} &= \Et_{3,1,1},                             & E_{3,2} &= \Et_{3,1,2} + \Et_{3,1,4} + \Et_{3,1,6}, \notag\\
  E_{3,3} &= \Et_{3,2,1} + \Et_{3,4,1} + \Et_{3,6,1}, & E_{3,4} &= \Et_{3,2,2} + \Et_{3,4,4} + \Et_{3,6,6}, \notag\\
  E_{3,5} &= \Et_{3,2,4} + \Et_{3,4,6} + \Et_{3,6,2}, & E_{3,6} &= \Et_{3,2,6} + \Et_{3,4,2} + \Et_{3,6,4}.
\end{align}
where $\Et_{3,\alpha,\beta}$ are the exceptional divisors on the cover.

This gives a total of $3\cdot 2+12\cdot 1+6\cdot 1=24$ exceptional divisors which agrees with $h_{\rm twist.}^{1,1}$ in Table~\ref{tab:ZN}. There are five classes of $\IC^3/\IZ_2$ fixed lines without fixed points on it, therefore, by~(\ref{eq:h21tw}), $h^{2,1}_{\rm twist.}=5$.

The divisors $D_{1,\alpha}$ and $D_{3,\gamma}$ are the same as those for the lattice $G_2 \times SU(3)^2$, only those in the in the $z^2$--direction are different:
\begin{align}
  D_{2,1} &= \Dt_{2,1} & D_{2,2} &= \Dt_{2,2} + \Dt_{2,4} + \Dt_{2,6} & D_{2,3} &= \Dt_{2,3} + \Dt_{2,5}.    
\end{align}
With this, the linear relations change as follows: The relations~(\ref{eq:Z6IrelD11}) and~(\ref{eq:Z6IrelD12}) for $D_{1,\alpha}$ become
\begin{eqnarray}
  \label{eq:Z6IrelD11a} 
  R_1&=&6\,D_{{1,1}}+\sum_{\gamma=1}^3E_{{1,\gamma}} +2\, \sum_{\mu=1}^2 \sum_{\gamma=1}^3 E_{{2,\mu,\gamma}}+3\, \sum_{\nu=1,2} E_{{3,\nu}}, \nonumber\\
  R_1&=&2D_{{1,2}}+\sum_{\nu=3}^6 E_{{3,\nu}},
\end{eqnarray}
while~(\ref{eq:Z6Irel13}) remains unchanged. The linear relations for $D_{2,\beta}$ are the same as those for $D_{1,\alpha}$: 
\begin{eqnarray}
  \label{eq:Z6Irels2a}
  R_2&=&6\,D_{{2,1}}+\sum_{\gamma=1}^3 E_{{1,\gamma}}+2\, \sum_{\mu=1,3} \sum_{\gamma=1}^3 E_{{2,\mu,\gamma}}+3\, \sum_{\nu=1,3} E_{{3,\nu}},\nonumber\\
  R_2&=&2D_{{2,2}}+\, \sum_{\nu=2,4,5,6} E_{{3,\nu}}, \nonumber \\
  R_2&=&3\,D_{{2,3}}+\sum_{\mu=2,4,5} \sum_{\gamma=1}^3 E_{{2,\mu,\gamma}}.
\end{eqnarray}
Finally, the relations for $D_{3,\gamma}$ are the same as in~(\ref{eq:Z6Irels3}). The polyhedron $\Delta^{(3)}$ is the same as in~(\ref{eq:Z6Ipoly}), so are the polyhedra for the $\IC^3/\IZ_3$ patches and the two $\IC^2/\IZ_2$ fixed lines with $\nu=1,2$ in Section~\ref{sec:Z6I_G2xSU3xSU3}. We only need the polyhedra for the additional $\IC^2/\IZ_2$ fixed lines. We drop again $v_7$ and $v_8$, and for the fixed line at $\nu=3$ we set $v_5=(0,6,0), v_9=(3,9,0)$, while for those at $\nu=4,5,6$ we set $v_4=(6,0,0), v_5=(0,6,0), v_9=(9,9,0)$. Performing the same steps as before we obtain the intersection ring of $X$ in the basis $\{R_i, E_{k\alpha\beta\gamma}\}$ (without the divisors $R_{i\jbar}$):
\begin{align}
  \label{eq:ringZ6Ia}
  R_1R_2R_3 &= 18, & R_3E_{3,1}^2 &= -2, & R_3E_{3,\nu}^2 &= -6, & E_{1,\gamma}^3 &= 8, \notag\\
  E_{1,\gamma}^2 E_{2,1,\gamma} &= 2, & E_{1,\gamma}E_{2,1,\gamma}^2 &= -4, & E_{2,1\gamma}^3 &= 8, & E_{2,\mu,\gamma}^3 &= 9, \notag\\
  E_{2,1,\gamma}E_{3,1}^2 &= -2, & E_{3,1}^3 &= 8,
\end{align}
for $\mu=2,\dots,5$, $\nu=2,\dots,6$, $\gamma=1,2,3$. 

For the topology of the divisors there are only a few changes with respect to the lattice $SU(3)^2\times G_2$. First of all, all divisors that were present in the resolved torus orbifold based on that lattice have the same topology here, with the underlying lattice $SU(3)\times G_2^2$. Then, there are new divisors $E_{3,\nu}$, $\nu=2,\dots,6$, instead of $E_{3,2}$. These are all of type~E\ref{item:E2}) with 3 representatives, hence their topology is that of $\IP^1\times T^2$. The divisor $D_{2,2}$ of the previous section is denoted $D_{2,3}$ here, and its topology is that of $\Bl{12}\IP^2$. The remaining new divisor is $D_{2,2}$ which has the same structure as $D_{1,2}$, therefore its topology is that of a $\IP^1\times T^2$. We want to point out that unlike in the case of the lattice $SU(3)^2\times G_2$, here all divisor classes of the divisors $D_{i\alpha}$ viewed as resolved $T^4$ orbifolds are realized as divisor classes in the resolved $T^6$ orbifold. This is due to the fact that the complex structure for the lattice $G_2^2\times SU(3)$ factorizes into the complex structure of three $T^2$, as we observed at the beginning of this subsection. For completeness, we display the second Chern classes in the basis $\{R_i, E_{k\alpha\beta\gamma}\}$:
\begin{align}
  \label{eq:c2Z6Ib}
  \ch_2\cdot E_{1,\gamma} &= -4, & \ch_2\cdot E_{2,1,\gamma} &= -4, & \ch_2 \cdot E_{2,\mu,\gamma} &= -6, & \ch_2\cdot E_{3,1} & = -4,\notag\\
  \ch_2\cdot E_{3,\nu} &= 0, & \ch_2 \cdot R_i &= 0, & \ch_2 \cdot R_3 = & 24. 
\end{align}


\subsubsection{The $\IZ_2 \times \IZ_2$ orbifold with one shift}
\label{sec:Z2xZ2one}

We add to the twist $\theta^2$ in~(\ref{eq;twisttwotwo}) a shift by half a lattice vector in the third coordinate, and consequently also to $\theta^a\theta^b$:
\begin{eqnarray}
  \label{eq;twistZ2xZ2one}
  \theta^1:\ (z^1,\, z^2,\, z^3)& \to& ( -z^1,  z^2,  -z^3),\cr
  \theta^2:\ (z^1,\, z^2,\, z^3)& \to& (  z^1, -z^2,  -z^3+\tfrac{1}{2}),\cr
  \theta^1\theta^2:\ (z^1,\, z^2,\, z^3)& \to& (-z^1,  -z^2,  z^3-\tfrac{1}{2}),\cr
  \theta^2\theta^1:\ (z^1,\, z^2,\, z^3)& \to& (-z^1,  -z^2,  z^3+\tfrac{1}{2}).
\end{eqnarray}
\begin{figure}[h!]
  \begin{center}
  \includegraphics[width=140mm]{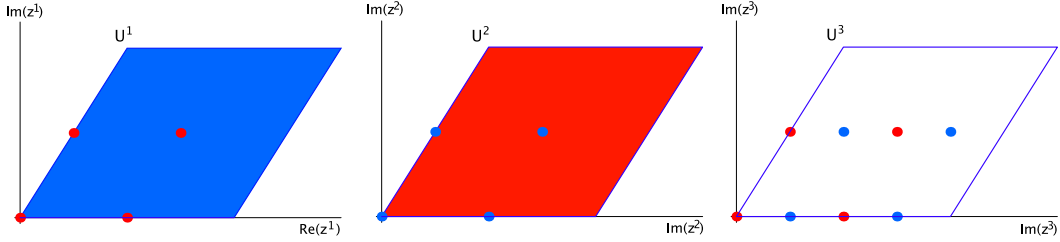}
  \caption{Fundamental regions for the $\IZ_2 \times \IZ_2$ orbifold with a shift in $z_3$}
  \label{fig:fztwotwoi}
  \end{center}
\end{figure}
Figure~\ref{fig:fztwotwoi} shows the fundamental regions of the three tori corresponding to $z^1,\,z^2,\,z^3$ and their fixed points in the different sectors. The twist $\theta^1$ has 16 fixed tori at $\zf{1}{\alpha}=0,1/2,\tau/2,(1+\tau)/2$, and $\zf{3}{\gamma}=0,1/2,\tau/2,(1+\tau)/2$. $\theta^2$ also has 16 fixed tori at $\zf{2}{\beta}=0,1/2,\tau/2,(1+\tau)/2$, $\zf{3}{\gamma}=1/4, 3/4,1/4+\tau/2,3/4+\tau/2$. $\theta^1\theta^2$ and $\theta^2\theta^1$ have no fixed points.

The equivalence classes of the fixed point set are described as follows: The fixed lines under $\theta^1$ fall into orbits of length two under $\theta^1\theta^2$: $\zf{3}{\gamma}\in\{0,1/2\}$ and $\zf{3}{\gamma}\in\{\tau/2, 1/2+\tau/2\}$. The fixed lines under $\theta^2$ fall into orbits of length two under $\theta^1$: $\zf{3}{\gamma}\in\{1/4,3/4\}$, and $\zf{3}{\gamma}\in\{1/4+\tau/2,3/4+\tau/2\}$.
\begin{table}[h!]
  \begin{center}
  \begin{tabular}{|c|c|c|c|}
    \hline
    Group el.& Order & Fixed Set& Conj. Classes \cr
    \hline 
    \noalign{\hrule}
    $\ \theta^1$ & 2 & 16\ {\rm fixed\ lines} &\  8\cr
    $\ \theta^2$ & 2 & 16\ {\rm fixed\ lines} &\  8\cr
    $\ \theta^1\theta^2$ & 2 &  -- &\  --\cr
    $\ \theta^2\theta^1$ & 2 &  -- &\  --\cr
    \hline
  \end{tabular}
  \caption{Fixed point sets for the $\IZ_2 \times \IZ_2$ orbifold with one shift.}
  \label{tab:fstwotwoi}
\end{center} 
\end{table}
Table~\ref{tab:fstwotwoi} summarizes the relevant data of the fixed sets. 
\begin{figure}[h!]
\begin{center}
\includegraphics[width=85mm]{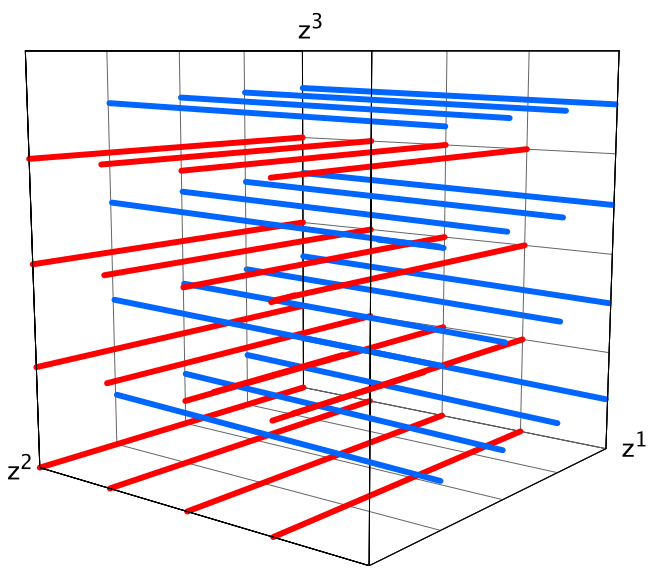}
\caption{Schematic picture of the fixed point set of the $\IZ_2 \times \IZ_2$ orbifold with one shift.}
\label{fig:ffixtwotwoi}
\end{center}
\end{figure}
Figure~\ref{fig:ffixtwotwoi} shows the configuration of the fixed point set in a schematic way.

Next, we resolve the singularities and determine all the divisors of the blown--up torus orbifold including their topologies. From Table~\ref{tab:fstwotwoi} and Figure~\ref{fig:ffixtwotwoi}, we see that there are no fixed points and all fixed lines are lines of $\IC^2/\IZ_2$ singularities. According to Appendix~\ref{sec:C2Zn} the resolution of these singularities contributes exceptional divisors $E_{1,\alpha,2\gamma-1}$, $E_{2,\beta,2\gamma}$, $\alpha,\beta=1,\dots,4$, and $\gamma=1,2$, where we have already formed the invariant combinations of the divisors on the cover. This results in 16 exceptional divisors. Together with the 3 inherited divisors $R_i$ we find $h^{1,1}=19$. Since none of these fixed lines has fixed points on it, we have $h^{2,1}_{\rm twist.}=16$, and therefore $h^{2,1}=19$.

From the local linear relations~(\ref{eq:lineqsctwo}), we find the following global linear relations:
\begin{align}
  \label{eq:Reqtwotwoi}
  R_{1} &\sim 2\,D_{1,\alpha}+\sum_{\gamma=1}^2 E_{1,\alpha,2\gamma-1}, &
  R_{2} &\sim 2\,D_{2, \beta}+\sum_{\gamma=1}^2 E_{2, \beta,2\gamma},\notag\\
  R_{3} &\sim 2\,D_{3,\gamma}+\sum_{\alpha=1}^4 E_{1,\alpha,2\gamma-1}, &
  R_{3} &\sim 2\,D_{3,\gamma}+\sum_{ \beta=1}^4 E_{2, \beta,2\gamma},
\end{align}
where $\alpha,\beta=1,\dots,4$, $\gamma=1,2$.
To compute the intersection ring, we need to determine the basis for the lattice $N$ in which the auxiliary polyhedron will live. From~(\ref{eq:Reqtwotwoi}) we see that $n_i=2$, $i=1,2,3$. Hence we can choose $m_1=m_2=2$, $m_3=1$, and the lattice basis is $f_1=(2,0,0)$, $f_2=(0,2,0)$, $f_3=(0,0,1)$. The lattice points of the polyhedron $\Delta^{(3)}$ for the local compactification of the $\IZ_2$ fixed lines in the $z^1$ direction are
\begin{align}
  \label{eq:Z2poly1}
  v_1 &= (-2,0,0), & v_2 &= (0,-2,0), & v_3 &= (0,0,-1), & v_4 &= (4,0,0), \notag\\
  v_5 &= (0,4,0), & v_6 &= (0,0,2) & v_7 &= (0,2,1),
\end{align}
while those for the $\IZ_2$ fixed lines in the $z^2$ direction are
\begin{align}
  \label{eq:Z2poly2}
  v_1 &= (-2,0,0), & v_2 &= (0,-2,0), & v_3 &= (0,0,-1), & v_4 &= (4,0,0), \notag\\
  v_5 &= (0,4,0), & v_6 &= (0,0,2) & v_7 &= (2,0,1),
\end{align}
both corresponding to the divisors $R_1,R_2,R_3,D_1,D_2,D_3,E_1$ in that order. 
From the intersection ring of these polyhedra and the linear relations~(\ref{eq:Reqtwotwoi}) we obtain the following nonvanishing intersection numbers of $X$ in the basis $\{R_i,E_{k\alpha\beta\gamma}\}$. We have thereby to take into account that the shift by half a lattice vector in~(\ref{eq;twistZ2xZ2one}) entails volume reduction by a factor of 2:
\begin{align}
  R_1R_2R_3 &=1, & R_1E_{2,\beta,2\gamma}^2 &=-1, & R_2E_{1,\alpha,2\gamma-1}^2&=-1.
\end{align}
The second Chern class is
\begin{align}
  \ch_2\cdot E_{1,\alpha,2\gamma-1} &= 0, & \ch_2 \cdot E_{2,\beta,2\gamma} &= 0, & \ch_2 \cdot R_i &= 12, & \ch_2 \cdot R_3 &= 0.
\end{align}


\subsubsection{The $\IZ_2 \times \IZ_2$ orbifold with two shifts}
\label{sec:Z2xZ2two}

We now also add to the twist $\theta^1$ in~(\ref{eq;twisttwotwo}) a shift by half a lattice vector in the second coordinate:
\begin{eqnarray}
  \label{eq;twistZ2xZ2two}
  \theta^1:\ (z^1,\, z^2,\, z^3)& \to& (\varepsilon\, z^1, \, z^2+\tfrac{1}{2}, \varepsilon\, z^3),\cr
  \theta^2:\ (z^1,\, z^2,\, z^3)& \to& (\, z^1, \, \varepsilon\,z^2, \varepsilon\, z^3+\tfrac{1}{2}),\cr
  \theta^1\theta^2:\ (z^1,\, z^2,\, z^3)& \to& (\varepsilon\, z^1, \, \varepsilon\,z^2+\tfrac{1}{2},  z^3-\tfrac{1}{2}),\cr
  \theta^2\theta^1:\ (z^1,\, z^2,\, z^3)& \to& (\varepsilon\, z^1, \, \varepsilon\,z^2+\tfrac{1}{2},  z^3-\tfrac{1}{2}),
\end{eqnarray}
\begin{figure}[h!]
  \begin{center}
  \includegraphics[width=140mm]{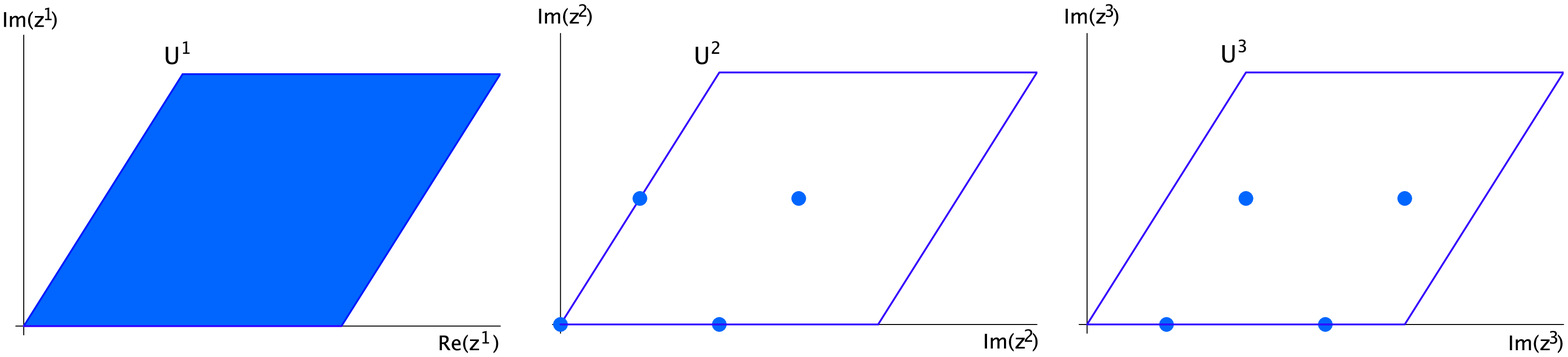}
  \caption{Fundamental regions for the $\IZ_2 \times \IZ_2$ orbifold with two shifts}
  \label{fig:fztwotwoii}
  \end{center}
\end{figure}
Figure~\ref{fig:fztwotwoii} shows the fundamental regions of the three tori corresponding to $z^1,\,z^2,\,z^3$ and their fixed points in the different sectors. The twists $\theta^1$ and $\theta^1\theta^2$ have no fixed points. $\theta^2$ has 16 fixed tori at $\zf{1}{\alpha}=0,1/2,\tau/2,(1+\tau)/2$, $\zf{3}{\gamma}=1/4, 3/4,1/4+\tau/2,3/4+\tau/2$. 

The fixed lines under $\theta^2$ fall again into orbits of length two under $\theta^1$: $\zf{3}{\gamma}\in\{1/4,3/4\}$, and $\zf{3}{\gamma}\in\{1/4+\tau/2,3/4+\tau/2\}$.
\begin{figure}[h!]
\begin{center}
\includegraphics[width=85mm]{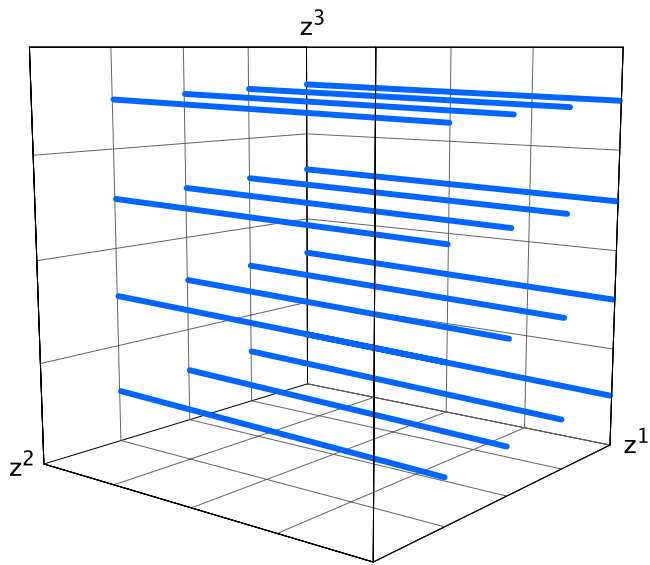}
\caption{Schematic picture of the fixed point set of the $\IZ_2 \times \IZ_2$ orbifold with two shifts.}
\label{fig:ffixtwotwoii}
\end{center}
\end{figure}
Figure~\ref{fig:ffixtwotwoii} shows the configuration of the fixed point set in a schematic way. 

The resolution of the lines of the lines of $\IC^2/\IZ_2$ singularities along the $z^1$ direction is the same in the previous subsection, but we only have 8 exceptional divisors $E_{1,\beta,\gamma}$, $\beta=1,\dots,4$ and $\gamma=1,2$. Together with the 3 inherited divisors $R_i$ we find $h^{1,1}=11$. Since none of these fixed lines has fixed points on it, we have $h^{2,1}_{\rm twist.}=8$, and therefore $h^{2,1}=11$. The intersection ring and the second Chern class reduce to:
\begin{align}
  R_1R_2R_3 &=1, & R_1E_{1,\beta,\gamma}^2&=-1, \notag\\
  \ch_2\cdot E_{1,\beta,\gamma} &= 0, & \ch_2 \cdot R_1 &= 12, & \ch_2 \cdot R_i &= 0.
\end{align}


\section{The orientifolds}
\label{sec:orientifolds}

Now that we have glued the local patches of our orbifolds together, we want to perform the orientifold projection.  This will tell us how many O--planes our model contains and how many D--branes we must introduce to balance the resulting charges.

\subsection{The orientifold actions}
\label{sec:Actions}

At the orbifold point, the orientifold projection is $\Omega\,I_6$, where $\Omega$ is the worldsheet orientation reversal and $I_6$ is an involution on the compactification manifold. In type IIB string theory with O3/O7--planes (instead of O5/O9), the holomorphic (3,0)--form $\Omega^{3,0}$ must transform as $\Omega^{3,0} \to -\Omega^{3,0}$. Therefore we choose
\begin{equation}
  \label{eq:I6}
  I_6:\ (z^1,z^2,z^3)\to (-z^1,-z^2,-z^3).
\end{equation}
Geometrically, this involution corresponds to taking a $\IZ_2$-quotient of the compactification manifold.

As long as we are at the orbifold point, all necessary information is encoded in (\ref{eq:I6}). To find the configuration of O3--planes, the fixed points under $I_6$ must be identified. On the covering space, $I_6$ always gives rise to 64 fixed points, i.e. 64 O3--planes. Some of them may be identified under the orbifold group $G$, such that there are less than 64 equivalence classes on the quotient. The O7--planes are found by identifying the fixed planes under the combined action of $I_6$ and the generators $\theta_{\IZ_2}$ of the $\IZ_2$ subgroups of $G$.
A point $x$ belongs to a fixed set, if it fulfills
\begin{equation}
  \label{eq:fixO}
  I_6\,\theta_{\IZ_2}\,x=x+a,\quad a\in \Lambda,
\end{equation}
where $\Lambda$ is the torus lattice.
Each $\IZ_2$ subgroup of $G$ gives rise to a stack of O7--planes. Therefore, there are none in the prime cases, one stack e.g. for $\IZ_{6-I}$ and three in the case of e.g. $\IZ_2\times \IZ_6$, which contains three $\IZ_2$ subgroups. The number of O7--planes per stack depends on the fixed points in the direction perpendicular to the O--plane and therefore on the particulars of the specific torus lattice.

Whenever $G$ contains a subgroup $H$ of odd order, some of the fixed point sets of $H$ will not be invariant under the global orientifold involution $I_6$ and will fall into orbits of length two under $I_6$. This can also happen for groups of even order giving rise to fixed tori with non--trivial volume factors, see e.g.~\cite{Ibanez:1987pj},~\cite{Erler:1992ki}. Some of these $I_6$--orbits may coincide with the $G$--orbits. In this case, no further effect arises. When $G$ contains in particular a $\IZ_2$ subgroup in each coordinate direction, all equivalence classes under $I_6$ and these subgroups coincide.
When certain fixed points or lines (which do not already form an orbit under $G$) are identified under the orientifold quotient, the second cohomology splits into an invariant and an anti--invariant part under $I_6$~\cite{Brunner:2003zm}:
$$H^{1,1}(X)=H^{1,1}_{+}(X)\oplus H^{1,1}_{-}(X).$$
The number of geometric moduli is effectively reduced by the orientifold quotient. The moduli associated to the exceptional divisors of the anti--invariant part are no longer geometric, but take the form~\cite{Grimm:2004uq},~\cite{Grimm:2005fa}
\begin{equation}
  \label{eq:newmoduli}
  G^a=C^a_2+S\,B^a_2.
\end{equation}
The only contributions to $H^{1,1}_{-}$ come from the twisted Kaehler moduli in $H^{1,1}_{\rm twist.}$. Table~\ref{tab:hminusone} gives the values of $h^{1,1}_{-}$ for all orbifolds given in Table~\ref{tab:ZN}. For a detailed discussion of these moduli see~\cite{pheno}.
Similarly, the third cohomology splits into an invariant and an anti--invariant part under $I_6$:
$$H^{2,1}(X)=H^{2,1}_{+}(X)\oplus H^{2,1}_{-}(X).$$
Since the action of $I_6$ lifts to $H^3$ such that $\Omega$ changes sign (in order to yield O3-- and O7--planes), the geometric moduli are now in $H^{2,1}_{-}$. In fact, there are no moduli in $H^{2,1}_{+}$. Again, the only contributions come from the twisted complex structure moduli in $H^{2,1}_{\rm twist.}$. In Table~\ref{tab:hminusone} we also indicate the values of $h^{2,1}_{+}$.
\begin{table}[h!]
\begin{center}
\begin{center}
{\small
\begin{tabular}{|l|c|c|c|c|c|c|c|}
\hline
$\ G$&Lattice &$h^{1,1}$&$h^{1,1}_{-}$&$h^{2,1}$&$h^{2,1}_{+}$ \\ [2pt]
\hline
$\  \IZ_3 $               &$SU(3)^3 $                                 &36 &13 & 0 & 0\cr
$\  \IZ_4  $              &$SU(4)^2  $                                &25 &6  & 1 & 0\cr
$\  \IZ_4 $               &$SU(2)\times SU(4)\times SO(5)$            &27 &4  & 3 & 2\cr
$\  \IZ_4$                &$SU(2)^2\times SO(5)^2 $                   &31 &0  & 7 & 6\cr
$\  \IZ_{6-I} $            & $(G_2\times SU(3)^{2})^{\flat}$            &25 &6  & 1 & 1\cr
$\  \IZ_{6-I}  $           &$ SU(3)\times G_2^2$                       &29 &6  & 5 & 5\cr
$\  \IZ_{6-II} $           &$ SU(2)\times SU(6)$                       &25 &6 & 1 & 0\cr
$\  \IZ_{6-II} $           &$SU(3)\times SO(8) $                       &29 &6 & 5 & 4\cr
$\  \IZ_{6-II} $           &$(SU(2)^2\times SU(3)\times SU(3))^{\sharp}$ &31&8 & 7 & 3\cr
$\  \IZ_{6-II}  $          &$ SU(2)^2\times SU(3)\times G_2$           &35 &8 & 11& 7\cr
$\  \IZ_7  $               &$ SU(7) $                                 &24 &9 & 0 & 0\cr
$\  \IZ_{8-I}  $           &$ (SU(4)\times SU(4))^*$                   &24 &5 & 0 & 0\cr
$\  \IZ_{8-I}  $           &$SO(5)\times SO(9)    $                    &27 &0 & 3 & 3\cr
$\  \IZ_{8-II} $           &$ SU(2)\times SO(10)  $                    &27 &4 & 3 & 2\cr
$\  \IZ_{8-II}  $          &$ SO(4)\times SO(9)$                       &31 &0 & 7 & 6\cr
$\  \IZ_{12-I}$            &$ E_6 $                                    &25 &6 & 1 & 1\cr
$\  \IZ_{12-I} $           &$ SU(3)\times F_4$                         &29 &6 & 5 & 5\cr
$\  \IZ_{12-II} $           &$SO(4)\times F_4$                         &31 &0 & 7 & 6\cr
$\  \IZ_2 \times\IZ_2$     &$SU(2)^6$                                 &51 &0 & 3 & 0\cr
$\  \IZ_2 \times\IZ_4$     &$SU(2)^2\times SO(5)^2$                   &61 &0 & 1 & 0\cr
$\  \IZ_2 \times\IZ_6 $    &$ SU(2)^2\times SU(3)\times G_2$          &51 &0 & 3 & 2\cr
$\  \IZ_2 \times\IZ_{6'}$  &$ SU(3)\times G_2^2$                       &36 &0 & 0 & 0\cr
$\  \IZ_3 \times\IZ_3  $   &$SU(3)^3$                                 &84 &37& 0 & 0\cr
$\  \IZ_3 \times\IZ_6 $    &$SU(3)\times G_2^2 $                      &73 &22& 1 & 1\cr
$\  \IZ_4 \times\IZ_4 $    &$SO(5)^3 $                                &90 & 0& 0 & 0\cr
$\  \IZ_6 \times\IZ_6  $   &$G_2^3$                                   &84 & 0& 0 & 0\cr
\hline 
\end{tabular}}
\end{center}
\caption{Twists, lattices, $h^{1,1}_{-}$ and $h^{2,1}_{+}$.}
\label{tab:hminusone}
\end{center}\end{table}

Now we want to discuss the orientifold action for the smooth Calabi--Yau manifolds $X$ resulting from the resolved torus orbifolds. For such a manifold $X$, we will denote its orientifold quotient $X/I_6$ by $B$ and the orientifold projection by $\pi:X \to B$. Away from the location of the resolved singularities, the orientifold involution retains the form~(\ref{eq:I6}). As explained above, the orbifold fixed points fall into two classes:
\renewcommand{\labelenumi}{O\theenumi)}
\begin{enumerate}
  \item The fixed point is invariant under $I_6$, i.e. its exceptional divisors are in $h^{1,1}_{+}$.
  \label{item:O1}
  \item The fixed point lies in an orbit of length two under $I_6$, i.e. is mapped to another fixed point. The invariant combinations of the corresponding exceptional divisors contribute to $h^{1,1}_{+}$, while the remaining linear combinations contribute to $h^{1,1}_{-}$. 
  \label{item:O2}
\end{enumerate}
\renewcommand{\labelenumi}{\arabic{\theenumi}}
The fixed points of class~O\ref{item:O1}) locally feel the involution: Let $\zf{}{\alpha}$ denote some fixed point. Since $\zf{}{\alpha}$ is invariant under~(\ref{eq:I6}),
\begin{equation}
  \label{eq:onei}
  (\zf{1}{\alpha}+\Delta z^1, \zf{2}{\alpha}+\Delta z^2, \zf{3}{\alpha}+\Delta z^3) \to (\zf{1}{\alpha}-\Delta z^1, \zf{2}{\alpha}-\Delta z^2, \zf{3}{\alpha}-\Delta z^3).
\end{equation}
In local coordinates centered around $\zf{}{\alpha}$, $I_6$ therefore acts as 
\begin{equation}
  \label{eq:locali}
  (z^1, z^2, z^3) \to (-z^1,-z^2, -z^3).
\end{equation}
In case~O\ref{item:O2}), the point $\zf{}{\alpha}$ is not fixed, but gets mapped to a different fixed point $\zf{}{\beta}$. So locally,
\begin{equation}\label{eq:oneii}
(\zf{1}{\alpha}+\Delta z^1, \zf{2}{\alpha}+\Delta z^2, \zf{3}{\alpha}+\Delta z^3) \to
(\zf{1}{\beta}-\Delta z^1, \zf{2}{\beta}-\Delta z^2, \zf{3}{\beta}-\Delta z^3).
\end{equation}
In the quotient, $\zf{}{\alpha}$ and $\zf{}{\beta}$ are identified, i.e. correspond the the same point.
In local coordinates centered around this point, $I_6$ therefore acts again as in (\ref{eq:locali}).

For the fixed lines, we apply the same prescription. The involution on fixed lines with fixed points on them is constrained by the involution on the fixed points.

What happens in the local patches after the singularities were resolved? 
A local involution $\Ic$ has to be defined in terms of the local coordinates, such that it agrees with the restriction of the global involution $I_6$ on $X$. Therefore, we require that $\Ic$ maps $z^i$ to $-z^i$. In addition to the three coordinates $z^i$ inherited from $\IC^3$, there are now also the new coordinates $y^k$ corresponding to the exceptional divisors $E_k$. For the choice of the action of $\Ic$ on the $y^k$ of an individual patch, there is some freedom. 

For simplicity we restrict the orientifold actions to be multiplications by $-1$ only. We do not take into account transpositions of coordinates or shifts by half a lattice vector. The latter have been considered in the context of toric Calabi--Yau hypersurfaces in~\cite{Berglund:1998va}. The allowed transpositions can be determined from the toric diagram of the local patch by requiring that the adjacencies of the diagram be preserved. We leave these cases to future work.

The only requirements ${\Ic}$ must fulfill are compatibility with the ${\IC}^*$--action of the toric variety, i.e. 
\begin{equation}
  \label{eq:Caction}
  (-z^1,-z^2,-z^3,(-1)^{\sigma_4}y^1,\dots,(-1)^{\sigma_n}y^n) = (\prod_{a=1}^r \lambda_1^{l_1^{(a)}} z^1,\dots,\prod_{a=1}^r \lambda_n^{l_n^{(a)}} y^n)
\end{equation}
where $l_i^{(a)}$ encode the linear relations~\eqref{eq:linrels} of the toric patch, and $\sigma_i\in \{0,1\}$. Moreover, we require that subsets of the set of solutions to~\eqref{eq:Caction} must not be mapped to the excluded set of the toric variety and vice versa.

The fixed point set under the combined action of ${\Ic}$ and the scaling action of the toric variety gives the configuration of O3-- and O7--planes in the local patches. Care must be taken that only these solutions which do not lie in the excluded set are taken into account. We also exclude solutions which do not lead to solutions of the right dimension, i.e. do not lead to O3/O7--planes.

On an individual patch, we can in principle choose any of the possible involutions on the local coordinates. In the global model however, the resulting solutions of the individual patches must be compatible with each other. While O7--planes on the exceptional divisors in the interior of the toric diagram are not seen by the other patches, O7--plane solutions which lie on the D--planes or on the exceptional divisors on a fixed line must be consistent with the solutions of all patches which lie in the same plane, respectively on the same fixed line. This is of course also true for different types of patches which lie in the same plane. 

It is in principle possible for examples with many interior points of the toric diagram to choose different orientifold involutions on the different patches which lead to solutions that are consistent with each other. We choose here the same involution on all patches, which for simple examples such as the $T^6/\IZ_4$ or $T^6/\IZ_6$ orbifolds is the only consistent possibility.

The solutions for the fixed sets under the combined action of ${\Ic}$ and the scaling action~\eqref{eq:Caction} give also conditions on the $\lambda_i$, in general they are set to $\pm 1$. If the corresponding toric diagram has points on its boundary, the O--plane solutions of the full patch descend to solutions on the restriction to the fixed lines that correspond to these points. For the restriction, we set the $\lambda_i$ which do not correspond to the Mori generators of the fixed line to the values of the $\lambda_i$ of the solution for the whole patch which lies on this fixed line. This agrees with the solutions of~\eqref{eq:Caction} on the toric variety corresponding to the fixed line.

A further global consistency requirement comes from the observation that the orientifold action commutes with the singularity resolution. A choice of the orientifold action on the resolved torus orbifold must therefore reproduce the orientifold action on the orbifold and yield the same fixed point set in the blow--down limit.

It turns out that it is not always possible to find an involution which reproduces the same O--plane configuration as at the orbifold point. Nevertheless we believe that the orientifold configuration in the resolved phase makes sense. It seems that in these examples the blow--up and the orientifold do not commute and that no smooth limit from the orientifold of the resolved Calabi--Yau manifold to the orientifold of the singular orbifold exists. It may also be that there is some further freedom in defining orientifold actions on orbifold CFTs which would commute with the blow--up. It would be very interesting to understand this point in more detail.

Given a consistent global orientifold action it might still happen that the model does not exist. This is the case if the tadpoles cannot be cancelled. While we will explain how to compute the tadpoles from the topological data in Section~\ref{sec:Tadpoles}, we do not consider the possibilities of their cancellation here. For a detailed discussion of the tadpole cancellation in some of our examples we refer to~\cite{pheno}. We would like to mention that tadpole cancellation conditions can generically be different in different regions of the moduli space of the Calabi--Yau orientifold. This has first been observed on the two sides of the conifold singularity of Calabi-Yau hypersurfaces in toric varieties in~\cite{Brunner:2004zd}. In~\cite{Denef:2005mm} a similiar observation was made for the torus orbifold $T^6/\IZ_2\times\IZ_2$ where the tadpole cancellation condition at large radius differed from the one at the orbifold point.

\subsection{The intersection ring}
\label{sec:Oring}

The intersection ring of the orientifold can be determined in two equivalent ways. The basis for both ways is the relation between the divisors on the Calabi--Yau manifold $X$ and the divisors on the orientifold $B$~\cite{Denef:2005mm}. The first observation is that the integral on $B$ is half the integral on $X$: 
\begin{equation}
  \label{eq:Ointegral}
  \int_B \widehat{S}_a \wedge \widehat{S}_b \wedge \widehat{S}_c = \frac{1}{2} \int_X S_a \wedge S_b \wedge S_c ,
\end{equation}
where the hat denotes the corresponding divisor on $B$. The second observation is that for a divisor $S_a$ on $X$ which is not fixed under $I_6$, we have $S_a = \pi^*\widehat{S}_a$. If, however, $S_a$ is fixed under $I_6$, we have to take $S_a=\frac{1}{2}\pi^*\widehat{S}_a$ because the volume of $S_a$ in $X$ is the same as the volume of $\widehat{S}_a$ on $B$. Applying these rules to the intersection ring obtained in Section~\ref{sec:IntersectionRing} immediately yields the intersection ring of $B$:  Triple intersection numbers between divisors which are not fixed under the orientifold involution become halved. If one of the divisors is fixed, the intersection numbers on the orientifold are the same as on the Calabi--Yau. If two (three) of the divisors are fixed, the intersection numbers on the orientifold must be multiplied by a factor of two (four).

The second way consists of applying these rules to the intersection ring of the local patches of the resolved singularities obtained in Section~\ref{sec:toric}, more precisely on the intersection ring of the auxiliary polyhedra $\Delta^{(3)}$ in Section~\ref{sec:IntersectionRing} and the global linear equivalences~(\ref{eq:Reqglobal}). This means that for each divisor which is fixed under $\Ic$, the corresponding coefficient in~(\ref{eq:Reqglobal}) is divided by 2. In the polyhedra, the distance to the origin of all those divisors which are fixed under the orientifold involution is halved. Then we solve the resulting system of equations for $\widehat{S}_a\widehat{S}_b\widehat{S}_c$ which we set up at the end of Section~\ref{sec:IntersectionRing}. Both methods give the same result.

\subsection{Global O--plane configuration and tadpole cancellation}
\label{sec:Tadpoles}

Those of the 64 O3--planes on the cover which are located away from the locations of the resolved patches resulting from the global involution descend to the orbifold of the resolved manifold. They are untouched by the process of resolving the singularities and the resulting modified local orientifold actions. 
The O3--plane solutions which coincide with orbifold fixed sets are replaced by the solutions of the corresponding resolved patch. The total number of O3--planes on the resolved orbifold quotient is obtained by counting the equivalence classes of O3--planes under the orbifold group and replacing those classes which coincide with resolved patches by the $O3$--plane solutions on these patches.
The O3--plane solutions are also reflected in the intersection ring. Take for example the solution $\{z^1=y^1=y^3=0\}$ given in (1) of (\ref{eq:Z6IIOplanes}). The corresponding intersection number is $D_1E_1E_3=\tfrac{1}{2}$, indicating the $\IZ_2$--singularity at the intersection point. Thus fractional intersection numbers indicate the presence of O3--planes. O3--planes which are located away from the fixed points and do not lie in the D--planes are reflected in the intersection numbers with the inherited divisors $R_i$, see for example $T^6/\IZ_3$ discussed in Appendix \ref{eq:Z3O}. If on the other hand the O3--planes lie in an O7--plane, their intersection numbers do not become fractional, since the effect of the orientifold involution is already captured by the O7--plane.

Since each O7--plane induces $-8$ units of D7--brane charge, a stack of 8 coincident D7--branes must be placed on top of each divisor fixed under the combination of the involution and the scaling action. 

For the D3--brane charge, the case is a bit more involved. The contribution from the O3--planes is
\begin{equation}
  \label{othreetadpole}
  Q_3(O3)=-{1\over 4}\times n_{O3},
\end{equation}
where $n_{O3}$ denotes the number of O3--planes. The D7--branes also contribute to the D3--tadpole:
\begin{equation}
  \label{dseventadpole}
  Q_3(D7)=-\frac{1}{2}\sum_a\,{n_{D7,a}\,\chi(S_a)\over 24},
\end{equation}
where $n_{D7,a}$ denotes the number of D7--branes in the stack located on the divisor $S_a$. As we have seen, the $S_a$ can be local $D$--divisors as well as exceptional divisors.
The last contribution to the D3--brane tadpole comes from the O7--planes:
\begin{equation}
  \label{oseventadpole}
  Q_3(O7)=-\frac{1}{2}\sum_a\,{\chi(S_a)\over 6}.
\end{equation}
So the total D3--brane charge that must be cancelled is
\begin{equation}
  \label{totaltaddpole}
  Q_{3,tot}=-{n_{O3}\over4}-\frac{1}{2}\sum_a\,{(n_{D7,a}+4)\,\chi(S_a)\over 24}.
\end{equation}
These are the values for the orientifold quotient, in the double cover this value must be multiplied by two.

\subsection{Examples}
\label{sec:OExamples}

\subsubsection{The $\IZ_{6-I}$ orbifold on $G_2\times SU(3)^2$}

We examine first the orbifold phase. We have 64 O3--planes from the action of $I_6$. From the $\IZ_2$--twist $I_6\,\theta^3$, we get one O7--plane at $z^3=0$. The 64 O3--planes fall into 22 conjugacy classes under the orbifold group. $(0,0,0)$ is alone in its equivalence class, while all other points are in orbits of length 3. 

Comparison with the fixed set configuration (Figure~\ref{fig:ffixedi}) shows that the O3--planes sit on top of the $\theta^3$--fixed lines. The fixed points located in the $z^3=0$ plane lie on an $O7$ plane, the others do not.

For this example we have $h^{1,1}_{-}=6$. The fixed points except for the one at $(0,0,0)$ fall into equivalence classes of length two under $I_6$. They partly coincide with the equivalence classes under the orbifold group. Two of the divisors in $H^{1,1}_{-}$ arise from the two $\IC^3/\IZ_{6-I}$--patches at $z^3\neq0$ which are being interchanged, the remaining four come from the $\IC^3/\IZ_3$ patches in these two planes. The fixed line without fixed point is invariant under the orientifold action, hence $h^{2,1}_+=1$.

We will now discuss the orientifold of the resolved orbifold. For the local involution $\Ic$ in the $\IC^3/\IZ_{6-I}$--patches, there are four possibilities which lead to solutions of the right dimensionality (i.e. O3-- and O7--planes):
\begin{eqnarray}
(1)\quad {\Ic}(z,y)&=&(-z^1,-z^2, -z^3, y^1,y^2,y^3),\cr
(2)\quad {\Ic}(z,y)&=&(-z^1,-z^2, -z^3, y^1,-y^2,-y^3),\cr
(3)\quad {\Ic}(z,y)&=&(-z^1,-z^2, -z^3, -y^1,y^2,-y^3),\cr
(4)\quad {\Ic}(z,y)&=&(-z^1,-z^2, -z^3, -y^1,-y^2,y^3).
\end{eqnarray}
They lead to the following two distinct solutions:
\begin{eqnarray}
(1),\ (3)\quad \{y^2=0\} \cup \{z^3=0\},\cr
(2),\ (4)\quad \{y^1=0\} \cup \{y^3=0\}.
\end{eqnarray}
Choosing the simplest possibility for $\Ic$, namely $(1)$, we have to solve
\begin{equation}
  \label{OZsixi}
  {(-z, y)=(\lambda_1\,z^1,\,\lambda_1\,z^2,\,\lambda_2\,z^3,\, {\lambda_3\over\lambda_2^2}\,y^1,\,{\lambda_2\over \lambda_3^2}\,y^2,\,{\lambda_3\over\lambda_1^2}\,y^3).}
\end{equation}
We find two solutions which are in the allowed set of the toric variety:
\begin{equation}{a)\quad y^2=0\ {\rm with}\ \lambda_1=\lambda_2=-1,\, \lambda_3=1.}\end{equation}
This corresponds to the whole exceptional divisor $E_2$ being fixed, therefore this gives an O7--plane. This solution does not lead to any global consistency conditions since the other patches do not see it.
\begin{equation}{b)\quad z^3=0\ {\rm with}\ \lambda_1=-1,\, \lambda_2= \lambda_3=1.}\end{equation}
This corresponds an O7--plane on the divisor $D_3$. This solution must be compatible with the solutions of the other patches which lie in the same plane.

We will now check the consistency of this solution with the restriction of the $\IC^3/\IZ_{6-I}$ patch to the $\IC^2/\IZ_2$ fixed line. The latter is described by the coordinates $z^1,\,z^2$ and $y^3$. Choosing e.g. $\lambda_2=\lambda_3=1$ in the scaling action produces the restriction. Since neither of the two coordinates appearing in the above solutions are contained in the fixed line, no restrictions on the $\lambda_i$ arise.
The equation we must solve for the fixed line is
\begin{equation}\label{OZsixitwo}{(-z^1, -z^2, y^3)=(\lambda_1\,z^1,\,\lambda_1\,z^2,\,\frac{1}{\lambda_1^2}\,y^3).}\end{equation}
It is trivially fulfilled by
$$\lambda_1=-1,$$
and therefore also does not lead to any more restrictions on the solutions for the $\IC^3/\IZ_{6-I}$ patch. The only thing left to check is the compatibility of the O--plane solutions of the other patches with the O7--planes on the $D_{3, \gamma}$. 
For this we must examine the $\IC^3/\IZ_3$--patches. The details of their resolution can be found in Appendix \ref{sec:Z3res}. 
There are two possible choices for the local involution:
\begin{eqnarray}
(1)\quad {\Ic}(z,y)&=&(-z^1,-z^2, -z^3, y),\cr
(2)\quad {\Ic}(z,y)&=&(-z^1,-z^2, -z^3, -y).
\end{eqnarray}
$(1)$ leads to the equation (see~(\ref{rescalesthree}))
\begin{equation}\label{OZthreea}{(-z^1,-z^2,-z^3, y)=(\lambda\,z^1,\,\lambda\,z^2,\,\lambda\,z^3,\, {1\over\lambda^3}\,y).}\end{equation}
with the solution 
$$y=0,\quad \lambda=-1.$$
$(2)$ leads to the equation
\begin{equation}\label{OZthree}{(-z^1,-z^2,-z^3, -y)=(\lambda\,z^1,\,\lambda\,z^2,\,\lambda\,z^3,\, {1\over\lambda^3}\,y).}\end{equation}
For this choice of involution, (\ref{OZthree}) is trivially fulfilled by
$$\lambda=-1,$$
without any restriction on the coordinates. 
Since both solutions do not lead to any further restriction of $z^3$, they are also consistent with $z^3=0$, i.e. an O7--plane on $D_3$.

For the fixed lines without fixed points on them, we simply solve
\begin{equation}\label{OZtwo}{(-z^1, -z^2,-z^3, y)=(\lambda\,z^1,\,\lambda\,z^2,\,z^3,\, {1\over\lambda^2}\,y).}\end{equation}
This gives one allowed solution, $z^3=0,\ \lambda=-1$, corresponding to an O7--plane at the locations of the fixed points in $z^3$ direction.

In total, there are three O7--planes on the $D_{3,\gamma}$--planes and three O7--planes on the $E_{2, \gamma}$ divisors. In comparison with the O--plane configuration at the orbifold point, we see that there is no continuous limit since the two O7--planes on $D_{3,2}$ and $D_{3,3}$ do not appear at the orbifold point.
Before the blow--up, we had 22 equivalence classes of O3--planes. The one at $(0,0,0)$ coincides with a fixed point and is not present after the blow--up, since no O3--brane solutions appear in the resolved patch. The others lie away from the local patches. Therefore we are only left with 21 O3--planes.

The modified intersection numbers are (cf. (\ref{eq:ringZ6I}))
\begin{align}
  \label{eq:ringZ6IO}
  R_1R_2R_3 &= 9, & R_3E_{3,1}^2 &= -1, & R_3E_{3,2}^2 &= -3, & E_{1,\gamma}^3 &= 4, \notag\\
  E_{1,\gamma}^2 E_{2,1,\gamma} &= 2, & E_{1,\gamma}E_{2,1,\gamma}^2 &= -8, & E_{2,1\gamma}^3 &=32, & E_{2,\mu,\gamma}^3 &= 9/2, \notag\\
  E_{2,1,\gamma}E_{3,1}^2 &= -2, & E_{3,1}^3 &= 3.
\end{align}

\subsubsection{The $\IZ_{6-I}$ orbifold on $G_2^2\times SU(3)$}

This case differs from $\IZ_{6-I}$ on $G_2\times SU(3)^2$ only in the number of $\IZ_2$ fixed lines. All that was said in the last subsection applies in this case as well. There are two differences: One arises for the O7--planes at the orbifold point, where we have four which fall into two equivalence classes. The other for the fixed lines without fixed points. There are now five of themwhich are invariant under the orientifold action, hence $h^{2,1}_+=5$.

\section{Conclusions}
\label{sec:Conclusions}

In the present paper we have provided a toolbox for studying smooth Calabi--Yau manifolds out of singular toroidal orbifolds. We have discussed how to determine the topology of all naturally arising divisors and how to compute the intersection ring, two essential ingredients for the calculation of the superpotential and the tree--level K\"ahler potential of the K\"ahler moduli. Furthermore, we have started to systematize the transition to the corresponding orientifold quotients. Explicit examples with $h^{1,1}_{-}\neq0$ and $h^{2,1}_{+}\neq0$ were discussed.

It should be stressed that the class of smooth manifolds obtained from maximally resolved toroidal orbifolds is one of the few classes of Calabi--Yau manifolds which are well--understood and allow a number of explicit calculations. We have provided a systematic approach for the transition to their orientifold quotients which deserves further study. The full details must be clarified in future works.

For the future, there is a variety of paths that could be pursued starting from the present state of knowledge. One possibility would be to attempt the construction of the corresponding mirror manifolds and their orientifolds. To our knowledge, only the mirror of $T^6/\IZ_2\times\IZ_2$ is fully understood. Continuing this aspect one should determine the variation of the Hodge structure and the period integrals of the resolved toroidal orbifolds. In particular, the period integrals associated to the twisted complex structure moduli have to be better understood in order to be able to turn on fluxes through the corresponding 3--cycles. Moreover, once this is known, the world--sheet instantons could be calculated in the topological string theory on this type of manifolds. 

Another possibility, is to find the construction of the Calabi--Yau fourfolds corresponding to the resolved threefolds analogous to the one used in Section 3 of~\cite{Denef:2005mm},  could be constructed, yielding the F--theory lifts for the type IIB models.


\vskip12pt

\noindent{\bf Acknowledgements:} We are grateful to Victor Batyrev, Ilka Brunner, Frederik Denef, Bogdan Florea, Antonella Grassi, Maximilian Kreuzer and Domenico Orlando for helpful discussions and correspondence. 
This work is supported in part by the
Deutsche Forschungsgemeinschaft as well as by the
EU-RTN network {\sl Constituents, Fundamental Forces and Symmetries
of the Universe} (MRTN-CT-2004-005104) and is a collaboration between the contractors Ludwig--Maximilians University at Munich and Universita di Torino to which the Universita del Piemonte Orientale, Alessandria, is associated. 

S.R. thanks the university of Munich for hospitality. The work of E. S. is supported by the Marie Curie Grant MERG--CT--2004--006374. E.S. thanks the Erwin Schroedinger Institute in Vienna, the Simons Workshop in Mathematics and Physics at Stony Brook, the Arnold Sommerfeld Center for Theoretical Physics and the Ludwig--Maximilians University at Munich for hospitality.

\break
\begin{appendix}



\section{Resolutions of the local orbifolds}
\label{sec:toricpatches}

In this appendix, we treat the resolutions of those ${\IC}^3/\IZ_N$, ${\IC}^3/\IZ_N\times\IZ_M$ and ${\IC}^2/\IZ_N$ orbifolds which occur in our examples and were not yet treated in the main text.

\subsection{Resolution of ${\IC}^3/\IZ_{3}$}
\label{sec:Z3res}

$\IZ_3$ acts as follows on ${\IC}^3$:
\begin{equation}
  \label{eq:twistthree}
  \theta:\ (z^1,\, z^2,\, z^3) \to (\varepsilon\, z^1, \varepsilon\, z^2, \varepsilon\, z^3),\quad \varepsilon=e^{2 \pi i/3}.
\end{equation}
To find the components of the $v_i$, we have to solve $(v_1)_i+(v_2)_i+(v_3)_i=0\ \mod\,3$. This leads to the following three generators of the fan (or some other linear combination thereof):
\begin{equation*}{v_1=(-1,-1,1),\ v_2=(1,0,1),\ v_3=(0,1,1).
}\end{equation*}
To resolve the singularity, we find that only $\theta$ fulfills (\ref{eq:criterion}). This leads to one  new generator:
\begin{equation*}{
w={1\over 3}\,v_1+{1\over 3}\,v_2+{1\over 3}\,v_3=(0,0,1).}
\end{equation*}
In this case, the triangulation is unique. 
\begin{figure}[h!]
\begin{center}
\includegraphics[width=120mm]{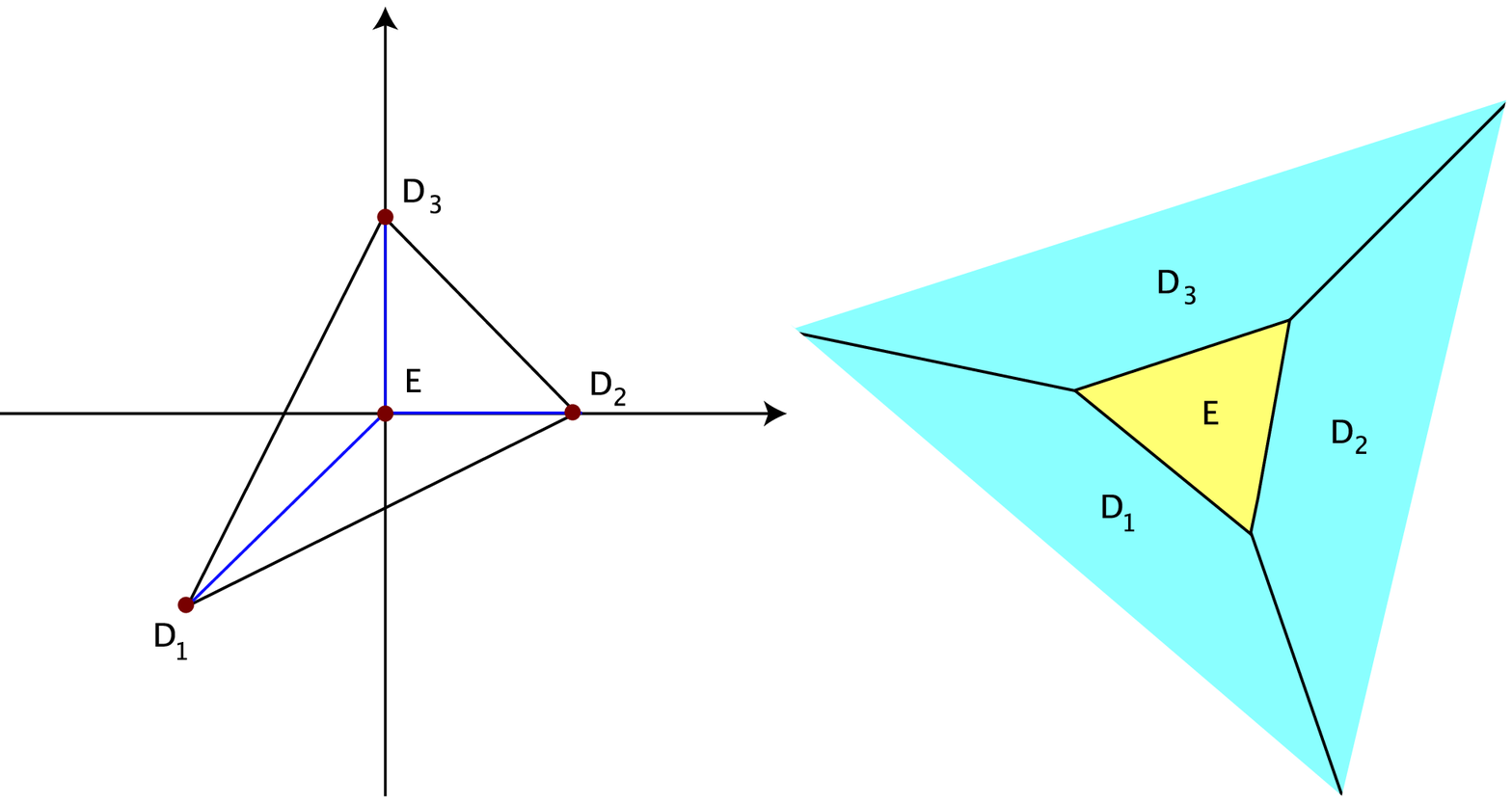}
\caption{Toric diagram of the resolution of ${\IC}^3/\IZ_{3}$ and dual graph}
\label{fig:frthree}
\end{center}
\end{figure}
Figure \ref{fig:frthree} shows the corresponding toric diagram and its dual graph.
We have now three three-dimensional cones: $(D_1,\,E,\,D_2)$, $(D_1,\,E,\,D_3)$ and $(D_2,\,E,\,D_3).$
Let us identify the blown--up geometry. The $\tilde U_i$ are
\begin{equation}
{\tilde U_1={z^2\over z^1},\quad \tilde U_2={z^3\over z^1},\quad \tilde U_3=z^1z^2z^3y.
}
\end{equation}
The rescaling that leaves the $\tilde U_i$ invariant is 
\begin{eqnarray}\label{rescalesthree}
&&(z^1,z^2,z^3,y)\to(\lambda\,z^1,\,\lambda\,z^2,\,\lambda\,z^3,\,\lambda^{-3}\,y).
\end{eqnarray}
Thus the blown--up geometry corresponds to
\begin{equation}
  \label{eq:blowupthree}
  X_{\tilde\Sigma}=(\IC^4\setminus F_{\tilde\Sigma})/\IC^*.
\end{equation}
The excluded set is $F_{\tilde\Sigma}=\{\,(z^1,z^2,z^3)=0\}$, the action of ${\IC}^{*}$ is given by (\ref{rescalesthree}).
It turns out that $X_{\tilde\Sigma}$ corresponds to the line bundle $\Oc(-3)$ over ${\IP}^2$. The exceptional divisor $E$ is identified with the zero section of this bundle.
(\ref{rescalesthree}) corresponds to the linear relation between our divisors
\begin{equation*}\label{linrelthree}{
D_1+D_2+D_3-3\,E=0.
}\end{equation*}
With this, we are ready to write down $(P\,|\,Q)$:
\begin{equation}{(P\,|\,Q)=\left( \begin{array}{cccccc}
D_1&1&-1&1&|&1\cr
D_2&1&0&1&|&1\cr
D_3&0&1&1&|&1\cr
E&0&0&1&|&\!\!-3\end{array}\right).
}
\end{equation}
This immediately yields the following linear equivalences:
\begin{equation}
\label{eq:lineqthree}
0 \sim 3\,D_i + E,\qquad i=1,\dots,3.
\end{equation}
The curve $C$ corresponding to the single column of $Q$ generates the Mori cone. We find that $C=\,D_1\cdot E=D_2\cdot E=D_3\cdot E$. Using~(\ref{eq:lineqthree}) we find e.g. $E^3=9$.

We will now discuss the topology of $E$. The star of $E$ is the whole toric diagram. Its Mori generator is exactly that of ${\IP}^2$, so this is the topology of $E$.

\subsection{Resolution of ${\IC}^3/\IZ_{4}$}
\label{sec:Z4res}

$\IZ_{4}$ acts as follows on ${\IC}^3$:
\begin{equation}
  \label{eq:twistfour}  
  \theta:\ (z^1,\, z^2,\, z^3) \to (\varepsilon\, z^1, \varepsilon\, z^2, \varepsilon^2\, z^3),\quad \varepsilon=e^{2 \pi i/4}.
\end{equation}
To find the components of the $v_i$, we have to solve $(v_1)_i+(v_2)_i+2\,(v_3)_i=0\ \mod\,4$. This leads to the following three generators of the fan:
\begin{equation*}{v_1=(2,0,1),\ v_2=(0,2,1),\ v_3=(-1,1,1).
}
\end{equation*}
To resolve the singularity, we find that $\theta$ and $\theta^2$ fulfill (\ref{eq:criterion}). This leads to two new generators:
\begin{eqnarray*}
w_1&=&{1\over 4}\,v_1+{1\over 4}\,v_2+{1\over 2}\,v_3=(0,1,1),\\ [2pt]
w_2&=&{1\over 2}\,v_1+{1\over 2}\,v_2=(1,1,1).
\end{eqnarray*}
In this case, there is again but one triangulation. 
\begin{figure}[h!]
  \begin{center}
  \includegraphics[width=120mm]{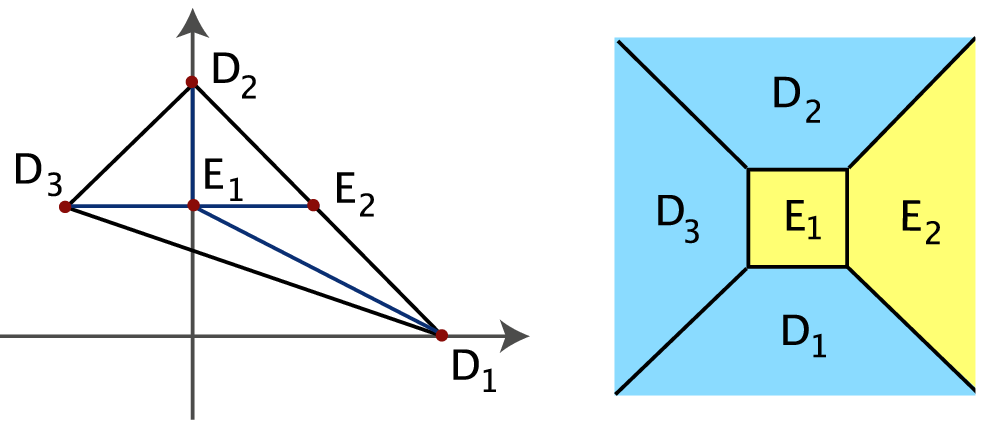}
  \caption{Toric diagram of the resolution of ${\IC}^3/\IZ_{4}$ and dual graph}
  \label{fig:frfour}
  \end{center}
\end{figure}
Figure~\ref{fig:frfour} shows the toric diagram and its dual graph.

The $\tilde U_i$ of the resolved geometries are
\begin{equation}{
\tilde U_1={(z^1)^2(z^3)^{-1}y^2},\quad \tilde U_2={(z^2)^2z^3y^1y^2},\quad \tilde U_3=z^1z^2z^2y^1y^2.
}\end{equation}
The rescalings that leave the $\tilde U_i$ invariant are
\begin{equation}\label{rescalesfour}{(z^1,\,z^2,\,z^3,\,y^1,\,y^2) \to (\lambda_1\,z^1,\,\lambda_1\,z^2,\,\lambda_2\,z^3,\, \frac{1}{\lambda_2^2}\,y^1,\,\frac{\lambda_2}{\lambda_1^2}\,y^2).
}
\end{equation}
According to (\ref{eq:toricvariety}), the new blown-up geometry is
\begin{equation*}{
X_{\tilde\Sigma}=\,({\IC}^{5}\setminus F_{\tilde\Sigma})/({\IC}^*)^2,
}
\end{equation*}
where the action of $({\IC}^*)^2$ is given by (\ref{rescalesfour}).

We have the following four three-dimensional cones: $(D_1,\,D_3,\,E_1),\,(D_1,\,E_1,\,E_2)$,\\ 
$(D_2,\,E_1,\,E_2),\,(D_2,\,D_3,\,E_1)$.
We identify the two generators of the Mori cone and write them for the columns of $Q$:
\begin{equation}{(P\,|\,Q)=\left(\begin{array}{cccccc}
2&0&1&|&1&0\cr
0&2&1&|&1&0\cr
\!\!\!-1&1&1&|&0&1\cr
0&1&1&|&0&\!\!\!-2 \cr
1&1&1&|&\!\!\!-2&1\end{array}\right).
}\end{equation}
{From} $Q$, we can determine the linear equivalences:
\begin{eqnarray}\label{lineqfour}
0&\sim& 4\,D_{{1}}+E_{{1}}+2\,E_{{2}},\cr
0&\sim& 4\,D_{{2}}+E_{{1}}+2\,E_{{2}},\cr
0&\sim& 2\,D_{{3}}+E_{{1}}.
\end{eqnarray}
There are four compact curves in our geometry, which are related to the $C_i$ as follows: $C_1=E_1\cdot E_2,\  C_2=D_1\cdot E_1=D_2\cdot E_1,\  E_1\cdot D_3=C_1+2\,C_2$. Furthermore, $E_1^3=8$.

From the Mori generators of the star of $E_1$, we find $E_1$ to be an ${\IF}_2$. $E_2$ corresponds to $\IP^1\times \IC$.

\subsection{Resolution of ${\IC}^3/\IZ_{6-II}$}
\label{sec:Z6IIres}

We are now going to present an example that allows several resolutions to illustrate the differences in the intersection numbers. 
$\IZ_{6-II}$ acts as follows on ${\IC}^3$:
\begin{equation*}
{\theta:\ (z^1,\, z^2,\, z^3) \to (\varepsilon\, z^1, \varepsilon^2\, z^2, \varepsilon^3\, z^3),\quad \varepsilon=e^{2 \pi i/6}.
}\end{equation*}
To find the components of the $v_i$, we have to solve $(v_1)_i+2\,(v_2)_i+3\,(v_3)_i=0\ \mod\,6$. This leads to the following three generators of the fan (or some other linear combination thereof):
\begin{equation*}{v_1=(-2,-1,1),\ v_2=(1,-1,1),\ v_3=(0,1,1).
}\end{equation*}

To resolve the singularity, we find that $\theta,\,\theta^2,\,\theta^3$ and $\theta^4$ fulfill (\ref{eq:criterion}). This leads to four  new generators:
\begin{eqnarray*}
w_1&=&{1\over 6}\,v_1+{2\over 6}\,v_2+{3\over 6}\,v_3=(0,0,1),\\ [2pt]
w_2&=&{2\over 6}\,v_1+{4\over 6}\,v_2=(0,-1,1),\\ [2pt]
w_3&=&{3\over 6}\,v_1+{3\over 6}\,v_3=(-1,0,1),\\ [2pt]
w_4&=&{4\over 6}\,v_1+{2\over 6}\,v_2=(-1,-1,1)
.\end{eqnarray*}
In this case, there are five  triangulations. 
\begin{figure}[h!]
  \begin{center}
  \includegraphics[width=140mm]{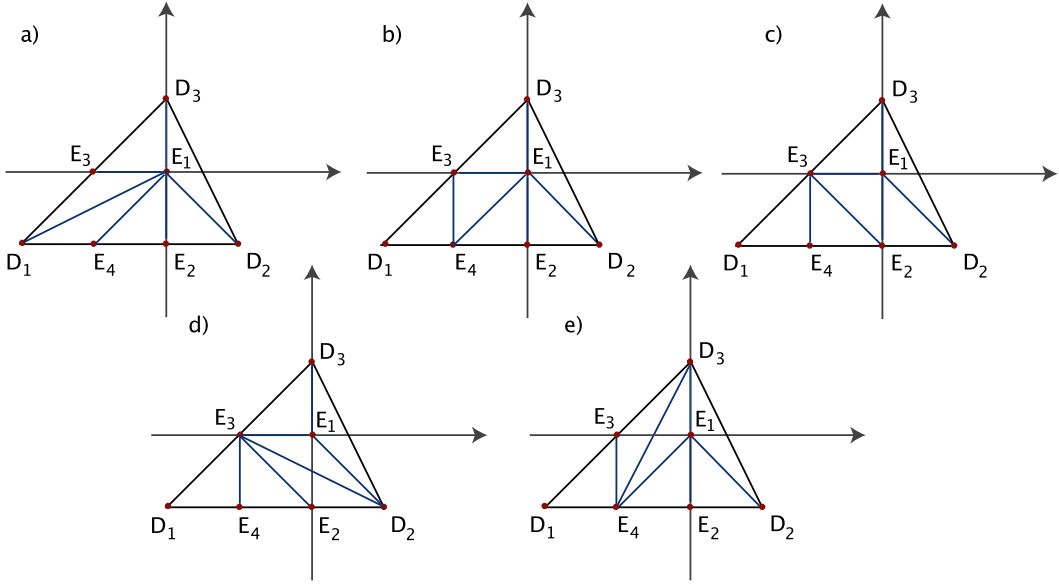}
  \caption{Toric diagrams of the resolutions of ${\IC}^3/\IZ_{6-II}$ and dual graphs}
  \label{fig:fsixii}
  \end{center}
\end{figure}
Figure~\ref{fig:fsixii} shows three of the corresponding toric diagrams and their dual graphs.
The remaining two not shown here are obtained by (i) taking the second case and flopping the curve $E_1\cdot E_3$ to $E_2\cdot D_2$ and (ii) taking the third case and flopping the curve $E_1\cdot E_2$ to $D_3\cdot E_4$. 

The $\tilde U_i$ of the resolved geometries are
\begin{eqnarray}
\tilde U_1&=&{z^2\over (z^1)^2y^3y^4},\quad \tilde U_2={z^3\over z^1z^2y^2y^4},\quad \tilde U_3=z^1z^2z^2y^1y^2y^3y^4.
\end{eqnarray}
The rescalings that leave the $\tilde U_i$ invariant are
\begin{equation}\label{rescalessixii}{(z^1,\,z^2,\,z^3,\,y^1,\,y^2,\,y^3,\,y^4) \to (\lambda_1\,z^1,\,\lambda_2\,z^2,\,\lambda_1\lambda_2\lambda_3\lambda_4\,z^3,\, {1\over\lambda_2^3\lambda_3^2\lambda_4}\,y^1,\,\lambda_3\,y^2,\,{\lambda_2\over \lambda_1^2\lambda_4}\,y^3, \lambda_4\,y^4).
}\end{equation}
According to (\ref{eq:toricvariety}), the new blown-up geometry is
\begin{equation*}{
X_{\tilde\Sigma}=\,({\IC}^{7}\setminus F_{\tilde\Sigma})/({\IC}^*)^4,
}\end{equation*}
where the action of $({\IC}^*)^4$ is given by (\ref{rescalessixii}). The five different resolutions of ${\IC}^3/\IZ_{6-II}$ only differ from each other by the excluded set. We must identify it for each case separately. So is for example $(z^1, y^1)=0$ in the excluded set for the cases b) and c), but not for a). We will not write down the three excluded sets explicitly. In what follows, we will only treat the cases depicted in Figure  \ref{fig:fsixii}.

\vskip0.5cm
\noindent{\it Case a)}
\vskip0.5cm

In this case, we have the following six three-dimensional cones: $(D_1,\,E_4,\,E_1)$, $(D_1,\,E_1,\,E_3)$, $(D_2,\,E_2,\,E_1),\,(D_2,\,E_1,\,D_3)$, $(D_3,\,E_1,\,E_3),\,(E_1,\,E_2,\,E_4)$.
We identify the four generators of the Mori cone and write them for the columns of $Q$:
\begin{equation}{(P\,|\,Q)=\left(\begin{array}{cccccccc}
\!\!-2&\!\!-1&1&|&1&\!\!\!-1&1&0\cr
1&\!\!-1&1&|&0&0&0&1\cr
0&1&1&|&1&0&0&0\cr
0&0&1&|&0&\!\!\!-1&0&0 \cr
0&\!\!-1&1&|&0&0&1&\!\!\!-2\cr
\!\!-1&0&1&|&\!\!\!-2&1&0&0\cr
\!\!-1&-1&1&|&0&1&\!\!\!-2&1\end{array}\right).
}\end{equation}
{From} $Q$, we can determine the linear equivalences:
\begin{eqnarray}\label{lineqsixii}
0&\sim& 6\,D_{{1}}+E_{{1}}+2\,E_{{2}}+3\,E_{{3}}+4\,E_{{4}},\cr
0&\sim& 3\,D_{{2}}+E_{{1}}+2\,E_{{2}}+E_{{4}},\cr
0&\sim& 2\,D_{{3}}+E_{{1}}+E_{{3}}.
\end{eqnarray}
There are six compact curves in our geometry, which are related to the $C_i$ as follows: $C_1=E_1\cdot E_3,\  C_2=E_1\cdot D_1,\  C_3=E_1\cdot E_4,\ C_4=E_1\cdot E_2,\ E_1\cdot D_3=C_1+3\,C_2+2\,C_3+C_4,\ E_1\cdot D_2=C_1+2\,C_2+C_3$. Furthermore, $E_1^3=6$.

\vskip0.5cm
\noindent{\it Case b)}
\vskip0.5cm

In this case, we have the following six three-dimensional cones: $(D_2,\,E_1,\,D_3)$, $(D_3,\,E_1,\,E_3)$, $(D_1,\,E_3,\,E_4),\,(E_4,\,E_2,\,E_3)$, $(E_1,\,E_2,\,E_3),\,(E_1,\,E_2,\,D_2)$.
We identify the four generators of the Mori cone and write them for the columns of $Q$:
\begin{equation}{(P\,|\,Q)=\left(\begin{array}{cccccccc}
\!\!-2&\!\!-1&1&|&0&0&0&1\cr
1&\!\!-1&1&|&0&0&1&0\cr
0&1&1&|&1&0&0&0\cr
0&0&1&|&\!\!\!-2&1&\!\!\!-1&0 \cr
0&\!\!-1&1&|&1&\!\!\!-1&\!\!\!-1&1\cr
\!\!-1&0&1&|&0&\!\!\!-1&1&0\cr
\!\!-1&-1&1&|&0&1&0&\!\!\!-2\end{array}\right).
}\end{equation}
The linear equivalences between the divisors remain the same as in case a).
There are again six compact curves in our geometry, which are related to the $C_i$ as follows: $C_1=E_1\cdot D_2=E_1\cdot E_3,\  C_2=E_2\cdot E_3,\  C_3=E_1\cdot E_2,\ C_4=E_3\cdot E_4,\ D_3\cdot E_1=C_1+C_3$.  Here, $E_1^3=8$.

\vskip0.5cm
\noindent{\it Case c)}
\vskip0.5cm

In this case, we have the following six three-dimensional cones: $(D_2,\,E_1,\,D_3)$, $(D_3,\,E_1,\,E_3)$, $(D_1,\,E_3,\,E_4),\,(E_4,\,E_1,\,E_3)$, $(E_1,\,E_2,\,E_4),\,(E_1,\,E_2,\,D_2)$. We identify the four generators of the Mori cone and write them for the columns of $Q$:
\begin{equation}{(P\,|\,Q)=\left(\begin{array}{cccccccc}
\!\!-2&\!\!-1&1&|&1&0&0&0\cr
1&\!\!-1&1&|&0&0&0&1\cr
0&1&1&|&0&0&1&0\cr
0&0&1&|&1&\!\!\!-1&\!\!\!-1&0 \cr
0&\!\!-1&1&|&0&1&0&\!\!\!-2\cr
\!\!-1&0&1&|&\!\!\!-1&1&\!\!\!-1&0\cr
\!\!-1&-1&1&|&\!\!\!-1&\!\!\!-1&1&1\end{array}\right).
}\end{equation}
The linear equivalences are the same as in case a).
There are again six compact curves in our geometry, which are related to the $C_i$ as follows: $C_1=E_3\cdot E_4,\  C_2=E_1\cdot E_4,\  C_3=E_1\cdot E_3,\ C_4=E_1\cdot E_2,\ E_1\cdot D_2=C_2+C_3, \ E_1\cdot D_3=2\,C_2+C_3+C_4$.  Here, $E_1^3=7$.

We come now to the topologies of the exceptional divisors. The only compact exceptional divisor is $E_1$. In triangulations b) and d), we recognize its star to be that of an ${\IF}_1$. In triangulation a) and c) it is that of an $\IF_1$ blown up in 2 and 1 points, respectively. Finally, in triangulation e) the topology of $E_1$ is that of a $\IP^2$. 

In triangulation a), all non-compact exceptional divisors have the topology of $\IP^1\times \IC$. Clearly, all triangulations are related via flops. In triangulation b), $E_2$ can only be described as being related to $E_2$ of a) via two flops. $E_3$ is $\IP^1\times \IC$ blown up in two points, while $E_4$ is $\IP^1\times \IC$. In triangulation c), $E_2$ is a $\IP^1\times \IC$, while the other two non-compact exceptional divisors are related to those of b) by flopping the curve $(E_2, E_3)$ to $(E_1, E_4)$.


\subsection{Resolution of ${\IC}^3/(\IZ_{2}\times\IZ_2)$}
\label{sec:localZ2xZ2}

$(\IZ_{2}\times\IZ_2)$ acts as follows on ${\IC}^3$:
\begin{eqnarray}
  \label{eq;twisttwotwo}
  \theta^1:\ (z^1,\, z^2,\, z^3)& \to& (\varepsilon\, z^1, \, z^2, \varepsilon\, z^3),\cr
  \theta^2:\ (z^1,\, z^2,\, z^3)& \to& (\, z^1, \, \varepsilon\,z^2, \varepsilon\, z^3),\cr
  \theta^1\theta^2:\ (z^1,\, z^2,\, z^3)& \to& (\varepsilon\, z^1, \, \varepsilon\,z^2,  z^3),
\end{eqnarray}
with $\varepsilon=e^{2 \pi i/2}$.
To find the components of the $v_i$, we have to solve 
\begin{eqnarray}
(v_1)_i+(v_3)_i&=&0\ \mod\,2,\cr
(v_2)_i+(v_3)_i&=&0\ \mod\,2,\cr
(v_1)_i+(v_3)_i&=&0\ \mod\,2.
\end{eqnarray}
This leads to the following three generators of the fan:
\begin{equation*}{v_1=(0,2,1),\ v_2=(0,0,1),\ v_3=(2,0,1).
}\end{equation*}
To resolve the singularity, we find that $\theta^1,\,\theta^2$  and $\theta^1\theta^2$ fulfill (\ref{eq:criterion}). This leads to three new generators:
\begin{eqnarray*}
w_1&=&(1,0,1),\ w_2=(1,1,1),\ w_3=(0,1,1).\end{eqnarray*}
In this case, there are four distinct triangulations. 
\begin{figure}[h!]
\begin{center}
\includegraphics[width=120mm]{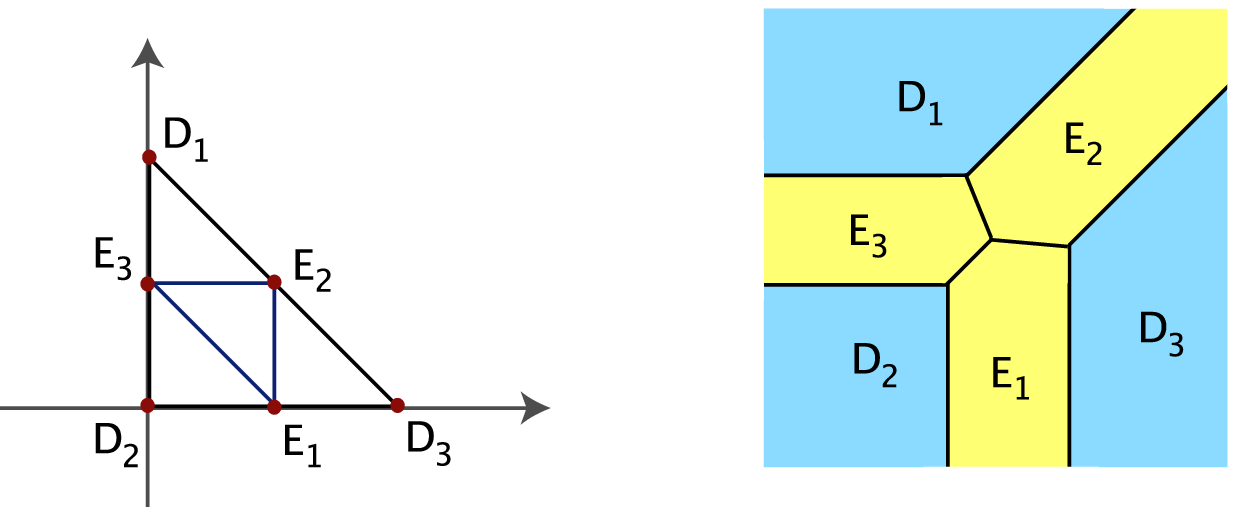}
\caption{Toric diagram of resolution of ${\IC}^3/\IZ_{2}\times\IZ_2$ and dual graph}\label{frtwotwo}
\end{center}
\end{figure}
Figure \ref{frtwotwo} shows two of them.
Let us identify the blown--up geometry. The $\tilde U_i$ are
\begin{eqnarray}\label{tildeUtwotwo}
\tilde U_1&=&{(z^3)^2y^1y^2},\cr
\tilde U_2&=&{(z^1)^2y^2y^3},\cr
\tilde U_3&=&z^1z^2z^3y^1y^2y^3.
\end{eqnarray}
The rescaling that leaves the $\tilde U_i$ invariant is 
\begin{eqnarray}\label{rescalestwotwo}&&(z^1,z^2,z^3,y^1,y^2,y^3) \to \cr
&&(\lambda_1\,z^1,\,\lambda_1\lambda_2\lambda_3\,z^2,\,\lambda_2\,z^3,\,{1\over \lambda_2^2\lambda_3}\,y^1, {\lambda_3}\,y^2,\,{1\over\lambda_1^2\lambda_3}\,y^3).
\end{eqnarray}
Thus the blown-up geometry corresponds to
\begin{equation*}{
X_{\tilde\Sigma}=({\IC}^{6}\setminus F_{\tilde\Sigma})/({\IC}^*)^3.
}\end{equation*}
The excluded sets differ for the different resolutions. We refrain from giving them explicitly. The action of $({\IC}^{*})^3$ is given by (\ref{rescalestwotwo}).

The $2\cdot 2=4$ three-dimensional cones are in this case $(D_1,E_2,E_3),\ (D_2,E_1,E_3)$,  $(D_3,E_1,E_2)$, and $(E_1,E_2,E_3)$.
We find three generators of the Mori cone and write them as columns of $Q$:
\begin{equation}{(P|\,Q)=\left(\begin{array}{ccccccccccc}
D_1&0&2&1&|&1&0&0\cr
D_2&0&0&1&|&0&1&0\cr
D_3&2&0&1&|&0&0&1\cr
E_1&1&0&1&|&1&\!\!\!-1&\!\!\!-1\cr
E_2&1&1&1&|&\!\!\!-1&1&\!\!\!-1\cr
E_3&0&1&1&|&\!\!\!-1&\!\!\!-1&1
\end{array}\right)}\end{equation}
This leads to the following linear equivalences between the divisors:
\begin{eqnarray}\label{lineqtwotwo}
0&\sim&2\,D_{{1}}+E_{{2}}+E_{{3}},\cr
0&\sim&2\,D_{{2}}+E_{{1}}+E_{3},\cr
0&\sim&2\,D_{{3}}+E_{{1}}+E_{{2}}.
\end{eqnarray}

From the intersection numbers, we find the following relations between the Mori generators and the nine compact curves of our geometry: $C_1=E_2\cdot E_3,\ C_2=E_1\cdot E_3,\ C_3=E_1\cdot E_2$.

We will now discuss the topologies of the exceptional divisors. They are all semi-compact and correspond to  $\IP^1\times \IC$.


\subsection{Resolution of ${\IC}^3/(\IZ_{2}\times\IZ_4)$}
\label{sec:localZ2xZ4}

$(\IZ_{2}\times\IZ_4)$ acts as follows on ${\IC}^3$:
\begin{eqnarray}
  \label{twisttwofour}
  \theta^1:\ (z^1,\, z^2,\, z^3)& \to& (\varepsilon^2\, z^1, \, z^2,  \varepsilon^2\, z^3),\cr
  \theta^2:\ (z^1,\, z^2,\, z^3)& \to& (\, z^1, \, \varepsilon\,z^2,  \varepsilon^3\, z^3),\cr
  \theta^1\theta^2:\ (z^1,\, z^2,\, z^3)& \to& (\varepsilon^2\, z^1, \,  \varepsilon\,z^2, \varepsilon\, z^3),
\end{eqnarray}
with $\varepsilon=e^{2 \pi i/4}$.
To find the components of the $v_i$, we have to solve
\begin{eqnarray}
  2\,(v_1)_i+2\,(v_3)_i&=&0\ \mod\,4,\cr
  (v_2)_i+3\,(v_3)_i&=&0\ \mod\,4,\cr
  2\,(v_1)_i+(v_2)_i+(v_3)_i&=&0\ \mod\,4.
\end{eqnarray}
This leads to the following three generators of the fan:
\begin{equation*}
  v_1=(1,-1,1),\ v_2=(-3,1,1),\ v_3=(1,1,1).
\end{equation*}
To resolve the singularity, we now have many more possibilities for  new vertices. We find that $\theta^1,\,\theta^2, \,(\theta^2)^2,\, (\theta^2)^3,\,\theta^1\theta^2$  and $\theta^1(\theta^2)^2$ fulfill  (\ref{eq:criterion}). This leads to six new generators:
\begin{eqnarray*}
  w_1&=&(1,0,1),\ w_2=(0,1,1),\ w_3=(-1,1,1),\ w_4=(-2,-1,1),\ w_5= (0,0,1),\cr
  w_6&=&(-1,0,1).
\end{eqnarray*}
In this case, there are 24 distinct triangulations.
\begin{figure}[h!]
\begin{center}
\includegraphics[width=120mm]{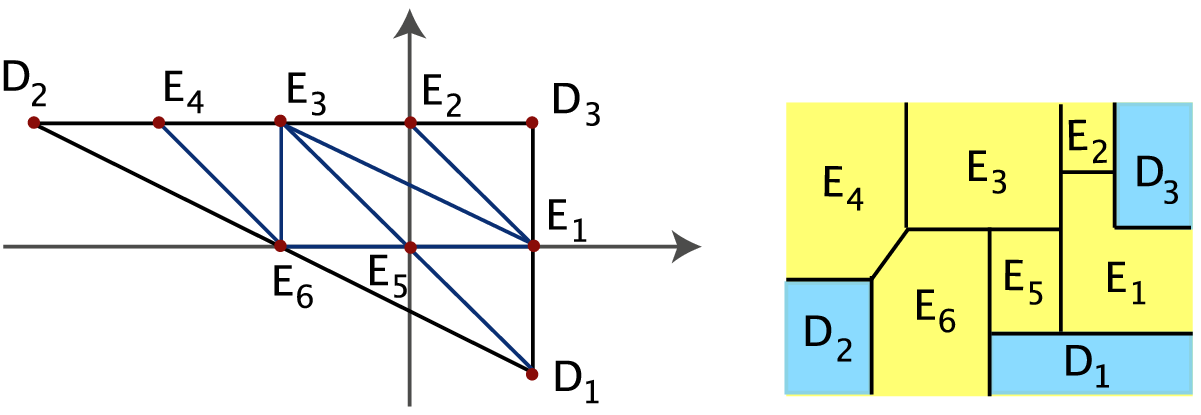}
\caption{Toric diagram of resolution of ${\IC}^3/\IZ_{2}\times\IZ_4$  and dual graph}\label{frtwofour}
\end{center}
\end{figure}
Figure \ref{frtwofour} shows one of them. It was chosen to be  compatible with the triangulation of $\IZ_2\times\IZ_2$.
Let us identify the blown--up geometry. The $\tilde U_i$ are
\begin{eqnarray}\label{tildeUtwofour}
  \tilde U_1&=&{z^1z^3y^1\over(z^2)^3y^3(y^4)^2y^6},\cr
  \tilde U_2&=&{z^2z^3y^2y^3y^4\over z^1},\cr
  \tilde U_3&=&z^1z^2z^3y^1y^2y^3y^4y^5y^6y^7.
\end{eqnarray}
The rescaling that leaves the $\tilde U_i$ invariant is
\begin{eqnarray}\label{rescalestwofour}(z^1,z^2,z^3,y^1,..,y^6) &\to&
(\lambda_1\,z^1,\,\lambda_2\,z^2,\,\lambda_3\,z^3,\,{\lambda_2^3 \lambda_4\lambda_5^2\lambda_6\over\lambda_1\lambda_3}\,y^1, {\lambda_1 \over \lambda_2\lambda_3\lambda_4\lambda_5}\,y^2,\cr
&&{\lambda_4}\,y^3,\lambda_5\,y^4,{\lambda_3\over \lambda_1\lambda_2^3 \lambda_4\lambda_5^2\lambda_6^2}\,y^5, \lambda_6\, y^6).
\end{eqnarray}
Thus the blown-up geometry corresponds to
\begin{equation*}{
X_{\tilde\Sigma}=({\IC}^{9}\setminus F_{\tilde\Sigma})/({\IC}^*)^6.
}\end{equation*}
The excluded sets differ for the different resolutions. We refrain  from giving them explicitly. The action of $({\IC}^{*})^6$ is given  by (\ref{rescalestwofour}).

The $2\cdot 4=8$ three-dimensional cones are in this case $ (D_2,E_4,E_6),\ (E_3,E_4,E_6)$,  $(E_3,E_5,E_6)$, $(D_3,E_1,E_2),\  (D_1,E_1,E_5),\ (D_1,E_5,E_6),\ (E_2,E_3,E_1),\ (D_3,E_1,E_2)$.
We find six generators of the Mori cone and write them as columns of  $Q$:
\begin{equation}{(P|\,Q)=\left(\begin{array}{ccccccccccc}
D_1&1&\!\!\!\-1&1&|&0&0&1&0&0&0\cr
D_2&\!\!\!-3&1&1&|&0&0&0&0&0&1\cr
D_3&1&1&1&|&1&0&0&0&0&0\cr
E_1&1&0&1&|&0&\!\!\!-1&0&1&0&0\cr
E_2&0&1&1&|&\!\!\!-2&1&0&0&0&0\cr
E_3&\!\!\!-1&1&1&|&1&\!\!\!-1&1&0&\!\!\!-1&1\cr
E_4&\!\!\!-2&1&1&|&0&0&0&0&1&\!\!\!-2\cr
E_5&0&0&1&|&0&1&\!\!\!-2&\!\!\!-2&1&0\cr
E_6&\!\!\!-1&0&1&|&0&0&0&1&\!\!\!-1&0
\end{array}\right)}\end{equation}
This leads to the following linear equivalences between the divisors:
\begin{eqnarray}\label{lineqtwofour}
0&\sim&2\,D_{{1}}+E_{{1}}+E_{{5}}+E_{{6}},\cr
0&\sim&4\,D_{{2}}+E_{{2}}+2\,E_{{3}}+3\,E_{{4}}+E_{5}+2\,E_{{6}},\cr
0&\sim&4\,D_{{3}}+2\,E_{{1}}+3\,E_{{2}}+2\,E_{{3}}+E_{{4}}+E_{{5}}.
\end{eqnarray}

From the intersection numbers, we find the following relations  between the Mori generators and the nine compact curves of our  geometry: $C_1=E_1\cdot E_2,\ C_2=E_1\cdot E_3,\ C_3=E_1\cdot E_5=E_5 \cdot E_6,\ C_4=E_3\cdot E_5=D_1\cdot E_5,\ C_5=E_3\cdot E_6,\ C_6=E_4 \cdot E_6$. Furthermore, $E_5^3=8$.

We will now discuss the topologies of the exceptional divisors. $E_5$  is an $\IF_1$. $E_2$ and $E_4$ are $\IP^1\times \IC$, $E_1,\ E_3$ and  $E_6$ are $\IP^1\times \IC$ with two blow-ups.


\subsection{Resolution of ${\IC}^3/(\IZ_{3}\times\IZ_3)$}
\label{sec:localZ3xZ3}

$(\IZ_{3}\times\IZ_3)$ acts as follows on ${\IC}^3$:
\begin{eqnarray}\label{twistthreethree}
\theta^1:\ (z^1,\, z^2,\, z^3)& \to& (\varepsilon\, z^1, \, z^2, \varepsilon^2\, z^3),\cr
\theta^2:\ (z^1,\, z^2,\, z^3)& \to& (\, z^1, \, \varepsilon\,z^2, \varepsilon^2\, z^3),\cr
\theta^1\theta^2:\ (z^1,\, z^2,\, z^3)& \to& (\varepsilon\, z^1, \, \varepsilon\,z^2, \varepsilon\, z^3),
\end{eqnarray}
with $\varepsilon=e^{2 \pi i/3}$.
To find the components of the $v_i$, we have to solve 
\begin{eqnarray}
(v_1)_i+2\,(v_3)_i&=&0\ \mod\,3,\cr
(v_2)_i+2\,(v_3)_i&=&0\ \mod\,3,\cr
(v_1)_i+(v_2)_i+(v_3)_i&=&0\ \mod\,3.
\end{eqnarray}
This leads to the following three generators of the fan:
\begin{equation*}{v_1=(-2,2,1),\ v_2=(-2,-1,1),\ v_3=(1,-1,1).
}\end{equation*}
To resolve the singularity, we now have many more possibilities for new vertices. We find that $\theta^1,\,(\theta^1)^2,\,\theta^2, \,(\theta^2)^2,\,\theta^1\theta^2, \,\theta^1(\theta^2)^2$  and $\,(\theta^1)^2\theta^2$ fulfill (\ref{eq:criterion}). This leads to seven new generators:
\begin{eqnarray*}
w_1&=&(0,0,1),\ w_2=(-1,1,1),\ w_3=(0,-1,1),\ w_4=(-1,-1,1),\ \cr
w_5&=&(-1,0,1),\ w_6=(-2,0,1),\ w_7=(-2,1,1)
.\end{eqnarray*}
In this case, there are 79 distinct triangulations. 
\begin{figure}[h!]
\begin{center}
\includegraphics[width=120mm]{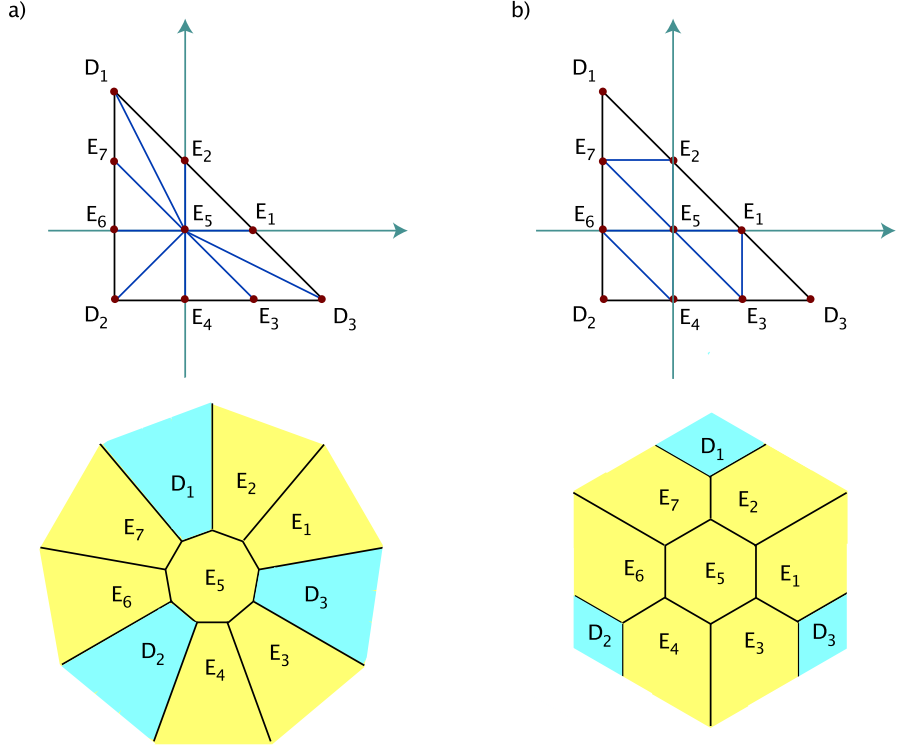}
\caption{Toric diagram of two of the resolutions of ${\IC}^3/\IZ_{3}\times\IZ_3$ and dual graphs}\label{frthreethree}
\end{center}
\end{figure}
Figure \ref{frthreethree} shows two of them.
Let us identify the blown--up geometry. The $\tilde U_i$ are
\begin{eqnarray}\label{tildeUthreethree}
\tilde U_1&=&{z^3\over(z^1)^2(z^2)^{2}y^2y^4y^5(y^6)^2(y^7)^2},\cr
\tilde U_2&=&{(z^1)^2y^2y^7\over z^2z^3y^3y^4},\cr
\tilde U_3&=&z^1z^2z^3y^1y^2y^3y^4y^5y^6y^7.
\end{eqnarray}
The rescaling that leaves the $\tilde U_i$ invariant is 
\begin{eqnarray}\label{rescalesthreethree}
(z^1,z^2,z^3,y^1,...,y^7) &\to&
(\lambda_1\,z^1,\,\lambda_2\,z^2,\,\lambda_1^2\lambda_2^2\lambda_3\lambda_4\lambda_5\lambda_6^2\lambda_7^2\,z^3,\,{1\over\lambda_1^{3}\lambda_3^2\lambda_5\lambda_6\lambda_7^2}\,y^1,\cr
&& \lambda_3\,y^2,\,{1\over \lambda_2^3\lambda_4^2\lambda_5\lambda_6^2\lambda_7}\,y^3,\lambda_4y^4,\lambda_5y^5,\lambda_6y^6,\lambda_7y^7).
\end{eqnarray}
Thus the blown-up geometry corresponds to
\begin{equation*}{
X_{\tilde\Sigma}=({\IC}^{10}\setminus F_{\tilde\Sigma})/({\IC}^*)^7.
}\end{equation*}
The excluded sets differ for the different resolutions. We refrain from giving them explicitly. The action of $({\IC}^{*})^7$ is given by (\ref{rescalesthreethree}).

Let us now give the intersection properties for the two resolutions shown in the figure.

\vskip0.5cm
\noindent{\it Case a)}
\vskip0.5cm

The $3\cdot 3=9$ three-dimensional cones are in this case $(D_3,E_1,E_5),\ (E_1,E_2,E_5)$, $(D_1,E_2,E_5)$, $(D_1,E_5,E_7),\ (E_5,E_6,E_7),\ (D_2,E_5,E_6),\ (D_2,E_4,E_5),\ (E_3,E_4,E_5)$, $(D_3,E_3,E_5)$.
We find nine generators of the Mori cone and write them as columns of $Q$:
\begin{equation}{(P|\,Q)=\left(\begin{array}{cccccccccccccc}
D_1&\!\!\!-2&2&1&|&0&0&1&\!\!\!-1&1&0&0&0&0\cr
D_2&\!\!\!-2&\!\!\!-1&1&|&0&0&0&0&0&1&\!\!\!-1&1&0\cr
D_3&1&\!\!\!-1&1&|&1&\!\!\!-1&0&0&0&0&0&0&1\cr
E_1&0&0&1&|&\!\!\!-2&1&1&0&0&0&0&0&0\cr
E_2&\!\!\!-1&1&1&|&1&0&\!\!\!-2&1&0&0&0&0&0\cr
E_3&0&\!\!\!-1&1&|&0&1&0&0&0&0&0&1&\!\!\!-2\cr
E_4&\!\!\!-1&\!\!\!-1&1&|&0&0&0&0&0&0&1&\!\!\!-2&1\cr
E_5&\!\!\!-1&0&1&|&0&\!\!\!-1&0&\!\!\!-1&0&0&\!\!\!-1&0&0\cr
E_6&\!\!\!-2&0&1&|&0&0&0&0&1&\!\!\!-2&1&0&0\cr
E_7&\!\!\!-2&1&1&|&0&0&0&1&\!\!\!-2&1&0&0&0
\end{array}\right)}\end{equation}
This leads to the following linear equivalences between the divisors:
\begin{eqnarray}\label{lineqthreethree}
0&\sim3&\,D_{{1}}+E_{{1}}+2\,E_{{2}}+E_{{5}}+E_{{6}}+2\,E_{{7}},\cr
0&\sim3&\,D_{{2}}+E_{{3}}+2\,E_{{4}}+E_{{5}}+2\,E_{6}+E_{{7}},\cr
0&\sim3&\,D_{{3}}+2\,E_{{1}}+E_{{2}}+2\,E_{{3}}+E_{{4}}+E_{{5}}.
\end{eqnarray}

From the intersection numbers, we find the following relations between the Mori generators and the nine compact curves of our geometry: $C_1=E_1\cdot E_5,\ C_2=D_3\cdot E_5,\ C_3=E_2\cdot E_5,\ C_4=D_1\cdot E_5,\ C_5=E_5\cdot E_7,\ C_6=E_5\cdot E_6,\ C_7=D_2\cdot E_5, \ E_4\cdot E_5=C_1+C_2+C_4-C_6-2\,C_7,\ E_3\cdot E_5=-C_1-2\,C_2+C_4+C_5+C_6+C_7$. Furthermore, $E_5^3=3$.

\vskip0.5cm
\noindent{\it Case b)}
\vskip0.5cm

Here, we have the following 9 tree-dimensional cones: $(D_1,E_2,E_7),\\
(E_5,E_6,E_7)$, $(E_2,E_5,E_7)$,  $(E_1,E_2,E_5),\,(D_2,E_4,E_6),\ (E_4,E_5,E_6),\\
(E_3,E_4,E_5),\ (E_1,E_3,E_5),\ (D_3,E_1,E_3)$.
$(P|\,Q)$ takes the following form:
\begin{equation}{(P|\,Q)=\left(\begin{array}{cccccccccccccc}
D_1&\!\!\!-2&2&1&|&1&0&0&0&0&0&0&0&0\cr
D_2&\!\!\!-2&\!\!\!-1&1&|&0&0&0&0&0&1&0&0&0\cr
D_3&1&\!\!\!-1&1&|&0&0&0&0&0&0&0&0&1\cr
E_1&0&0&1&|&0&0&0&1&\!\!\!-1&0&0&1&\!\!\!-1\cr
E_2&\!\!\!-1&1&1&|&\!\!\!-1&1&0&\!\!\!-1&1&0&0&0&0\cr
E_3&0&\!\!\!-1&1&|&0&0&0&0&1&0&1&\!\!\!-1&\!\!\!-1\cr
E_4&\!\!\!-1&\!\!\!-1&1&|&0&0&1&0&0&\!\!\!-1&\!\!\!-1&1&0\cr
E_5&\!\!\!-1&0&1&|&1&\!\!\!-1&\!\!\!-1&\!\!\!-1&\!\!\!-1&1&\!\!\!-1&\!\!\!-1&1\cr
E_6&\!\!\!-2&0&1&|&0&1&\!\!\!-1&0&0&\!\!\!-1&1&0&0\cr
E_7&\!\!\!-2&1&1&|&\!\!\!-1&\!\!\!-1&1&1&0&0&0&0&0\end{array}\right)}\end{equation}
The linear equivalences remain of course the same. The relations between the nine compact curves and the $C_i$ are $C_1=E_2\cdot E_7,\ C_2=E_5\cdot E_7,\ C_3=E_5\cdot E_6,\ C_4=E_2\cdot E_5,\ C_5=E_1\cdot E_5,\ C_6=E_4\cdot E_6,\ C_9=E_1\cdot E_3,\ E_4\cdot E_5=-C_3+C_4+C_5,\ E_3\cdot E_5=C_2+C_3-C_5.$ Here, $E_5^3=6$.

We will now discuss the topologies of the exceptional divisors.  The only compact exceptional divisor is $E_5$. In the triangulation a), the star of $E_5$ corresponds to the whole toric diagram. Unfortunately, one cannot read off the topology directly from the Mori generators. The triangulations a) and b) are connected by three flop transitions: $(E_1,E_3)\to(D_3,E_5),\ (E_2,E_7)\to (D_1,E_5)$ and $(E_4,E_6)\to (D_2,E_5)$. For the triangulation b), $D_1, D_2$ and $D_3$ are not part of the star. Unfortunately, the Mori generators of this star aren't helpful either. If we perform two flop-transitions, we end up in a very simple case. We flop the curve $(E_5, E_7)$ to $(E_2, E_6)$, furthermore, we flop $(E_3, E_5)$ to $(E_1, E_4)$. Thus, we arrive at a star which contains only $E_1, E_2, E_4, E_5, E_6$; it corresponds to an ${\IF}_0$. So both triangulations are birationally equivalent to ${\IF}_0$. Therefore, $h^{(1,0)}=h^{(2,0)}=0$, since the $h^{(p,0)}$ are birational invariants.

In triangulation a), all non-compact exceptional divisors have the topology of $\IP^1\times \IC$. 
In triangulation b), they are all $\IP^1\times \IC$ blown up in one point.

\subsection{Resolution of ${\IC}^2/\IZ_n$-type orbifolds}
\label{sec:C2Zn}

Here, we treat the patches of the form ${\IC}^2/\IZ_n$, associated to fixed lines. We will be rather brief. As mentioned already in the main text, these singularities are rational double points of type $A_{n-1}$. The toric diagrams of their resolutions (Hirzebruch-Jung sphere trees) are lines of length $n$ with $n+1$ points on it, each at distance one from its neighbors.

The action of $\IZ_n$ on ${\IC}^2$ is:
\begin{equation}\label{twistctwo}{\theta:\ (z^1,\, z^2) \to (\varepsilon\, z^1, \varepsilon^{n-1}\, z^2),\quad \varepsilon=e^{2 \pi i/n}.
}\end{equation}
To find the components of the $v_i$, we have to solve $(v_1)_i+(n-1)\,(v_2)_i=0\ \mod\,n$. This leads to the following two generators of the fan:
\begin{equation}{v_1=(n-1,1),\quad v_2=(-1,1).
}\end{equation}
All $\theta^i$, $i=1,...,n-1$ fulfill (\ref{eq:criterion}), so we get $n-1$ new generators:
\begin{equation}{
w_i=\frac{i}{n}v_1 + \frac{n-i}{n}v_2 = (i-1,1), \qquad i=1,\dots,n-1}\end{equation}
The $\tilde U_i$ are
 \begin{equation}\label{tildeUctwo}{\tilde U_1={(z^1)^{n-1}y^2(y^3)^2...\,(y^{n-1})^{n-2}\over z^2},\quad \tilde U_2=z^1z^2y^1y^2...\,y^{n-1}.
 }\end{equation}
 The rescaling that leaves the $\tilde U_i$ invariant is 
 \begin{eqnarray}\label{rescalesctwo}
&& (z^1,z^2,y^1,y^2, ..., y^{n-1}) \to \cr
 &&(\lambda_1\,z^1,\,\lambda_1^{n-1}\lambda_2\lambda_3^2...\lambda_{n-1}^{n-2}\,z^2,\,(\lambda_1^{n}\lambda_2^{2}\lambda_3^3...\lambda_{n-1}^{n-1})^{-1}\,y^1,\lambda_2\,y^2, \lambda_3\,y^3,...,\lambda_{n-1}\,y^{n-1}).\ \ \ \ \ \ \ \ \ \ 
 \end{eqnarray}
The blown--up geometry is
\begin{equation}\label{blowupcn}{
X_{\tilde\Sigma}=({\IC}^{n+1}\setminus F_{\tilde\Sigma})/\left({\IC}^*\right)^{n-1},
}\end{equation}
where the $\left({\IC}^{*}\right)^{n-1}$ action is determined in the following $(P|Q)$--matrix
\begin{equation}{(P\,|\,Q)=\left(\begin{array}{ccccccccccc}
D_1    &n-1&1&|& 1& 0&0&...&0& 0& 0\cr
E_1    &  0&1&|&-2& 1&0&...&0& 0& 0\cr
E_2    &  1&1&|& 1&-2&1&...&0& 0& 0\cr
...    &   &1&|&  &  & &...& &  &  \cr
E_{n-2}&  1&1&|& 0& 0&0&...&1&-2& 1\cr
E_{n-1}&  1&1&|& 0& 0&0&...&0& 1&-2\cr
D_2    & -1&1&|& 0& 0&0&...&0& 0& 1\cr
\end{array}\right)}\end{equation}
We observe that the Q matrix is nothing but the Cartan matrix for $A_{n-1}$. With it, we obtain the following linear equivalences:
\begin{equation}
  \label{eq:lineqsctwo}
  E_1 \sim 2D_1, \qquad 2E_i \sim E_{i-1} + E_{i+1}, i=2,...,n-2, \qquad E_{n-1} \sim 2D_2.
\end{equation}



\section{The $\IZ_3$ orbifold}
\label{sec:Z3}

\subsection{The resolved orbifold}
\label{sec:Z3orb}

\begin{figure}[h!]
  \begin{center}
  \includegraphics[width=140mm]{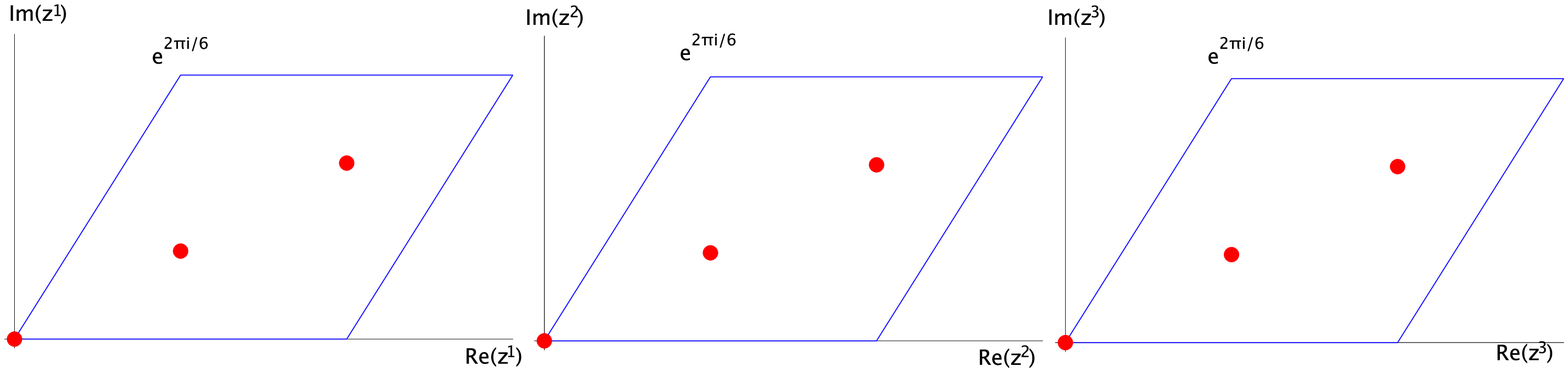}
  \caption{Fundamental regions for the $\IZ_3$ orbifold}
  \label{fig:ffuthree}
  \end{center}
\end{figure}

\begin{figure}[h!]
  \begin{center}
  \includegraphics[width=85mm]{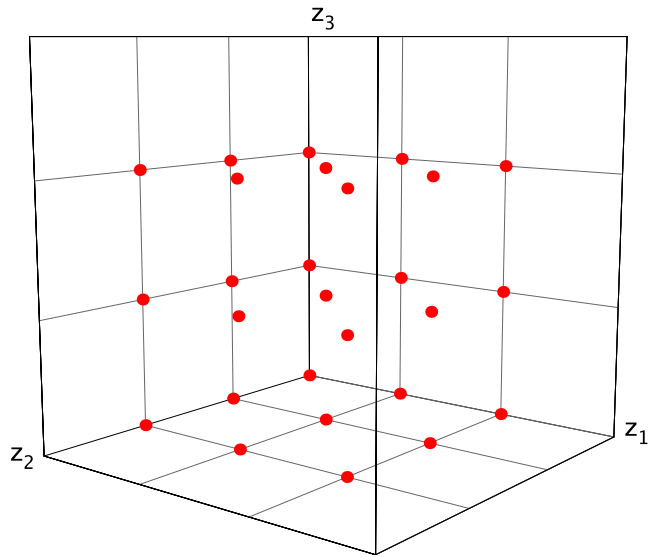}
  \caption{Schematic picture of the fixed set configuration of the $\IZ_3$ orbifold}
  \label{fig:ffixthree}
  \end{center}
\end{figure}

This is a prime orbifold and therefore an easy case. It is well understood and we provide this example just to display some of the techniques in detail. The action of the twist $\theta$ was given in (2.45) and the resulting complex structure in (2.47) of~\cite{Lust:2005dy}.
The $T^6$ factorizes into $(T^2)^3$. Figure~\ref{fig:ffuthree} shows the fundamental regions of the three tori corresponding to $z^1,\,z^2,\,z^3$ and their fixed points. 
The fundamental regions are all the same in this case and the three fixed points of the  $\IZ_3$--twist are $z^1_{{\rm fixed},1}=z^2_{{\rm fixed},1}=z^3_{{\rm fixed},1}=0,\ z^1_{{\rm fixed},2}=z^2_{{\rm fixed},2}=z^3_{{\rm fixed},2}=1/\sqrt3\,e^{\pi i/6}$, and $z^1_{{\rm fixed},3}=z^2_{{\rm fixed},3}=z^3_{{\rm fixed},3}=1+i/\sqrt3$. Table~\ref{tab:fsthree} summarizes the relevant data of the fixed point set.
\begin{table}[h!]
  \begin{center}
  \begin{tabular}{|l|c|l|c|}
    \hline
    \ Group el.&\ Order &\ Fixed Set& Conj. Classes\cr
    \hline
    \ $\theta$ & 3      &\ 27 fixed  points & 27\cr
    \hline
  \end{tabular}
  \caption{Fixed point set for $\IZ_3$.}
  \label{tab:fsthree}
  \end{center}
\end{table}
Figure~\ref{fig:ffixthree} shows the configuration of the fixed point set in a schematic way.

On each of the 27 isolated $\IZ_3$ fixed points we put a local patch, which each contributes one compact exceptional divisor. We denote these divisors by $E_{\alpha\beta\gamma}$, where $\alpha$ denotes the 3 fixed points on the $z^1$--axis, $\beta$ those on the $z^2$--axis and $\gamma$ those of the $z^3$--axis. 
There are 9 fixed planes for this example, i.e. $z^i=\zf{i}{\alpha}, \ z^j,z^k$ free. To each of these we associate a divisor which we denote them by $D_{i\alpha}, \ i,\alpha=1,2,3$. $D_{1\alpha}$ are the divisors associated to $z^1=\zf{1}{\alpha}$, where $\alpha$ labels the fixed point. Furthermore, there are 9 divisors $R_{i\jbar}$ which are inherited from $T^6$. Since there are no fixed lines in this example, $h^{2,1}_{\rm twist.}=0$.

From the linear relations~(\ref{eq:lineqthree}) for the $\IC^3/\IZ_3$ patch, we obtain the global linear relations~(\ref{eq:Reqglobal}):
\begin{equation}
  \label{eq:Reqthree}
  R_{i} \sim 3\,D_{i, \alpha}+\sum_{\beta,\gamma=1}^3 E_{{\alpha\beta\gamma}}.
\end{equation}
The calculation of the intersection ring is very simple, since all exceptional divisors neither intersect each other nor the $R_{i\jbar}$, and we do not need to work with the polyhedra. From~(\ref{eq:R1R2R3}) and the subsequent discussion as well as Appendix~\ref{sec:Z3res} we find that the only non-trivial intersection numbers of $X$ in the basis $\{R_i, R_{i\jbar}, E_{\alpha\beta\gamma}\}$ are 
\begin{align}
  \label{eq:Z3ring}
  R_1R_2R_3 & = 9, & R_1R_{2\bar3}R_{3\bar2} &= -9, & R_{1\bar3}R_2R_{3\bar1} &= -9, & R_{1\bar2}R_{2\bar1}R_3&=-9,\notag\\
  R_{1\bar2}R_{2\bar3}R_{3\bar1} &=9, & R_{1\bar3}R_{3\bar2}R_{2\bar1} &=9, & E_{\alpha\beta\gamma}^3 &= 9.
\end{align}
As noted in Appendix~\ref{sec:Z3res}, the $E_{\alpha\beta\gamma}$ have the topology of a ${\IP}^2$. All divisors $D_{i\alpha}$ have exactly the same properties, therefore it is enough to look at one of them and determine its Euler number. After removing the 9 fixed points from the $T^4$ the $\IZ_3$ action is free, hence the quotient has Euler number $(0-9)/3 = -3$. Resolving the singularities corresponds to gluing in 9 $P^1$. Therefore, $\chi(D) = -3 + 2\cdot9 = 15$. From~(\ref{eq:Reqthree}) and the intersection numbers given above, we find $D_{i\alpha}^3 = -3$, and using~(\ref{eq:S3}) its holomorphic Euler characteristic is $\chi(\Oc_{D_{i\alpha}}) = 1$. The $R_i$ and $R_{i\jbar}$ are $T^4$. Applying~(\ref{eq:c2.S}) to both $D_{i\alpha}$ and $E_{\alpha\beta\gamma}$ and then plugging into~(\ref{eq:Reqthree}) we can also determine the second Chern class to be
\begin{align}
  \label{eq:Z3c2}
  \ch_2\cdot R_i &= 0 & \ch_2\cdot E_{\alpha\beta\gamma} &= -6.
\end{align}
Another way to obtain this result and also the second Chern class on the $R_{i\jbar}$ goes as follows: We denote the covering torus $T^6$ by $A$ and the orbifold projection by $\pi:A\to X_0$, where $X_0$ is the unresolved orbifold. The resolution of the singularities is $\phi:X \to X_0$. We choose $H_0$ to be a general very ample divisor which does not pass through the singularities, i.e. $H_0$ is a linear combination of the $R_i$ and $R_{i\jbar}$. If we let $H = \phi^*H_0$ and $\tilde H = \pi^*H_0$ then we have that $\pi:\tilde H \to H$ is a $3:1$ cover. Therefore $\ch_2(\tilde H) = 3 \ch_2(H)$ and $\tilde H^3 = 3 H^3$. Applying~\eqref{eq:embedding} to both $H \subset X$ and $\tilde H \subset A$, and using~\eqref{eq:c2.S} as well as the fact that $\ch_2(A)\cdot \tilde H = 0$ we find that $\ch_2\cdot H = 0$ and hence also
\begin{equation}
  \label{eq:Z3c2b}
  \ch_2 \cdot R_{i\jbar} = 0.
\end{equation}

\subsection{The orientifold}
\label{eq:Z3O}

The O--plane configuration at the orbifold point is very simple, we have 64 O3--planes, located at the $I_6$ fixed points in each direction. They fall into 22 conjugacy classes, apart from $z_{\rm fixed}=(0,0,0)$, which is invariant, all other points fall into orbits of length 3. Only at $z_{\rm fixed}=(0,0,0)$, a fixed point coincides with one of the O3--planes.

All the exceptional divisors $E_{\alpha\beta\gamma}$ except $E_{111}$ fall into orbits of length two under $I_6$. Therefore, this example has $h^{1,1}_{-}=13$.

We now consider the orientifold of the resolved orbifold. There are two possible choices for the local involution leading to O3-- and O7--planes:
\begin{eqnarray}
(1)\quad {\Ic}(z,y)&=&(-z^1,-z^2, -z^3, y),\cr
(2)\quad {\Ic}(z,y)&=&(-z^1,-z^2, -z^3, -y).
\end{eqnarray}
We choose $(1)$ and have to solve (see~(\ref{rescalesthree}))
\begin{equation}
  \label{eq:othree}
  (-z^1,-z^2,-z^3,y)=(\lambda\, z^1, \lambda\,z^2, \lambda\,z^3, \lambda^{-3}\,y),
\end{equation}
leading to the solution $y=0,\ \lambda=-1$. This corresponds to an O7--plane located on the exceptional divisor $E$. Since only the divisor $E$ located at the fixed points itself appears in the solution, no further global consistency conditions need to be considered. Since only at $(0,0,0)$ a fixed point of the orbifold group coincides with a fixed point of $I_6$, we see that the O3--plane present at this point in the orbifold phase has disappeared.

The O3--planes away from the patches that we found in the orbifold phase are also present in the resolved case since they lie away from the resolved patches. We can see them by looking at the intersection ring of the orientifold and interpreting certain intersection numbers as number of O3--planes as discussed in Section~\ref{sec:Oring}. There are three cases. The fixed points of the orientifold action that lie at $z=(0,0,\frac{1}{2}),\, (0,0,\frac{\tau}{2},\, (0,0,\frac{1}{2}(1+\tau))$ form an equivalence class and correspond to the intersection of $D_{1,1} = \{ z^1=0 \},\, D_{2,1} = \{ z^2 =0 \}$, and $R_3 = \{z^3=c, c\not = 0,\,\frac{1}{3},\,\frac{2}{3} \}$. This intersection number is $D_{1,1}D_{2,1}R_3 = \frac{1}{2}$, hence we have a single O3--plane sitting there. By permuting the coordinates we get a total of 3 O3--planes. Then, there are the fixed points that lie at $z=(0,\frac{1}{2},\frac{1}{2}),\, (0,\frac{1}{2},\frac{\tau}{2},\dots, (0,\frac{1}{2}(1+\tau),\frac{\tau}{2}, (0,0,\frac{1}{2}(1+\tau))$. The corresponding intersection number is $D_{1,1}R_2R_3 = \frac{3}{2}$, hence we have 3 O3--planes. Taking into account the permutations of the coordinates yields a total of 9 O3-planes. Finally, the remaining orientifold fixed points correspond to the intersection $R_1R_2R_3 =\frac{9}{2}$, thus there are 9 more O3--planes. This makes a grand total of 21 O3--planes which agrees with the number of conjugacy classes.


\section{The $\IZ_{4}$ orbifolds}

\subsection{The lattice $SU(4)^2$}
\label{sec:Z4onSU4xSU4}

\subsubsection{Complex structure and fixed sets}

On the root lattice of $SU(4)^2$, the twist $Q$ has the following action:
\begin{eqnarray}
Q\ e_1&=&e_2,\quad Q\ e_2=e_3,\quad Q\ e_3=-e_1-e_2-e_3,\cr 
Q\ e_4&=&e_5,\quad Q\ e_5=e_6 ,\quad Q\ e_6=-e_4-e_5-e_6\ .\end{eqnarray}
The twist $Q$ allows for seven independent real deformations of the metric $g$
and five real deformations  of the
anti--symmetric tensor $b$. These results follow from solving the equations
$Q^tg\,Q=g$ and $Q^tb\,Q=b$:
{\arraycolsep2pt
\begin{equation}{
g\!=\!\left(\!\!\begin{array}{cccccc}
R_1^2&R_1^2\cos\theta_{23}&x&R_1R_2\cos\theta_{36}&y&R_1R_2\cos\theta_{34}\cr
R_1^2\cos\theta_{23}&R_1^2&R_1^2\cos\theta_{23}&R_1R_2\cos\theta_{35}&R_1R_2\cos\theta_{36}&y\cr
x&R_1^2\cos\theta_{23}&R_1^2&R_1R_2\cos\theta_{34}&R_1R_2\cos\theta_{35}&R_1R_2\cos\theta_{36}\cr
R_1R_2\cos\theta_{36}&R_1R_2\cos\theta_{35}&R_1R_2\cos\theta_{34}&R_2^2&R_2^2\cos\theta_{56}&z\cr
y&R_1R_2\cos\theta_{36}&R_1R_2\cos\theta_{35}&R_2^2\cos\theta_{56}&R_2^2&R_2^2\cos\theta_{56}\cr
R_1R_2\cos\theta_{34}&y&R_1R_2\cos\theta_{36}&z&R_2^2\cos\theta_{56}&R_2^2\end{array}\!\!\right),
}\end{equation}}
with $x=-R_1^2(1+2\,\cos\theta_{23})$, $y=-R_1R_2\,(\cos\theta_{34}+\cos\theta_{35}+\cos\theta_{36})$, $z=-R_2^2(1+2\,\cos\theta_{56})$ and the seven real parameters $R_1^2,\ R_2^2,\ \theta_{23}$, $\theta_{34},\ \theta_{35},\ \theta_{36},\,\theta_{56}$.  
For $b$ we find
\begin{equation}{
b=\left(\begin{array}{cccccc}
0&b_1&0&b_5&-b_3-b_4-b_5&b_3\cr
-b_1&0&b_1&b_4&b_5&-b_3-b_4-b_5\cr
0&-b_1&0&b_3&b_4&b_5\cr
-b_5&-b_4&-b_3&0&b_2&0\cr
b_3+b_4+b_5&-b_5&-b_4&-b_2&0&b_2\cr
-b_3&b_3+b_4+b_5&-b_5&0&-b_2&0\end{array}\right)}\end{equation}
with the five real parameters $b_1,\ b_2,\ b_3,\ b_4,\ b_5$. We see that we get 5 untwisted K\"ahler moduli and one untwisted complex structure modulus in this orbifold.

\begin{figure}[h!]
  \begin{center}
  \includegraphics[width=85mm]{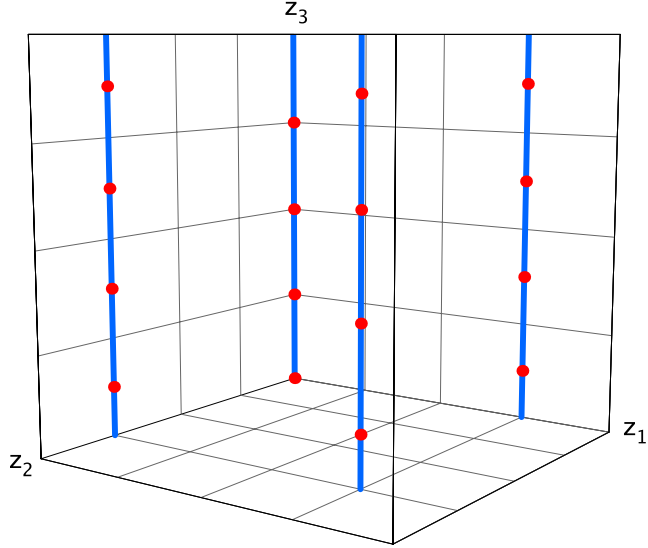}
  \caption{Schematic picture of the fixed set configuration of $\IZ_{4}$ on $SU(4)^2$}
  \label{fig:ffixfoura}
  \end{center}
\end{figure}

With the methods discussed in \cite{Lust:2005dy},
we arrive at the following complex structure:
\begin{eqnarray}
z^1&=&\frac{1}{\sqrt2}\,(x^1+i\,x^2-x^3),\cr
z^2&=&\frac{1}{\sqrt2}\,(x^4+i\,x^5-x^6),\cr
z^3&=&\frac{1}{2\sqrt{u_2}}\,[x^1-x^2+x^3+{\Uc}\,(x^4-x^5+x^6)],
\end{eqnarray}
with 
\begin{equation}
\Uc=-\frac{R_2}{2\,R_1}\sec\theta_{23}(\cos\theta_{34}+\cos\theta_{36}+i\,\sqrt{-(\cos\theta_{34}+\cos\theta_{36})^2+4\,\cos\theta_{23}\cos\theta_{56}}).
\end{equation}
The five untwisted real 2--forms that are invariant under this orbifold twist are
\begin{eqnarray}
\omega_1&=&dx^1\wedge dx^2+dx^2\wedge dx^3,\cr 
\omega_2&=&dx^1\wedge dx^4-dx^1\wedge dx^5+dx^2\wedge dx^5-dx^2\wedge dx^6+dx^3\wedge dx^6,\cr
\omega_3&=&-dx^1\wedge dx^5+dx^1\wedge dx^6-dx^2\wedge dx^6+dx^3\wedge dx^4,\cr
\omega_4&=&-dx^1\wedge dx^5+dx^2\wedge dx^4-dx^2\wedge dx^6+dx^3\wedge dx^5,\cr
\omega_5&=&dx^4\wedge dx^5+dx^5\wedge dx^6.
\end{eqnarray}
In order to determine the fixed point set of this twist, we need to look only at the $\theta$-, and $\theta^2$-twisted sectors. There are four $\IC^2/\IZ_2$ fixed lines which lie at $\zf{1}{\alpha}=0,\frac{1}{2}(1+i)$, $\alpha=1,2$, and $\zf{2}{\beta}=0,\frac{1}{2}(1+i)$, $\beta=1,2$. On each of them there are four $\IC^3/\IZ_4$ fixed points. At $(\zf{1}{1},\zf{2}{1})=(0,0)$ they are at $\zf{3}{2\gamma-1} = 0,\frac{1}{2},\frac{U}{2},\frac{1}{2}(1+U)$, $\gamma=1,\dots,4$. For the remaining values of $(\alpha,\beta)$ they are at $\zf{3}{2\gamma} = \frac{1}{4},\frac{3}{4},\frac{1}{4}(1+2U),\frac{1}{4}(3+2U)$, $\gamma=1,\dots,4$. Here, $U = \Uc$, see~\cite{Lust:2005dy}. Table~\ref{fsfoura} summarizes the relevant data of the fixed point set. The invariant subtorus under $\theta^2$ is $(x^3,0,x^3,x^6,0,x^6)$, corresponding to the $z^3$ coordinate being invariant.
\begin{table}[h!]\begin{center}
\begin{tabular}{|c|c|c|c|}
\hline
Group el.& Order & Fixed Set& Conj. Classes \cr
\hline
\noalign{\hrule}\noalign{\hrule}
$\theta$& 4     &\ 16 fixed points & 16\cr
$\theta^2$& 2     &\ 4 fixed lines & 4\cr
\hline
\end{tabular}
\caption{Fixed point set for $\IZ_{4}$ orbifold on  $SU(4)^2$.}\label{fsfoura}
\end{center}\end{table}
Figure~\ref{fig:ffixfoura} shows the configuration of the fixed point set in a schematic way.
This orbifold has special properties due to the $SU(4)^2$ lattice which leads to a non--standard volume factor for the fixed torus in $z^3$--direction (see~\cite{Ibanez:1987pj},~\cite{Erler:1992ki}) and changed periodicities of the real lattice. Instead of the usual real lattice shift of one unit, the shift is $1/2$ for the coordinates entering $z^3$. The fixed points in $z^3$ direction do not all lie in the same four $D_3$ planes as usual but the points at $(z^1, z^2)\neq (0,0)$ are shifted by $1/4$.

\subsubsection{The resolved orbifold}
\label{sec:Z4orbSU4SU4}

As we have seen, there are 16 local $\IZ_{4}$--patches which sit in groups of four on one of the four $\IZ_2$--fixed lines. From Appendix~\ref{sec:Z4res} we see that each of the $\IZ_{4}$ patches contributes one exceptional divisor and so does each of the fixed lines, therefore we get $16\cdot1+4\cdot1=20$ exceptional divisors in total. According the labeling of the fixed points, we denote them by $E_{1,\alpha\beta\gamma}$ and $E_{2,\alpha,\beta}$, $\alpha,\beta=1,2$, $\gamma = 1,3,5,7$ for $(\alpha,\beta)=(1,1)$ and $\gamma=2,4,6,8$ otherwise. Since there are no fixed lines without fixed points on them in this example, there are no twisted complex structure moduli.

\begin{figure}[h!]
\begin{center}
\includegraphics[width=50mm]{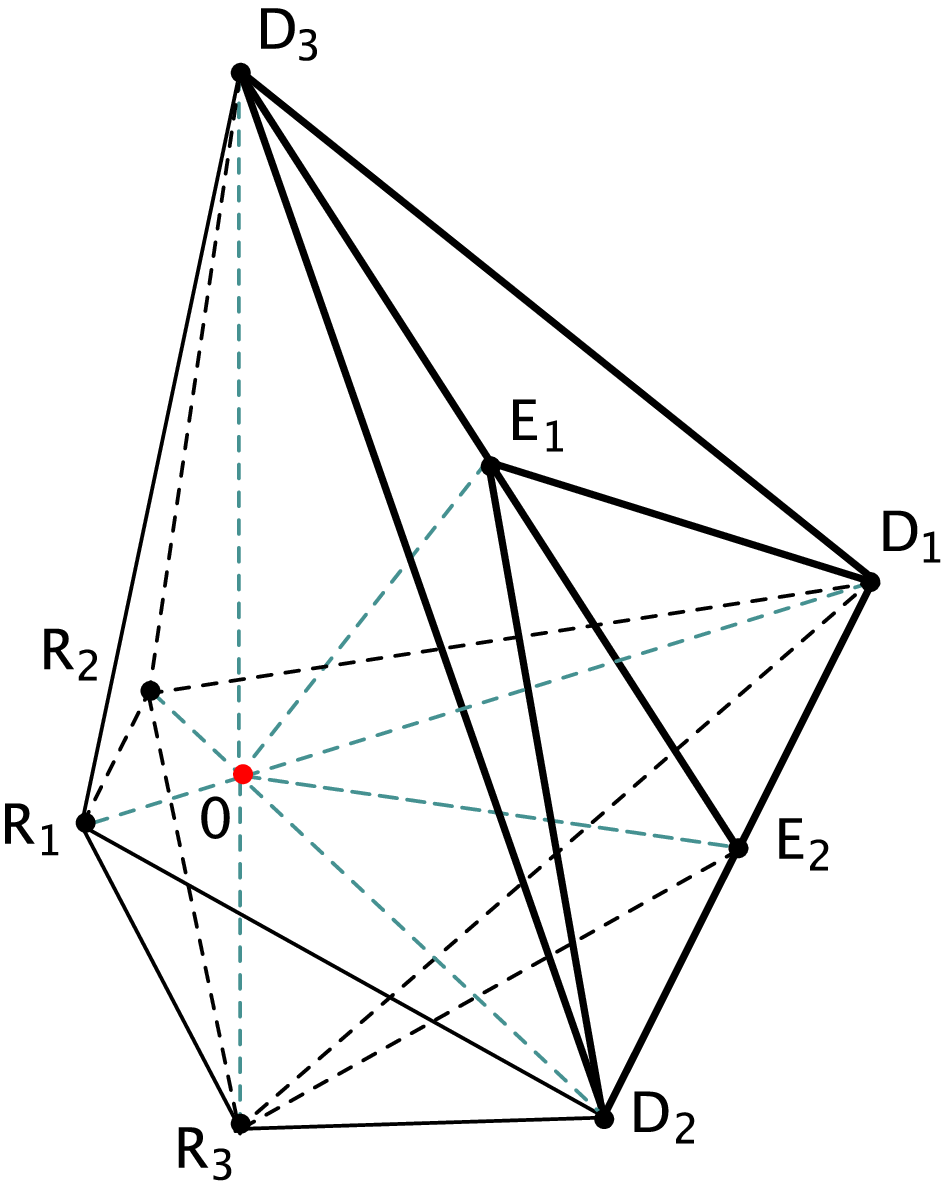}
\caption{The polyhedron $\Delta^{(3)}$ describing the local compactification of the resolution of $\IC^3/\IZ_4$.}
\label{fig:Z4-cpt}
\end{center}
\end{figure}

From the local linear relations~(\ref{lineqfour}) for the $\IC^3/\IZ_4$ patch, we find the following global linear relations~\eqref{eq:Reqglobal}:
\begin{eqnarray}
  \label{eq:Reqfour}
  R_{1} &\sim& 4\,D_{1,1}+\sum_{\gamma=1}^4 \left(E_{1,1,1,2\gamma-1}+E_{1,1,2,2\gamma}\right) + 2\sum_{\beta=1,2}E_{2,1,\beta},\cr
  R_{1} &\sim& 4\,D_{1,2}+\sum_{\beta=1,2}  \sum_{\gamma=1}^4 E_{1,2,\beta,2\gamma}  + 2\sum_{\beta=1,2}  E_{2,2,\beta},\cr
  R_{2} &\sim& 4\,D_{2,1}+\sum_{\gamma=1}^4 \left(E_{1,1,1,2\gamma-1}+E_{1,2,1,2\gamma}\right) + 2\sum_{\alpha=1,2}E_{2,\alpha,1},\cr
  R_{2} &\sim& 4\,D_{2,2}+\sum_{\alpha=1,2} \sum_{\gamma=1}^4 E_{1,\alpha,2,2\gamma} + 2\sum_{\alpha=1,2} E_{2,\alpha,2},\cr
  R_{3} &\sim& 2\,D_{3, 2\gamma-1}+E_{1,1,1,2\gamma-1}, \cr
  R_{3} &\sim& 2\,D_{3, 2\gamma}+E_{1,1,2\gamma} + \sum_{\beta=1,2} E_{1,2,\beta,2\gamma}, 
\end{eqnarray}
where $\gamma=1,...,4$.
To compute the intersection ring, we need to determine the basis for the lattice $N$ in which the auxiliary polyhedron will live. From~(\ref{eq:Reqfour}) we see that $n_1=n_2=4$, and $n_3=2$. Hence we can choose $m_1=m_2=m_3=2$, and the lattice basis is $f_1=(2,0,0)$, $f_2=(0,2,0)$, $f_3=(0,0,2)$. The lattice points of the polyhedron $\Delta^{(3)}$ for the local compactification of the $\IZ_4$ fixed points are
\begin{align}
  \label{eq:Z4poly}
  v_1 &= (-2,0,0), & v_2 &= (0,-2,0), & v_3 &= (0,0,-2), & v_4 &= (8,0,0), & v_5 &= (0,8,0),\notag\\
  v_6 &= (0,0,4), & v_7 &= (2,2,2), & v_8 &= (4,4,0),
\end{align}
corresponding to the divisors $R_1,R_2,R_3,D_1,D_2,D_3,E_1,E_2$ in that order. The polyhedron is shown in Figure~\ref{fig:Z4-cpt}. 
From the intersection ring of the 16 polyhedra and the linear relations~(\ref{eq:Reqfour}) we obtain the following nonvanishing intersection numbers of $X$ in the basis $\{R_i,E_{k\alpha\beta\gamma}\}$ (without the divisors $R_{i\jbar}$):
\begin{align}
  R_1R_2R_3 &=8, & R_{3}E_{2,\alpha\beta}^2 &=-2, & E_{1,\alpha\beta\gamma}E_{2,\alpha\beta}^2&=-2, & E_{1,\alpha,\beta,\gamma}^3&=8, & E_{2,\alpha,\beta}^3&=8.
\end{align}
Next, we discuss the topology of the divisors. The topology of the exceptional divisors $E_{1\alpha\beta\gamma}$ has already been determined in Appendix~\ref{sec:Z4res} to be that of an $\IF_2$. According to Section~\ref{sec:Topology}, the divisors $E_{2,\alpha\beta}$ are of type E\ref{item:E3}) with a single line ending on them (cf. Figure~\ref{fig:frfour}), hence their topology is that of an $\IF_0$. The restriction of the $\IZ_4$ action to $z^1=\zf{1}{\alpha}$ is $\frac{1}{4}(1,2)$ and has eight classes of fixed points. Each of them gets replaced by a $\IP^1$, hence the Euler number of $D_{1,\alpha}$ is $(0-16)/4+8\cdot 2=12$ and the topology of $D_{1,\alpha}$ can be viewed as that of $\Bl{8}\IF_n$. The same holds for $D_{2,\beta}$. For $D_{3,\gamma}$ we start with the restriction of the $\IZ_4$ action to $z^3=\zf{3}{\gamma}$, which is $\frac{1}{4}(1,1)$ on $T^4$. The Euler number of $D_{3,\gamma}$ minus the 16 fixed points is $(0-16)/4=-4$. The fixed points fall into 10 classes. For odd $\gamma$ 3 of these classes are not resolved and the corresponding singularities are replaced by points. At the remaining fixed points we glue in a $\IP^1$ as usual. Therefore the Euler number of $D_{3,\gamma}$, $\gamma$ odd, is $-4+3\cdot 1 + 7\cdot 2 =13$. For even $\gamma$ only one of the ten classes is not blown up and the Euler number is $-4+1+9\cdot 2=15$. We conclude that the topology of $D_{3,\gamma}$ is $\Bl{9}\IF_0$ and $\Bl{11}\IF_0$ for $\gamma$ odd and even, respectively. The topology of $R_1$ and $R_2$ is that of a $T^4$ and $R_3$ is a K3. The second Chern class is
\begin{align}
  \ch_2\cdot E_{1,\alpha\beta\gamma} &= -4, & \ch_2 \cdot E_{2,\alpha\beta} &= -4, & \ch_2 \cdot R_i &= 0, & \ch_2 \cdot R_3 &= 24.
\end{align}

\subsubsection{The orientifold}
\label{sec:Z4orbSU4SU4_O}

The O--plane configuration at the orbifold point is very simple, we have 64 O3--planes, located at the $I_6$ fixed points in each direction. They fall into 22 conjugacy classes, apart from $(0,0,0)$, which is invariant. Under the combination $I_6\,\theta^2$, one O7--plane is fixed located at $z^3=0$ in the $(z^1,z^2)$--plane.

The $h^{1,1}_{-}=6$ of this example is due to the 12 $\IC^3/\IZ_4$ patches which are shifted in the $z^3$--direction by $1/4$, which are mapped to each other by $I_6$ pairwise on each fixed line.

Now we discuss the orientifold of the resolved case. There are two possibilities for $\Ic$ on the $\IC^3/\IZ_4$ patch which lead to a solution with O3-- and O7--planes: 
\begin{eqnarray}(1)\quad {\Ic}(z,y)&=&(-z^1,-z^2,-z^3,y^1,y^2),\label{ZfourOi}\cr
(2)\quad {\Ic}(z,y)&=&(-z^1,-z^2,-z^3,-y^1,-y^2).
\end{eqnarray}
For the choice (1), we solve
\begin{equation}\label{ofouri}{(-z^1,-z^2,-z^3,y^1,y^2)=(\lambda_1\, z^1, \lambda_1\,z^2, \lambda_2\,z^3, \frac{1}{\lambda_2^2}\,y^1,\frac{\lambda_2}{\lambda_1^2}\,y^2),}
\end{equation}
leading to the following two solutions: 
$$y^2=0,\ \lambda_1=\lambda_2=-1,\ \ \ z^3=0,\ \lambda_1=-1,\lambda_2=1.$$ 
This corresponds to an O7--plane located on $E_2$ and one on $D_3$. 
The choice (2) leads to
$$y^1=0,\ \lambda_1=\lambda_2=-1.$$
We will concentrate on the choice (1) for $\Ic$. The restriction of the scaling action to the $\IC^2/\IZ_2$ fixed line which is compatible with the solution $y^2=0$ consists of $D_1,\, D_2$ and $E_2$ is given by setting $\lambda_2=-1$:
\begin{equation}
(-z^1, -z^2, -z^3,y^1,y^2)=(\lambda_1\, z^1, \lambda_1\,z^2,-z^3, y^1,-\frac{1}{\lambda_1^2}\,y^2).
\end{equation}
This obviously leads to the solution
$$y^2=0,\ \lambda_1=-1.$$
In the resolved case, we have thus 4 O7--planes on the $E_{2\alpha\beta}$, and 8 O7--planes on the 8 $D_{3\gamma}$. Four of the $\IC^3/\IZ_4$ patches coincide with location of O3--planes before the blow--up. Since no O3--plane solutions appear in the resolved patches, these O3--planes are not present in the resolved case and we are left with 18 O3--planes located away from the resolved patches.

The modified intersection intersection numbers are
\begin{eqnarray}
&&R_{1}R_{2}R_{3}=4,\ R_{3}E_{2,\alpha\beta}^2=-4,\cr
&&E_{1,\alpha\beta\gamma}E_{2,\alpha\beta}^2=-4,\ E_{1,\alpha,\beta,\gamma}^3=4,\ E_{2,\alpha,\beta}^3=32.
\end{eqnarray}


\subsection{The lattice $SU(2)\times SO(5)\times SU(4)$}
\label{sec:Z4onSU4xSU2xSO5}

\subsubsection{Complex structure and fixed sets}

\begin{figure}[h!]
\begin{center}
\includegraphics[width=85mm]{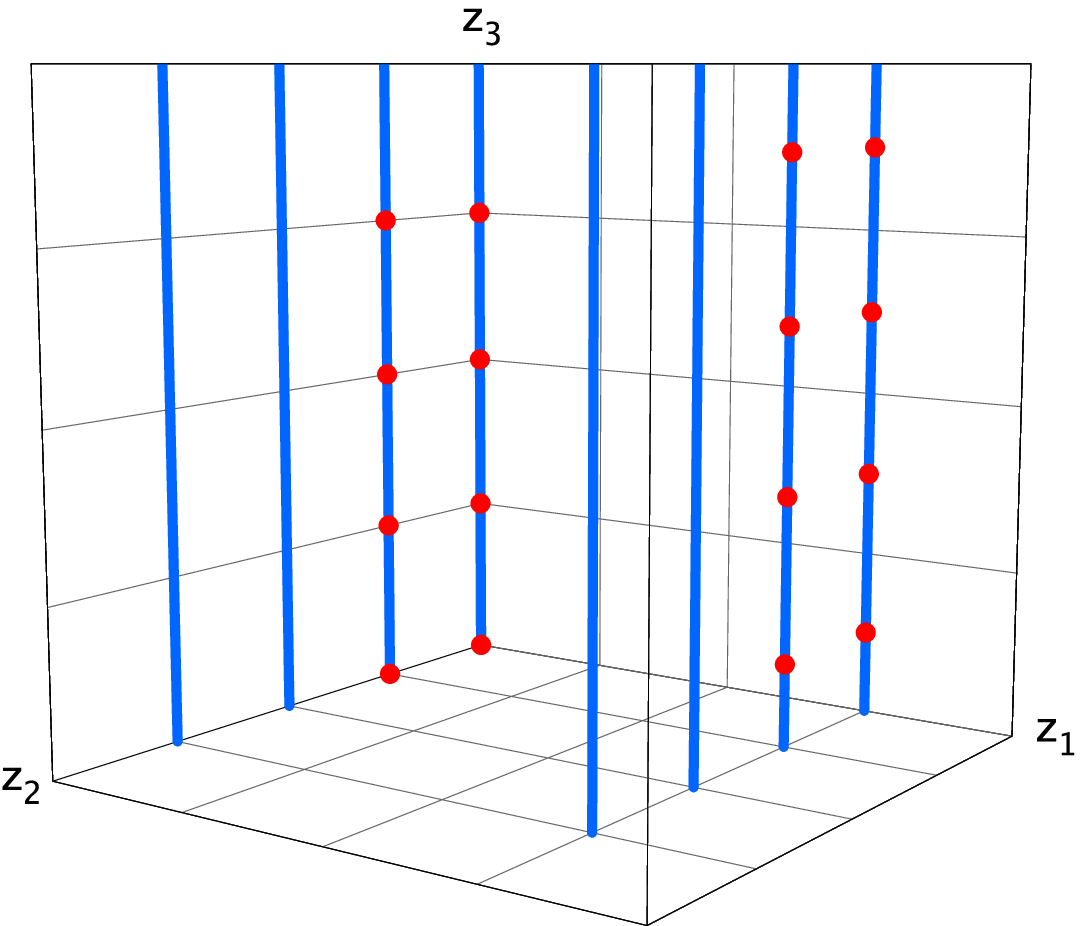}
\caption{Schematic picture of the fixed set configuration of $\IZ_{4}$ on $SU(2)\times SO(5)\times SU(4)$}
\label{fig:ffixfouraa}
\end{center}
\end{figure}

The twist $Q$ has the following action on the root lattice of $SU(2)\times SO(5)\times SU(4)$:
\begin{eqnarray}
Q\ e_1&=&e_2,\quad Q\ e_2=e_3,\quad Q\ e_3=-e_1-e_2-e_3,\cr 
Q\ e_4&=&e_4+2\,e_5,\quad Q\ e_5=-e_4-e_5 ,\quad Q\ e_6=-e_6\ .\end{eqnarray}
The form of metric and anti-symmetric tensor follow from solving the equations
$Q^tg\,Q=g$ and $Q^tb\,Q=b$:
{\arraycolsep1pt
\begin{equation}{
g\!=\!\!\left(\!\!\begin{array}{cccccc}
R_1^2&R_1^2\cos\theta_{23}&x&-R_1R_2\cos\theta_{34}&-R_1R_2\cos\theta_{35}&R_1R_3\cos\theta_{36}\cr
R_1^2\cos\theta_{23}&R_1^2&R_1^2\cos\theta_{23}&y&-y&-R_1R_3\cos\theta_{36}\cr
x&R_1^2\cos\theta_{23}&R_1^2&R_1R_2\cos\theta_{34}&R_1R_2\cos\theta_{35}&R_1R_2\cos\theta_{36}\cr
-R_1R_2\cos\theta_{34}&y&R_1R_2\cos\theta_{34}&2\,R_2^2&-R_2^2&0\cr
-R_1R_2\cos\theta_{35}&-y&R_1R_2\cos\theta_{35}&-R_2^2&R_2^2&0\cr
R_1R_3\cos\theta_{36}&-R_1R_3\cos\theta_{36}&R_1R_3\cos\theta_{36}&0&0&R_3^2\end{array}\!\!\right),
}\end{equation}}
with $x=-R_1^2(1+2\,\cos\theta_{23})$, $y=R_1R_2\,(\cos\theta_{34}+2\,\cos\theta_{35})$. The seven real parameters $R_1^2,\ R_2^2,\ R_3^2, \theta_{23}$, $\theta_{34},\ \theta_{35},\ \theta_{36}$.  
For $b$ we find
\begin{equation}{
b=\left(\begin{array}{cccccc}
0&b_1&0&-b_2&-b_3&b_4\cr
-b_1&0&b_1&b_2+2\,b_3&-b_2-b_3&-b_4\cr
0&-b_1&0&b_2&b_3&b_4\cr
b_2&-b_2-2\,b_3&-b_2&0&b_5&0\cr
b_3&b_2+b_3&-b_3&-b_5&0&0\cr
-b_4&b_4&-b_4&0&0&0\end{array}\right)}\end{equation}
with the five real parameters $b_1,\ b_2,\ b_3,\ b_4,\ b_5$. We see that we get 5 untwisted K\"ahler moduli and one untwisted complex structure modulus in this orbifold.
With the methods discussed in \cite{Lust:2005dy},
we arrive at the following complex structure:
\begin{eqnarray}
z^1&=&\frac{1}{\sqrt2}\,(x^1+i\,x^2-x^3),\cr
z^2&=&x^4+\left(\frac{1}{2}-\frac{i}{2}\right)\,x^5,\cr
z^3&=&\frac{1}{2\sqrt{2\,u_2}}\,(x^1-x^2+x^3+2\,{\Uc}\,x^6).
\end{eqnarray}
with 
\begin{equation}
\Uc=-\frac{R_3}{2\,R_1}\sec\theta_{23}(\cos\theta_{36}+i\,\sqrt{-\cos\theta_{23}-\cos\theta_{36}^2}).
\end{equation}
The five untwisted real 2--forms that are invariant under this orbifold twist are
\begin{eqnarray}
\omega_1&=&dx^1\wedge dx^2+dx^2\wedge dx^3,\cr 
\omega_2&=&-dx^1\wedge dx^4+dx^2\wedge dx^4-dx^2\wedge dx^5+dx^3\wedge dx^4,\cr
\omega_3&=&-dx^1\wedge dx^5+2\,dx^2\wedge dx^4-dx^2\wedge dx^5+dx^3\wedge dx^5,\cr
\omega_4&=&dx^1\wedge dx^6-dx^2\wedge dx^6+dx^3\wedge dx^6,\cr
\omega_5&=&dx^4\wedge dx^5.
\end{eqnarray}
The fixed point set consists of eight $\IC^2/\IZ_2$ fixed lines which lie at $\zf{1}{\alpha}=0,\frac{1}{2}(1+i)$, $\alpha=1,2$, and $\zf{2}{\beta}=0,\frac{1}{2},\frac{i}{2},\frac{1}{2}(1+i)$, $\beta=1,..,4$. On four of them there are four $\IC^3/\IZ_4$ fixed points. For $(\zf{1}{1},\zf{2}{\beta})=(0,0),(0,\frac{1}{2})$ they are at $\zf{3}{2\gamma-1} = 0,\frac{1}{2},\frac{U}{2},\frac{1}{2}(1+U)$, $\gamma=1,\dots,4$, while for $(\zf{1}{2},\zf{2}{\beta})=(\frac{1}{2}(1+i),0),(\frac{1}{2}(1+i),\frac{1}{2})$  they are at $\zf{3}{2\gamma} = \frac{1}{4},\frac{3}{4},\frac{1}{4}(1+2U),\frac{1}{4}(3+2U)$, $\gamma=1,\dots,4$. Here, $U = \Uc$, see~\cite{Lust:2005dy}. Table \ref{fsfouraa} summarizes the relevant data of the fixed sets. The invariant subtorus under $\theta^2$ is $(x^3,0,x^3,0,0,x^6)$, corresponding to the $z^3$ coordinate being invariant.
\begin{table}[h!]\begin{center}
\begin{tabular}{|c|c|c|c|}
\hline
Group el.& Order &Fixed Set& Conj. Classes \cr
\hline
\noalign{\hrule}\noalign{\hrule}
$\theta$& 4     &16  fixed points & 16\cr
$\theta^2$& 2   &8 fixed lines & 6\cr
\hline
\end{tabular}
\caption{Fixed point set for $\IZ_{4}$ orbifold on  $SU(2)\times SO(5)\times SU(4)$.}\label{fsfouraa}
\end{center}\end{table}
Figure~\ref{fig:ffixfouraa} shows the configuration of the fixed point set in a schematic way. 
Since in this case only one $SU(4)$ factor is present in the lattice, the periodicity change occurs only in one real direction and only the fixed points with $z^1\neq0$ are shifted by $1/4$.

\subsubsection{The resolved orbifold}
\label{sec:Z4orbSU4SU2SO5}

As we have seen, there are 16 local $\IC^3/\IZ_{4}$ patches which sit in groups of four on a $\IC^2/\IZ_2$ fixed line. They yield the same exceptional divisors as in Section~\ref{sec:Z4onSU4xSU4}, however, according the different labeling of the fixed points here, we denote them by $E_{1,\alpha\beta\gamma}$ and $E_{2,\alpha,\beta}$, $\alpha,\beta=1,2$, $\gamma = 1,3,5,7$ for $\alpha=1$ and $\gamma=2,4,6,8$ for $\alpha=2$. The remaining four $\IC^2/\IZ_2$ fixed lines fall into two equivalence classes $E_{2,\alpha,3}=\Et_{2,\alpha,3}+\Et_{2,\alpha,4}$, $\alpha=1,2$, where $\Et_{2,\alpha,\beta}$ are the divisors on the cover. Therefore we get $16\cdot1+6\cdot1=22$ exceptional divisors in total. These two classes of $\IC^2/\IZ_2$ fixed lines do not have fixed points on them, so by~\eqref{eq:h21tw} we have $h^{2,1}_{\rm twist.}=2$.

From the local linear relations~(\ref{lineqfour}), we find the following global relations:
\begin{eqnarray}
  \label{eq:Reqfour2}
  R_{1} &\sim& 4\,D_{1,1}+\sum_{\beta=1,2} \sum_{\gamma=1}^4 E_{1,1,\beta,2\gamma-1}  + 2\sum_{\beta=1}^3 E_{2,1,\beta},\cr
  R_{1} &\sim& 4\,D_{1,2}+\sum_{\beta=1,2} \sum_{\gamma=1}^4 E_{1,2,\beta,2\gamma}    + 2\sum_{\beta=1}^3 E_{2,2,\beta},\cr
  R_{2} &\sim& 4\,D_{2,\beta}+\sum_{\gamma=1}^4 \left(E_{1,1,\beta,2\gamma-1}+E_{1,2,\beta,2\gamma}\right) + 2\sum_{\alpha=1,2}E_{2,\alpha,\beta},\cr
  R_{2} &\sim& 2\,D_{2, 3} + \sum_{\alpha=1}^2 E_{2,\alpha,3},\cr
  R_{3} &\sim& 2\,D_{3, 2\gamma-1}+\sum_{\beta=1,2} E_{1,1,\beta,2\gamma-1}, \cr
  R_{3} &\sim& 2\,D_{3, 2\gamma}  +\sum_{\beta=1,2} E_{1,2,\beta,2\gamma}, 
\end{eqnarray}
where $\alpha,\beta=1,2,\ \gamma=1,...,4$.
The polyhedron for the $\IZ_2$ fixed lines is obtained from~(\ref{eq:Z4poly}) by dropping $v_7$. We obtain the following nonvanishing intersection numbers of $X$ in the basis $\{R_i,E_{k\alpha\beta\gamma}\}$ (without the divisors $R_{i\jbar}$):
\begin{align}
  R_1R_2R_3 &=8, & R_{3}E_{2,\alpha\beta}^2 &=-2, & E_{1,\alpha\beta\gamma}E_{2,\alpha\beta}^2&=-2, \notag\\
  E_{1,\alpha,\beta,\gamma}^3&=8, & E_{2,\alpha,\beta}^3&=8, & E_{2,\alpha,3}^2R_3 &= -4,
\end{align}
for $\alpha,\beta=1,2$ and all $\gamma$. The topology of those divisors which were already present in the model in Appendix~\ref{sec:Z4orbSU4SU4} does not change except for $D_{3\gamma}$. The difference is that for both even and odd $\gamma$ two of the ten classes of fixed points of $T^4/\IZ_4$ are not resolved and the corresponding singularities are replaced by points. The Euler number therefore is $-4 + 2\cdot 1 + 8\cdot 2 = 14$ and the topology of $D_{3,\gamma}$ is $\Bl{10}\IF_0$. According to Section~\ref{sec:Topology} the new divisors $E_{2,\alpha,3}$ are of type E\ref{item:E2}), hence their topology is $\IP^1\times T^2$. By a similar argument as for $D_{1,2}$ in Section~\ref{sec:Z6I_G2xSU3xSU3} we find that the topology of $D_{2,3}$ is also that of a $\IP^1 \times T^2$. Finally, the second Chern class is
\begin{align}
  \ch_2\cdot E_{1,\alpha\beta\gamma} &= -4, & \ch_2 \cdot E_{2,\alpha\beta} &= -4, & \ch_2\cdot E_{2,\alpha,3} &= 0, & \ch_2 \cdot R_i &= 0, \notag\\ \ch_2 \cdot R_3 &= 24.
\end{align}

\subsubsection{The orientifold}
\label{sec:Z4orbSU4SU2SO5_O}

In this example, we have at the orbifold point 64 O3--planes which fall into 28 conjugacy classes under the orbifold group. Under $I_6\,\theta^2$, we have two O7--planes in the $(z^1,z^2)$--plane located at $z^3=0$ and $z^3=\tfrac{1}{2}U$.
$h^{1,1}_{-}=4$ is due to the 8 $\IC^3/\IZ_4$ patches which are shifted in the $z^3$--direction by $1/4$ and are mapped pairwise onto each other under $I_6$. There are two classes of fixed lines with fixed points which are invariant under the orientifold action, hence $h^{2,1}_+=2$.

The local involution on the resolved patches is the same as in Appendix~\ref{sec:Z4orbSU4SU4_O}. Since 8 of the O3--planes coincide with the local patches, they are not present in the resolved case. This means that we are left with 20 O3--planes, which are located away from the fixed points. Choosing the involution (1) of~(\ref{ZfourOi}), we have 4 O7--planes on the $E_{2\alpha\beta}$ and 8 on the $D_{3\gamma}$--planes.

The modified intersection numbers are
\begin{align}
  R_1R_2R_3 &=4, & R_{3}E_{2,\alpha\beta}^2 &=-4, & E_{1,\alpha\beta\gamma}E_{2,\alpha\beta}^2&=-4, \notag\\
  E_{1,\alpha,\beta,\gamma}^3&=4, & E_{2,\alpha,\beta}^3&=32, & E_{2,\alpha,3}^2R_3 &= -16,
\end{align}
for $\alpha,\beta=1,2$ and all $\gamma$.


\subsection{The lattice $SU(2)^2\times SO(5)^2$}
\label{sec:Z4onSU2xSU2xSU5xSU5}

\subsubsection{Complex structure and fixed sets}

\begin{figure}[h!]
\begin{center}
\includegraphics[width=85mm]{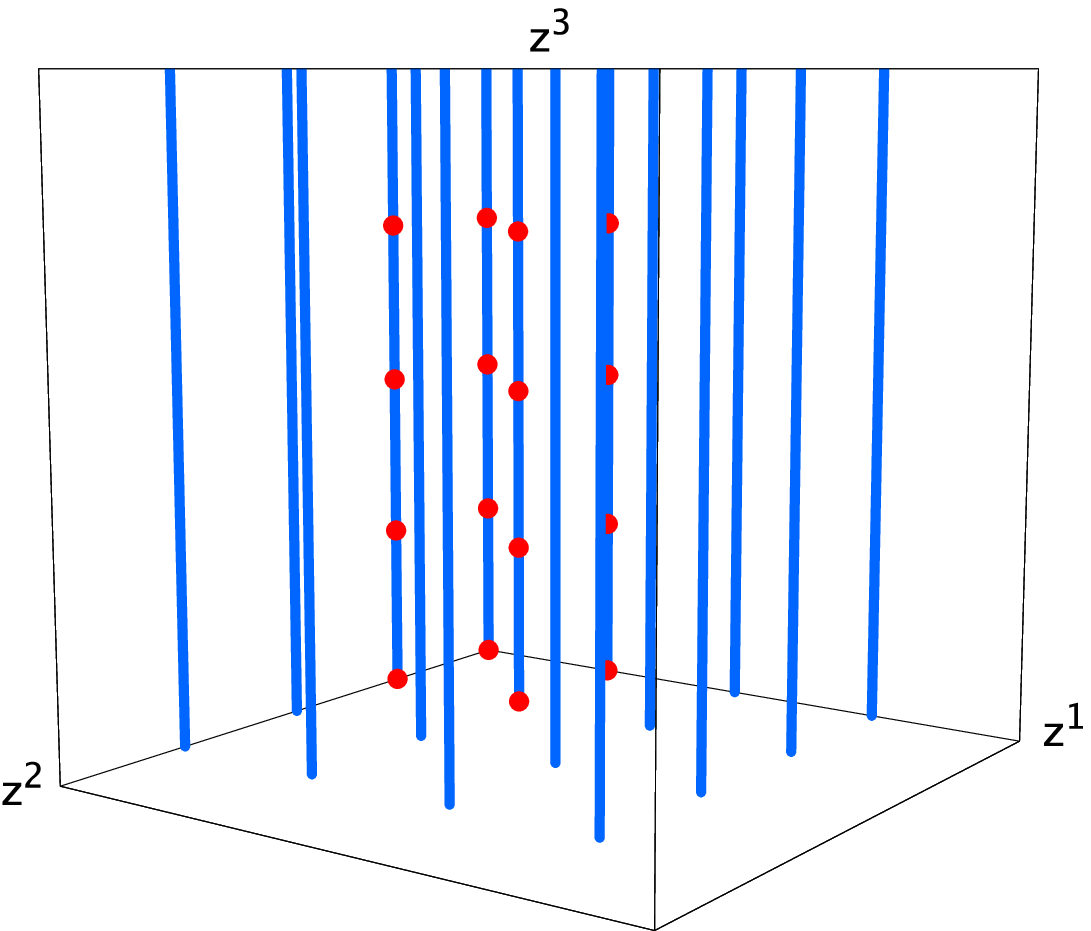}
\caption{Schematic picture of the fixed set configuration of $\IZ_{4}$ on $SU(2)^2\times SO(5)^2$}\label{ffixfouraaa}
\end{center}
\end{figure}

On the root lattice of $SU(4)^2$, the twist $Q$ has the following action:
\begin{eqnarray}
Q\ e_1&=&e_1+2\,e_2,\quad Q\ e_2=-e_1-e_2,\quad Q\ e_3=e_3+2\,e_4,\cr 
Q\ e_4&=&-e_3-e_4,\quad Q\ e_5=-e_5 ,\quad Q\ e_6=-e_6\ .\end{eqnarray}
The form of metric and anti-symmetric tensor follow from solving the equations
$Q^tg\,Q=g$ and $Q^tb\,Q=b$:
{\arraycolsep3pt
\begin{equation}{
g\!=\!\left(\!\!\begin{array}{cccccc}
2R_1^2&-R_1^2&2R_1R_2\cos\theta_{24}&x&0&0\cr
-R_1^2&R_1^2&R_1R_2\cos\theta_{23}&R_1R_2\cos\theta_{24}&0&0\cr
2R_1R_2\cos\theta_{24}&R_1R_2\cos\theta_{23}&2\,R_2^2&-R_2^2&0&0\cr
x&R_1R_2\cos\theta_{24}&-R_2^2&R_2^2&0&0\cr
0&0&0&0&R_3^2&R_3R_4\cos\theta_{56}\cr
0&0&0&0&R_3R_4\cos\theta_{56}&R_4^2\end{array}\!\!\right),
}\end{equation}}
with $x=-R_1R_2\,(\cos\theta_{23}+2\,\cos\theta_{24})$. The seven real parameters $R_1^2,\ R_2^2,\ R_3^2, \ R_4^2,\ \theta_{23}$, $\theta_{24},\ \theta_{56}$.  
For $b$ we find
\begin{equation}{
b=\left(\begin{array}{cccccc}
0&b_1&2\,b_4&-b_4-2\,b_5&0&0\cr
-b_1&0&b_4&b_5&0&0\cr
-2\,b_4&-b_4&0&b_2&0&0\cr
-b_4-2\,b_5&-b_5&-b_2&0&0&0\cr
0&0&0&0&0&b_3\cr
0&0&0&0&-b_3&0\end{array}\right)}\end{equation}
with the five real parameters $b_1,\ b_2,\ b_3,\ b_4,\ b_5$. We see that we get 5 untwisted K\"ahler moduli and one untwisted complex structure modulus in this orbifold.
With the methods discussed in \cite{Lust:2005dy},
we arrive at the following complex coordinates:
\begin{eqnarray}
z^1&=&x^1+\left(\frac{1}{2}-\frac{i}{2}\right)\,x^2,\cr
z^2&=&x^3+\left(\frac{1}{2}-\frac{i}{2}\right)\,x^4,\cr
z^3&=&\frac{1}{\sqrt{2\,{\rm Im}\,{\Uc}}}\,(x^5+{\Uc}\,x^6),
\end{eqnarray}
with  $\Uc=\frac{R_4}{R_3}\,e^{i\theta_{56}}$.
The five untwisted real 2--forms that are invariant under this orbifold twist are
\begin{eqnarray}
\omega_1&=&dx^1\wedge dx^2,\quad \omega_2=dx^3\wedge dx^4,\quad \omega_3=dx^5\wedge dx^6,\cr
\omega_4&=&-dx^1\wedge dx^4+dx^2\wedge dx^3,\cr
\omega_5&=&2\,dx^1\wedge dx^5-2\,dx^1\wedge dx^4+dx^2\wedge dx^4.
\end{eqnarray}
The fixed point set consists of 16 $\IC^2/\IZ_2$ fixed lines which lie at $\zf{1}{\alpha}=0,\frac{1}{2},\frac{i}{2},\frac{1}{2}(1+i)$, $\alpha=1,\dots,4$, and $\zf{2}{\beta}=0,\frac{1}{2},\frac{i}{2},\frac{1}{2}(1+i)$, $\beta=1,..,4$. On those with $\alpha=1,2$, $\beta=1,2$ there are four $\IC^3/\IZ_4$ fixed points which lie at $\zf{3}{\gamma} = 0,\frac{1}{2},\frac{U}{2},\frac{1}{2}(1+U)$, $\gamma=1,\dots,4$. Here, $U = \Uc$, see~\cite{Lust:2005dy}. Table \ref{fsfouraaa} summarizes the relevant data of the fixed sets. The invariant subtorus under $\theta^2$ is $(0,0,0,0,x^5,x^6)$, corresponding to the $z^3$ coordinate being invariant.
\begin{table}[h!]\begin{center}
\begin{tabular}{|c|c|c|c|}
\hline
Group el.& Order &Fixed Set& Conj. Classes \cr
\hline
\noalign{\hrule}\noalign{\hrule}
$\theta$&4    &16 fixed points & 16\cr
$\theta^2$& 2    &16 fixed lines & 10\cr
\hline
\end{tabular}
\caption{Fixed point set for $\IZ_{4}$ orbifold on  $SU(2)^2\times SO(5)^2$.}
\label{fsfouraaa}
\end{center}\end{table}
Figure \ref{ffixfouraaa} shows the configuration of the fixed set point set in a schematic way. 

\subsubsection{The resolved orbifold}
\label{sec:Z4orbSU2SU52}

As we have seen, there are 16 local $\IC^3/\IZ_{4}$ patches which sit in groups of four on a $\IC^2/\IZ_2$ fixed line. They yield the same exceptional divisors as in Section~\ref{sec:Z4onSU4xSU2xSO5}, however, according the different labeling of the fixed points here, we denote them by $E_{1,\alpha\beta\gamma}$ and $E_{2,\alpha,\beta}$, $\alpha,\beta=1,2$, $\gamma = 1,\dots,4$. The remaining 12 $\IC^2/\IZ_2$ fixed lines fall into six equivalence classes. 
The invariant divisors are 
\begin{align}
  E_{2,1} &= \Et_{2,1,1}, & E_{2,2} &= \Et_{2,1,2}, & E_{2,3} &= \Et_{2,1,3} + \Et_{2,1,4}, \notag\\ 
  E_{2,4} &= \Et_{2,2,1}, & E_{2,5} &= \Et_{2,2,2}, & E_{2,6} &= \Et_{2,2,3} + \Et_{2,2,4}, \notag\\
  E_{2,7} &= \Et_{2,3,1} + \Et_{2,4,1}, & E_{2,8} &= \Et_{2,3,2} + \Et_{2,4,2}, & E_{2,9} &= \Et_{2,3,3} + \Et_{2,4,4}, \notag\\
  E_{2,10} &= \Et_{2,3,4} + \Et_{2,4,3}.
\end{align}
where $\Et_{2,\alpha,\beta}$ are the representatives on the cover. Therefore we get $16\cdot1+10\cdot1=26$ exceptional divisors in total. The latter six classes of $\IC^2/\IZ_2$ fixed lines do not have fixed points on them, so by~\eqref{eq:h21tw} we have $h^{2,1}_{\rm twist.}=6$.

From the local linear relations~(\ref{lineqfour}), we find the following global relations:
\begin{align}
  \label{eq:Reqfour3}
  R_{1} &= 4\,D_{1, \alpha}+\sum_{\beta=1}^2\sum_{\gamma=1}^4 E_{1,{\alpha\beta\gamma}}+2\sum_{\mu=1}^3 E_{2,3\alpha-3+\mu}, && \alpha=1,2, \notag\\
  R_{1} &= 2\,D_{1, 3} + \sum_{\mu=7}^{10} E_{2,\mu},\notag\\
  R_{2} &= 4\,D_{2, \beta}+\sum_{\alpha=1}^2\sum_{\gamma=1}^4 E_{1,{\alpha\beta\gamma}}+2\sum_{\mu=0,3,6}E_{2,\mu+\beta},&& \beta=1,2, \notag\\
  R_{2} &= 2\,D_{2, 3} + \sum_{\mu=3,6,9,10} E_{2,\mu},\notag\\
  R_{3} &= 2\,D_{3, \gamma}+\sum_{\alpha=1}^2 \sum_{\beta=1}^2 E_{1,{\alpha\beta\gamma}}, && \gamma=1,\dots,4, 
\end{align}
The polyhedron the $\IZ_2$ fixed lines are is obtained from~(\ref{eq:Z4poly}) by dropping $v_7$. We obtain the following nonvanishing intersection numbers of $X$ in the basis $\{R_i,E_{k\alpha\beta\gamma}\}$ (without the divisors $R_{i\jbar}$):
\begin{align}
  R_1R_2R_3 &=8, & R_{3}E_{2,\mu}^2 &=-2, & E_{1,\alpha\beta\gamma}E_{2,\mu}^2&=-2, & \notag\\
  E_{1,\alpha,\beta,\gamma}^3&=8, & E_{2,\mu}^3&=8, & & \notag\\
  \intertext{for $\mu=1,2,4,5$ and} 
  E_{2,\mu}^2R_3 &= -4,
\end{align}
for $\mu=3,6,\dots,10$ and $\alpha,\beta=1,2$. The topology of those divisors which were already present in the model in Appendix~\ref{sec:Z4orbSU4SU2SO5} does not change except for $D_{3\gamma}$. The difference is that all of the ten classes of fixed points of $T^4/\IZ_4$ are resolved. The Euler number therefore is $-4 + 10\cdot 2 = 16$ and the topology of $D_{3,\gamma}$ is $\Bl{12}\IF_0$. All the new divisors $E_{2,\mu}$ and $D_{3,\beta}$ have the topology of a $\IP^1 \times T^2$. Finally, the second Chern class is
\begin{align}
  \ch_2\cdot E_{1,\alpha\beta\gamma} &= -4, & \ch_2 \cdot R_i &= 0, & \ch_2 \cdot R_3 &= 24, & \ch_2 \cdot E_{2,\mu} &= -4, \notag\\
  \intertext{for $\mu=1,2,4,5$ and} 
  \ch_2\cdot E_{2,\mu} &= 0,
\end{align}
for $\mu=3,6,\dots,10$.

\subsubsection{The orientifold}
\label{sec:Z4orbSU2SO52_O}

In this example, we have at the orbifold point 64 O3--planes which fall into 40 conjugacy classes under the orbifold group. Under $I_6\,\theta^2$, we have four O7--planes in the $(z^1,z^2)$--plane located at $z^3=0,\,\tfrac{1}{2},\,\tfrac{1}{2}\,{\Uc}^3$ and $z^3=\tfrac{1}{2}\,(1+{\Uc}^3)$.
$h^{1,1}_{-}=0$ since all $\IC^3/\IZ_4$ patches are fixed under $I_6$. There are six classes of fixed lines with fixed points which are invariant under the orientifold action, hence $h^{2,1}_+=6$.

The local involution on the resolved patches is the same as for the $SU(4)^2$--lattice in Appendix~\ref{sec:Z4orbSU4SU4_O}. Since all of the local patches coincide with the locations of the O3--planes, these 16 O3--planes are not present in the resolved case. This means that we are left with 24 O3--planes, which are located away from the fixed points. Choosing the involution (1) of (\ref{ZfourOi}), we have 4 O7--planes on the $E_{2\alpha\beta}$ and 4 on the $D_{3\gamma}$--planes.

The modified intersection numbers are
\begin{align}
  R_1R_2R_3 &=4, & R_{3}E_{2,\mu}^2 &=-4, & E_{1,\alpha\beta\gamma}E_{2,\mu}^2&=-4, & \notag\\
  E_{1,\alpha,\beta,\gamma}^3&=4, & E_{2,\mu}^3&=32, & & \notag\\
  \intertext{for $\mu=1,2,4,5$ and} 
  E_{2,\mu}^2R_3 &= -16,
\end{align}
for $\mu=3,6,\dots,10$ and $\alpha,\beta=1,2$.


\section{The $\IZ_{6-II}$ orbifolds}

\subsection{The lattice $SU(2)\times SU(6)$}
\label{sec:Z6IIonSU2xSU6}

\subsubsection{The resolved orbifold}

\begin{figure}[h!]
  \begin{center}
  \includegraphics[width=85mm]{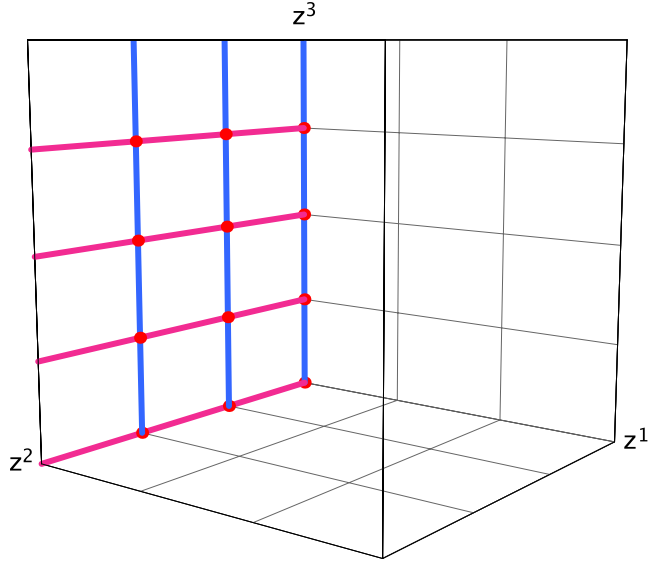}
  \caption{Schematic picture of the fixed set configuration of $\IZ_{6-II}$ on $SU(2)\times SU(6)$}
  \label{fig:ffixsixiiaa}
  \end{center}
\end{figure}

As for $\IZ_{6-I}$ in~\ref{sec:Z6I_G2xSU3xSU3}, we need to look again only at the $\theta$-, $\theta^2$- and $\theta^3$-twisted sectors. The action of the twist $\theta$ on the lattice $SU(2)\times SU(6)$ was given in (A.29) and the resulting complex structure in (A.36) of~\cite{Lust:2005dy} with $\Uc^3=U^3$. We denote the fixed points in each direction as follows: $\zf{1}{1}=0,\ \zf{2}{1}=0,\ \zf{2}{2}=\frac{1}{\sqrt3} \,e^{\pi i/6},\ \zf{2}{3}=1+i/\sqrt3,\ \zf{3}{1}=0,\ \zf{3}{2}=\frac{1}{2}\, U^3,\ \zf{3}{3}\frac{1}{2},\ \zf{3}{4}=\frac{1}{2}(1+U^3)$. They are left invariant under all twists, so that all the conjugacy classes contain a single element. Table~\ref{tab:fssixiia} summarizes the relevant data of the fixed point sets. The invariant subtori under $\theta^2$ and $\theta^3$ are $(x^5,0,x^5,0,x^5,x^6)$ and $(x^4,x^5,0,x^4,x^5,0)$ corresponding to complex coordinates $z^3$ and $z^2$ being invariant, respectively,
\begin{table}[h!]
  \begin{center}
  \begin{tabular}{|c|c|c|c|}
    \hline
    Group el.& Order & Fixed Set & Conj. Classes \cr
    \hline
    \noalign{\hrule}\noalign{\hrule}  
    $\theta  $ & 6     &12\ {\rm fixed\ points} &\ 12\cr
    $\theta^2$ & 3      &3\ {\rm fixed\ lines} &\ 3\cr
    $\theta^3$ & 2      &4\ {\rm fixed\ lines} &\ 4\cr
    \hline
  \end{tabular}
  \caption{Fixed point set for $\IZ_{6-II}$ orbifold on $SU(2)\times SU(6)$.}
  \label{tab:fssixiia}
  \end{center}
\end{table}
while Figure~\ref{fig:ffixsixiiaa} shows the configuration of the fixed point sets in a schematic way.

We see that in this model, there are 3 lines fixed by the order 3 element and 4 lines fixed by the order 2 element. They intersect in 12 fixed points at which we have a $\IC^3/\IZ_{6-II}$ patch. From the toric diagram of its resolution in Figure~\ref{fig:fsixii}, we get one compact divisor $E_1$ each, which we label them $E_{1,\alpha\beta\gamma}$, with $\alpha=1,\,\beta=1,2,3,\ \gamma=1,...,4$. The divisor $E_3$ is identified with the exceptional divisor on the resolved $\IC^2/\IZ_2$ patch, therefore there are four of them: $E_{3,\alpha\gamma}$. Of the divisors $E_2,\ E_4$, we get three each: $E_{2,\alpha\beta},\ E_{4,\alpha\beta}$. They are identified with the two exceptional divisors of the $\IC^2/\IZ_3$  patch. Since in this lattice $\alpha=1$ is the only value, we suppress the label $\alpha$ in the following. In total, there are $12\cdot 1+3\cdot 2+4\cdot 1=22$ exceptional divisors. 
There are 8 fixed planes with their associated divisors: $D_1,\,D_{2\beta},\ \beta=1,2,3,\, D_{3\gamma},\ \gamma=1,...,4$. On this lattice, there are no fixed lines without fixed points on them, hence by~\eqref{eq:h21tw} we have $h^{2,1}_{\rm twist.}=0$.

\begin{figure}[h!]
\begin{center}
\includegraphics[width=50mm]{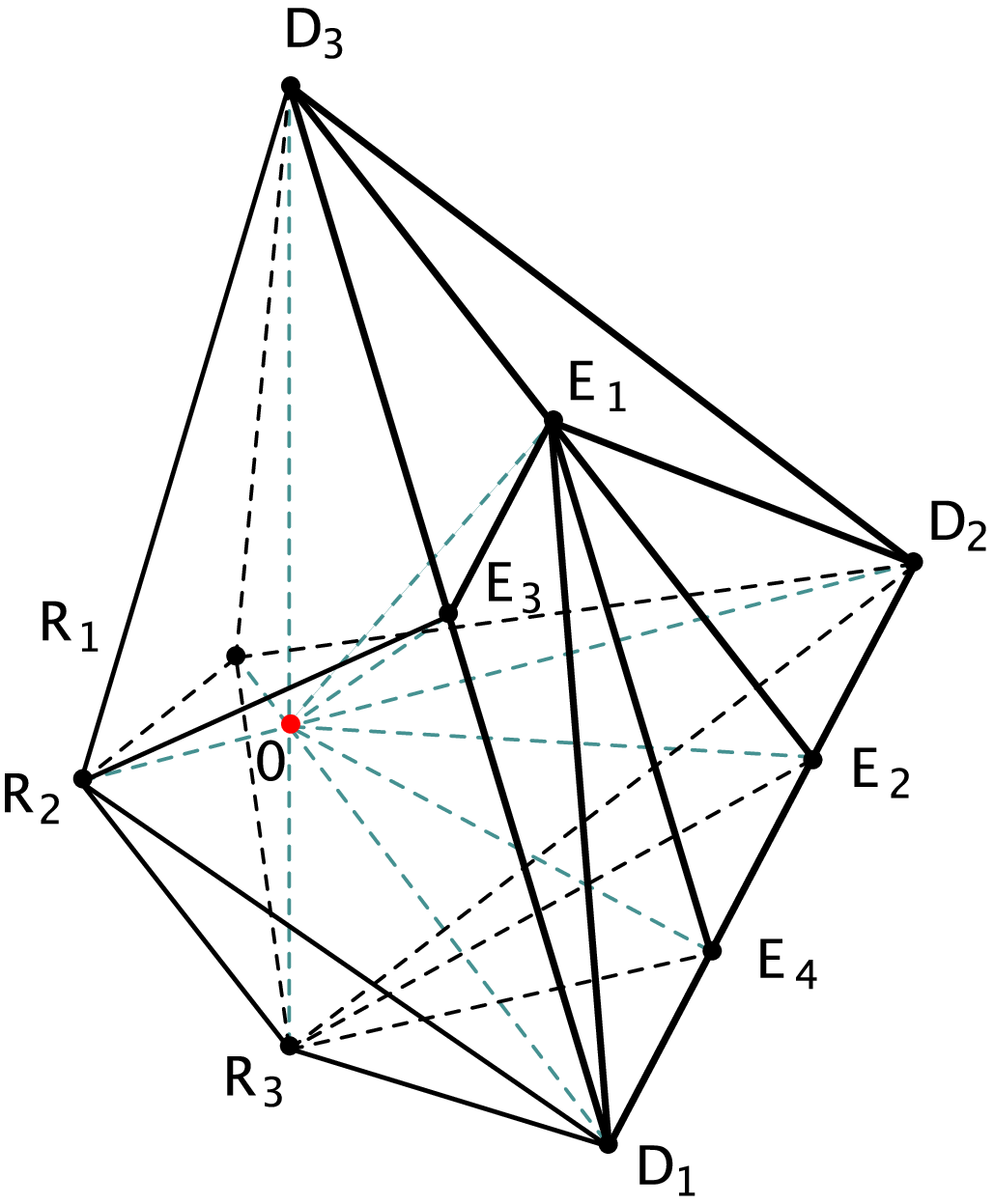}
\caption{The polyhedron $\Delta^{(3)}$ describing the local compactification of the resolution of $\IC^3/\IZ_{6-II}$.}
\label{fig:Z6II-cpt}
\end{center}
\end{figure}

From the local linear equivalences~(\ref{lineqsixii}) for the $\IC^3/\IZ_{6-II}$ patch, the global relations~\eqref{eq:Reqglobal} are constructed:
\begin{align}
  \label{eq:globalrelsixiia}
  R_1&=6\,D_{{1}}+3\,\sum_{\gamma=1}^4 E_{{3,\gamma}}+\sum_{\beta=1}^3\sum_{\gamma=1}^4E_{{1,\beta\gamma}}+\sum_{\beta=1}^3[\,2\,E_{{2,\beta}}+4\,E_{{4,\beta}}],\cr
  R_2&=3\,D_{{2,\beta}}+\sum_{\gamma=1}^4E_{{1,\beta\gamma}}+2\,E_{{2,\beta}}+E_{{4,\beta}}, && \beta=1,2,3 \ \ \ \ \ 
\end{align}
\begin{align}\nonumber
  R_3&=2\,D_{{3,\gamma}}+\sum_{\beta=1}^3E_{{1,\beta\gamma}}+E_{{3,\gamma}}, && \gamma=1,\dots,4.
\end{align}
To compute the intersection ring, we determine the basis for the lattice $N$ from~\eqref{eq:globalrelsixiia} to be $f_1=(1,0,0)$, $f_2=(0,2,0)$, $f_3=(0,0,3)$. The lattice points of the polyhedron $\Delta^{(3)}$ for the local compactification of the $\IZ_{6-II}$ fixed points are
\begin{align}
  \label{eq:Z6poly}
  v_1 &= (-1,0,0), & v_2 &= (0,-2,0), & v_3 &= (0,0,-3), & v_4 &= (6,0,0), & v_5 &= (0,6,0),\notag\\
  v_6 &= (0,0,6), & v_7 &= (1,2,3), & v_8 &= (2,4,0), &v_9&=(3,0,3), & v_{10} &= (4,2,0), 
\end{align}
corresponding to the divisors $R_1,R_2,R_3,D_1,D_2,D_3,E_1,E_2,E_3,E_4$ in that order. The polyhedron is shown in Figure~\ref{fig:Z6II-cpt}. 
From the intersection ring of the 12 polyhedra and the linear relations~\eqref{eq:globalrelsixiia} we obtain the following nonvanishing intersection numbers of $X$ in the basis $\{R_i,E_{k\alpha\beta\gamma}\}$:
\begin{align}
  R_1R_2R_3&=6, & R_2E_{3,\gamma}^2 &= -2, & R_3E_{2,\beta}^2 &= -2, & R_3E_{4,\beta} &= -2, \notag\\
  R_3E_{2,\beta}E_{4,\beta} &= 1, & E_{1,\beta\gamma}^3&=6, & E_{2,\beta}^3& =8, & E_{3,\gamma}^3&=8, \notag\\
  E_{4,\beta}^3&=8, & E_{1,\beta\gamma}E_{2\beta}^2 &= -1, & E_{1,\beta\gamma}E_{3\gamma}^2 &= -1, &  E_{1,\beta\gamma}E_{4\beta}^2 &= -1, \notag\\
  E_{1,\beta\gamma}E_{2,\beta}E_{4\beta} &= 1, &  E_{2,\beta}^2E_{4,\beta} &= -2.
\end{align}
Next, we discuss the topology of the divisors. For the moment we still only consider the triangulation a). The topology of the exceptional divisors has been determined in Appendix~\ref{sec:Z6IIres}: $E_{1,\beta\gamma}$ was found to be $\Bl{2}\IF_1$. The remaining exceptional divisors are all of type~E\ref{item:E3}), hence the basic topology is $\IF_0$. Looking at Figure~\ref{fig:fsixii} we see that in the toric diagram of triangulation a) there is only one line ending in each of $E_2$, $E_3$, and $E_4$, which corresponds to the exceptional curve of $\IF_0$. Therefore there is no additional blow--up, and each of $E_{2\beta}$, $E_{3\gamma}$, and $E_{4\beta}$ has the topology of a $\IF_0$. 

The topology of $D_1$ is seen as follows: The action of $\frac{1}{6}(2,3)$ on $T^4$ factorizes and the topology of $D_1$ minus the fixed point set is that of $(T^2/\IZ_3 \setminus \{3 \ {\rm pts} \}) \times (T^2/\IZ_2 \setminus \{4 \ {\rm pts} \})$. Looking again at the toric diagram in Figure~\ref{fig:fsixii} we see that there is one line ending in $D_1$, hence the 12 singular points are replaced by a $\IP^1$. The topology of $D_1$ therefore is that of $\Bl{12}\IF_0$, and its Euler number is 16. (For the other triangulations there is no line ending in $D_1$, hence there is no blow--up and the topology is that of $\IF_0$.) For $D_{2\beta}$ the action of $\frac{1}{6}(1,3)$ on $T^4$ yields the $\IZ_3$ fixed line with 3 $\IZ_{6-II}$ fixed points on top of it that we see in Figure~\ref{fig:ffixsixiiaa}. In addition, there are 2 more $\IZ_3$ fixed lines which fall into an orbit of length 2 under the residual $\IZ_2$ action, as well as 12 $\IZ_2$ fixed points which fall into 4 orbits of length 3 under the residual $\IZ_3$ action. The latter two sets are not realized in the $T^6$ orbifold for this lattice. The Euler number of $D_{2\beta}$ minus the fixed point set is $(0-3\cdot 0-12)/6 = -2$. The blow--up process glues in a $T^2 \times F$ at the class of the $\IZ_3$ fixed lines without fixed points, where $F$ are two $\IP^1$ intersecting in a point. There is no contribution to the Euler number from this space. The last fixed line is replaced by $T^2 \times F$ minus 4 points,  upon which there is still a free $\IZ_2$ action. Its Euler number is therefore $(0-4)/2 = -2$. For the 4 $\IZ_{6-II}$ fixed points on this fixed line we see that in the corresponding toric diagram there is one line ending in $D_2$. (For triangulation e) there are two lines.) For a single fixed point this contributes $\chi(\IP^1)=2$ to the Euler number. At the each of the four classes of $\IZ_2$ fixed points we also glue in a $\IP^1$. Adding everything up, the Euler number of $D_{2\beta}$ is $-2 + 0 -2 + 4\cdot 2 + 4\cdot 2 = 12$ which can be viewed as the result of a blow--up of $\IF_0$ is 8 points. For $D_{3\gamma}$ the computation is similar. The action of $\frac{1}{6}(1,2)$ on $T^4$ yields the $\IZ_2$ fixed line with 4 $\IZ_{6-II}$ fixed points on top of it that we see in Figure~\ref{fig:ffixsixiiaa}. In addition, there are 3 more $\IZ_2$ fixed lines which fall into an orbit of length 3 under the residual $\IZ_3$ action, as well as 6 $\IZ_3$ fixed points which fall into 3 orbits of length 2 under the residual $\IZ_2$ action. The latter two sets again are not realized in the $T^6$ orbifold for this lattice. The Euler number of $D_{2\beta}$ minus the fixed point set is $(0-4\cdot 0-6)/6 = -1$. The blow--up process glues in a $T^2 \times \IP^1$ at the class of the $\IZ_2$ fixed line without fixed points. There is no contribution to the Euler number from this space. The last fixed line is replaced by $T^2 \times \IP^1$ minus 3 points, upon which there is still a free $\IZ_3$ action. Its Euler number is therefore $(0-3)/3 = -1$. For the 3 $\IZ_{6-II}$ fixed points on this fixed line we see that in the corresponding toric diagram there is one line ending in $D_2$. (For triangulation d) there are two lines.) For a single fixed point this contributes $\chi(\IP^1)=2$ to the Euler number. At the each of the three classes of $\IZ_3$ fixed points we glue in two $\IP^1$ intersecting in one point whose Euler number is $2\cdot 2-1$. Adding everything up, the Euler number of $D_{3\gamma}$ is $-1 + 0 -1 + 3\cdot 2 + 3\cdot 3 = 13$ which can be viewed as the result of a blow--up of $\IF_0$ in 9 points.

The divisor $R_1$ does not intersect any fixed lines, therefore it simply has the topology of $T^4$. The divisors $R_2$ and $R_3$, on the other hand, have the topology of a K3. In Table~\ref{tab:TopZ6II} we have summarized the topology of all the divisors for all triangulations. All the Euler numbers and types of surfaces we have determined above together with~(\ref{eq:ringZ6I}) agree with Noethers formula~(\ref{eq:S3}). 
\begin{table}[h!]
  \begin{center}
  $
  \begin{array}{|c|c|c|c|c|c|c|c|c|c|}
    \hline
    {\rm Triang.} & E_{1,\beta\gamma} & E_{2,\beta}   & E_{3,\gamma}   & E_{4,\beta}   & D_1          & D_{2,\beta}  & D_{3,\gamma} & R_1 & R_2,R_3 \\  
    \hline
    {\rm a) }     & \Bl{2}\IF_1       & \IF_0         & \IF_0          & \IF_0         & \Bl{12}\IF_0 & \Bl{8}\IF_n  & \Bl{9}\IF_n  & T^4 & {\rm K3} \\
    {\rm b) }     & \IF_1             & \Bl{4}\IF_0   & \Bl{6}\IF_0    & \IF_0         & \IF_0        & \Bl{8}\IF_n  & \Bl{9}\IF_n  & T^4 & {\rm K3}\\ 
    {\rm c) }     & \Bl{1}\IF_1       & \IF_0         & \Bl{3}\IF_0    & \Bl{4}\IF_0   & \IF_0        & \Bl{8}\IF_n  & \Bl{9}\IF_n  & T^4 & {\rm K3}\\
    {\rm d) }     & \IF_1             & \IF_0         & \IF_0          & \Bl{8}\IF_0   & \IF_0        & \Bl{8}\IF_n  & \Bl{12}\IF_n & T^4 & {\rm K3}\\
    {\rm e) }     & \IP^2             & \IF_0         & \Bl{9}\IF_0    & \IF_0         & \IF_0        & \Bl{12}\IF_n & \Bl{9}\IF_n  & T^4 & {\rm K3}\\
    \hline
  \end{array}
  $
  \end{center}
  \caption{The topologies of the divisors for all triangulations.}
  \label{tab:TopZ6II}
\end{table}
With the knowledge of the Euler numbers and the intersection ring we can determine the second Chern class $\ch_2$ on the basis $\{R_i,E_{k\alpha\beta\gamma}\}$ by~(\ref{eq:c2.S}) (for triangulation a)):
\begin{align}
  \label{eq:c2Z6II}
  \ch_2\cdot E_{1\beta\gamma} &= 0, & \ch_2\cdot E_{2\beta} &= -4, & \ch_2 \cdot E_{3\gamma} &= -4, & \ch_2\cdot E_{4\beta} & = -4,\notag\\
  \ch_2 \cdot R_1 &= 0, & \ch_2 \cdot R_i &= 24. 
\end{align}
Since the second Chern class is a linear form on $H^2(X,\IZ)$ we can apply it to each of the linear relations in~(\ref{eq:Z6IrelD11}) to~(\ref{eq:Z6Irels3}) and again find complete agreement.

\subsubsection{The orientifold}
\label{sec:Z6IISU2xSU6_O}

At the orbifold point, there are 64 O3--planes which fall into 16 conjugacy classes under the orbifold group. The combination $I_6\,\theta^3$ gives rise to one O7--plane at $z^2=0$ in the $(z^1,z^3)$--plane.

The fixed sets located at $z^2=0$ are invariant under the global involution $I_6$, those located at $z^2=\frac{1}{3}$ are mapped to $z^2=\frac{2}{3}$ and vice versa. Consequently, the twelve divisors $E_{1,\beta\gamma},\, E_{2,\beta}$ and $E_{4,\beta}$ for $\beta=2,3$ are not invariant. We can form six invariant linear combinations: $\frac{1}{2}(E_{1,2\gamma}+E_{1,3\gamma}),\ \frac{1}{2}(E_{2,2}+E_{2,3})$ and $\frac{1}{2}(E_{4,2}+E_{4,3})$. With a minus sign instead of a plus sign, the combinations are anti-invariant, therefore $h^{1,1}_{-}=6$.

We now discuss the orientifold for the resolved case.
On the homogeneous coordinates $y^k$, several different local actions are possible. We give the eight possible actions which only involve sending coordinates to their negatives: 
\begin{align}
  \label{eq:invlocal}
  (1)\quad\Ic(z,y) &= (-z^1,-z^2,-z^3,y^1,y^2,y^3,y^4) \notag\\
 (2)\quad \Ic(z,y) &= (-z^1,-z^2,-z^3,y^1,y^2,-y^3,-y^4) \notag\\
  (3)\quad\Ic(z,y) &= (-z^1,-z^2,-z^3,y^1,-y^2,y^3,-y^4) \notag\\
  (4)\quad\Ic(z,y) &= (-z^1,-z^2,-z^3,y^1,-y^2,-y^3,y^4) \notag\\
  (5)\quad\Ic(z,y) &= (-z^1,-z^2,-z^3,-y^1,y^2,y^3,-y^4) \notag\\
  (6)\quad\Ic(z,y) &= (-z^1,-z^2,-z^3,-y^1,y^2,-y^3,y^4) \notag\\
 (7)\quad \Ic(z,y) &= (-z^1,-z^2,-z^3,-y^1,-y^2,y^3,y^4) \notag\\
 (8)\quad \Ic(z,y) &= (-z^1,-z^2,-z^3,-y^1,-y^2,-y^3,-y^4) 
\end{align}
In the orbifold limit, (\ref{eq:invlocal}) reduces to $I_6$. The combination of~(\ref{eq:invlocal}) and the scaling action of the resolved patch~(\ref{rescalessixii}) has the following fixed point sets:
\begin{eqnarray}
  \label{eq:Z6IIOplanes}
 (1),\,(8) &&\{z^2=0\} \cup \{y^4=0\} \cup \{z^1=y^1=y^3=0\} \cup \{z^3=y^1=y^3=0\}, \notag\\
 (2),\,(7) &&\{y^3=0\} \cup \{z^1=y^1=y^4=0\} \cup \{z^2=z^3=y^3=0\}\notag\\
 && \cup\ \{z^2=y^1=y^2=0\} \cup \{y^1=y^2=y^4=0\}, \notag\\
 (3),\,(6) &&\{z^1=0\} \cup \{z^3=0\} \cup \{y^2=0\}, \notag\\
  (4),\,(5) &&\{y^1=0\}.
  \end{eqnarray}
Note that the eight possible involutions only lead to four distinct fixed sets (but to different values for the $\lambda_i$).

We focus for the moment on the third possibility. 
With the scaling action
\begin{equation}\label{rescalessixiia}{(z^1,\,z^2,\,z^3,\,y^1,\,y^2,\,y^3,\,y^4) \to (\frac{\lambda_1\lambda_3}{\lambda_4}\,z^1,\,\lambda_2\,z^2,\,\lambda_3\,z^3,\, {1\over\lambda_4}\,y^1,\,\frac{\lambda_1}{\lambda_2^2}\,y^2,\,{\lambda_4\over \lambda_3^2}\,y^3, {\lambda_2\lambda_4\over \lambda_1^2}\,y^4)
}\end{equation}
we get the solutions
\begin{eqnarray}
{\rm (i)}. \quad z^1&=&0,\ \ \lambda_1=\lambda_2=\lambda_3=-1,\ \lambda_4=1,\cr
{\rm (ii)}.\quad z^3&=&0,\ \ \lambda_1=\lambda_2=-1,\ \lambda_3=\lambda_4=1,\cr
{\rm (iii)}.\quad y^2&=&0,\ \ \lambda_1=\lambda_4=1,\ \lambda_2=\lambda_3=-1.
\end{eqnarray}
This corresponds to an O7--plane wrapped on $D_{1}$, one on each of the four $D_{3,\gamma}$ and one wrapped on each of the two invariant $E_{2,\beta}$. No O3--plane solutions occur.

$\lambda_1$ and $\lambda_2$ correspond to the two Mori generators of the $\IZ_3$--fixed line. We restrict to it by setting $\lambda_3=-1,\ \lambda_4=1$ in accordance with solution (i) and (ii) which are seen by this fixed line. The scaling action thus becomes
\begin{equation}\label{rescalessixiizthree}{(z^1,\,z^2,\,z^3,\,y^1,\,y^2,\,y^3,\,y^4) \to (-\lambda_1\,z^1,\,\lambda_2\,z^2,-z^3,y^1,\,\frac{\lambda_1}{\lambda_2^2}\,y^2, y^3, {\lambda_2\over \lambda_1^2}\,y^4).
}\end{equation}
$y^1$ and $y^3$ do not appear in the fixed line, and the restriction makes sense only directly at the fixed point, i.e. for $z^3=0$. With this scaling action and the involution (3), we again reproduce the solutions (i) and (ii).

$\lambda_3$ corresponds to the Mori generator of the $\IZ_2$ fixed line. We restrict to it by setting $
\lambda_1=\lambda_2=-1,\ \lambda_4=1$. 
The scaling action becomes
\begin{equation}\label{rescalessixiiaztwo}
(z^1,\,z^2,\,z^3,\,y^1,\,y^2,\,y^3,\,y^4) \to (-\lambda_3\,z^1,-\,z^2,\,\lambda_3\,z^3,\,y^1,\,-y^2,\,{1\over \lambda_3^2}\,y^3, -\,y^4),
\end{equation}
which together with the involution (3) again reproduces the solutions (i) and (iii).
Global consistency is ensured since we only have one kind of patch on which we choose the same involution for all patches.

The four patches at $z^1=z^2=0$ coincide with locations of O3--planes at the orbifold point. Since no O3--plane solutions occur for our choice of involution, we are left with 12 O3--plane in the resolved case, which are located away from the $\IC^3/\IZ_{6-II}$ patches.

The modified intersection numbers are 
\begin{align}
  R_1R_2R_3&=3, & R_2E_{3,\gamma}^2 &= -1, & R_3E_{2,\beta}^2 &= -4, & R_3E_{4,\beta} &= -1, \notag\\
  R_3E_{2,\beta}E_{4,\beta} &= 1, & E_{1,\beta\gamma}^3&=3, & E_{2,\beta}^3& =32, & E_{3,\gamma}^3&=4, \notag\\
  E_{4,\beta}^3&=4, & E_{1,\beta\gamma}E_{2\beta}^2 &= -2, & E_{1,\beta\gamma}E_{3\gamma}^2 &= -1/2, &  E_{1,\beta\gamma}E_{4\beta}^2 &= -1/2, \notag\\
  E_{1,\beta\gamma}E_{2,\beta}E_{4\beta} &= 1, &  E_{2,\beta}^2E_{4,\beta} &= -4.
\end{align}


\subsection{The lattice $SU(3)\times SO(8)$}
\label{sec:Z6IIonSU3xSO8}

\subsubsection{The resolved orbifold}

\begin{figure}[h!]
\begin{center}
\includegraphics[width=85mm]{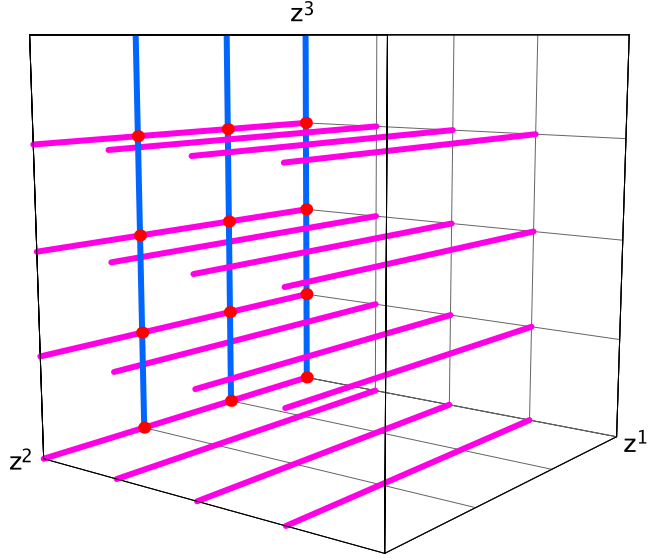}
\caption{Schematic picture of the fixed set configuration of $\IZ_{6-II}$ on $SU(3)\times SO(8)$}\label{ffixsixiiaaa}
\end{center}
\end{figure}

Here, the analysis of the fixed point set is very similar to the previous example and we will only point out the differences. The action of the twist $\theta$ on the lattice $SU(3)\times SO(8)$ was given in (A.48) and the resulting complex structure in (A.51) of~\cite{Lust:2005dy}. As for the fixed point set, the only change occurs in the $z^1$- direction. Apart from $\zf{1}{1}=0$, we now have $\zf{1}{2}=\half({1\over\sqrt3}\,e^{\pi i/6}),\ \zf{1}{4}=\half,\ \zf{1}{6}=\half(1+{1\over\sqrt3}\,e^{\pi i/6})$, at which we have further $\IZ_2$ fixed lines in the $z^2$ direction. In addition, the order three element maps these points as $\half({1\over\sqrt3}\,e^{\pi i/6}) \to \frac{1}{2} \to \half(1+{1\over\sqrt3}\,e^{\pi i/6}) \to \half({1\over\sqrt3}\,e^{\pi i/6})$. The resulting conjugacy classes are
\begin{align}
  \label{eq:classessixiiii}
  \alpha=1:\; & \notag\\
  \gamma=1:\; & (0, z^2, 0)\quad \gamma=2:\;\ (0,z^2, \tfrac{1}{2})\quad \gamma=3:\;\ (0,z^2,\tfrac{1}{2} U^3)\quad \gamma=4:\; \ (0,z^2,\tfrac{1}{2}(1+U^3))\notag\\
  \alpha=2:\; & \notag\\
  \gamma=1:\; & (\tfrac{1}{2}, z^2, 0),\ (\tfrac{1}{2\sqrt3}\,e^{\pi i/6}, z^2,0),\ (\tfrac{1}{2}\,(1+\tfrac{1}{\sqrt3}\,e^{\pi i/6}), z^2, 0)\notag\\
  \gamma=2:\; & (\tfrac{1}{2}, z^2,  \tfrac{1}{2}),\ (\tfrac{1}{2\sqrt3}\,e^{\pi i/6}, z^2, \tfrac{1}{2}),\ (\tfrac{1}{2}\,(1+\tfrac{1}{\sqrt3}\,e^{\pi i/6}), z^2,  \tfrac{1}{2})\notag\\
  \gamma=3:\; & (\tfrac{1}{2}, z^2,\tfrac{1}{2} U^3),\ (\tfrac{1}{2\sqrt3}\,e^{\pi i/6}, z^2,\tfrac{1}{2} U^3),\ (\tfrac{1}{2}\,(1+\tfrac{1}{\sqrt3}\,e^{\pi i/6}), z^2,\tfrac{1}{2} U^3)\notag\\
  \gamma=4:\; & (\tfrac{1}{2}, z^2, \tfrac{1}{2}(1+U^3)),\ (\tfrac{1}{2\sqrt3}\,e^{\pi i/6}, z^2,\tfrac{1}{2}(1+U^3)),\ (\tfrac{1}{2}\,(1+\tfrac{1}{\sqrt3}\,e^{\pi i/6}), z^2, \tfrac{1}{2}(1+U^3)).
\end{align}
Table~\ref{fssixiaaa} summarizes the relevant data of the fixed point set. The invariant subtori under $\theta^2$ and $\theta^3$ are $(x^3+x^4,0,x^3,x^4,0,0)$ and $(0,0,0,0,x^5,x^6)$, respectively.
\begin{table}[h!]\begin{center}
\begin{tabular}{|c|c|c|c|}
\hline
Group el.& Order &Fixed Set& Conj. Classes \cr
\hline
\noalign{\hrule}\noalign{\hrule}
$ \theta$& 6   &12  fixed points & 12\cr
$\theta^2$& 3   &3  fixed lines & 3\cr
$\theta^3$& 2 &16  fixed lines & 8\cr
\hline
\end{tabular}
\caption{Fixed point set for $\IZ_{6-II}$ orbifold on $SU(3)\times SO(8)$}\label{fssixiaaa}
\end{center}\end{table}
Figure \ref{ffixsixiiaaa} shows the configuration of the fixed point set in a schematic way.

Here we have again 12 $\IC^3/\IZ_{6-II}$ patches which each sit at the intersection of two fixed lines, and in addition 3 fixed lines originating from the order 3 element as well as 8 classes of fixed lines from the order two element. The fixed points yield the same exceptional divisors $E_{1,\beta\gamma}, E_{2,\beta}, E_{3,\alpha\gamma}, E_{4,\beta}$, $\alpha=1$, $\beta=1,2,3$, $\gamma=1,\dots,4$, as in Appendix~\ref{sec:Z6IIonSU2xSU6}. Moreover, there are four exceptional divisors $E_{3,2\gamma}$ coming from the additional $\IC^2/\IZ_2$ fixed lines. These are the invariant combinations $E_{3,2\gamma}=\sum_{\alpha=2,4,6} \Et_{3,\alpha\gamma}$, where $\Et_{3,\alpha\gamma}$ are the representatives on the cover. This gives a total of $12\cdot 1+3\cdot 2+8\cdot 1=26$ exceptional divisors. On this lattice there are four classes of $\IC^2/\IZ_2$ fixed lines without fixed points on them, therefore by~\eqref{eq:h21tw} we have $h^{2,1}_{\rm twist.}=4$. We have 9 fixed planes with their associated divisors: $D_{1\alpha},\ \alpha=1,2,\,D_{2\beta},\ \beta=1,2,3,\, D_{3\gamma},\ \gamma=1,...,4$. Here, $D_{1,2}$ is the invariant combination $D_{1,2}=\sum_{\alpha=2,4,6} \Dt_{1\alpha}$ of the representatives $\Dt_{1\alpha}$ on the cover.

The global linear relations are the same as those in~\eqref{eq:globalrelsixiia} except for a new relation involving $R_1$ and $D_{1,2}$, as well as an additional term involving $E_{3,2,\gamma}$ in the relations for $R_3$:
\begin{align}
  \label{eq:globalrelsixiib}
  R_1&=6\,D_{{1,1}}+3\,\sum_{\gamma=1}^4 E_{{3,1,\gamma}}+\sum_{\beta=1}^3\sum_{\gamma=1}^4 E_{{1,\beta\gamma}}+2\sum_{\beta=1}^3 \left( 2\,E_{{2,\beta}}+4\,E_{{4,\beta}}\right), \cr
R_1&=2\,D_{{1,2}}+\sum_{\gamma=1}^4 E_{{3,2,\gamma}},\cr
R_2&=3\,D_{{2,\beta}}+\sum_{\gamma=1}^4E_{{1,\beta\gamma}}+2\,E_{{2,\beta}}+E_{{4,\beta}}, \qquad \beta=1,2,3, \cr
  R_3&=2\,D_{{3,\gamma}}+\sum_{\beta=1}^3E_{{1,\beta\gamma}}+\sum_{\alpha=1}^2 E_{{3,\alpha\gamma}}, \quad\qquad \gamma=1,\dots,4.
\end{align}
We obtain the following nonvanishing intersection numbers of $X$ in the basis $\{R_i,E_{k\alpha\beta\gamma}\}$:
\begin{align}
  R_1R_2R_3&=6, & R_2E_{3,1,\gamma}^2 &= -2, & R_2E_{3,2,\gamma}^2 &= -6, & R_3E_{2,\beta}^2 &= -2, \notag\\
  R_3E_{4,\beta} &= -2, & R_3E_{2,\beta}E_{4,\beta} &= 1, & E_{1,\beta\gamma}^3&=6, & E_{2,\beta}^3& =8, \notag\\
  E_{3,1,\gamma}^3&=8, & E_{4,\beta}^3&=8, & E_{1,\beta\gamma}E_{2\beta}^2 &= -2, & E_{1,\beta\gamma}E_{3,1,\gamma}^2 &= -2, \notag\\
  E_{1,\beta\gamma}E_{4,\beta}^2 &= -2, & E_{1,\beta\gamma}E_{2,\beta}E_{4,\beta} &= 1, &  E_{2,\beta}^2E_{4,\beta} &= -2,
\end{align}
For the topology of the divisors there are only a few changes with respect to the lattice $SU(2)\times SU(6)$. First of all, the topology of the divisors $E_{1,\beta\gamma}$, $E_{2,\beta}$, $E_{4,\beta}$, $D_{2,\beta}$, and $D_{3,\gamma}$ are the same as in Table~\ref{tab:TopZ6II}. The divisors $E_{3,1,\gamma}$ and $D_{1,1}$ have the same topology as $E_{3,\gamma}$ and $D_1$, respectively, in that table. The topology of the new divisors $E_{3,2\gamma}$ and $D_{1,2}$ are as follows: The divisors are of type~E\ref{item:E2}) with 3 representatives, hence their topology is that of $\IP^1 \times T^2$. The topology of each representative of $D_{1,2}$ minus the fixed point set, viewed as a $T^4$ orbifold, is that of a $T^2 \times (T^2/\IZ_2 \setminus \{ 4\ \rm{pts} \})$. They are permuted under the residual $\IZ_3$ action and the 12 points fall into 3 orbits of length 1 and 3 orbits of length 3. Hence, the topology of the class is still that of a $T^2 \times (T^2/\IZ_2 \setminus \{ 4 \ \rm{pts} \})$. After the blow--up it is therefore a $\IP^1 \times T^2$. For both, $E_{3,2\gamma}$ and $D_{1,2}$, the topology is obviously independent of the choice of resolution of $\IC^3/\IZ_{6-II}$. For completeness, we display the second Chern classes in the basis $\{R_i, E_{k\alpha\beta\gamma}\}$ (for triangulation a)):
\begin{align}
  \label{eq:c2Z6IIb}
  \ch_2\cdot E_{1\beta\gamma} &= 0, & \ch_2\cdot E_{2\beta} &= -4, & \ch_2 \cdot E_{3,1\gamma} &= -4, & \ch_2\cdot E_{3,2\gamma} &= 0, & \notag\\
  \ch_2\cdot E_{4\beta} & = -4, &\ch_2 \cdot R_1 &= 0, & \ch_2 \cdot R_i &=  24. 
\end{align}

\subsubsection{The orientifold}
\label{sec:Z6IISU3xSO8_O}

At the orbifold point, there are 64 O3--planes which fall into 24 conjugacy classes under the orbifold group. Under $I_6\,\theta^3$ we find four O7--planes in the $(z^1,z^3)$--plane which fall into two conjugacy classes: The one at $z^2=0$ and those at $z^2=\frac{1}{2},\, \frac{\tau}{2},\, \frac{1}{2}(1+\tau)$. $h^{(1,1)}_{-}=6$ is obtained in the same way as in Appendix~\ref{sec:Z6IISU2xSU6_O}. Similarly, everything we found for the resolved orbifold in that appendix applies here as well. There are four classes of fixed lines with fixed points which are invariant under the orientifold action, hence $h^{2,1}=4$.


\subsection{The lattice $SU(2)^2\times SU(3)^2$}
\label{Z6IIonSU2xSU2xSU3xSU3}

\subsubsection{The resolved orbifold}

\begin{figure}[h!]
\begin{center}
\includegraphics[width=85mm]{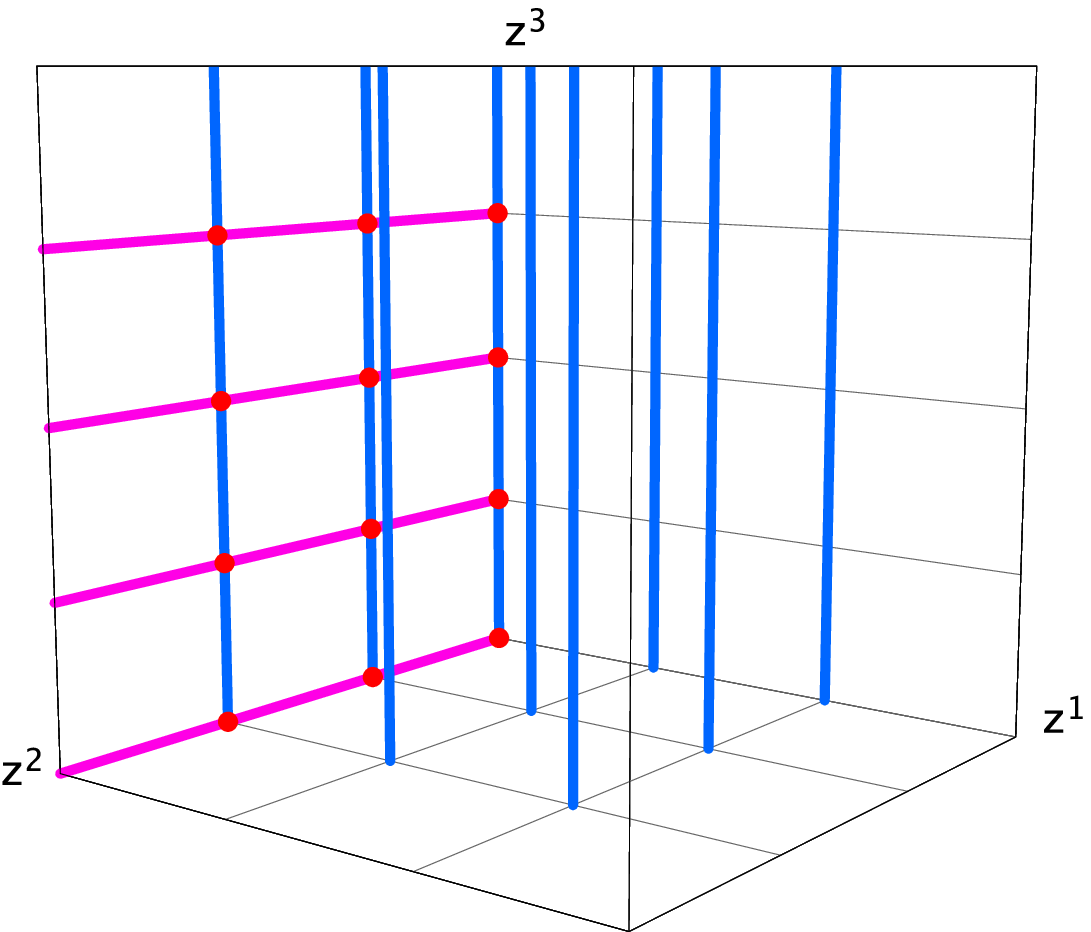}
\caption{Schematic picture of the fixed set configuration of $\IZ_{6-II}$ on $SU(2)^2\times SU(3)^2$}
\label{ffixsixiig}
\end{center}
\end{figure}

Here, the analysis of the fixed point set is very similar to the example in Appendix~\ref{sec:Z6IIonSU3xSO8} and we will only point out the differences. The action of the twist $\theta$ on the lattice $SU(2)^2\times SU(3)^2$ was given in (A.41) and the resulting complex structure in (A.44) of~\cite{Lust:2005dy}. The only change is that instead of $\IC^2/\IZ_2$ fixed lines in the $z^2$ direction, we now have $\IC^2/\IZ_3$ fixed lines in the $z^3$ direction which lie, apart from $\zf{1}{1}=0$, at $\zf{1}{3}=1/3$ and $\zf{1}{5}=2/3$. In addition, the order two element maps the latter two points into each other. The resulting conjugacy classes are
\begin{align}
  \label{eq:classsixii}
  \mu=1:\; &(0,0,z^3) & \mu=2:\; & (0,\tfrac{1}{\sqrt3}\,e^{\pi i/6}, z^3)\notag\\
  \mu=3:\; &(0,1+\tfrac{i}{\sqrt3}, z^3) & \mu=4:\; &(\tfrac{1}{3}, 0, z^3),\ (\tfrac{2}{3}, 0, z^3)\notag\\
  \mu=5:\; &(\tfrac{1}{3}, \tfrac{1}{\sqrt3}\,e^{\pi i/6}, z^3),\ (\tfrac{2}{3}, \tfrac{1}{\sqrt3}\,e^{\pi i/6}, z^3) & \mu=6:\; &(\tfrac{1}{3}, 1+\tfrac{i}{\sqrt3}, z^3),\ (\tfrac{2}{3}, 1+\tfrac{i}{\sqrt3}, z^3).
\end{align}
Table~\ref{fssixiiaaa} summarizes the important data of the fixed sets. The invariant subtori under $\theta^2$ and $\theta^3$ are $(x^1,x^2,0,0,0,0)$ and $(0,0,x^5-x^6,-x^6,x^5,x^6)$, respectively.
\begin{table}[h!]\begin{center}
\begin{tabular}{|c|c|c|c|}
\hline
Group el.& Order & Fixed Set& Conj. Classes \cr
\noalign{\hrule}\noalign{\hrule}
$\theta$& 6    &12 fixed points & 12\cr
$ \theta^2$&3      &9  fixed lines &6\cr
$ \theta^3$&2      &4 fixed lines &4\cr
\hline
\end{tabular}
\caption{Fixed point set for $\IZ_{6-II}$ on $SU(2)^2\times SU(3)^2$.}
\label{fssixiig}
\end{center}\end{table}
Figure~\ref{ffixsixiig} shows the configuration of the fixed sets in a schematic way. 

The fixed point set yields the same exceptional divisors $E_{1,\beta\gamma}, E_{2,\alpha\beta}, E_{3,\gamma}, E_{4,\alpha\beta}$, $\alpha=1$, $\beta=1,2,3$, $\gamma=1,\dots,4$, as in Appendix~\ref{sec:Z6IIonSU2xSU6}. Moreover, there are three pairs of exceptional divisors $E_{2,3\beta}$, $E_{4,3\beta}$ coming from the additional $\IC^2/\IZ_3$ fixed lines. These are the invariant combinations $E_{i,3\beta}=\sum_{\alpha=3,5} \Et_{i,\alpha\gamma\beta}$, $i=2,4$, where $\Et_{i,\alpha\beta}$ are the representatives on the cover. This gives a total of $12\cdot 1+6\cdot 2+4\cdot 1=28$ exceptional divisors. On this lattice there are three classes of $\IC^2/\IZ_3$ fixed lines without fixed points on them, therefore by~\eqref{eq:h21tw} we have $h^{2,1}_{\rm twist.}=6$. We have 9 fixed planes with their associated divisors: $D_{1\alpha},\ \alpha=1,3,\,D_{2\beta},\ \beta=1,2,3,\, D_{3\gamma},\ \gamma=1,...,4$. Here, $D_{1,3}$ is the invariant combination $D_{1,3}=\sum_{\alpha=3,5} \Dt_{1\alpha}$ of the representatives $\Dt_{1\alpha}$ on the cover.

The global linear relations are the same as those in~\eqref{eq:globalrelsixiia} except for a new relation involving $R_1$ and $D_{1,3}$, as well as additional terms involving $E_{i,3\beta}$, $i=2,4$ in the relations for $R_2$:
\begin{align}
  \label{eq:globalrelsixiic}
  R_1&=6\,D_{{1,1}}+3\,\sum_{\gamma=1}^4 E_{{3,1,\gamma}}+\sum_{\beta=1}^3\sum_{\gamma=1}^4 E_{{1,\beta\gamma}}+\sum_{\beta=1}^3 \left( 2\,E_{{2,1,\beta}}+4\,E_{{4,1,\beta}}\right), \cr
  R_1&=3\,D_{{1,3}}+\sum_{\beta=1}^3 \left( E_{{2,3,\beta}} + 2\,E_{4,3,\beta} \right), \cr
  R_2&=3\,D_{{2,\beta}}+\sum_{\gamma=1}^4E_{{1,\beta\gamma}}+\sum_{\alpha=1,3} \left(2\,E_{{2,\alpha\beta}}+ E_{{4,\alpha\beta}}\right), \qquad \beta = 1,2,3 \cr
  R_3&=2\,D_{{3,\gamma}}+\sum_{\beta=1}^3E_{{1,\beta\gamma}}+ E_{{3,1,\gamma}}, \qquad \gamma=1,\dots,4.
\end{align}
For the nonvanishing intersection numbers of $X$ in the basis $\{R_i,E_{k\alpha\beta\gamma}\}$ we find:
\begin{align}
  R_1R_2R_3&=6, & R_2E_{3,1,\gamma}^2 &= -2, & R_3E_{2,1,\beta}^2 &= -2, & R_3E_{2,3,\beta}^2 &= -4, \notag\\
  R_3E_{4,1,\beta} &= -2, & R_3E_{4,3,\beta} &= -4, & R_3E_{2,1,\beta}E_{4,1,\beta} &= 1, & R_3E_{2,3,\beta}E_{4,3,\beta} &= 2, \notag\\  
  E_{1,\beta\gamma}^3&=6, & E_{2,1,\beta}^3& =8, & E_{3,1,\gamma}^3&=8, & E_{4,1,\beta}^3&=8, \notag\\
  E_{1,\beta\gamma}E_{2,1,\beta}^2 &= -1, & E_{1,\beta\gamma}E_{3,1,\gamma}^2 &= -1, & E_{1,\beta\gamma}E_{4,1,\beta}^2 &= -1, & E_{1,\beta\gamma}E_{2,1,\beta}E_{4,1,\beta} &= 1, \notag\\
  E_{2,1,\beta}^2E_{4,1,\beta} &= -2,
\end{align}
The topology of the divisors $E_{1,\beta\gamma}$, $E_{3,1,\gamma}$, $D_{2,\beta}$, and $D_{3,\gamma}$ are the same as in Table~\ref{tab:TopZ6II}. The divisors $E_{2,1,\beta}$, $E_{4,1,\beta}$, and $D_{1,1}$ have the same topology as $E_{2,\beta}$, $E_{4,\beta}$, and $D_1$, respectively, in that table. The topology of the new divisors $E_{2,3,\beta}$, $E_{4,3,\beta}$, and $D_{1,3}$ are $\IP^1 \times T^2$ and $\Bl{8}\IF_n$, respectively, independent of the choice of resolution of $\IC^3/\IZ_{6-II}$. For completeness, we display the second Chern classes in the basis $\{R_i, E_{k\alpha\beta\gamma}\}$ (for triangulation a)):
\begin{align}
  \label{eq:c2Z6IIc}
  \ch_2\cdot E_{1\beta\gamma} &= 0, & \ch_2\cdot E_{2,1\beta} &= -4, & \ch_2\cdot E_{2,3\beta} &= 0, & \ch_2 \cdot E_{3,\gamma} &= -4, & \notag\\
  \ch_2\cdot E_{4,1\beta} & = -4, &\ch_2\cdot E_{4,3\beta} & = 0, &\ch_2 \cdot R_1 &= 0, & \ch_2 \cdot R_i &=  24. 
\end{align}

\subsubsection{The orientifold}
\label{sec:Z6IISU2xSU3_O}

At the orbifold point, there are 64 O3--planes which fall into 16 conjugacy classes under the orbifold group, as for the lattice $SU(2)xSU(6)$ in Appendix~\ref{sec:Z6IISU2xSU6_O}. Under $I_6\,\theta^3$ we find four O7--planes in the $(z^1,z^3)$--plane which fall into two conjugacy classes: The one at $z^2=0$ and those at $z^2=\frac{1}{2},\, \frac{\tau}{2},\, \frac{1}{2}(1+\tau)$. $h^{(1,1)}_{-}=8$ is obtained in the same way as in Appendix~\ref{sec:Z6IISU2xSU6_O} with two additional divisors coming from the new divisors $E_{2,3,\beta}$, $E_{4,3,\beta}$. Since these divisors come from fixed lines without fixed point, and latter also contribute to the twisted complex structure moduli according to the discussion in Section~\ref{sec:TwistedCplx}, some of these moduli are projected out by the induced orientifold action on $H^{2,1}_{+}$. In fact, since the three fixed lines at $(\zf{1}{2},\zf{2}{\beta})$, $\beta=1,2,3$ are identified with those at $(\zf{1}{3},\zf{2}{\beta})$, we find that $h^{2,1}_{+}=3$. Similarly, everything we found for the resolved orbifold in that appendix applies here as well.

\subsection{The lattice $SU(2)^2\times SU(3)\times G_2$}
\label{sec:Z6IIonSU2xSU2xSU3xG2}

\subsubsection{The resolved orbifold}

\begin{figure}[h!]
\begin{center}
\includegraphics[width=140mm]{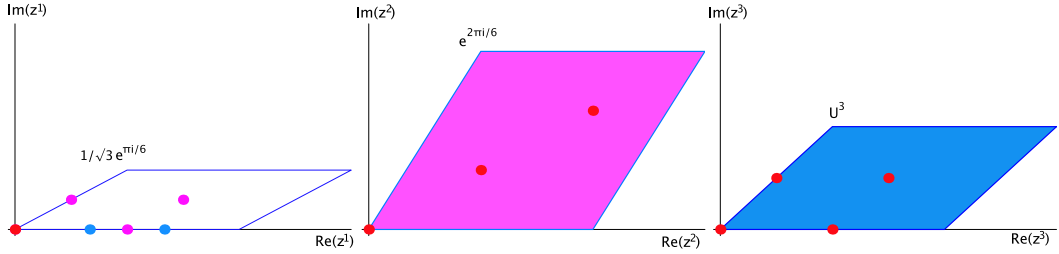}
\caption{Fundamental regions for the $\IZ_{6-II}$ orbifold on $SU(2)^2\times SU(3)\times G_2$}\label{ffusixi}
\end{center}
\end{figure}

\begin{figure}[h!]
\begin{center}
\includegraphics[width=85mm]{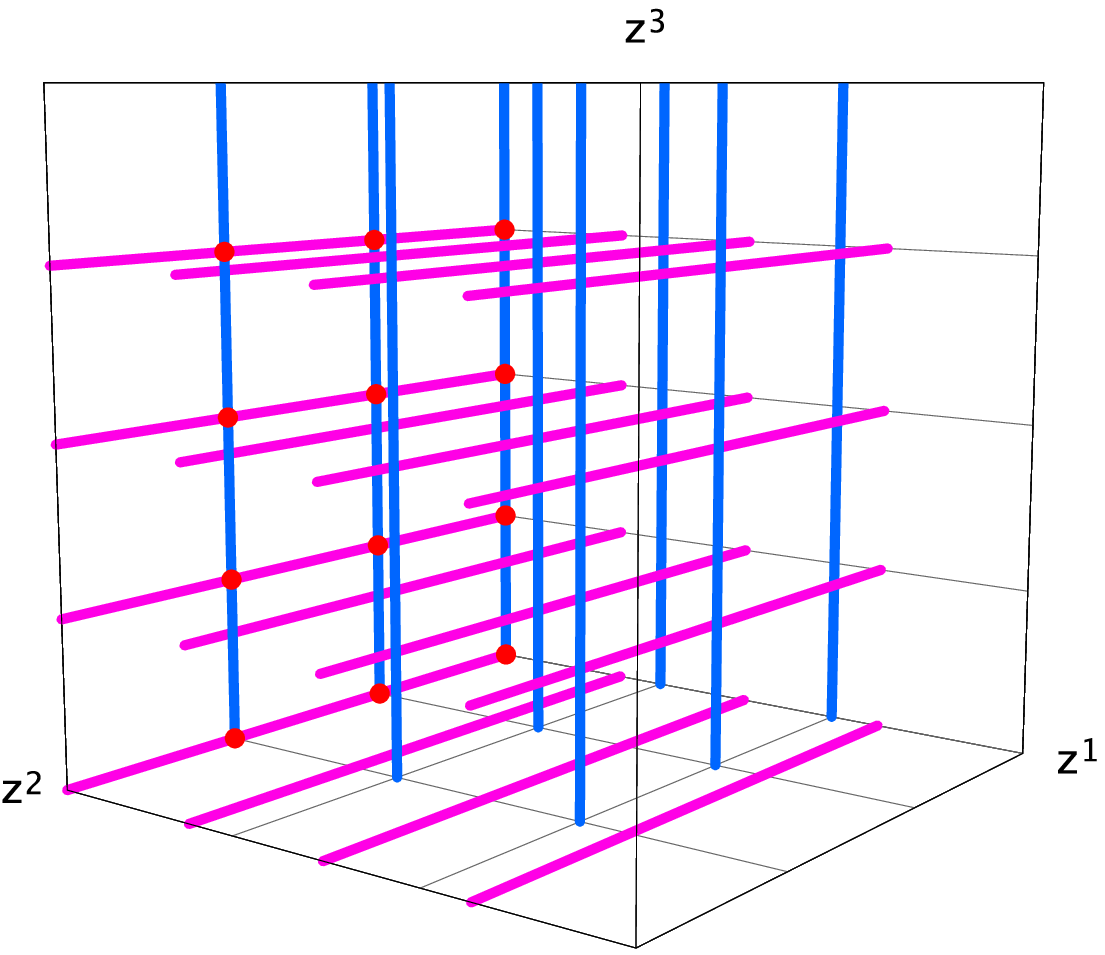}
\caption{Schematic picture of the fixed set configuration of $\IZ_{6-II}$ on $SU(2)^2\times SU(3)\times G_2$}
\label{ffixsixiia}
\end{center}
\end{figure}

Here, the analysis of the fixed point set is a combination of those in the Appendices~\ref{sec:Z6IIonSU3xSO8} and~\ref{Z6IIonSU2xSU2xSU3xSU3}. The action of the twist $\theta$ on the lattice $SU(2)^2\times SU(3)\times G_2$ was given in (A.22) and the resulting complex structure in (A.25) of~\cite{Lust:2005dy}. The main difference to the lattices in the Appendices~\ref{sec:Z6IIonSU2xSU6} to~\ref{Z6IIonSU2xSU2xSU3xSU3} is that the torus now factorizes into $(T^2)^3$. Figure~\ref{ffusixi} shows the fundamental regions of the three tori corresponding to $z^1,\,z^2,\,z^3$ and their fixed points. 
For $z^1$, we have 0 as the fixed point of the $\IZ_6$--twist, $0,1/3,2/3$ as the fixed points of the $\IZ_3$--twist arising in the second twisted sector and the four fixed points of the $\IZ_2$--twist arising in the third twisted sector. For $z^2$ we get the usual three fixed points of the $\IZ_3$--twist, namely $0, 1/\sqrt3 \,e^{\pi i/6}$ and $1+i/\sqrt3$, and for $z^3$ we find the four fixed points $0,\half, \half U^3, \half(1+U^3)$. Therefore, apart from $\zf{1}{1}=0$, we now have both $\zf{1}{2}=\half({1\over\sqrt3}\,e^{\pi i/6}),\ \zf{1}{4}=\half,\ \zf{1}{6}=\half(1+{1\over\sqrt3}\,e^{\pi i/6})$, at which we have further $\IZ_2$ fixed lines in the $z^2$ direction, and $\zf{1}{3}=1/3$ and $\zf{1}{5}=2/3$, at which we have further $\IZ_3$ fixed lines in the $z^3$ direction. The conjugacy classes of these fixed lines were given in~\eqref{eq:classessixiiii} and~\eqref{eq:classsixii}. Table~\ref{fssixiiaaa} summarizes the relevant data of the fixed point set. The invariant subtori under $\theta^2$ and $\theta^3$ are $(0,0,0,0,x^5,x^6)$ and $(0,0,x^3,x^4,0,0)$, respectively. 
\begin{table}[h!]\begin{center}
\begin{tabular}{|c|c|c|c|}
\hline
Group el.& Order & Fixed Set& Conj. Classes \cr
\noalign{\hrule}\noalign{\hrule}
$\theta$& 6    &12 fixed points &\ 12\cr
$ \theta^2$&3      &9  fixed lines &\ 6\cr
$ \theta^3$&2      &16 fixed lines &\ 8\cr
\hline
\end{tabular}
\caption{Fixed point set for $\IZ_{6-II}$ on $SU(2)^2\times SU(3)\times G_2$.}\label{fssixiiaaa}
\end{center}\end{table}
Figure \ref{ffixsixiia} shows the configuration of the fixed sets in a schematic way.

The fixed point set yields the same exceptional divisors $E_{1,\beta\gamma}, E_{2,\alpha\beta}, E_{3,\alpha\gamma}, E_{4,\alpha\beta}$, $\alpha=1,2,3$, $\beta=1,2,3$, $\gamma=1,\dots,4$, as in the Appendices~\ref{sec:Z6IIonSU3xSO8} and~\ref{Z6IIonSU2xSU2xSU3xSU3}. Here, $\alpha=1,2$ for $E_{3,\alpha\gamma}$ and $\alpha=1,3$ for $E_{2,\alpha\beta}$ and $E_{4,\alpha\beta}$. This gives a grand total of $12\cdot 1+6\cdot 2+8\cdot 1=32$ exceptional divisors. On this lattice, there are four classes of $\IC^2/\IZ_2$ fixed lines and three classes of $\IC^2/\IZ_3$ fixed lines without fixed points on them, therefore by~\eqref{eq:h21tw}, we have $h^{2,1}_{\rm twist.}=10$.

The global linear relations are obtained by combining~\eqref{eq:globalrelsixiib} and~\eqref{eq:globalrelsixiic}:
\begin{align}
  \label{eq:globalrelsixiid}
  R_1&=6\,D_{{1,1}}+3\,\sum_{\gamma=1}^4 E_{{3,1,\gamma}}+\sum_{\beta=1}^3\sum_{\gamma=1}^4 E_{{1,\beta\gamma}}+\sum_{\beta=1}^3 \left( 2\,E_{{2,1,\beta}}+4\,E_{{4,1,\beta}}\right), \cr
  R_1&=2\,D_{{1,2}}+\sum_{\gamma=1}^4 E_{{3,2,\gamma}}, \cr
  R_1&=3\,D_{{1,3}}+\sum_{\beta=1}^3 \left( E_{{2,3,\beta}} + 2\,E_{4,3,\beta} \right),
\end{align}
\begin{align}\nonumber
  R_2&=3\,D_{{2,\beta}}+\sum_{\gamma=1}^4E_{{1,\beta\gamma}}+\sum_{\alpha=1,3} \left(2\,E_{{2,\alpha\beta}}+ E_{{4,\alpha\beta}}\right), \qquad \beta = 1,2,3 \cr
  R_3&=2\,D_{{3,\gamma}}+\sum_{\beta=1}^3E_{{1,\beta\gamma}}+ \sum_{\alpha=1}^2 E_{{3,\alpha,\gamma}}, \qquad \gamma=1,\dots,4.
\end{align}
We obtain the following nonvanishing intersection numbers of $X$ in the basis $\{R_i,E_{k\alpha\beta\gamma}\}$:
\begin{align}
  R_1R_2R_3&=6, & R_2E_{3,1,\gamma}^2 &= -2, & R_2E_{3,2,\gamma}^2 &= -6,& R_3E_{2,1,\beta}^2 &= -2, \notag\\
  R_3E_{2,3,\beta}^2 &= -4, & R_3E_{4,1,\beta} &= -2, & R_3E_{4,3,\beta} &= -4, & R_3E_{2,1,\beta}E_{4,1,\beta} &= 1, \notag\\ 
  R_3E_{2,3,\beta}E_{4,3,\beta} &= 2, & E_{1,\beta\gamma}^3&=6, & E_{2,1,\beta}^3& =8, & E_{3,1,\gamma}^3&=8, \notag\\
  E_{4,1,\beta}^3&=8, & E_{1,\beta\gamma}E_{2,1,\beta}^2 &= -1, & E_{1,\beta\gamma}E_{3,1,\gamma}^2 &= -1, & E_{1,\beta\gamma}E_{4,1,\beta}^2 &= -1, \notag\\
  E_{1,\beta\gamma}E_{2,1,\beta}E_{4,1,\beta} &= 1, & E_{2,1,\beta}^2E_{4,1,\beta} &= -2,
\end{align}
The topology of all the divisors has already been determined in one of the Appendices~\ref{sec:Z6IIonSU2xSU6} to~\ref{Z6IIonSU2xSU2xSU3xSU3}. The second Chern class in the basis $\{R_i, E_{k\alpha\beta\gamma}\}$ (for triangulation a)) reads:
\begin{align}
  \label{eq:c2Z6IId}
  \ch_2\cdot E_{1\beta\gamma} &= 0, & \ch_2\cdot E_{2,1\beta} &= -4, & \ch_2\cdot E_{2,3\beta} &= 0, & \ch_2 \cdot E_{3,1\gamma} &= -4, & \notag\\
  \ch_2 \cdot E_{3,2\gamma} &= 0, & \ch_2\cdot E_{4,1\beta} & = -4, &\ch_2\cdot E_{4,3\beta} & = 0, &\ch_2 \cdot R_1 &= 0, & \notag\\
  \ch_2 \cdot R_i &=  24. 
\end{align}

\subsubsection{The orientifold}

At the orbifold point, there are 64 O3--planes which fall into 24 conjugacy classes under the orbifold group as for the lattice $SU(3)\times SO(8)$ in Appendix~\ref{sec:Z6IISU3xSO8_O}. Under $I_6\,\theta^3$ we find four O7--planes in the $(z^1,z^3)$--plane which fall into two conjugacy classes: The one at $z^2=0$ and those at $z^2=\frac{1}{2},\, \frac{\tau}{2},\, \frac{1}{2}(1+\tau)$. $h^{(1,1)}_{-}=8$ is obtained in the same way as in Appendix~\ref{Z6IIonSU2xSU2xSU3xSU3} with two additional divisors coming from the new divisors $E_{2,3,\beta}$, $E_{4,3,\beta}$. Since these divisors come from fixed lines without fixed point, and latter also contribute to the twisted complex structure moduli according to the discussion in Section~\ref{sec:TwistedCplx}, some of these moduli are projected out by the induced orientifold action on $H^{2,1}_{+}$. In fact, the three fixed lines at $(\zf{1}{2},\zf{2}{\beta})$, $\beta=1,2,3$ are identified with those at $(\zf{1}{3},\zf{2}{\beta})$. Moreover, there are four classes of $\IC^2/\IZ_2$ fixed lines without fixed points, therefore we find that $h^{2,1}_{+}=7$. Similarly, everything we found for the resolved orbifold in Appendix~\ref{sec:Z6IISU2xSU6_O} applies here as well.


\section{The $\IZ_2\times\IZ_4$ orbifold}

\subsection{Complex structure and fixed sets}

The torus factorizes into $(T^2)^3$ under the combined twists, where the first of the $T^2$ is not constrained. The twists act on the lattice basis:
\begin{eqnarray}
Q_1\, e_1&=&- e_1,\quad Q_1\,e_2=-e_2,\quad Q_1\,e_3=e_3,\quad Q_1\,e_4=e_4,\cr
Q_1\,e_5&=&-e_5,\quad Q_1\,e_6=-e_6,\cr
Q_2\, e_1&=& e_1,\quad Q_2\,e_2=e_2,\quad Q_2\,e_3=e_3+2\,e_4,\quad Q_2\,e_4=-e_3-e_4,\cr
Q_2\,e_5&=&e_5+2\,e_6,\quad Q_2\,e_6=-e_5-e_6.
\end{eqnarray}
The combined twist $Q_3$ has the form
\begin{eqnarray}
Q_3\, e_1&=& -e_1,\quad Q_3\,e_2=-e_2,\cr
Q_3\,e_3&=&e_3+2\,e_4,\quad Q_3\,e_4=-e_3-e_4,\cr
Q_3\,e_5&=&-e_5-2\,e_6,\quad Q_3\,e_6=e_5+e_6.
\end{eqnarray}
We require the metric to be invariant under all three twists, i.e. we impose the three conditions $Q_i^Tg\,Q_i=g,\quad i=1,2,3$. This leads to the following solution:
\begin{equation}{g=\left(\begin{array}{cccccc}
R_1^2&R_1R_2\,\cos\theta_{12}&0&0&0&0\cr
R_1R_2\,\cos\theta_{12}&R_1^2&0&0&0&0\cr
0&0&2\,R_3^2&-R_3^2&0&0\cr
0&0&-R_3^2&R_3^2&0&0\cr
0&0&0&0&2\,R_5^2&-R_5^2\cr
0&0&0&0&-R_5^2&R_5^2\end{array}\right).}
\end{equation}
From the solution for $b$ we see that we have three K\"ahler moduli and three untwisted complex structure moduli. 
With the methods discussed in \cite{Lust:2005dy}, we get the following complex structure:
\begin{equation}
z^1={1\over\sqrt{2\,{\rm Im}\,{\Uc}}}\,(x^1+{\Uc}\, x^2),\quad
z^2=x^3-{1\over2}(1-i)\,x^4,\quad
z^3=x^5-{1\over2}(1-i)\,x^6,\end{equation}
with ${\Uc}=R_2/R_1\, e^{i\theta_{12}}$.

\begin{figure}[h!]
\begin{center}
\includegraphics[width=140mm]{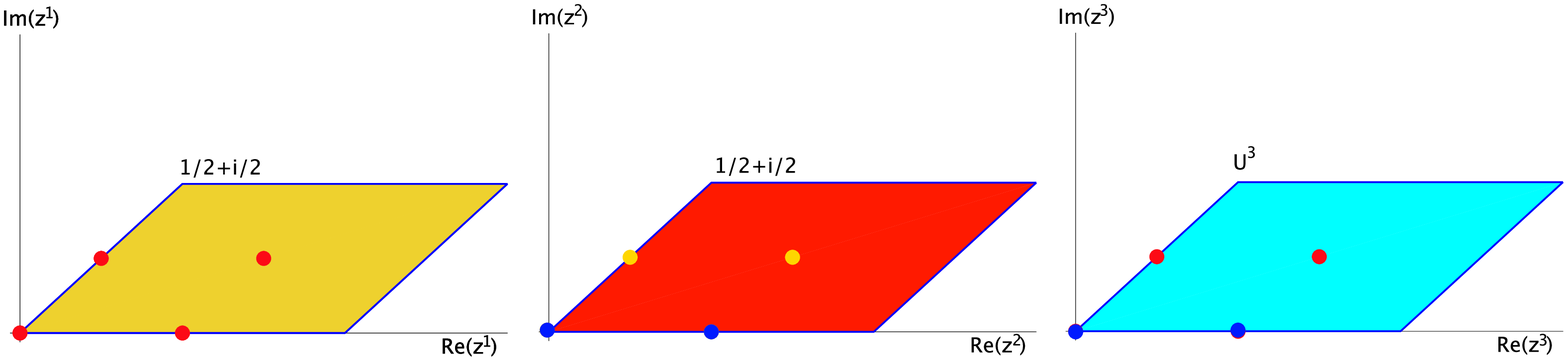}
\caption{Fundamental regions for the $\IZ_2\times\IZ_4$ orbifold}\label{ffutwofour}
\end{center}
\end{figure}

\begin{figure}[h!]
\begin{center}
\includegraphics[width=85mm]{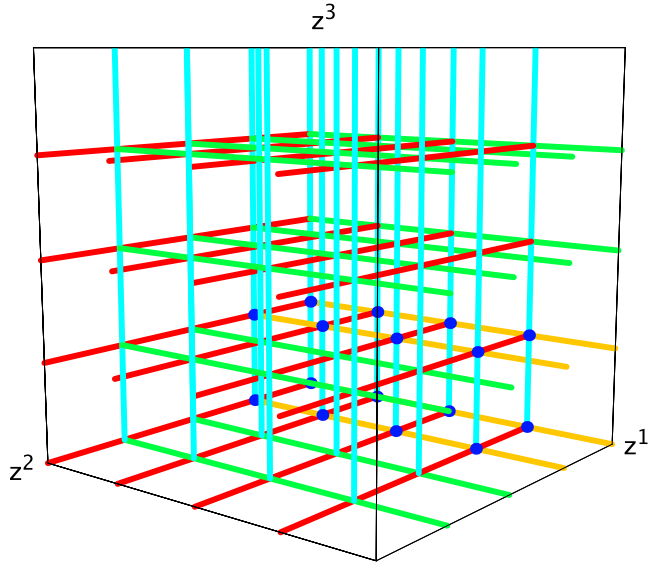}
\caption{Schematic picture of the fixed set configuration of $\IZ_2\times \IZ_4$}\label{ffixtwofour}
\end{center}
\end{figure}

In order to determine the fixed point set, we need to examine the $\theta^1,\,\theta^2,\,(\theta^2)^2,\,\theta^1\theta^2$, and $\theta^1(\theta^2)^2$ twists. Figure~\ref{ffutwofour} shows the fundamental regions of the three tori corresponding to $z^1,\,z^2,\,z^3$ and their fixed points. 
In each of them, we get the usual four fixed points of the $\IZ_2$--twist, $\zf{1}{\alpha}=0,\frac{1}{2},\frac{U}{2},\frac{1}{2}(1+U)$, $\alpha=1,\dots,4$, and $\zf{2}{\beta},\zf{3}{\gamma}=0,\frac{1}{2},\frac{i}{2},\frac{1}{2}(1+i)$, $\beta,\gamma=1,\dots,4$. At each pair $(\zf{i}{\alpha},\zf{j}{\beta})$ we have a $\IC^2/\IZ_2$ fixed line. In addition, at the fixed points with $\beta,\gamma=1,2$, and $\alpha=1,\dots,4$ there are $\IC^3/(\IZ_2\times\IZ_4)$ singularities while the singularities at the remaining fixed points are of the form $\IC^3/(\IZ_2\times\IZ_2)$. The latter, as well as the corresponding fixed lines, at $\beta=3, 4$ and $\gamma=3, 4$ are mapped into each other by an order two twist, respectively, and fall into conjugacy classes. Table~\ref{fstwofour} summarizes the relevant data of the fixed point set.
\begin{table}[h!]\begin{center}
\begin{tabular}{|c|c|c|c|}
\hline
Group el.& Order &Fixed Set& Conj. Classes \cr
\hline
\noalign{\hrule}\noalign{\hrule}
$\theta^1   $&2     &16  fixed lines& 12\cr
$\theta^2   $&4  &4  fixed lines&4\cr
$ (\theta^2)^2   $&2      &16  fixed lines& 10\cr
$ \theta^1\theta^2   $&8      &16  fixed points& 16\cr
$ \theta^1(\theta^2)^2   $&2  &16  fixed lines& 12\cr
\hline
\end{tabular}
\caption{Fixed point set for $\IZ_2\times \IZ_4$.} \label{fstwofour}
\end{center}\end{table}
Figure \ref{ffixtwofour} shows the configuration of the fixed sets in a schematic way.

\subsection{The resolved orbifold}

From the 16 $\IC^3/\IZ_2\times\IZ_4$ patches, see Figure~\ref{frtwofour}, we get for each an exceptional divisor, $E_{5,\alpha\beta\gamma},\ \alpha=1,\dots,4,\ \beta,\gamma=1,2$. From the 4 $\IC^2/\IZ_4$ fixed lines in the $z^1$ direction, we get three each: $E_{2,\beta\gamma},\, E_{3,\beta\gamma},\, E_{4,\beta\gamma},\ \beta,\gamma=1,2$. From the $12+12+(10-4)=30$ $\IC^2/\IZ_2$--fixed lines we get one for each: $E_{1,\alpha\gamma'}, \, E_{6,\alpha\beta'},\ i=\alpha=1,\dots,4,\, \beta',\gamma'=1,2,3$ in the $z^2$, respectively $z^3$ direction and in the $z^1$ direction $E_{3,\mu},\ \mu=3,6,\dots,10$. Since these divisors and the divisors $E_{3,\beta\gamma}$ have similar properties we collect them and denote the latter by $E_{3,\mu}$, $\mu=1,2,4,5$. This adds up to $16\cdot1+4\cdot3+30\cdot1=58$ exceptional divisors. The intersection points of three $\IZ_2$ fixed lines are locally described by the resolved $\IC^3/\IZ_2\times\IZ_2$ patches in Appendix~\ref{sec:localZ2xZ2}. Since in this example, there are no fixed lines without fixed points on them, $h^{2,1}_{\rm twist.}=0$.

Since the $\IC^3/\IZ_2\times\IZ_2$ fixed points fall into conjugacy classes, we have to form the corresponding invariant divisors:
\begin{align}
  \label{eq:Z2xZ4invdivs}
  D_{2,1} &= \Dt_{2,1}, & D_{2,2} &= \Dt_{2,2}, & D_{2,3} &= \Dt_{2,3} + \Dt_{2,4}, \notag\\
  D_{3,1} &= \Dt_{3,1}, & D_{3,2} &= \Dt_{3,2}, & D_{3,3} &= \Dt_{3,3} + \Dt_{3,4}, \notag\\
  E_{1,\alpha,1} &= \Et_{1,\alpha,1}, & E_{1,\alpha,2} &= \Et_{1,\alpha,2}, & E_{1,\alpha,3} &= \Et_{1,\alpha,3} + \Et_{1,\alpha,4}, \notag\\
  E_{6,\alpha,1} &= \Et_{6,\alpha,1}, & E_{6,\alpha,2} &= \Et_{6,\alpha,2}, & E_{6,\alpha,3} &= \Et_{6,\alpha,3} + \Et_{6,\alpha,4} ,\notag\\  
  E_{3,1} &= \Et_{3,1,1}, & E_{3,2} &= \Et_{3,1,2}, & E_{3,3} &= \Et_{3,1,3} + \Et_{3,1,4}, \notag\\
  E_{3,4} &= \Et_{3,2,1}, & E_{3,5} &= \Et_{3,2,2}, & E_{3,6} &= \Et_{3,2,3} + \Et_{3,2,4}, \notag\\
  E_{3,7} &= \Et_{3,3,1} + \Et_{3,4,1}, & E_{3,8} &= \Et_{3,3,2} + \Et_{3,4,2}, & E_{3,9} &= \Et_{3,3,3} + \Et_{3,4,4}, \notag\\
  E_{3,10} &= \Et_{3,3,4} + \Et_{3,4,3},
\end{align}
where $\Dt_{2,\beta}$, $\Dt_{3,\gamma}$, $\Et_{1,\alpha,\gamma}$, $\Et_{6,\alpha,\beta}$, $\Et_{3,\beta\gamma}$ are the divisors on the cover. From the local linear relations~(\ref{lineqtwofour}) and~(\ref{lineqtwotwo}) we arrive at the following global relations:
\begin{eqnarray}
  \label{globalreltwofour}
  R_1&\sim&2\,D_{1,\alpha}+\sum_{\gamma=1}^3 E_{1,\alpha\gamma}+\sum_{\beta,\gamma=1,2} E_{5,\alpha\beta\gamma}+\sum_{\beta=1}^3 E_{6,\alpha\beta},\quad \alpha=1,..,4,\cr
  R_2&\sim&4\,D_{2,\beta}+\sum_{\gamma=1,2}[\,E_{2,\beta\gamma}+2\,E_{3,\beta\gamma}+3\,E_{4,\beta\gamma}]+\sum_{\alpha=1}^4\sum_{\gamma=1,2}E_{5,\alpha\beta\gamma}\cr
     &    &+2\sum_{\alpha=1}^4E_{6,\alpha\beta}+2\,E_{3,\mu},\ \ \beta=1,2,\cr
  R_2&\sim&2\,D_{2,3}+\sum_{\alpha=1}^4E_{6,\alpha3}+\sum_{\mu=7}^{10} E_{3,\mu},\cr
  R_3&\sim&4\,D_{3,\gamma}+2\sum_{\alpha=1}^4E_{1,\alpha\gamma}+\sum_{\beta=1,2}[\,3\,E_{2,\beta\gamma}+2\,E_{3,\beta\gamma}+E_{4,\beta\gamma}]+\sum_{\alpha=1}^4\sum_{\beta=1,2}E_{5,\alpha\beta\gamma}\cr
     &    &+2\,E_{3,\mu},\quad \gamma=1,2,\cr
  R_3&\sim&2\,D_{3,3}+\sum_{\alpha=1}^4E_{1,\alpha3}+\sum_{\mu=3,6,9,10} E_{3,\mu},
\end{eqnarray}
where we set in the second line $\mu=3,6$ for $\beta=1,2$, respectively, and in the fourth line $\mu=7,8$ for $\gamma=1,2$, respectively. Furthermore for compactness of the display in these two lines, we have written $E_{3,\mu}$, $\mu=1,\dots,4$ instead of $E_{3,\beta\gamma}$, $\beta,\gamma=1,2$ according to their origin on the cover as in~(\ref{eq:Z2xZ4invdivs}).
\begin{figure}[h!]
\begin{center}
\includegraphics[width=50mm]{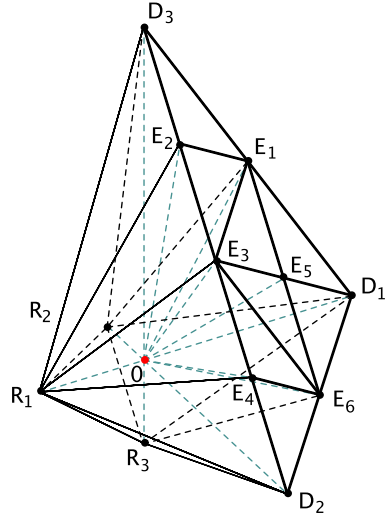}
\caption{The polyhedron $\Delta^{(3)}$ describing the local compactification of the resolution of $\IC^3/\IZ_2\times\IZ_4$.}
\label{fig:Z2Z4-cpt}
\end{center}
\end{figure}
We obtain the following nonvanishing intersection numbers of $X$ in the basis $\{R_i,E_{k\alpha\beta\gamma}\}$:
\begin{align}
  R_1R_2R_3 &=4, & R_1E_{2\beta\gamma}^2 &=-2, & R_1E_{2\beta\gamma}E_{3\mu} &=1, & R_1E_{3\mu'}^2 &=-2, \notag\\
  R_1E_{3\mu}E_{4\beta\gamma} &= 1, & R_1E_{4\beta\gamma}^2 &=-2, & R_2E_{1\alpha\gamma}^2 &=-2, & R_2E_{1\alpha3}^2 &= -4, \notag\\
  R_3E_{6\alpha\beta}^2 &=-2, & R_3E_{6\alpha3}^2 &= -4, & E_{1\alpha\gamma}^3 &=3, & E_{1\alpha3}^3 &=4, \notag\\
  E_{1\alpha\gamma'}^2E_{3\mu'} &= -1, & E_{1\alpha\gamma}^2E_{6\alpha3} &= -1, & E_{1\alpha3}^2E_{6\alpha\beta} &= -1, & E_{1\alpha3}^2E_{6\alpha3} &= -2, \notag\\ 
  E_{1\alpha\gamma}E_{2\beta\gamma}^2 &=-2, & E_{1\alpha\gamma}E_{2\beta\gamma}E_{3\mu} &= 1, & E_{1\alpha\gamma'}E_{3\mu'}^2 &= -1, & E_{1\alpha\gamma}E_{3\mu}E_{5\alpha\beta\gamma} &= 1, \notag\\
  E_{1\alpha3}E_{3\mu_1}E_{6\alpha\beta} &=1, & E_{1\alpha\gamma}E_{3\mu_2}E_{6\alpha3} &=1, & E_{1\alpha3}E_{3\mu_3}E_{6\alpha3} &=1, & E_{1\alpha\gamma}E_{5\alpha\beta\gamma}^2 &= -2, \notag\\
  E_{1\alpha3}E_{6\alpha\beta}^2 &=-1, & E_{1\alpha\gamma}E_{6\alpha3}^2 &=-1, & E_{1\alpha3}E_{6\alpha3}^2 &=-2, & E_{2\beta\gamma}^3 &= 8, \notag\\
  E_{2\beta\gamma}^2E_{3\mu} &= -4, & E_{2\beta\gamma}E_{3\mu}^2 &= 2, & E_{3\mu''}^3 &= 4, & E_{3\mu}^2E_{4\beta\gamma} &=2, \notag\\
  E_{3\mu'}^2E_{6\alpha\beta'} &= -1, & E_{3\mu}E_{4\beta\gamma}^2 &= -4, & E_{3\mu}E_{4\beta\gamma}E_{6\alpha\beta} &= 1, & E_{3\mu}E_{5\alpha\beta\gamma}^2 &=-2, \notag\\
  E_{3\mu}E_{5\alpha\beta\gamma}E_{6\alpha\beta} &=1, & E_{3\mu'}E_{6\alpha\beta'}^2 &= -1, & E_{4\beta\gamma}^3 &= 8, & E_{4\beta\gamma}^2E_{6\alpha\beta} &= -2, \notag\\
  E_{5\alpha\beta\gamma}^3 &= 8, & E_{5\alpha\beta\gamma}^2E_{6\alpha\beta} & =-2, & E_{6\alpha\beta}^3 &=3, & E_{6\alpha3}^3 &= 4,
\end{align}
where $\alpha=1,\dots,4$, $\beta,\gamma=1,2$, $\beta',\gamma'=1,\dots,3$, $\mu=1,2,4,5$, $\mu'=1,\dots,10$, $\mu''=3,6,\dots,10$, $\mu_1 = 3,6$, $\mu_2=7,8$, and $\mu_3=9,10$. The intersection numbers involving $\beta,\beta',\gamma,\gamma',\mu$, and $\mu'$ only have the given value for appropriate values of the labels, otherwise they vanish. As an example, $R_1E_{2\beta\gamma}E_{3\mu} =1$ has to be understood as $R_1E_{2,1,1}E_{3,1} = 1, R_1E_{2,1,2}E_{3,2} =1, R_1E_{2,2,1}E_{3,4} =1, R_1E_{2,2,2}E_{3,5} = 1$. Which values of $\mu$ fit to the values of $\beta,\gamma$ can be determined by looking at the first summand in the definition of $E_{3,\mu}$ in~(\ref{eq:Z2xZ4invdivs}). 

Next, we discuss the topology of the divisors. The topology of the exceptional divisors $E_{5,\alpha\beta\gamma}$ was determined in Appendix~\ref{sec:localZ2xZ4} to be an $\IF_1$. According to Section~\ref{sec:Topology} the remaining divisors are of type E\ref{item:E3}). Since the there is only one line ending at $E_{2\beta\gamma}$ and $E_{4,\beta\gamma}$ in the toric diagram in Figure~\ref{frtwofour}, both of them have the topology of an $\IF_0$. The divisors $E_{1,\alpha\gamma}$ and $E_{6,\alpha\beta}$ have two $\IC^3/\IZ_2\times\IZ_4$ and one $\IC^3/\IZ_2\times\IZ_2$ fixed points lying on them, with three and two lines ending at them in the toric diagram in the respective Figures~\ref{frtwofour} and~\ref{frtwotwo}. Hence, their topology is $\Bl{5}\IF_0$. The divisors $E_{3\mu}$, $\mu=1,2,4,5$ have four $\IC^3/\IZ_2\times\IZ_4$ fixed points lying on them, with three lines ending at them in the corresponding toric diagram. Therefore their topology is that of $\Bl{8}\IF_0$. Similarly, the divisors $E_{1,\alpha3}$, $E_{3\mu}$, $\mu=3,6,\dots,10$, and $E_{6,\alpha3}$ have four $\IC^3/\IZ_2\times\IZ_2$ patches lying on them, with two lines ending at them in the toric diagram. Therefore their topology is that of $\Bl{4}\IF_0$. The topology of the divisors $D_{i\alpha}$ is obtained from that of $T^2\setminus \{\textrm{4 pts} \} \times T^2\setminus \{\textrm{4 pts} \}$. Since in Figure~\ref{frtwofour} there is one line ending at $D_1$ and none at $D_2$ and $D_3$, the topology of $D_{1\alpha}$ is that of $\Bl{4}\IF_0$, while $D_{2\beta'}$ and $D_{3\gamma'}$ have the topology of $\IF_0$. Finally, the $R_i$ are K3 surfaces.

Finally, the second Chern class is
\begin{align}
  \ch_2\cdot E_{1,\alpha\gamma} &= 6, & \ch_2\cdot E_{1,\alpha3} &= 4, & \ch_2\cdot E_{2,\beta\gamma} &= -4, & \ch_2\cdot E_{3\mu} &= 12, \notag\\
  \ch_2\cdot E_{3\mu"} &= 4 & \ch_2\cdot E_{4,\beta\gamma} &= -4, & \ch_2\cdot E_{5,\alpha\beta\gamma} &= -4, & \ch_2\cdot E_{6\alpha\beta} &= 6, \notag\\
  \ch_2\cdot E_{6,\alpha3} &=4, & \ch_2 \cdot R_i &= 24,
\end{align}
for $\mu=1,2,4,5$ and $\mu"=3,6,\dots,10$.

\subsection{The orientifold}

At the orbifold point, we get the following configuration of O--planes:
The fixed points under $I_6$ are 64 O3--planes which fall into 40 conjugacy classes under the orbifold group. From the combination $I_6\,\theta^1$ arise four O7--planes in the $(z^1,z^3)$--plane. They fall into the conjugacy classes $z^2=0,\,z^2=1/2$ and $z^2=1/2\,\tau,\ 1/2\,(1+\tau)$. Under $I_6\,(\theta^2)^2$ are four O7--planes fixed in the $(z^2, z^3)$--plane, which are all in separate conjugacy classes. Under $I_6\,\theta^1(\theta^2)^3$, another four O7--planes arise, this time in the $(z^1,z^2)$--plane in the three conjugacy classes $z^3=0,\,z^3=1/2$ and $z^3=1/2\,\tau,\ 1/2\,(1+\tau)$.

For this example, $h^{(1,1)}_{-}=0$ since all the fixed points under the orbifold group lie on $\IZ_2$ fixed points and are therefore invariant under $I_6$.

After the blow--up, we have to deal with two different patches, the $\IC^3/\IZ_2\times\IZ_2$--patch and the $\IC^3/\IZ_2\times\IZ_4$--patch. For both cases, we choose the simplest possibility for $\Ic$, namely sending $z^i\to -z^i$ while leaving the $y^i$ unchanged.
The fixed sets under the combination of the scaling action of $\IC^3/\IZ_2\times\IZ_4$ (\ref{rescalestwofour}) and $\Ic$ are
\begin{itemize}
\item $z^1=0,\ \lambda_1=1,\,\lambda_2=\lambda_3=-1,\,\lambda_4=\lambda_5=\lambda_6=1,$
\item $z^2=0,\ \lambda_1=-1,\,\lambda_2=1,\,\lambda_3=-1,\ \lambda_4=\lambda_5=\lambda_6=1,$
\item $z^3=0,\ \lambda_1=\lambda_2=-1,\, \lambda_3=\lambda_4=\lambda_5=\lambda_6=1,$
\item $y^3=0,\ \lambda_1=\lambda_2=\lambda_3=\lambda_4=-1,\,\lambda_5=\lambda_6=1.$
\end{itemize}
The fixed sets under the combination of the scaling action of $\IC^3/\IZ_2\times\IZ_2$ (\ref{rescalestwofour}) and $\Ic$ are
\begin{itemize}
\item $z^1=0,\ \lambda_1=\lambda_2=-1,\,\lambda_3=1,$
\item $z^2=0,\ \lambda_1=1,\,\lambda_2=-1,\,\lambda_3=1,$
\item $z^3=0,\ \lambda_1=-1,\,\lambda_2=\lambda_3=1.$
\end{itemize}
These solutions correspond to O7--planes on the four $D_1$--planes, on the three equivalence classes of $D_2$ and $D_3$--planes and on the four exceptional divisors $E_{3,1},\,E_{3,2},\,E_{3,4}$ and $E_{3,5}$ which arise from the resolution of the four $\IZ_4$ fixed lines. This amounts to a total of 14 O7--planes. In the blown down limit, the O7--planes on the $E_3$ disappear and we recover the 10 O7--planes of the orbifold limit. There are no O3--plane solutions in the local patches, and since such a patch sits at every location of an O3--plane in the orbifolds limit, no O3--planes arise in the blown up case.

Intersection numbers involving divisors not fixed under the orientifold involution are halved:
\begin{align}
  R_1R_2R_3 &=2, & R_1E_{2\beta\gamma}^2 &=-1, & R_1E_{3\mu''}^2 &=-1, \notag\\
   R_1E_{4\beta\gamma}^2 &=-1, & R_2E_{1\alpha\gamma}^2 &=-1, & R_2E_{1\alpha3}^2 &= -2, \notag\\
  R_3E_{6\alpha\beta}^2 &=-1, & R_3E_{6\alpha3}^2 &= -2, & E_{1\alpha\gamma}^3 &=3/2, \notag \\
  E_{1\alpha3}^3 &=2, & E_{1\alpha\gamma'}^2E_{3\mu''} &= -1/2, & E_{1\alpha\gamma}^2E_{6\alpha3} &= -1/2,\notag \\
   E_{1\alpha3}^2E_{6\alpha\beta} &= -1/2, & E_{1\alpha3}^2E_{6\alpha3} &= -1, &
  E_{1\alpha\gamma}E_{2\beta\gamma}^2 &=-1, \notag \\
  E_{1\alpha\gamma'}E_{3\mu''}^2 &= -1/2, &
  E_{1\alpha3}E_{3\mu_1}E_{6\alpha\beta} &=1/2, & E_{1\alpha\gamma}E_{3\mu_2}E_{6\alpha3} &=1/2,\notag \\
   E_{1\alpha3}E_{3\mu_3}E_{6\alpha3} &=1/2, & E_{1\alpha\gamma}E_{5\alpha\beta\gamma}^2 &= -1, &
  E_{1\alpha3}E_{6\alpha\beta}^2 &=-1/2,\notag \\
   E_{1\alpha\gamma}E_{6\alpha3}^2 &=-1/2, & E_{1\alpha3}E_{6\alpha3}^2 &=-1, & E_{2\beta\gamma}^3 &= 4, \notag\\
  E_{3\mu''}^3 &= 2, &
  E_{3\mu''}^2E_{6\alpha\beta'} &= -1/2, &
   E_{3\mu''}E_{6\alpha\beta'}^2 &= -1/2, \notag \\
   E_{4\beta\gamma}^3 &= 4, & E_{4\beta\gamma}^2E_{6\alpha\beta} &= -1, &
  E_{5\alpha\beta\gamma}^3 &= 4, \notag \\
  E_{5\alpha\beta\gamma}^2E_{6\alpha\beta} & =-1, & E_{6\alpha\beta}^3 &=3/2, & E_{6\alpha3}^3 &= 2.
\end{align}
The intersection numbers involving the $E_{3,\mu}, \ \mu=1,2,4,5$, which are fixed are
\begin{align}
R_1E_{3\mu}E_{4\beta\gamma} &= 1, & R_1E_{2\beta\gamma}E_{3\mu} &=1, & E_{1\alpha\gamma'}^2E_{3\mu} &= -1,\notag \\
 E_{1\alpha\gamma}E_{2\beta\gamma}E_{3\mu} &= 1, & E_{1\alpha\gamma'}E_{3\mu}^2 &= -2,& E_{1\alpha\gamma}E_{3\mu}E_{5\alpha\beta\gamma} &= 1,\notag \\
  E_{2\beta\gamma}^2E_{3\mu} &= -4, & E_{2\beta\gamma}E_{3\mu}^2 &= 4, & E_{3\mu}^2E_{4\beta\gamma} &=4,\notag \\
  E_{3\mu}^2E_{6\alpha\beta'} &= -2,& E_{3\mu}E_{4\beta\gamma}^2 &= -4, & E_{3\mu}E_{4\beta\gamma}E_{6\alpha\beta} &= 1,\notag\\
  E_{3\mu}E_{5\alpha\beta\gamma}^2 &=-2, & E_{3\mu}E_{5\alpha\beta\gamma}E_{6\alpha\beta} &=1, & E_{3\mu}E_{6\alpha\beta'}^2 &= -1. 
\end{align}


\section{The $\IZ_3\times\IZ_3$ orbifold}

\subsection{The resolved orbifold}

\begin{figure}[h!]
\begin{center}
\includegraphics[width=140mm]{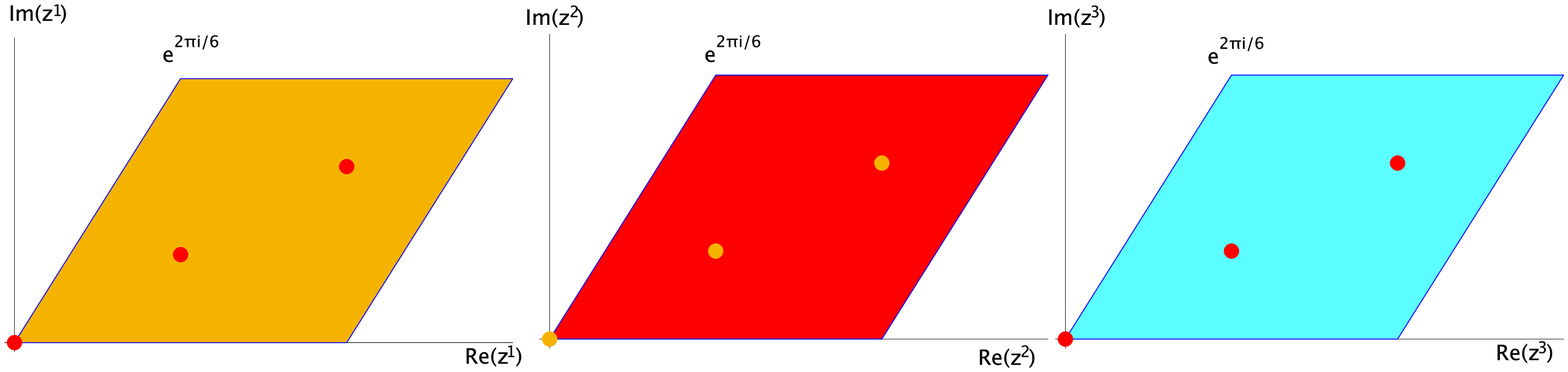}
\caption{Fundamental regions for the $\IZ_3\times\IZ_3$ orbifold}\label{ffuthreethree}
\end{center}
\end{figure}

\begin{figure}[h!]
\begin{center}
\includegraphics[width=85mm]{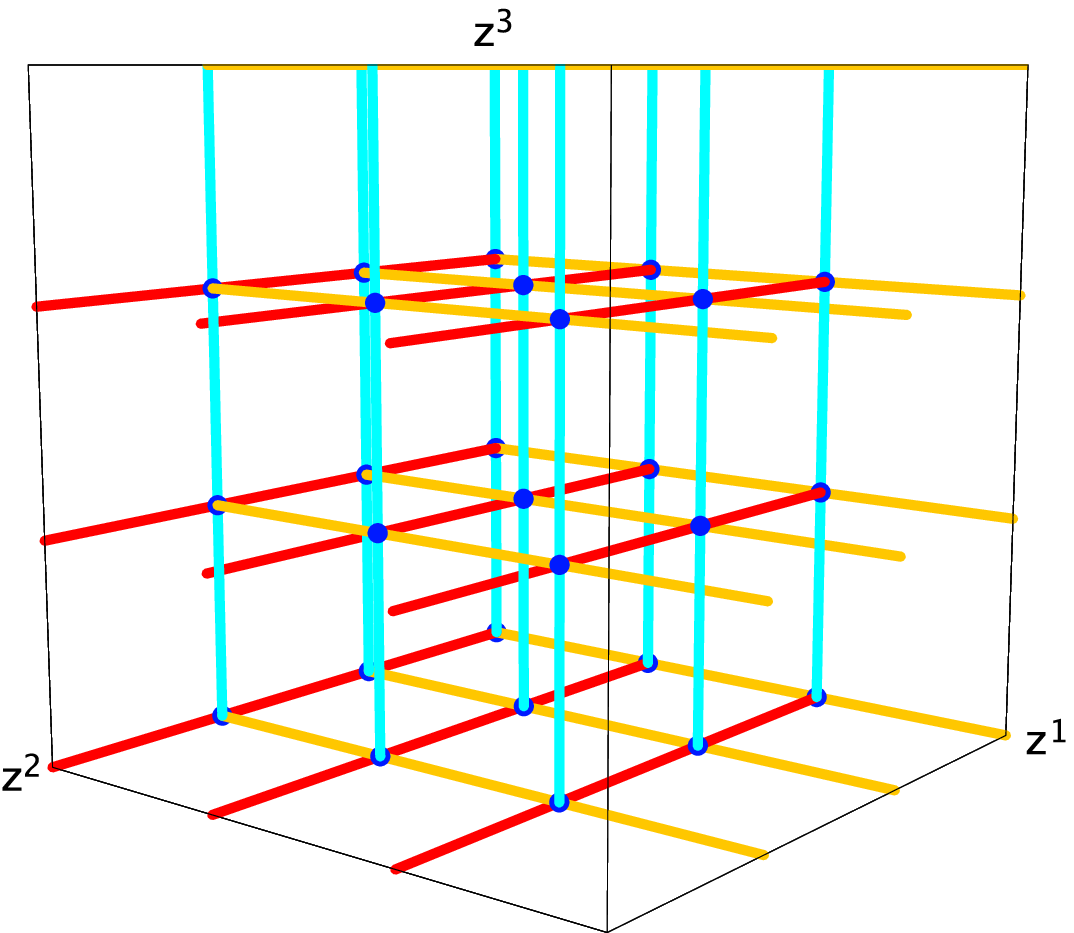}
\caption{Schematic picture of the fixed set configuration of $\IZ_3\times \IZ_3$}\label{fig:ffixthreethree}
\end{center}
\end{figure}

This is a combination of two prime orbifolds, therefore the conjugacy classes are in one-to-one correspondence with the fixed points. We need to examine the $\theta^1,\ \theta^2,\ \theta^1\theta^2$ and $\theta^1(\theta^2)^2$--twists ($\,(\theta^1)^2\theta^2$ is the anti-twist of $\theta^1(\theta^2)^2$). The action of the twists $\theta^1$ and $\theta^2$ was given in (A.84) and the resulting complex structure in (A.88) of~\cite{Lust:2005dy}. The torus factorizes into $(T^2)^3$ and Figure~\ref{ffuthreethree} shows the fundamental regions of the three tori corresponding to $z^1,\,z^2,\,z^3$ and their fixed points.
In each of them, we get  the usual three fixed points of the $\IZ_3$--twist, namely $\zf{1}{1}=\zf{2}{1}=\zf{3}{1}=0,\ \zf{1}{2}=\zf{2}{2}=\zf{3}{2}=\frac{1}{\sqrt3}e^{\pi i/6}$, and $\zf{1}{3}=\zf{2}{3}=\zf{3}{3}=1+\frac{i}{\sqrt3}$. At each pair $(\zf{i}{\alpha},\zf{j}{\beta}), i,j,\alpha,\beta=1,2,3$ we have a $\IC^2/\IZ_3$ fixed line. These fixed lines intersect in $\IC^3/(\IZ_3\times\IZ_3)$ fixed points. Table~\ref{fsthreethree} summarizes the relevant data of the fixed point set. 
\begin{table}[h!]\begin{center}
\begin{tabular}{|c|c|c|c|}
\hline
Group el.& Order & Fixed Set& Conj. Classes \cr
\hline
\noalign{\hrule}\noalign{\hrule}
$ \theta^1   $&3      &9  fixed lines& 9\cr
$\theta^2   $&3   &9  fixed lines& 9\cr
$ \theta^1\theta^2   $&$ 9 $    &27  fixed points&27\cr
$ \theta^1(\theta^2)^2   $&3      &9  fixed lines& 9\cr
\hline
\end{tabular}
\caption{Fixed point set for $\IZ_3\times\IZ_3$.}\label{fsthreethree}
\end{center}\end{table}
Figure \ref{fig:ffixthreethree} shows the configuration of the fixed sets in a schematic way.

Every ${\IC}^2/\IZ_3$ fixed line contributes two exceptional divisors, namely $E_{1,\alpha \gamma}, E_{2,\alpha \gamma}$, $E_{3,\beta\gamma}, E_{4,\beta\gamma}$, $E_{6,\alpha\beta}, E_{7,\alpha\beta}$, $\alpha,\beta,\gamma=1,2,3$. The resolved $\IC^3/(\IZ_3\times \IZ_3)$ patches yield one additional exceptional divisor $E_{5,\alpha\beta\gamma}$, $\alpha,\beta,\gamma=1,2,3$, besides the two on each of the sides of the toric diagram in Figure~\ref{frthreethree}. The latter are identified with those of the fixed lines, on whose intersections the fixed points sit. So we have $9\cdot2+9\cdot2+9\cdot2+27=81$ exceptional divisors. Since in this example there are no fixed lines without fixed points on them, we have $h^{2,1}_{\rm twist.}=0$. Finally, in each of the coordinate planes we have 3 fixed planes with associated divisors $D_{i\alpha},\ \alpha, i=1,2,3$. 

\begin{figure}[h!]
\begin{center}
\includegraphics[width=50mm]{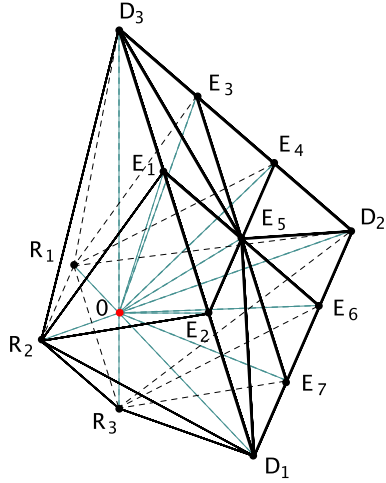}
\caption{The polyhedron $\Delta^{(3)}$ describing the local compactification of the resolution of $\IC^3/\IZ_3\times\IZ_3$.}
\label{fig:Z3Z3a-cpt}
\end{center}
\end{figure}

We do the calculation of the intersection ring for triangulation a) in Figure~\ref{frthreethree}. From the local linear equivalences~(\ref{lineqthreethree}) we arrive at the following global relations:
\begin{eqnarray}
  \label{eq:globalrelZ3Z3}
  R_1&\sim&3\,D_{{1,\alpha}}+\sum_{\beta=1}^3( E_{{6,\alpha\beta}}+2\,E_{{7,\alpha\beta}})+\sum_{\gamma=1}^3(E_{{1, \alpha\gamma}}+2\,E_{{2, \alpha\gamma}})+\sum_{\beta,\gamma=1}^3E_{{5,\alpha\beta\gamma}},\cr
  R_2&\sim&3\,D_{{2,\beta}}+\sum_{\alpha=1}^3(2\,E_{{6,\alpha\beta}}+E_{{7,\alpha\beta}})+\sum_{\gamma=1}^3(2\,E_{{4,\beta\gamma}}+E_{{3,\beta\gamma}})+\sum_{\alpha,\gamma=1}^3 E_{{5,\alpha\beta\gamma}},\cr  
  R_3&\sim&3\,D_{{3,\gamma}}+\sum_{\alpha=1,}^3(2\,E_{{1,\alpha\gamma}}+E_{{2,\alpha\gamma}})+\sum_{\beta=1}^3(2\,E_{{3,\beta\gamma}}+E_{{4,\beta\gamma}})+\sum_{\alpha,\beta=1}^3 E_{{5,\alpha\beta\gamma}}.
\end{eqnarray}
The basis for the lattice $N$ from~\eqref{eq:globalrelZ3Z3} to be $f_1=(3,0,0)$, $f_2=(0,1,0)$, $f_3=(0,0,1)$. The lattice points of the polyhedron $\Delta^{(3)}$ for the local compactification of the $\IZ_3\times\IZ_3$ fixed points are
\begin{align}
  \label{eq:Z3Z3apoly}
  v_1 &= (-3,0,0), & v_2 &= (0,-1,0), & v_3 &= (0,0,-1), & v_4 &= (9,0,0), & v_5 &= (0,3,0), \notag\\
  v_6 &= (0,0,3), & v_7 &= (1,0,2), & v_8 &= (2,0,1), & v_9 &= (0,1,2), & v_{10} &= (0,2,1), \notag\\
  v_{11} &= (1,1,1), & v_{12} &= (1,2,0), & v_{13} &= (2,1,0), 
\end{align}
corresponding to the divisors $R_1,R_2,R_3,D_1,D_2,D_3,E_1,\dots,E_7$ in that order. The polyhedron is shown in Figure~\ref{fig:Z3Z3a-cpt}. 
We obtain the following nonvanishing intersection numbers of $X$ in the basis $\{R_i,E_{k\alpha\beta\gamma}\}$:
\begin{align}
  R_{1}R_{2}R_{3}&=3, & R_1E_{3,\beta\gamma}^2&=-2, & R_1E_{3,\beta\gamma}E_{4,\beta\gamma}&=1, \notag\\
  R_1E_{4,\beta\gamma}^2&=-2, & R_2E_{1,\alpha\gamma}^2&=-2, & R_2E_{1,\alpha\gamma}E_{2,\alpha\gamma}&=1, \notag\\
  R_2E_{2,\alpha\gamma}^2&=-2, & R_3E_{6,\alpha\beta}^2&=-2, & R_3E_{6,\alpha\beta}E_{7,\alpha\beta} &=1, &\notag\\
  R_3E_{7,\alpha\beta}^2&=-2, & E_{1,\alpha\gamma}^3&=8, & E_{1,\alpha\gamma}^2E_{2,\alpha\gamma}&=-1,\notag\\
  E_{1,\alpha\gamma}E_{2,\alpha\gamma}^2&=-1, & E_{1,\alpha\gamma}E_{2,\alpha\gamma}E_{5,\alpha\beta\gamma}&=1, & E_{1,\alpha\gamma}^2E_{5,\alpha\beta\gamma}&=-2, &\notag\\
  E_{2,\alpha\gamma}^3&=8, & E_{2,\alpha\gamma}^2E_{5,\alpha\beta\gamma}&=-2, & E_{3,\beta\gamma}^3&=8, \notag\\
  E_{3,\beta\gamma}^2E_{4,\beta\gamma}&=-1, & E_{3,\beta\gamma}E_{4,\beta\gamma}^2&=-1, & E_{3,\beta\gamma}E_{4,\beta\gamma}E_{5,\alpha\beta\gamma}&=1, &\notag\\
  E_{3,\beta\gamma}^2E_{5,\alpha\beta\gamma}&=-2, & E_{4,\beta\gamma}^3&=8, & E_{4,\beta\gamma}^2E_{5,\alpha\beta\gamma}&=-2, \notag\\
  E_{5,\alpha\beta\gamma}^3&=3, & E_{5,\alpha\beta\gamma}E_{6,\alpha\beta}^2&=-2, & E_{5,\alpha\beta\gamma}E_{6,\alpha\beta}E_{7,\alpha\beta}&=1, \notag\\ 
  E_{5,\alpha\beta\gamma}E_{7,\alpha\beta}^2&=-2, & E_{6,\alpha\beta}^3&=8, & E_{6,\alpha\beta}^2E_{7,\alpha\beta}&=-1, \notag\\
  E_{6,\alpha\beta}E_{7,\alpha\beta}^2&=-1 & E_{7,\alpha\beta}^3&=8.
\end{align}

Next, we discuss the topology of the divisors. The topology of the exceptional divisors $E_{5,\alpha\beta\gamma}$ was determined in Appendix~\ref{sec:localZ3xZ3} to be a $\Bl{5}\IF_0$. According to Section~\ref{sec:Topology} the remaining divisors are of type E\ref{item:E3}). Since the there is only one line ending at all the other exceptional divisors in the toric diagram in Figure~\ref{frthreethree}, all of them have the topology of an $\IF_0$. The topology of the divisors $D_{i\alpha}$ is obtained from that of $T^2\setminus \{\textrm{3 pts} \} \times T^2\setminus \{\textrm{3 pts} \}$. Since in Figure~\ref{frthreethree} there is one line ending at all the $D_i$, the topology of $D_{i\alpha}$ is that of a $\Bl{9}\IF_0$. Finally, the $R_i$ are K3 surfaces.

Finally, the second Chern class is
\begin{align}
  \ch_2\cdot E_{1,\alpha\gamma} &= -4, & \ch_2\cdot E_{2,\alpha\gamma} &= -4, & \ch_2\cdot E_{3,\beta\gamma} &= -4, & \ch_2\cdot E_{4,\beta\gamma} &= -4, \notag\\
  \ch_2\cdot E_{5\alpha\beta\gamma} &= 6 & \ch_2\cdot E_{6,\alpha\beta} &= -4, & \ch_2\cdot E_{7,\alpha\beta} &= -4, & \ch_2\cdot R_i &= 24.
\end{align}

\subsection{The orientifold}

At the orbifold point, there are 64 O3--planes which fall into 18 conjugacy classes under the orbifold action. Since $\IZ_3\times\IZ_3$ has no $\IZ_2$ subgroups, no O7--planes appear.

For this example, $h^{1,1}_{-}=37$. 13 of these divisors come from the $E_{5\alpha\beta\gamma}$, which except for the one located at $(0,0,0)$ fall into conjugacy classes of length two under $I_6$. The remaining 24 come from the two exceptional divisors of the $\IC^2/\IZ_3$ fixed lines, which except for those three which pass the origin also fall into equivalence classes of length two.

Now we discuss the orientifold for the resolved case. For the local involution, we choose the simplest possibility, i.e. $z^i\to -z^i$ while on the $y^i$ nothing happens. Looking for the fixed points of the combination of the involution and the scaling action of the resolved patch (\ref{rescalesthreethree}), we find 
$$y^5=0,\quad \lambda_1= \lambda_2= \lambda_5=-1,\  \lambda_3= \lambda_4= \lambda_6= \lambda_7=1,$$
i.e. an O7--plane wrapped on $E_5$. In total, we arrive at 14 O7--planes wrapped on the remaining linear combinations of the $E_{5\alpha\beta\gamma}$s in $H^{1,1}_{+}$.
Only the resolved patch at $(0,0,0)$ coincides with an O3--plane location at the orbifold point. This O3--plane is not present in the resolved case. We are therefore left with 17 equivalence classes of O3--planes which are located away from the resolved patches. In this case, the O--plane configuration of the orbifold point is reproduced in the blown--down limit.

In the global relations (\ref{eq:globalrelZ3Z3}), the coefficient of $E_5$ is changed to $1/2$ and the intersection numbers change as follows:

\begin{align}
  R_{1}R_{2}R_{3}&=\frac{3}{2}, & R_1E_{3,\beta\gamma}^2&=-1, & R_1E_{3,\beta\gamma}E_{4,\beta\gamma}&=\frac{1}{2}, \notag\\
  R_1E_{4,\beta\gamma}^2&=-1, & R_2E_{1,\alpha\gamma}^2&=-1, & R_2E_{1,\alpha\gamma}E_{2,\alpha\gamma}&=\frac{1}{2}, \notag\\
  R_2E_{2,\alpha\gamma}^2&=-1, & R_3E_{6,\alpha\beta}^2&=-1, & R_3E_{6,\alpha\beta}E_{7,\alpha\beta} &=\frac{1}{2}, &\notag\\
  R_3E_{7,\alpha\beta}^2&=-1, & E_{1,\alpha\gamma}^3&=4, & E_{1,\alpha\gamma}^2E_{2,\alpha\gamma}&=-\frac{1}{2},\notag\\
  E_{1,\alpha\gamma}E_{2,\alpha\gamma}^2&=-\frac{1}{2}, & E_{1,\alpha\gamma}E_{2,\alpha\gamma}E_{5,\alpha\beta\gamma}&=1, & E_{1,\alpha\gamma}^2E_{5,\alpha\beta\gamma}&=-2, &\notag\\
  E_{2,\alpha\gamma}^3&=4, & E_{2,\alpha\gamma}^2E_{5,\alpha\beta\gamma}&=-2, & E_{3,\beta\gamma}^3&=4, \notag\\
  E_{3,\beta\gamma}^2E_{4,\beta\gamma}&=-\frac{1}{2}, & E_{3,\beta\gamma}E_{4,\beta\gamma}^2&=-\frac{1}{2}, & E_{3,\beta\gamma}E_{4,\beta\gamma}E_{5,\alpha\beta\gamma}&=1, &\notag\\
  E_{3,\beta\gamma}^2E_{5,\alpha\beta\gamma}&=-2, & E_{4,\beta\gamma}^3&=4, & E_{4,\beta\gamma}^2E_{5,\alpha\beta\gamma}&=-2, \notag\\
  E_{5,\alpha\beta\gamma}^3&=12, & E_{5,\alpha\beta\gamma}E_{6,\alpha\beta}^2&=-2, & E_{5,\alpha\beta\gamma}E_{6,\alpha\beta}E_{7,\alpha\beta}&=1, \notag\\ 
  E_{5,\alpha\beta\gamma}E_{7,\alpha\beta}^2&=-2, & E_{6,\alpha\beta}^3&=4, & E_{6,\alpha\beta}^2E_{7,\alpha\beta}&=-\frac{1}{2}, \notag\\
  E_{6,\alpha\beta}E_{7,\alpha\beta}^2&=-\frac{1}{2} & E_{7,\alpha\beta}^3&=4.
\end{align}

\end{appendix}

\bibliographystyle{my-h-elsevier}

\end{document}